\documentclass[preprint2]{aastex}
\usepackage{amssymb}
\usepackage{amsmath}
\usepackage{comment}
\newcommand\ghrs{GHRS}
\newcommand\fuse{\emph{FUSE}}
\newcommand\hst{\emph{HST}}
\newcommand\stis{STIS}
\providecommand{\cm[1]}{\ensuremath{\,{\rm cm}^{#1}}}
\providecommand{\kms}{\ensuremath{\,{\rm km\,s}^{-1}}}
\providecommand{\mA}{\ensuremath{\,\mbox{m\AA}}}
\providecommand{\Ang}{\ensuremath{\,\mbox{\AA}}}
\providecommand{\kpc}{\ensuremath{\,\mathrm{kpc}}}
\providecommand{\Lya}{\ensuremath{\mathrm{Ly}\alpha}} 
\providecommand{\Lyb}{\ensuremath{\mathrm{Ly}\beta}} 
\providecommand{\HH}{\ensuremath{\mathrm{H}_{2}}}
\providecommand{\Cthr}{\ensuremath{\mathrm{C}^{+3}}}
\providecommand{\zabs}{\ensuremath{z_{abs}}}
\providecommand{\zciv}{\ensuremath{z_{1548}}}
\providecommand{\zem}{\ensuremath{z_{em}}}

\providecommand{\dvabs}{\ensuremath{\delta v_{abs}}}
\providecommand{\dvciv}{\ensuremath{\delta v_{\rm C\,IV}}}
\providecommand{\EWr}{\ensuremath{W_{r}}}
\providecommand{\sigEWr}{\ensuremath{\sigma_{\EWr}}}
\providecommand{\EWlin[1]}{\ensuremath{\EWr{}_{,\mathrm{#1}}}}
\providecommand{\sigEWlin[1]}{\ensuremath{\sigma_{\EWr,\mathrm{#1}}}}
\providecommand{\EWo}{\ensuremath{W_{obs}}}
\providecommand{\sigEWo}{\ensuremath{\sigma_{\EWo}}}
\providecommand{\fx}{\ensuremath{f_{\lambda}}}
\providecommand{\sigfx}{\ensuremath{\sigma_{\fx}}}
\providecommand{\wvr}{\ensuremath{\lambda_{r}}}
\providecommand{\fval}{\ensuremath{f_{osc}}}
\providecommand{\wvlo}{\ensuremath{\lambda_{l}}}
\providecommand{\wvhi}{\ensuremath{\lambda_{h}}}
\providecommand{\N[1]}{\ensuremath{N_{#1}}}
\providecommand{\NCIV}{\ensuremath{N(\Cthr)}}
\providecommand{\sigNCIV}{\ensuremath{\sigma_{N(\Cthr)}}}
\providecommand{\logCIV}{\ensuremath{\log \NCIV}}
\providecommand{\logCIVlim}{\ensuremath{\log \N{lim}(\Cthr)}}
\providecommand{\logN}{\ensuremath{\log N}}
\providecommand{\siglogN}{\ensuremath{\sigma_{\logN}}}
\providecommand{\OmCIV}{\ensuremath{\Omega_{\Cthr}}}
\providecommand{\sigOmCIV}{\ensuremath{\sigma_{\Omega}}}
\providecommand{\ff[1]}{\ensuremath{f(#1)}}
\providecommand{\aff[1]}{\ensuremath{\alpha_{#1}}}
\providecommand{\kff[1]}{\ensuremath{k_{#1}}}
\providecommand{\Num}{\ensuremath{\mathcal{N}}}
\providecommand{\Dz}{\ensuremath{\Delta z}}
\providecommand{\DXp}{\ensuremath{\Delta X}}
\providecommand{\sigDX}{\ensuremath{\sigma_{\DXp}}}
\providecommand{\DX[1]}{\ensuremath{\DXp(#1)}}
\providecommand{\ud}{\ensuremath{\mathrm{d}}}
\providecommand{\dNLyadz}{\ensuremath{\ud \Num_{\mbox{Ly}\alpha}/\ud z}}
\providecommand{\dNCIVdz}{\ensuremath{\ud \Num_{\mathrm{C\,IV}}/\ud z}}

\providecommand{\dNCIVdX}{\ensuremath{\ud \Num_{\mathrm{C\,IV}}/\ud X}}
\newcommand\ie{\emph{i.e.},\ }
\newcommand\eg{\emph{e.g.},\ }

\begin{document}


\title{The Last Eight-Billion Years of Intergalactic \ion{C}{4} Evolution}

\author{Kathy L. Cooksey,\altaffilmark{1} Christopher
  Thom,\altaffilmark{2} J. Xavier Prochaska,\altaffilmark{1,3} and Hsiao-Wen
  Chen\altaffilmark{2}}

\altaffiltext{1}{Department of Astronomy; University of California;
  1156 High St., Santa Cruz, CA 95064;
  kcooksey@ucolick.org\protect\renewcommand{\thefootnote}{\fnsymbol{footnote}}\footnotemark[1]\renewcommand{\thefootnote}{\arabic{footnote}}
}
\altaffiltext{2}{Department of Astronomy; University of Chicago; 5640
  S. Ellis Ave., Chicago, IL 60637;
  cthom@oddjob.uchicago.edu,\protect\renewcommand{\thefootnote}{\fnsymbol{footnote}}\footnotemark[2]\renewcommand{\thefootnote}{\arabic{footnote}}\protect~hchen@oddjob.uchicago.edu}
\altaffiltext{3}{UCO/Lick Observatory;
  University of California; 1156 High St., Santa Cruz, CA 95064;
  xavier@ucolick.org}

\shorttitle{Intergalactic \ion{C}{4}} \shortauthors{Cooksey et al.}

\slugcomment{Draft 2: \today}


\begin{abstract}
  We surveyed the \hst\ UV spectra of 49 low-redshift quasars for $z <
  1$ \ion{C}{4} candidates, relying solely on the characteristic
  wavelength separation of the doublet. After consideration of the
  defining traits of \ion{C}{4} doublets (\eg consistent line
  profiles, other associated transitions, etc.), we defined a sample
  of 38 definite (group G = 1) and five likely (G = 2) doublets with
  rest equivalent widths \EWr\ for both lines detected at
  $\ge3\sigEWr$. We conducted Monte-Carlo completeness tests to
  measure the unblocked redshift (\Dz) and co-moving pathlength (\DXp)
  over which we were sensitive to \ion{C}{4} doublets of a range of
  equivalent widths and column densities. The absorber line density of
  (G = 1+2) doublets is $\dNCIVdX = 4.1^{+0.7}_{-0.6}$ for $\logCIV
  \ge 13.2$, and \dNCIVdX\ has not evolved significantly since
  $z=5$. The best-fit power-law to the G = 1 frequency distribution of
  column densities $\ff{\NCIV} \equiv k(\NCIV/\N{0})^{\aff{N}}$ has
  coefficient $k = 0.67^{+0.18}_{-0.16} \times 10^{-14}\cm{2}$ and
  exponent $\aff{N} = -1.50^{+0.17}_{-0.19}$, where $\N{0} =
  10^{14}\cm{-2}$. Using the power-law model of \ff{\NCIV}, we measured
  the \Cthr\ mass density relative to the critical density: $\OmCIV =
  (6.20^{+1.82}_{-1.52}) \times 10^{-8}$ for $13 \le \logCIV \le
  15$. This value is a $2.8\pm0.7$ increase in \OmCIV\ compared to the
  error-weighted mean from several $1 < z < 5$ surveys for \ion{C}{4}
  absorbers. A simple linear regression to \OmCIV\ over the age of the
  Universe indicates that \OmCIV\ has slowly but steadily increased from
  $z = 5 \rightarrow 0$, with $\ud\OmCIV / \ud t_{age} =
  (0.42\pm0.2)\times10^{-8}\,{\rm Gyr}^{-1}$.
\end{abstract}



\section{Introduction}\label{sec.intro}

\renewcommand{\thefootnote}{\fnsymbol{footnote}}
\footnotetext[1]{Present address: Department of
    Physics, Massachusetts Institute of Technology, 77 Massachusetts
    Ave., Cambridge, MA 02139; kcooksey@space.mit.edu}  
\footnotetext[2]{Present address: Space Telescope
    Science Institute, 3700 San Martin Dr., Baltimore, MD 21218;
    cthom@stsci.edu} 
\renewcommand{\thefootnote}{\arabic{footnote}}


With the inception of echelle spectrometers on 10\,m-class optical
telescopes, observers unexpectedly discovered that a significant
fraction of the $z > 1.5$ intergalactic medium (IGM; a.k.a. the \Lya\
forest) was enriched \citep{cowieetal95, tytleretal95}. The \ion{C}{4}
absorption lines have proven to be valuable transitions for studying
the enrichment of the IGM since: (1) their rest wavelengths
$\lambda\lambda1548.20,1550.77\Ang$ are redward of \Lya\
$\lambda1215.67\Ang$; (2) they redshift into the optical for $\zciv
\gtrsim 1.5$; (3) they constitute a doublet with characteristic rest
wavelength separation ($2.575\Ang$ or $498\kms$); and (4) have an
equivalent width ratio $2:1$ for $\EWlin{1548}:\EWlin{1550}$ in the
unsaturated regime.

Quantitative studies based on the \ion{C}{4} doublet measured the
enrichment level to be $\approx10^{-2}$ to $10^{-4}$ the chemical
abundance of the Sun over the range $1.8 \lesssim z \lesssim 5$
\citep{songaila01, boksenbergetal03ph, schayeetal03}. No viable model
of Big Bang nucleosynthesis can explain this observed level of
enrichment; therefore, the metals observed have been produced in stars
and transported to the \Lya\ forest. The mechanisms typically invoked
include ``primary'' enrichment by some of the earliest stars at $z>6$
\citep[\eg][]{madauetal01,wiseandabel08} or ``contemporary'' injection
through galactic feedback processes from $z \approx 6 \rightarrow 2$
\citep[\eg][]{scannapiecoetal02, oppenheimeranddave06}.

Numerous high-redshift surveys have shown that the mass density of
triply-ionized carbon relative to the critical density of the Universe
$\OmCIV=\rho_{\Cthr}/\rho_{c,0}$ has {\it not} evolved substantially
from $z=5$ ($\approx 1$\,Gyr after the Big Bang) to $z=1.5$
\citep[$\approx 4$\,Gyr; \eg][]{songaila01, boksenbergetal03ph,
  pettinietal03, schayeetal03, songaila05}. The studies used a variety
of techniques, from traditional absorption line surveys
\citep[\eg][]{songaila01,boksenbergetal03ph}, where hundreds of
\ion{C}{4} doublets were analyzed, to variants on the pixel optical
depth (POD) method \citep[\eg][]{schayeetal03,songaila05}, which
statistically correlate the amount of flux absorbed to the \ion{C}{4}
mass density. \citet{songaila01} pioneered modern \OmCIV\ studies, and
her results for the range $1.5 < z < 5$ have been confirmed by
subsequent surveys: $1.6\times10^{-8} \lesssim \OmCIV \lesssim
3\times10^{-8}$. (We adjust all \OmCIV\ values quoted in this paper to
our adopted cosmology: $H_{0}=70\kms\,{\rm Mpc}$, $\Omega_{\rm M} =
0.3$, and $\Omega_{\Lambda} = 0.7$ and to sample doublets with $13 \le
\logCIV \le 15$; see Appendix \ref{appdx.adjOmCIV} for more details.)

Recent studies have focused on increasing the statistics on \ion{C}{4}
absorption at $z>5$, where \citet{songaila01} only detected one
absorber. These studies had to await the development of near-infrared
spectrographs and the discovery of $z\approx 6$ QSOs. With only one or
two sightlines, \citet{ryanweberetal06} and \citet{simcoe06},
respectively, measured the $5.4 \lesssim z \lesssim 6.2$ \ion{C}{4}
mass density to be consistent with the previous $1.5 < z < 5$ surveys.
However, the most recent work by \citet{ryanweberetal09} and
\citet{beckeretal09} observed that \OmCIV\ at $5.2 \lesssim z
\lesssim 6.2$ is a factor of $\approx4$ smaller than \OmCIV\ at $z<5$.
Although these two studies included two to three times as many
sightlines as \citet{simcoe06}, the results are based on $\lesssim 3$
detected \ion{C}{4} doublets. Obviously, the small number statistics
at $z>5$ leave the \OmCIV\ measurements more susceptible to cosmic
variance.

Cosmological hydrodynamic simulations have been used to understand the
interplay between metallicity, feedback, the ionizing background,
etc., and the evolution of \OmCIV.  \citet{aguirreetal01} and
\citet{springelandhernquist03} argued that most of the metals observed
in the IGM were distributed by galactic winds at $3\le z\le10$, and
observations at $z\approx2.5$ offer empirical support
\citep[\eg][]{simcoeetal04}.  \citet{oppenheimeranddave06} evolved
cosmological hydrodynamic simulations from $z=6 \rightarrow 2$ with a
range of prescriptions for galactic winds that enrich the IGM. They
found that the increasing cosmic metallicity from $z=5 \rightarrow 2$
balanced the decreasing fraction of carbon traced by the \ion{C}{4}
transition (\ie changing ionization state of the IGM). Thus, they
neatly reproduced the observed lack of \OmCIV\ evolution. In their
momentum-driven winds simulation (their favored `{\it vzw}' model),
\OmCIV\ increased from $z=6 \rightarrow 5$, consistent with the
measurements available at the time \citep{songaila01, ryanweberetal06,
  simcoe06} and with the more recent results \citep{ryanweberetal09,
  beckeretal09}.

In \citet{oppenheimeranddave08}, the authors included feedback from
asymptotic giant branch (AGB) stars, in addition to a new method for
deriving the velocity dispersion, $\sigma$, of galaxies, which defined
the momentum-driven wind speed. They compared the evolution of \OmCIV\
from $z=2 \rightarrow 0$ in the simulations with the old and new
$\sigma$-derived winds with AGB feedback and the new $\sigma$-derived
winds without AGB feedback. In all three simulations, \OmCIV\ did not
evolve from $z=3 \rightarrow 1$ (2\,Gyr to 6\,Gyr after the Big
Bang). In the simulation with the new $\sigma$-derived winds and AGB
feedback, \OmCIV\ increased by 70\% from $z=1 \rightarrow 0$---to
$\OmCIV \approx 7\times10^{-8}$---over the last 8\,Gyr of the cosmic
enrichment cycle. The AGB feedback increased the star formation rate,
which increased the mass of carbon in the IGM. This predicted increase
in \OmCIV\ at $z < 1$ can and should be tested by empirical
observation.

The \ion{C}{4} mass density at $z<1.5$ has not been studied as
extensively as at high redshift, where the \ion{C}{4} doublet is
redshifted into optical passbands. At low redshift, ultraviolet
spectrographs on space-based telescopes are required. Recent studies
\citep[\eg][]{fryeetal03,danforthandshull08} have leveraged 
high-resolution, UV echelle spectra to examine the low-redshift IGM:
the {\it Hubble Space Telescope} Space Telescope Imaging Spectrograph
and Goddard High-Resolution Spectrograph (\hst\ \stis\ and \ghrs,
respectively), supplemented by spectra from the {\it Far Ultraviolet
  Spectrograph Explorer} (\fuse). Through an analysis of nine quasar
sightlines observed with the \stis\ E140M grating, \citet{fryeetal03}
found six \ion{C}{4} doublets and measured $\OmCIV \approx
12\times10^{-8}$ for $z<0.1$ from a preliminary analysis.  With an
expanded survey of 28 sightlines with E140M spectra,
\citet{danforthandshull08} detected 24 \ion{C}{4} doublets in 28
sightlines and measured $\OmCIV = (7.8\pm1.5)\times10^{-8}$ for
$z<0.12$.  These low-redshift \OmCIV\ measurements are consistent with that
predicted by the $\sigma$-derived winds with AGB feedback simulation
of \citet{oppenheimeranddave08}.

These initial studies focused on \ion{C}{4} absorbers at $z \lesssim
0.1$ and did not present comprehensive analysis of their survey
completeness, nor did they take advantage of the full set of \hst\
archival data. There are currently 69 sightlines in the \hst\ archives
with UV spectra, where the \ion{C}{4} doublet can be detected at
$z<1$. We have conducted a large survey for \ion{C}{4} systems in
these sightlines. We analyzed the $z<1$ data in a consistent
fashion, which allowed for a uniform comparison throughout the
eight-billion year interval. We introduced robust search algorithms and
constrained the frequency distribution \ff{\NCIV} for the full sample.
Finally, we compared our results with the other low- and high-redshift
surveys, focusing on the evolution of \OmCIV.

The paper is organized as follows: we present the spectra, the
reduction procedures, and the measurements in \S\ \ref{sec.data}; our
sample selection is described in \S\ \ref{sec.selec}; \S\
\ref{sec.cmplt} outlines the completeness tests; we analyze the
frequency distribution and the \ion{C}{4} mass density in \S\
\ref{sec.analysis}; the final discussion is provided in \S\
\ref{sec.disc}; and \S\ \ref{sec.summ} is a summary.


\begin{deluxetable}{llllllr@{\ }lllllp{1.8cm}p{2.5cm}}
\rotate
\tablecolumns{14}
\tablewidth{0pc}
\tablecaption{OBSERVATION SUMMARY \label{tab.obssumm}}
\tabletypesize{\scriptsize}
\tablehead{
\colhead{(1)} & \colhead{(2)} & \colhead{(3)} & \colhead{(4)} & \colhead{(5)} & 
\colhead{(6)} & \multicolumn{2}{c}{(7)} & \colhead{(8)} & \colhead{(9)} & \colhead{(10)} & 
\colhead{(11)} & \colhead{(12)} & \colhead{(13)} \\
\colhead{Target} & \colhead{RA} & \colhead{Dec} & 
\colhead{$\zem$} & \colhead{Instr.} & \colhead{Grating} & 
\multicolumn{2}{c}{$R$ (FWHM)} & 
\colhead{S/N} & \colhead{$\lambda_{\rm min}$} & \colhead{$\lambda_{\rm max}$} & 
\colhead{$t_{exp}$} & \colhead{PID} & \colhead{Notes} \\
 & \colhead{(J2000)} & \colhead{(J2000)} & & & & 
\multicolumn{2}{c}{\hspace{\stretch{1}}(\!\kms)} & 
\colhead{(pix$^{-1}$)} & \colhead{(\AA)} & \colhead{(\AA)} & 
\colhead{(ks)}  & & 
}
\startdata
                   MRK335 &     00:06:19 &    +20:12:10 & 0.0258 & STIS &  E140M &   45800 & $(   7)$ &    7 & 1141 & 1710 &    17 & 9802 &  \\
                     &              &              &        & GHRS &  G160M &   20000 & $(  15)$ &   34 & 1222 & 1258 &    15 & 3584 &  \\
                     &              &              &        & {\it FUSE} &   LWRS &   20000 & $(  15)$ & $\le  10$ &  904 & 1188 &    99 & P101 &  \\
               Q0026+1259 &     00:29:13 &    +13:16:04 & 0.1450 & GHRS &  G270M &   20000 & $(  15)$ &    5 & 2785 & 2831 &     5 & 3755 & Exclude (coverage) \\
                     &              &              &        & {\it FUSE} &   LWRS &   20000 & $(  15)$ & $\le   2$ &  904 & 1188 &    20 & Q206 & \\
                  TONS180 &     00:57:20 &   --22:22:56 & 0.0620 & STIS &  G140M &   12700 & $(  24)$ &   25 & 1244 & 1298 &     3 & 7345 & Exclude (coverage) \\
                     &              &              &        &      & G230MB &    9450 & $(  32)$ &    9 & 2758 & 2912 &     1 & 9128 & Exclude (coverage) \\
                     &              &              &        & {\it FUSE} &   LWRS &   20000 & $(  15)$ & $\le   5$ &  904 & 1188 &    25 & D028; P101 & \\
               PG0117+213 &     01:20:17 &    +21:33:46 & 1.4930 & STIS &  E230M &   30000 & $(  10)$ &    8 & 2278 & 3072 &    42 & 8673 &  \\
                  TONS210 &     01:21:51 &   --28:20:57 & 0.1160 & STIS &  E140M &   45800 & $(   7)$ &    6 & 1141 & 1710 &    22 & 9415 &  \\
                     &              &              &        &      &  E230M &   30000 & $(  10)$ &    4 & 1988 & 2782 &     5 & 9415 &  \\
                     &              &              &        &      & G230MB &    9450 & $(  32)$ &    0 & 2759 & 2913 &     2 & 9128 & Exclude ($R$) \\
                     &              &              &        & {\it FUSE} &   LWRS &   20000 & $(  15)$ & $\le   8$ &  904 & 1188 &    53 & P107 & \\
\enddata
\tablecomments{
The targets and spectra included in the current survey.
The targets, their coordinates, and their redshifts are listed in Columns 1--4.
The instrument and grating of the archived observations (Columns 5 and 6) are listed with the resolution $R$ (and FWHM in \!\kms), exposure time $t_{exp}$, and the proposal identification number (PID) from the MAST query (Columns 7, 11, and 12).
The signal-to-noise ratio S/N was measured from the normalized spectra (Column 8).
Columns 9 and 10 are the wavelength coverage of each spectrum.
The status of the spectra (\ie whether it was excluded and why) is listed in Column 13.
Table \ref{tab.obssumm} is published in its entirety in the electronic edition of the {\it Astrophysical Journal}.  A portion is shown here for guidance regarding its form and content.
}
\end{deluxetable}

\section{Data, Reduction, and Measurements}\label{sec.data}

To assemble our target list, we searched the {\it HST} \stis\ and
\ghrs\ spectroscopic archives for objects with target descriptions
including the terms, \eg QSO, quasar, Seyfert, etc. Our final list
included 69 objects with redshifts $0.001<\zem<2.8$. We retrieved all
available spectra from the Multimission Archive at Space
Telescope\footnote{See
  http://archive.stsci.edu/.} (MAST), including supplementary \fuse\
data, when available (see Table \ref{tab.obssumm}\footnote{We have
  adopted the target name that MAST used. So note: B0312--770 is also
  PKS0312--770; QSO--123050+011522 is also Q1230--0115; and PHL1811 is
  also known as FJ2155--092.}). 
We reduced (in the case of the \fuse\ spectra), co-added, and
normalized the spectra with similar algorithms as described in
\citet{cookseyetal08}. The reduced \hst\ spectra were retrieved from
MAST directly.

One goal of this study was to search the {\it entire} \hst\ archive
for \ion{C}{4} absorption; thus, we initially included the full
archival dataset. However, we excluded targets that only had
un-normalized spectra with signal-to-noise ratio ${\rm S/N} < 2\,{\rm
  pix}^{-1}$. The signal-to-noise ratio was measured by fitting a
Gaussian to the histogram of flux \fx\ and noise \sigfx\ per pixel,
clipping the highest and lowest outliers and demanding $f_\lambda >
0$. 
Ten sightlines do not meet the S/N criterion: QJ0640--5031; TON34;
HE1104--1805A; HE1122--1648; NGC4395; Q1331+170; IR2121--1757;
HDFS--223338--603329-QSO; AKN564; and PG2302+029. They are indicated
by `Exclude (low S/N)' in Table \ref{tab.obssumm}. Q1331+170 has a
damped \Lya\ system, with molecular hydrogen lines, at $z = 1.7765$
\citep{cuietal05} that made the E230M spectra exceptionally difficult
to analyze.\footnote{Q1331+170 also had a known \ion{Mg}{2} system at
  $z=0.7450$ \citep{ellisonetal03}, which likely had blended
  \ion{C}{4} absorption.}  Therefore, Q1331+170 was also excluded,
though it had ${\rm S/N} = 5\,{\rm pix}^{-1}$.

In addition, we excluded the spectra for which higher-resolution
spectra covered the same wavelength range (noted as `Exclude
(overlap)' in Table \ref{tab.obssumm}). Typically, we excluded spectra
with resolution $R<20,000$ (${\rm FWHM} > 15\kms$). There were ten
targets that had no \ion{C}{4} coverage, given the spectral wavelength
range and \zem: Q0026+1259, TONS180, PKS0558--504, PG1004+130,
HE1029+140, Q1230+0947, \linebreak[4]PG1307+085, MARK290, Q1553+113,
and \linebreak[4]PKS2005-489.

\begin{figure}[!hbt]
\begin{center}
  \includegraphics[height=0.47\textwidth,angle=90]{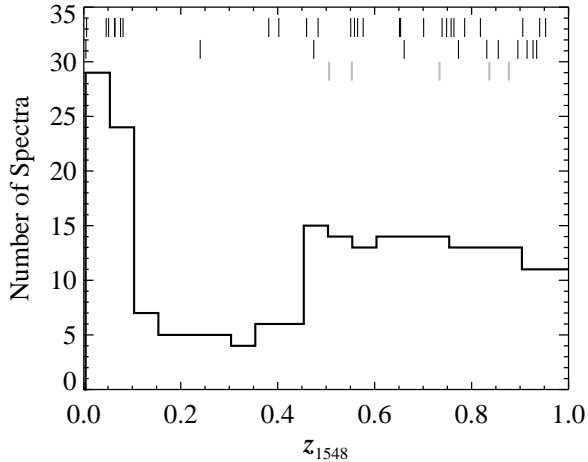}
\end{center}
  \caption[Schematic of redshift coverage for the current survey.]
  {Schematic of redshift coverage for the current survey. The number
    of spectra with coverage of the \ion{C}{4} doublet is shown as a
    function of \zciv, the redshift of the 1548 line (black
    histogram). The \stis\ E140M spectra covered $\zciv \lesssim
    0.1$. The \stis\ E230M spectra typically covered $0.4 \lesssim
    \zciv < 1$.  The redshift range $0.1 \lesssim \zciv \lesssim 0.4$
    was covered by some E230M spectra as well as the medium-resolution
    gratings and \ghrs\ spectra. The redshifts of the doublets
    detected with $\EWr \ge 3\,\sigEWr$ in both lines are shown with
    the hashes across the top. The top and middle rows indicate the
    redshifts of the 27 unsaturated (top) and 11 saturated (middle)
    doublets in the definite group (G = 1; see \S\
    \ref{subsec.smpl}). The bottom row shows the redshift of the five
    unsaturated likely (G = 2) doublets.
    \label{fig.gz}
  }
\end{figure}

Ultimately, 49 targets had spectra with usable wavelength coverage and
S/N ratios.  All reduced, co-added, and normalized spectra are
available online,\footnote{See
  http://www.ucolick.org/$\sim$xavier/HSTCIV/ for the
    normalized spectra, the continuum fits, the \ion{C}{4} candidate
    lists, and the Monte-Carlo completeness limits for all sightlines
    as well as the completeness test results for the full data
    sample.} even those not explicitly searched in this paper. The
number of spectra covering the \ion{C}{4} doublet redshift range is
shown schematically in Figure \ref{fig.gz}.

\subsection {\hst\ \stis\ and \ghrs}\label{subsec.hst}

The \hst\ spectra drive the target selection because its UV
instruments, \stis\ and \ghrs, have the wavelength coverage to detect
\ion{C}{4} in the $z<1$ Universe. We preferred higher-resolution data
over lower-resolution data to resolve the \ion{C}{4} doublets and
better distinguish them from coincident \Lya\ features.  The \stis\
echelle gratings were the preferred set-up in most cases, since they
were high resolution and covered a large wavelength range.

All \stis\ observations were reduced with CalSTIS 2.23 on UT
06-October-2006 with On-the-Fly Reprocessing and co-added with
XIDL\footnote{See http://www.ucolick.org/$\sim$xavier/IDL/.} COADSTIS
\citep{cookseyetal08}. The \stis\ long-slit spectra were co-added by
re-binning the individual exposures to the same wavelength array and
then combining the error-weighted flux. For the echelle data, the
observations for each order were co-added separately, in the same
manner as the long-slit spectra. Then all of the orders were co-added
into a single spectrum.

The \stis\ echelle gratings E140M and E230M contributed the most to
the total pathlength of this survey. All 49 targets have at least one
echelle spectrum. The E140M grating covered the wavelength range
$1140\Ang \lesssim \lambda \lesssim 1710\Ang$ or
$\zciv\lesssim0.1$ and had a resolution of $R=45,000$ (${\rm FWHM} =
7\kms$). The E230M grating covered $\approx\!800\Ang$ per tilt over
the range $1570\Ang \lesssim \lambda \lesssim 3110\Ang$ or
$\zciv\lesssim1$. Typically, the observations covered $2280\Ang
\lesssim \lambda \lesssim 3070\Ang$ or $0.5\lesssim \zciv \lesssim
1$. The E230M grating had resolution $R=30,000$ ($10\kms$).  The only
other \stis\ gratings included in the \ion{C}{4} search were G140M and
G230M, which have wavelength coverage similar to E140M and E230M, respectively,
albeit with less total coverage. For more information about
\stis, see \citet{mobasher02}.

The calibrated \ghrs\ spectra were retrieved from MAST. They were co-added like
the \stis\ long-slit spectra. Observations with central wavelengths that varied by
less than 5\% of the total wavelength coverage of that grating were co-added
into a single spectrum.  This avoided the problem of automatic scaling to the
highest-S/N spectrum.

The \ghrs\ ECH-A, ECH-B, G160M, G200M, and G270M gratings were used in
the final \ion{C}{4} search. The ECH-A and ECH-B gratings had
resolution $R=100,000$ (${\rm FWHM} = 3\kms$). The G160M, G200M, and
G270M gratings had resolution $R=20,000$ ($15\kms$).  We excluded all
\ghrs\ spectra where there existed \stis\ coverage since, in general,
the \stis\ data have higher resolution and larger wavelength coverage.
For more information about \ghrs, see
\citet{brandtetal94}.

\subsection {\emph{Far Ultraviolet Spectroscopic
    Explorer}}\label{subsec.fuse}

The raw \fuse\ observations were retrieved from MAST. We reduced the
spectra with a modified CalFUSE\footnote{See
  ftp://fuse.pha.jhu.edu/fuseftp/calfuse/.} v3.2 pipeline \citep[see
][]{cookseyetal08}. To summarize, the partially processed data from
the exposures (\ie intermediate data file or IDF) were combined before
the extraction window centroid was determined. Then, the bad pixel
masks were generated, and the spectra were optimally extracted.

The observations were co-added with \linebreak[4]FUSE\_REGISTER. The
\fuse\ wavelength solutions were {\it not} shifted to match the \stis\
wavelength solution. The \fuse\ observations were only searched for
absorption lines (\eg \Lyb, \ion{C}{3}, \ion{O}{6}) to supplement
candidate \ion{C}{4} systems. The blind doublet search made allowance
for offsets in redshift, due to multi-phase absorbers or misaligned
wavelength solutions, when it assigned additional absorption lines to
the candidate \ion{C}{4} systems (see \S\ \ref{subsec.detec} for more
details).

We excluded \fuse\ spectra with S/N ratios too low for continuum
fitting: TON28, NGC4395, QSO--123050+011522, Q1230+0947, NGC5548,
PG1444+407, and PG1718+481.

\subsection{Continuum Fitting}\label{subsec.conti}

The continuum for each spectrum was fit semi-automatically with a
B-spline. ``Semi-automatically'' refers to the subjective nature of
continuum fitting. The best B-spline was based on the input
parameters, primarily, the low- and high-sigma clips and the
breakpoint spacing. However, the authors varied these parameters until
they agreed the ``best'' B-spline was a good match to un-absorbed regions.

The breakpoint spacing was typically $\approx\!4\Ang$ to $6\Ang$. The
breakpoint spacing was refined to increase in regions where the flux
was not varying substantially and to decrease in regions where it
was. Regions of great variation were determined by comparing the
median flux of each bin to the mean variance-weighted flux of all
bins, where the bins are determined by the initial breakpoint
spacing. If the difference between the median flux in a bin and the
mean flux were more than one standard deviation of the flux bins, the
breakpoint spacing was decreased. For most spectra, the low- and
high-sigma clips were $2\sigma$ and $2.5\sigma$, respectively. The
regions masked out by the sigma clipping were increased by two pixels
on both sides.

The B-spline was iteratively fit to the variance-weighted flux in the
bins, with the clipped regions masked out, until the percent
difference across the continuum fit was less than $0.001\%$ compared
to the previous (converged). For several spectra (\eg
the NGC galaxies), the semi-automatic continuum fit was adjusted by
hand.

As mentioned previously, the ``best-fit'' continuum was a subjective
judgment. Several, slightly different continuum fits would have
satisfied the authors. To gauge the systematic error introduced by the
subjective nature of continuum fitting, we measured the differences
due to changing the continuum from the semi-automatic fit to one
generated ``by hand'' for one sightline \citep[see ][]{cookseyetal08}.
The root-mean-square fractional difference in the observed equivalent
width and column density were $\lesssim10\%$.

\subsection{Redshift, Equivalent Width, and Column
  Densities}\label{subsec.meas}

We measured redshifts from the optical depth-weighted central
wavelengths; equivalent widths from simple boxcar summation; and
column densities from the apparent optical depth method \cite[AODM;
][]{savageandsembach91}. To minimize the effect of spurious, outlying
pixels on the measurements, the flux was trimmed: $-\sigfx \le \fx \le
1+\sigfx$. 
Outlying flux pixels were set to the appropriate
extrema. The trimming affected measurements in the lowest S/N regions
of the spectra, typically the edges. It also assisted the
identification of absorption lines in the noisy regions by the
automated feature search (see \S\ \ref{subsec.detec}), from which the
candidate \ion{C}{4} doublets were drawn. When the final \ion{C}{4}
sample was defined and the wavelength bounds visually confirmed or
changed, this flux trimming had negligible effect. For optical depth
measurements, the minimum flux was set, so that:
$\fx \ge (0.2\sigfx > 0.05)$. This prevented the optical depth, $\tau
= \ln(1/\fx)$, from being overwhelmingly large for the saturated
pixels. These cases were then reported as lower limits to \NCIV.

We measured redshifts from the mean optical depth-weighted central
wavelengths of the absorption lines and the rest wavelength \wvr\ of
the transition. The wavelengths per pixel $\lambda_{i}$ were weighted
by their optical depth per pixel $\tau_{i} = \ln(1/\fx{}_{i})$ within
the bounds of the absorption line, defined by the wavelength range:
$\wvlo \le \lambda_{i} \le \wvhi$. Thus, the pixels with the strongest
absorption dominated the redshift estimate:
\begin{equation}
  1+\zabs = 
  \frac{\displaystyle \sum_{\wvlo}^{\wvhi} \lambda_{i} \ln\bigg(\frac{1}{\fx{}_{i}}\bigg)}
  {\displaystyle \wvr \sum_{\wvlo}^{\wvhi} \ln\bigg(\frac{1}{\fx{}_{i}}\bigg)} {\rm .} \label{eqn.zabs}
\end{equation}

The rest equivalent width \EWr\ were measured with a boxcar summation
over the wavelength bounds of the feature:
\begin{eqnarray}
  \EWr & = & \frac{1}{(1+\zabs)}\sum_{\wvlo}^{\wvhi} (1-\fx{}_{i}) 
  \delta\lambda_{i} \label{eqn.ewr} \\
  \sigEWr^{2} & = & \frac{1}{(1+\zabs)^{2}}\sum_{\wvlo}^{\wvhi} 
  \sigfx{}_{i}^{2} \delta\lambda_{i}^{2} {\rm ,} \nonumber
\end{eqnarray}
where $\delta \lambda_{i}$ (\AA) is the wavelength pixel scale of the
spectrum. The second equation is the variance of the \EWr\ measurement
from error propagation. The observed equivalent width \EWo\ and error
\sigEWo\ were measured with the previous equations but without the
$1+\zabs$ factor.

Most of the column densities were measured with the apparent optical
depth method \citep[AODM; ][]{savageandsembach91}:
\begin{eqnarray}
  \N{\rm AOD} & = & \frac{10^{14.5762}}{\fval\wvr} 
  \sum_{\wvlo}^{\wvhi} \ln\bigg(\frac{1}{\fx{}_{i}}\bigg)\delta v_{\lambda,i}\,(\!\cm{-2})
  \label{eqn.aodm} \\
  \sigma_{\N{\rm AOD}}^{2} & = &
  \bigg(\frac{10^{14.5762}}{\fval\wvr}\bigg)^{2}
  \sum_{\wvlo}^{\wvhi}\bigg(\frac{ \sigfx{}_{i}}{ \fx{}_{i}}\delta
    v_{\lambda,i}\bigg)^{2}\,(\!\cm{-4}) \nonumber\\
  \delta v_{\lambda,i} & = & c\bigg(\frac{\delta \lambda_{i}}
  {\wvr(1+\zabs)}\bigg) {\rm ,} \nonumber 
\end{eqnarray}
where \fval\ (unitless) is the oscillator strength of the transition
with rest wavelength \wvr\ (\AA) and $\delta v_{\lambda,i}$ (\!\kms)
is the velocity pixel scale of the spectrum. The atomic data and
sources are tabulated in \citet{prochaskaetal04}. The AOD profiles
were used as a diagnostic (see \S\ \ref{subsec.flag}) and were
constructed from the un-summed versions of the above equations,
smoothed over three pixels (see Figure \ref{fig.civcand}).

In some cases, only a column density limit could be set, when the AODM
resulted in a measurement $\N{\rm AOD} < 3\sigma_{\N{\rm AOD}}$
(resulting in an upper limit) or the line was saturated (lower
limit). In low-S/N spectra, some lines satisfied both criteria (\eg
$\zciv=0.40227$ doublet in the PKS0454--22 sightline). In which case,
we counted the line as an estimate of the upper limit if $\EWr <
3\,\sigEWr$ and as a lower limit otherwise. Since the column density
from the AODM was a poor measurement in the aforementioned instances,
we estimated the column density by assuming the \EWr\ reflects the
column density from the linear portion of the curve of growth
(COG). Then, we use whichever column density measurement resulted in
the more extreme limit. For example, if the COG column density was
lower than the AODM column density for a saturated, $\ge 3\sigEWr$
feature, the COG column density was used. 

For analyses pertaining to the equivalent width, we used \EWlin{1548}
from the \ion{C}{4} 1548 line. For analyses relating to the column
density \NCIV\ we either used the error-weighted average of $\N{1548}$
and $\N{1550}$ when both were measurements or constrained the value
based on the limits of the two doublet lines. The greater lower limit
(or the smaller upper limit) was used. In a few cases, $\N{1548}$ and
$\N{1550}$ constrained a \NCIV\ range (see the bracketed values in
Table \ref{tab.civ}). For the column density analyses, we took the
average of the limits and used the difference between the average and
the values as the errors. For the $\zciv=0.38152$ \ion{C}{4} doublet
in the PKS0454--22 sightline, the two line limits did not overlap; we
increased/decreased the limits by $1\sigma$ to constrain \NCIV.

For several sightlines (\eg the \fuse\ spectra or the two E230M grating tilts
for PG1634+706), there were overlapping spectra. We quote measurements
from the spectrum where \EWr\ was measured with the higher estimated
significance $\EWr/\sigEWr$.

\section{Sample Selection}\label{sec.selec}

We conducted a ``blind'' survey for \ion{C}{4} doublets, where
candidate \ion{C}{4} absorbers and any associated lines were
identified exclusively by the characteristic wavelength separation of
the \ion{C}{4} doublet and the measured redshift of the \ion{C}{4}
1548 line. This eliminated any bias associated with identifying \Lya\
absorbers first and then searching for associated \ion{C}{4} doublets.
Also, for several sightlines, we did not have the wavelength coverage
to search for \Lya. Though \ion{C}{4} systems frequently show strong
\Lya\ absorption \citep{ellisonetal99,simcoeetal04}, \ion{C}{4}
systems with weak \Lya\ absorption do occur \citep{schayeetal07}, and
they might occur with higher frequency at $z<1$ if the IGM is more
highly enriched. Once we searched for the candidate \ion{C}{4}
doublets, we used other diagnostics (see \S\ \ref{subsec.flag}) and
visual inspection to define our final sample.

\subsection{Automatic Line Detection and Blind Doublet
  Search}\label{subsec.detec}

First, we searched for absorption features\footnote{We refer to
  absorption ``features'' to indicate the results from the automated
  feature-finding program, which also returned absorption line-like
  artifacts in the spectrum (\eg regions with poor continuum fit,
  multiple pixels with spuriously low flux) in addition to real
  absorption {\it lines}.} with observed equivalent widths $\EWo \ge
3\,\sigEWo$ with an automated procedure \citep{cookseyetal08}. We
convolved the flux with a Gaussian with width equal to the FWHM of the
instrument to yield $f_{G}$. Then, adjacent convolved pixels with
significance $\ge1\sigma_{f_{G}}$ were grouped into absorption
features. The observed equivalent width for all features were measured
with the boxcar extraction window defined by the wavelength limits,
\wvlo\ and \wvhi. The final product was a list of candidate absorption
features with $\EWo \ge 3\,\sigEWo$, defined by their flux
decrement-weighted wavelength centroid and wavelength limits.

Weighting with the flux decrement $1-\fx$, where $\fx < 1$, was better
behaved than weighting with the optical depth $\tau =
\ln(1/\fx)$. For strong absorption lines, the two weighting methods
result in comparable values. However, for weak absorption lines and
noisy absorption features, the flux decrement-weighted centroid was
closer to the central wavelength, as defined by the bounds $\wvlo \le
\lambda \le \wvhi$

There was no attempt to separate blends at this level. This method
automatically recovered nearly all of the features we would have
identified visually, but it was not sensitive to broad, shallow
features, \ie $b \gtrsim 50\kms$.  However, \citet{rauchetal96}
measured the distribution of Doppler parameters of
\ion{C}{4} doublets at high redshift to be $5\kms \le b \le 20\kms$.  This gives
us reason to expect that our algorithms would miss very few (if any)
systems.

\begin{figure*}[!hbt]
  \begin{center}$
    \begin{array}{cc}
      \includegraphics[width=0.4\textwidth]{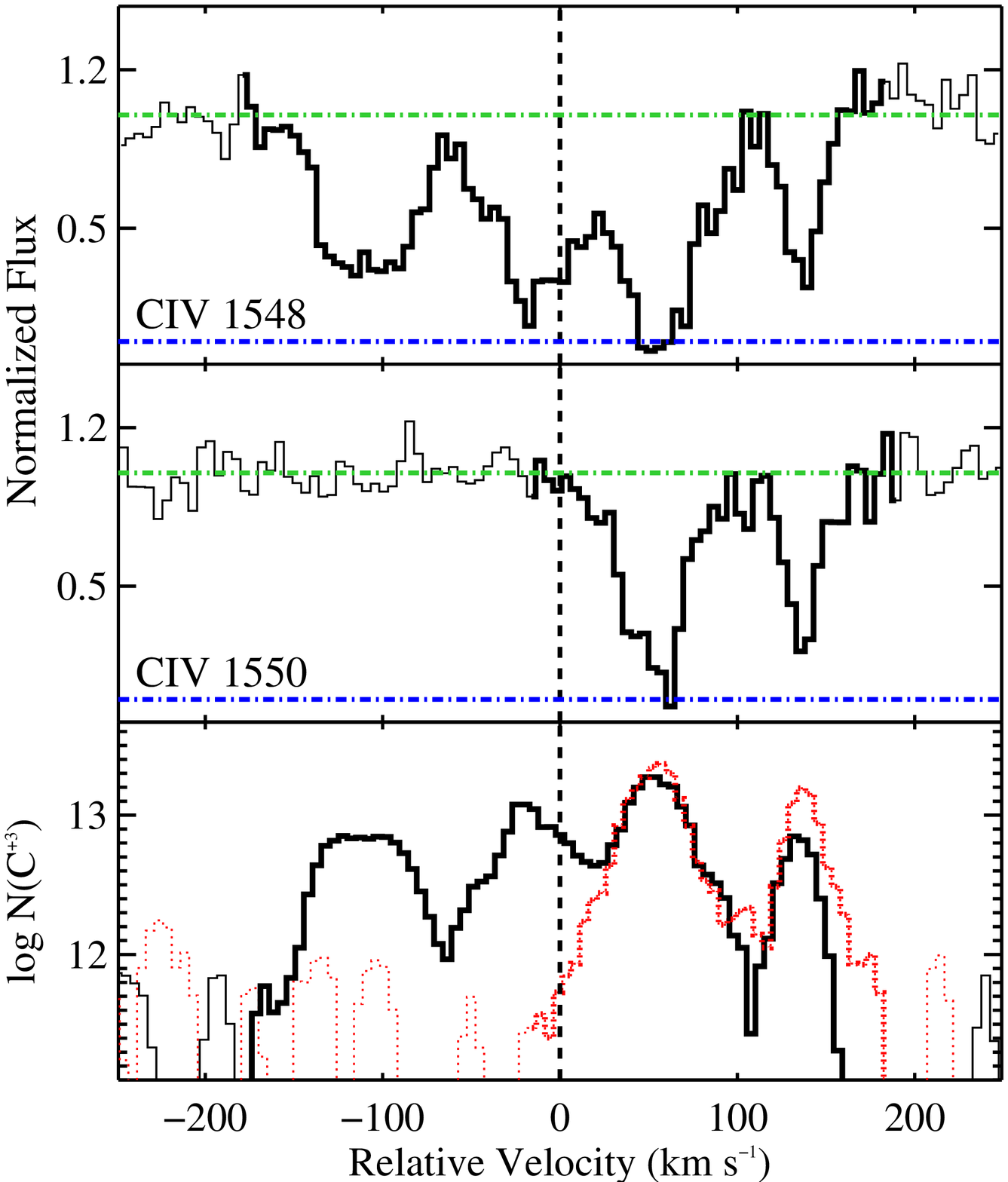} & 
      \includegraphics[width=0.4\textwidth]{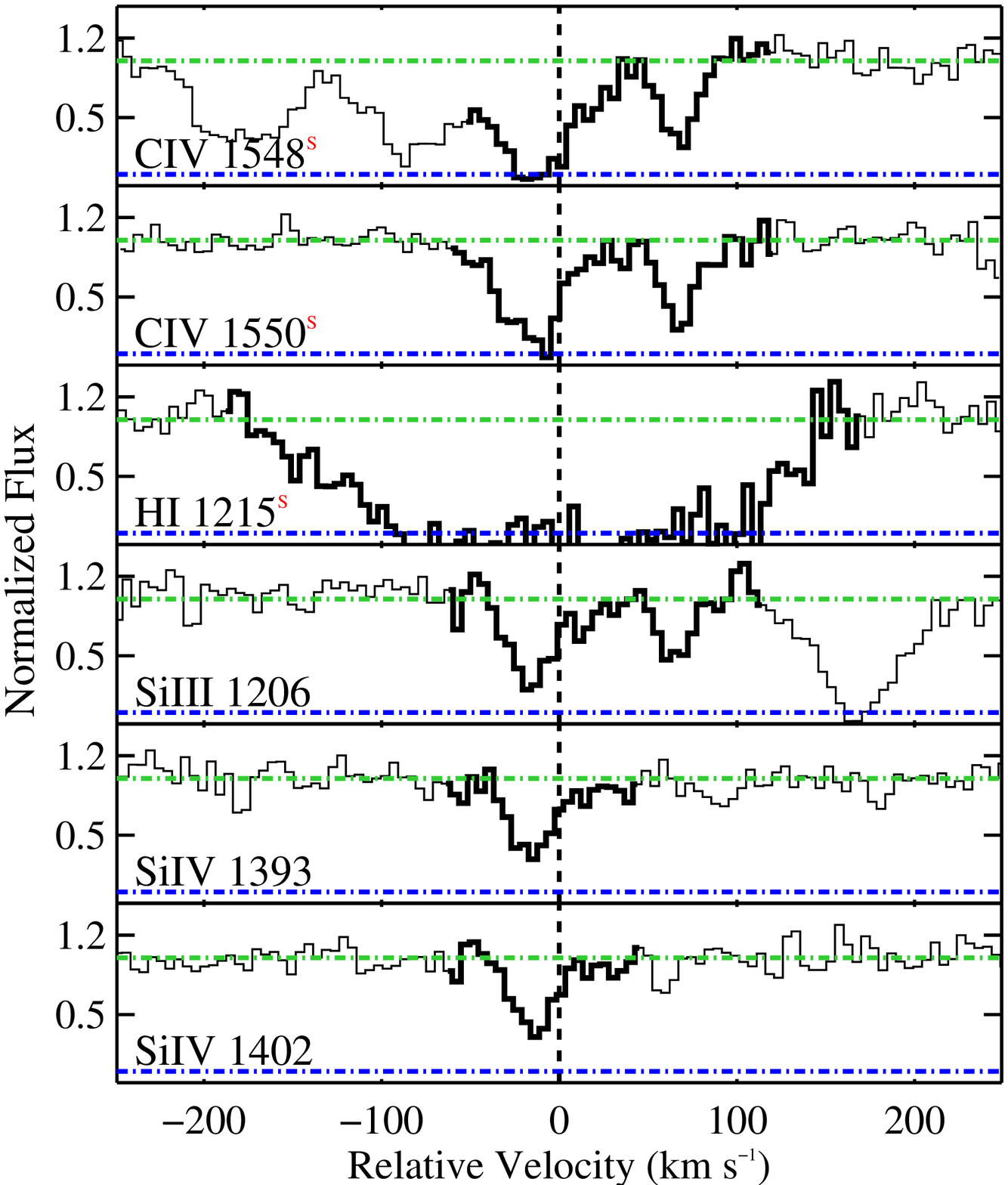} 
    \end{array}$
  \end{center}
  \caption[Example of an initial {\it candidate} \ion{C}{4}
    doublet and the final $\zciv = 0.91440$ system in the PG1630+377
    sightline.]
    {Example of an initial {\it candidate} \ion{C}{4} doublet and the
      final $\zciv = 0.91440$ system in the PG1630+377 sightline. Two
      automatically-detected absorption features were paired as a
      candidate \ion{C}{4} doublet based on their wavelength
      separation (left panel).  The wavelength bounds of the candidate
      1550 line were expanded to match the range defined by the
      candidate 1548 line. Several automatically-detected features
      were added to the candidate \ion{C}{4} system as common
      transitions, based on their observed wavelengths and the
      redshift of the candidate 1548 line. Finally, we visually
      inspected the candidate, deemed it a definite system, and
      modified the transitions and wavelength bounds (right panel).
      The bottom panel on the left shows the AOD profile of the 1548
      and 1550 lines (black solid and red dotted lines, respectively;
      see \S\ \ref{subsec.flag}). For a description of the velocity
      plots, see Figure \ref{fig.g1}.
    \label{fig.civcand}
  }
\end{figure*}

The automatically-detected features were paired into candidate
\ion{C}{4} doublets based purely on the characteristic wavelength
separation ($2.575\Ang$ or $498\kms$ in the rest frame). Every feature
with $\lambda_{cent}$ between $\lambda_{1548}$ and
$\lambda_{1548}(1+\zem)$ was assumed to be \ion{C}{4} 1548
absorption.\footnote{Actually, we searched for doublets at higher
  redshifts than \zem\ as a secondary check to the adopted values of
  \zem. We did not find any \ion{C}{4} doublets with $\zciv > \zem$.}
If there were an automatically-detected feature between the wavelength
limits $(\wvlo,\wvhi)$ at the location of the \ion{C}{4} 1550 line, it
was assumed to be \ion{C}{4} 1550 absorption (see Figure
\ref{fig.civcand}). Otherwise, the region that would have included
\ion{C}{4} 1550 was used to give an estimate of the upper limit of the
column density and equivalent width. The \ion{C}{4} 1550 region was
set by the wavelength bounds of \ion{C}{4} 1548:
\begin{eqnarray}
  \wvlo{}_{,1550} & = & \wvlo{}_{,1548} \bigg( 
  \frac{\wvr{}_{,1550}}{\wvr{}_{,1548}} \bigg) \\
  \wvhi{}_{,1550} & = & \wvhi{}_{,1548} \bigg( 
  \frac{\wvr{}_{,1550}}{\wvr{}_{,1548}} \bigg) {\rm .} \nonumber
\end{eqnarray}
Similarly, the wavelength bounds of the candidate doublet were
adjusted so that they were aligned (\eg $v_{l,1548} =
v_{l,1550}$).

The doublet search was then performed in reverse, where
automatically-detected features were assumed to be \ion{C}{4} 1550
lines, if not already included. The corresponding region for the
\ion{C}{4} 1548 absorption yielded an estimate for the upper limit of
the column density and equivalent width.

Automatically-detected features at the expected locations, defined by
the redshift of the \ion{C}{4} 1548 lines, of several common
transitions were added to the candidate \ion{C}{4} systems. The common
transitions were: \Lya; \Lyb; \ion{C}{3} 977; \ion{O}{6} 1031, 1037;
\ion{Si}{2} 1260; \ion{Si}{3} 1206; \ion{Si}{4} 1393, 1402; \ion{N}{3}
989; and \ion{N}{5} 1238, 1242. If there were no
automatically-detected candidate \Lya\ absorption, the region where
the \Lya\ line would have been was included as an estimate of the
upper limit of the column density and equivalent width, if the
wavelength coverage existed.

There were some automatically-detected features with $\lambda_{1548}
\le \lambda \le \lambda_{1548}(1+\zem)$ that were more than $498\kms$
wide, which is the characteristic separation of the \ion{C}{4}
doublet.  For example, there is a damped \Lya\ system in the spectrum
of PG1206+459 at $\zabs=0.92677$, which has a strong, multi-component
\ion{C}{4} doublet that is $633\kms$ wide, and the doublet lines 1548
and 1550 are blended with each other (see Appendix
\ref{appdx.plots}). These cases were visually evaluated and included
as candidate \ion{C}{4} doublets, since no automated attempt was made
to separate blended absorption features. The search for common
transitions then proceeded as described previously.

\begin{figure}[!hbt]
  \begin{center}
 \epsscale{0.8} \plotone{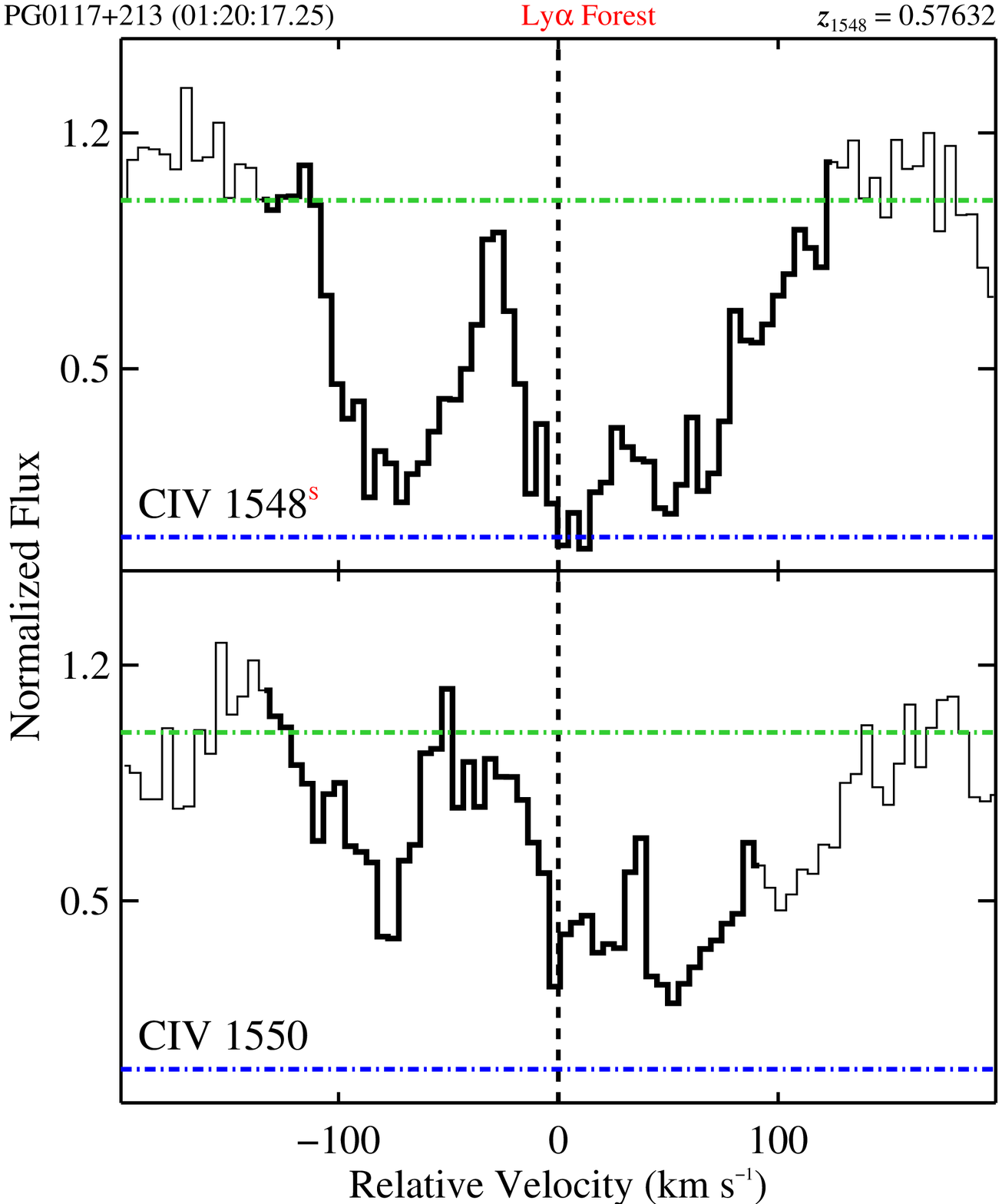}
 \end{center}
  \caption[Velocity plot of G = 1 \ion{C}{4} system in the PG0117+213
    sightline.]
    {Velocity plot of G = 1 \ion{C}{4} system in the PG0117+213
      sightline. The regions of spectra around each absorption line
      are aligned in velocity space with respect to the rest
      wavelength of the transition and $\zciv=0.49907$. Saturated
      transitions are indicated with the (red) `S.' The regions used
      to measure \EWr\ and \logCIV\ are shown by the dark outline. The
      flux at zero and unity are shown with the dash-dot lines (blue
      and green, respectively); the vertical dashed line indicates
      $v=0\kms$ corresponding to the optical depth-weighted velocity
      centroid of the \ion{C}{4} 1548 transition.  (The velocity plots
      for all G = 1 and G = 2 absorbers are in Appendix
      \ref{appdx.plots}.)
    \label{fig.g1}
  }
\end{figure}

Finally, the redshifts, rest equivalent widths, column densities, and
errors were measured for all transitions in the candidate \ion{C}{4}
systems (see \S\ \ref{subsec.meas}). The measured quantities for the
candidate doublets are listed in Table \ref{tab.cand}.


\begin{deluxetable}{rlllllllll}
\tablewidth{0pc}
\tablecaption{\ion{C}{4} CANDIDATES SUMMARY \label{tab.cand}}
\tabletypesize{\scriptsize}
\tablehead{ 
\colhead{(1)} & \colhead{(2)} & \colhead{(3)} & \colhead{(4)} & \colhead{(5)} & 
\colhead{(6)} & \colhead{(7)} & \colhead{(8)} & \colhead{(9)} & \colhead{(10)} \\ 
\colhead{$z_{1548}$} & \colhead{\dvabs} & 
\colhead{$\lambda_{r}$} & \colhead{\wvlo} & 
\colhead{\wvhi} & \colhead{\EWr} & \colhead{\sigEWr} & 
\colhead{\logN} & \colhead{\siglogN} & 
\colhead{Flag} \\
 & \colhead{(\!\kms)} & \colhead{(\AA)} & \colhead{(\AA)} & 
\colhead{(\AA)} & \colhead{(\!\mA)} & \colhead{(\!\mA)} & & & 
}
\startdata
\hline \\[-1ex]
\multicolumn{10}{c}{MRK335 ($\zem=0.026$)} \\[1ex]
\hline
-0.00162 & \nodata & 1548.19 & 1545.33 & 1545.97 & $<     48$ & \nodata & $< 13.09$  & \nodata & 157 \\
 &   -11.7 & 1550.77 & 1547.90 & 1548.54 &     286 &      21 & $> 14.39$  & \nodata &  \\
-0.00092 & \nodata & 1548.19 & 1546.49 & 1547.03 &     153 &      20 & 13.72 &  0.07 & 335 \\
 &   -18.0 & 1550.77 & 1549.06 & 1549.61 & $<     59$ & \nodata & $< 13.50$  & \nodata &  \\
 0.00000 & \nodata & 1548.19 & 1547.90 & 1548.54 &     286 &      21 & $> 14.09$  & \nodata & 495 \\
 &     4.2 & 1550.77 & 1550.47 & 1551.12 &     183 &      31 & 14.15 &  0.09 &  \\
\hline \\[-1ex]
\multicolumn{10}{c}{PG0117+213 ($\zem=1.493$)} \\[1ex]
\hline
 0.47166 & \nodata & 1548.19 & 2278.01 & 2278.98 &     691 &      78 & $> 14.23$\tablenotemark{a} & \nodata & 288 \\
 &     3.7 & 1550.77 & 2281.80 & 2282.77 & $<    119$ & \nodata & $< 13.94$  & \nodata &  \\
 0.47372 & \nodata & 1548.19 & 2280.33 & 2282.56 & $<    191$ & \nodata & $<  0.30$  & \nodata & 128 \\
 &   -23.5 & 1550.77 & 2284.12 & 2286.36 &     544 &      73 & $> 14.66$  & \nodata &  \\
 0.47605 & \nodata & 1548.19 & 2283.79 & 2286.36 &     522 &      79 & $> 14.35$  & \nodata & 386 \\
 &   -50.0 & 1550.77 & 2287.59 & 2290.16 &     978 &      71 & $> 15.05$  & \nodata &  \\
\hline \\[-1ex]
\multicolumn{10}{c}{TONS210 ($\zem=0.116$)} \\[1ex]
\hline
-0.00073 & \nodata & 1548.19 & 1546.84 & 1547.40 &     155 &      20 & $> 13.85$  & \nodata & 367 \\
 &     2.1 & 1550.77 & 1549.41 & 1549.98 & $<     64$ & \nodata & $< 13.63$  & \nodata &  \\
-0.00001 & \nodata & 1548.19 & 1547.64 & 1548.67 &     480 &      25 & $> 14.32$  & \nodata & 495 \\
 &    -7.9 & 1550.77 & 1550.21 & 1551.24 &     328 &      36 & 14.42 &  0.06 &  \\
 0.00090 & \nodata & 1548.19 & 1549.43 & 1549.73 &      74 &      23 & $< 13.23$  & \nodata & 269 \\
 &   -18.0 & 1550.77 & 1552.01 & 1552.31 & $<     38$ & \nodata & $<  0.30$  & \nodata &  \\
\enddata
\tablenotetext{a}{\logN\ measured by assuming \EWr\ results from the linear portion of the COG.}
\tablecomments{
Summary of C\,IV doublet candidates by target and redshift of C\,IV 1548.
Upper limits are $2\sigma$ limits for both \EWr\ and \logN.
The binary flag is described in Table \ref{tab.flag} and \S\ \ref{subsec.flag}.
Table \ref{tab.cand} is published in its entirety in the electronic edition of the {\it Astrophysical Journal}.  A portion is shown here for guidance regarding its form and content.
}
\end{deluxetable}

\subsection{Machine-Generated Diagnostics}\label{subsec.flag}

We assembled several thousand candidate \ion{C}{4} systems by the
process described in the previous section. We then used
machine-generated diagnostics to assist in identifying the true
\ion{C}{4} systems visually. The candidate systems were assigned flags
based on nine criteria that leveraged our knowledge of the \ion{C}{4}
doublet and other desirable characteristics (see Table
\ref{tab.flag}). The criteria were the following: \ion{C}{4} 1548 and
1550 had $\EWr \ge3\,\sigEWr$ (flags = 256 and 128, respectively); the
\EWr\ ratio of 1548 to 1550 was between $1:1$ and $2:1$, accounting
for the error (64, see Equation \ref{eqn.ratio}); the redshifts of the
doublet lines were within $10\kms$ (32, see Equation \ref{eqn.dvciv});
there was a candidate \Lya\ absorber detected with $\EWr \ge
3\,\sigEWr$ (16); \ion{C}{4} 1548 was outside the \Lya\ and the \HH\
forest (8 and 4, respectively); the apparent optical depth profiles of
the doublet were in agreement (2); and other candidate transitions
beside \Lya\ were detected with $\EWr \ge 3\,\sigEWr$ (1).

The equivalent width ratio of \ion{C}{4} 1548 to 1550 was measured as
follows:
\begin{eqnarray}
  R_{W} & = & \frac{\EWlin{1548}}{\EWlin{1550}} \label{eqn.ratio} \\
  \sigma_{R_{W}}^{2} & = &
  R_{W}^{2} \bigg(\Big(\frac{\sigEWr{}_{,1548}}{\EWlin{1548}}\Big)^{2} + 
  \Big(\frac{\sigEWr{}_{,1550}}{\EWlin{1550}}\Big)^{2}\bigg) {\rm ,} \nonumber
\end{eqnarray}
where the variance $\sigma_{R_{W}}^{2}$ resulted from the propagation
of errors. The actual diagnostic flag = 64 applied to doublets where
$1-\sigma_{R_{W}} \le R_{W} \le 2+\sigma_{R_{W}}$. The upper limit was
set by the characteristic ratio for an unsaturated doublet, and the
expected ratio decreases for increasingly saturated doublets until
$\EWlin{1548} = \EWlin{1550}$, hence, the lower limit.

The redshift of the 1548 line $z_{1548}$ must equal the redshift of
the 1550 line $z_{1550}$ for an un-blended \ion{C}{4} doublet,
detected with infinite S/N and infinite resolution. To accommodate
blending and the range of S/N ratio and resolution of the spectra, the
binary diagnostic flag = 32 applied to doublets with $\vert \dvciv
\vert \le10\kms$, where:
\begin{equation}
  \dvciv = c\bigg( \frac{z_{1550}-z_{1548}}{1+z_{1548}}\bigg) {\rm .} \label{eqn.dvciv}
\end{equation}

A candidate system was outside the \Lya\ forest when
$\lambda_{1548}(1+\zciv) \ge \lambda_{\alpha}(1+\zem)$ (flag = 8).
Similarly, a candidate was outside the \HH\ forest when
$\lambda_{1548}(1+\zciv) \ge \lambda_{\mbox{\HH}}(1+\zem)$ (flag = 4),
where $\lambda_{\mbox{\HH}} = 1138.867\Ang$ for \HH\ B0-0P(7), the
last \HH\ transition considered \citep[see ][for atomic and molecular
line data]{prochaskaetal04}.

The profiles of the 1548 and 1550 lines were considered to have matched when
they agreed within $1\sigma$ for $68.3\%$ of the range \wvlo\ to
\wvhi\ (flag = 2). The profiles were defined by the apparent optical
depth (AOD) column densities, similar to Equation \ref{eqn.aodm}:
\begin{eqnarray}
  \N{i} & = & \frac{10^{14.5762}}{\fval\wvr} 
  \ln\bigg(\frac{1}{\fx{}_{i}}\bigg)\delta v_{\lambda,i}\,(\!\cm{-2}) \\
  \sigma_{\N{i}}^{2} & = &
  \bigg(\frac{10^{14.5762}}{\fval\wvr}
  \frac{ \sigfx{}_{i}}{ \fx{}_{i}}\delta
    v_{\lambda,i}\,(\!\cm{-2})\bigg)^{2}{\rm ,} \nonumber
\end{eqnarray}
where $i$ indicates the pixels between \wvlo\ and \wvhi. The profiles
were smoothed over three pixels to minimize the effects of spurious
pixels and noisy spectra. \ion{C}{4} doublets with the diagnostic flag
= 2 have:
\begin{equation}
\lvert \N{i,1548} - \N{i,1550} \rvert \le 
\sqrt{\sigma_{\N{i,1548}}^{2} + \sigma_{\N{i,1550}}^{2}} 
\end{equation}
for more than $68.3\%$ of the range \wvlo\ to \wvhi. The AOD profiles
were computed only for the candidate \ion{C}{4} doublets.

The flags are binary and sum uniquely. They are ordered so that
doublets satisfying more of the diagnostic criteria have higher flag
values. For the current study, we focus on the doublets with both
lines detected at $\ge 3\sigEWr$, or ${\rm flag} \ge 384$. 

\begin{deluxetable}{rp{6cm}}
  \tablewidth{0pt}
  \tabletypesize{\scriptsize}
  \tablecaption{\ion{C}{4} DIAGNOSTIC FLAGS\label{tab.flag} }
  \tablehead{ \colhead{Flag} & \colhead{Description} } \startdata
  256 & $\EWlin{1548} \ge 3\sigEWlin{1548}$ \\
  128 & $\EWlin{1550} \ge 3\sigEWlin{1548}$\\
  64 & $1-\sigma_{R_{W}} \le R_{W} \le 2+\sigma_{R_{W}}$ \\
  32 & $\vert \dvciv \vert \le 10\kms$ \\
  16 & Candidate \Lya\ with $\EWlin{1215} \ge 3\sigEWlin{1215}$ \\
  8 & C\,IV 1548 outside \Lya\ forest  \\
  4 & C\,IV 1548 outside \HH\ forest  \\
  2 & $\ge68.3\%$ AOD profile per element (pixel) in 1-$\sigma$ agreement \\
  1 & Other candidate absorption lines with $\EWr\ge3\sigEWr$ \\
  \enddata
  \tablecomments{Binary diagnostic flags for C\,IV doublets. (For more
    information, see \S\ \ref{subsec.flag}.) }
\end{deluxetable}


\begin{deluxetable}{rllllllllll}
\tablewidth{0pc}
\tablecaption{\ion{C}{4} SYSTEMS SUMMARY \label{tab.sys}}
\tabletypesize{\scriptsize}
\tablehead{ 
\colhead{(1)} & \colhead{(2)} & \colhead{(3)} & \colhead{(4)} & \colhead{(5)} & 
\colhead{(6)} & \colhead{(7)} & \colhead{(8)} & \colhead{(9)} & \colhead{(10)} & 
\colhead{(11)} \\
\colhead{$z_{1548}$} & \colhead{\dvabs} & 
\colhead{$\lambda_{r}$} & \colhead{\wvlo} & 
\colhead{\wvhi} & \colhead{\EWr} & \colhead{\sigEWr} & 
\colhead{\logN} & \colhead{\siglogN} & 
\colhead{G} & \colhead{Flag} \\ 
 & \colhead{(\!\kms)} & \colhead{(\AA)} & \colhead{(\AA)} & 
\colhead{(\AA)} & \colhead{(\!\mA)} & \colhead{(\!\mA)} & & & & 
}
\startdata
\hline \\[-1ex]
\multicolumn{11}{c}{PG0117+213 ($\zem=1.493$)} \\[1ex]
\hline
0.51964 & \nodata & 1548.19 & 2352.30 & 2353.07 &    94 &    24 & 13.47 &  0.11 &  2 & 288 \\
 &    -3.5 & 1550.77 & 2356.21 & 2356.98 & $<   45$ & \nodata & $< 13.39$ & \nodata &      &   \\
0.57632 & \nodata & 1548.19 & 2439.35 & 2441.47 &   728 &    28 & $> 14.56$ & \nodata &  1 & 450 \\
 &    12.6 & 1550.77 & 2443.41 & 2445.24 &   442 &    25 & 14.52 &  0.03 &      &   \\
\hline \\[-1ex]
\multicolumn{11}{c}{PKS0232--04 ($\zem=1.440$)} \\[1ex]
\hline
0.73910 & \nodata & 1548.19 & 2691.50 & 2693.40 &   372 &    34 & 14.16 &  0.08 &  1 & 450 \\
 &    18.7 & 1550.77 & 2695.98 & 2697.88 &   359 &    41 & $> 14.43$ & \nodata &      &   \\
0.86818 & \nodata & 1548.19 & 2891.94 & 2892.69 &    69 &    22 & $< 13.13$ & \nodata &  2 & 358 \\
 &    -5.6 & 1550.77 & 2896.75 & 2897.50 & $<   37$ & \nodata & $< 13.32$ & \nodata &      &   \\
\hline \\[-1ex]
\multicolumn{11}{c}{PKS0405--123 ($\zem=0.573$)} \\[1ex]
\hline
0.36071 & \nodata & 1548.19 & 2106.28 & 2107.05 &    86 &    17 & 13.37 &  0.09 &  1 & 383 \\
 &     6.4 & 1550.77 & 2109.79 & 2110.55 & $<   35$ & \nodata & $< 13.27$ & \nodata &      &   \\
 &    53.0 & 1215.67 & 1653.80 & 1655.51 &   769 &    35 & $> 14.47$ & \nodata &      &   \\
 &    15.7 & 1025.72 & 1395.47 & 1396.23 &   270 &    10 & $> 14.93$ & \nodata &      &   \\
 &     8.7 &  977.02 & 1329.24 & 1329.69 &   142 &     9 & $> 13.68$ & \nodata &      &   \\
 &    19.9 & 1206.50 & 1641.62 & 1641.94 &    73 &    12 & 12.70 &  0.09 &      &   \\
\hline \\[-1ex]
\multicolumn{11}{c}{PKS0454--22 ($\zem=0.534$)} \\[1ex]
\hline
0.20645 & \nodata & 1548.19 & 1867.44 & 1868.12 &   142 &    24 & 13.64 &  0.08 &  2 & 366 \\
 &     0.9 & 1550.77 & 1870.54 & 1871.23 &    81 &    26 & 13.70 &  0.14 &      &   \\
0.24010 & \nodata & 1548.19 & 1919.35 & 1920.52 &   644 &    57 & $> 14.20$\tablenotemark{a} & \nodata &  1 & 494 \\
 &     7.7 & 1550.77 & 1922.54 & 1923.71 &   495 &    54 & $> 14.39$\tablenotemark{a} & \nodata &      &   \\
0.27797 & \nodata & 1548.19 & 1978.18 & 1979.04 &   274 &    52 & $> 13.83$\tablenotemark{a} & \nodata &  1 & 366 \\
 &    -9.6 & 1550.77 & 1981.47 & 1982.33 & $<  104$ & \nodata & $< 14.20$ & \nodata &      &   \\
0.38152 & \nodata & 1548.19 & 2138.54 & 2139.20 &    96 &    32 & $< 13.55$ & \nodata &  1 & 494 \\
 &     1.2 & 1550.77 & 2142.10 & 2142.76 &   110 &    27 & $> 13.73$\tablenotemark{a} & \nodata &      &   \\
 &     3.8 & 1215.67 & 1679.22 & 1679.74 & $<  466$ & \nodata & $< 14.23$\tablenotemark{a} & \nodata &      &   \\
0.40227 & \nodata & 1548.19 & 2170.55 & 2171.37 &   177 &    29 & $< 13.72$ & \nodata &  1 & 494 \\
 &     1.7 & 1550.77 & 2174.16 & 2174.99 &   140 &    31 & $> 13.84$\tablenotemark{a} & \nodata &      &   \\
 &    -5.4 & 1215.67 & 1704.35 & 1704.99 & $<  458$ & \nodata & $< 14.23$\tablenotemark{a} & \nodata &      &   \\
0.42955 & \nodata & 1548.19 & 2212.93 & 2213.58 &    81 &    25 & $< 13.20$ & \nodata &  2 & 366 \\
 &     9.0 & 1550.77 & 2216.61 & 2217.26 & $<   50$ & \nodata & $< 13.44$ & \nodata &      &   \\
 &    -1.6 & 1215.67 & 1737.63 & 1738.14 & $<  269$ & \nodata & $< 13.97$ & \nodata &      &   \\
 &     2.0 &  977.02 & 1396.51 & 1396.92 & $<  259$ & \nodata & $< 13.90$\tablenotemark{a} & \nodata &      &   \\
0.47436 & \nodata & 1548.19 & 2281.82 & 2283.46 &   645 &    37 & $> 14.55$ & \nodata &  1 & 511 \\
 &    -6.2 & 1550.77 & 2285.62 & 2287.25 &   524 &    33 & $> 14.75$ & \nodata &      &   \\
 &   105.2 & 1215.67 & 1790.54 & 1795.92 &  2768 &   327 & $> 14.71$\tablenotemark{a} & \nodata &      &   \\
 &     7.3 & 1206.50 & 1778.44 & 1779.41 &   581 &   126 & $> 13.43$\tablenotemark{a} & \nodata &      &   \\
 &     8.1 & 1260.42 & 1857.67 & 1859.04 &   617 &    31 & $> 14.03$ & \nodata &      &   \\
 &    -0.0 & 1393.76 & 2054.25 & 2055.56 &   635 &    52 & $> 13.84$\tablenotemark{a} & \nodata &      &   \\
 &    -1.7 & 1402.77 & 2067.54 & 2068.85 &   475 &    56 & $> 14.02$\tablenotemark{a} & \nodata &      &   \\
0.48328 & \nodata & 1548.19 & 2295.80 & 2296.92 &   246 &    27 & 14.04 &  0.08 &  1 & 511 \\
 &    -3.4 & 1550.77 & 2299.61 & 2300.74 &   183 &    29 & 14.12 &  0.09 &      &   \\
 &    28.5 & 1215.67 & 1802.80 & 1803.92 &   599 &   105 & $> 14.04$\tablenotemark{a} & \nodata &      &   \\
 &    -0.2 & 1206.50 & 1789.01 & 1790.12 &   631 &   124 & $> 13.47$\tablenotemark{a} & \nodata &      &   \\
 &     0.4 & 1260.42 & 1869.08 & 1869.97 &   233 &    21 & $> 13.52$ & \nodata &      &   \\
\hline \\[-1ex]
\multicolumn{11}{c}{HE0515--4414 ($\zem=1.710$)} \\[1ex]
\hline
0.50601 & \nodata & 1548.19 & 2330.91 & 2332.37 &   448 &    28 & $> 14.36$ & \nodata &  2 & 482 \\
 &    -0.8 & 1550.77 & 2334.79 & 2335.94 &   333 &    21 & 14.39 &  0.04 &      &   \\
0.73082 & \nodata & 1548.19 & 2679.29 & 2679.99 &    27 &     9 & 12.86 &  0.14 &  2 & 354 \\
 &    -0.1 & 1550.77 & 2683.74 & 2684.44 & $<   17$ & \nodata & $< 12.96$ & \nodata &      &   \\
0.94042 & \nodata & 1548.19 & 3003.24 & 3005.31 &   365 &    20 & $> 14.27$ & \nodata &  1 & 499 \\
 &     2.1 & 1550.77 & 3008.24 & 3010.31 &   242 &    23 & 14.24 &  0.05 &      &   \\
 &    32.9 & 1215.67 & 2357.83 & 2360.83 &  1052 &    29 & $> 14.68$ & \nodata &      &   \\
 &    20.0 & 1206.50 & 2340.48 & 2342.09 &   190 &    18 & 13.08 &  0.05 &      &   \\
 &    10.9 & 1260.42 & 2445.09 & 2446.77 &   122 &    19 & 13.00 &  0.07 &      &   \\
 &     6.9 & 1393.76 & 2703.77 & 2705.23 &   233 &    13 & 13.58 &  0.03 &      &   \\
 &     7.3 & 1402.77 & 2721.26 & 2722.73 &    56 &    11 & 13.15 &  0.08 &      &   \\
\hline \\[-1ex]
\multicolumn{11}{c}{HS0624+6907 ($\zem=0.370$)} \\[1ex]
\hline
0.06351 & \nodata & 1548.19 & 1646.18 & 1646.78 &   106 &    10 & 13.54 &  0.04 &  1 & 503 \\
 &    -7.7 & 1550.77 & 1648.92 & 1649.52 &    61 &    10 & 13.55 &  0.07 &      &   \\
 &   -16.2 & 1215.67 & 1292.28 & 1293.28 &   594 &    11 & $> 14.43$ & \nodata &      &   \\
 &   -20.7 & 1025.72 & 1090.36 & 1091.20 &   357 &    23 & $> 15.09$ & \nodata &      &   \\
 &    -1.6 & 1206.50 & 1282.91 & 1283.30 &   150 &     9 & 13.05 &  0.04 &      &   \\
0.07574 & \nodata & 1548.19 & 1665.24 & 1665.72 &    53 &     6 & 13.19 &  0.05 &  1 & 503 \\
 &     5.4 & 1550.77 & 1668.01 & 1668.49 &    32 &     8 & 13.23 &  0.11 &      &   \\
 &     0.9 & 1215.67 & 1307.45 & 1308.15 &   300 &    10 & $> 14.09$ & \nodata &      &   \\
 &     1.3 & 1025.72 & 1103.16 & 1103.68 &   127 &    28 & 14.49 &  0.12 &      &   \\
\hline \\[-1ex]
\multicolumn{11}{c}{HS0747+4259 ($\zem=1.897$)} \\[1ex]
\hline
0.83662 & \nodata & 1548.19 & 2842.55 & 2844.22 &   303 &    25 & $> 14.11$ & \nodata &  2 & 481 \\
 &    -0.7 & 1550.77 & 2847.27 & 2848.95 &   232 &    30 & 14.20 &  0.06 &      &   \\
 &   -19.8 & 1215.67 & 2232.02 & 2233.33 & $<   86$ & \nodata & $< 13.39$ & \nodata &      &   \\
\hline \\[-1ex]
\multicolumn{11}{c}{HS0810+2554 ($\zem=1.510$)} \\[1ex]
\hline
0.83135 & \nodata & 1548.19 & 2834.27 & 2836.40 &   774 &    49 & $> 14.63$ & \nodata &  1 & 499 \\
 &     2.8 & 1550.77 & 2838.99 & 2841.12 &   724 &    53 & $> 14.89$ & \nodata &      &   \\
 &    62.4 & 1215.67 & 2225.35 & 2228.26 &  1242 &   106 & $> 14.61$ & \nodata &      &   \\
 &    36.6 & 1206.50 & 2209.27 & 2210.36 &   380 &    62 & $> 13.25$\tablenotemark{a} & \nodata &      &   \\
 &    13.1 & 1393.76 & 2551.82 & 2553.29 &   262 &    30 & 13.67 &  0.08 &      &   \\
 &    13.0 & 1402.77 & 2568.32 & 2569.80 &   271 &    31 & 13.91 &  0.06 &      &   \\
0.87687 & \nodata & 1548.19 & 2905.00 & 2906.32 &   219 &    55 & $> 13.73$\tablenotemark{a} & \nodata &  2 & 486 \\
 &    -7.8 & 1550.77 & 2909.83 & 2911.15 &   162 &    49 & $< 13.94$ & \nodata &      &   \\
 &   -24.5 & 1215.67 & 2281.20 & 2282.09 & $<   98$ & \nodata & $<  0.30$ & \nodata &      &   \\
\hline \\[-1ex]
\multicolumn{11}{c}{PG0953+415 ($\zem=0.234$)} \\[1ex]
\hline
0.06807 & \nodata & 1548.19 & 1653.40 & 1653.77 &   136 &    26 & $> 13.53$\tablenotemark{a} & \nodata &  1 & 383 \\
 &    -2.7 & 1550.77 & 1656.15 & 1656.52 & $<   50$ & \nodata & $< 13.63$ & \nodata &      &   \\
 &     3.7 & 1215.67 & 1298.15 & 1298.74 &   281 &     8 & $> 14.08$ & \nodata &      &   \\
 &   -12.3 & 1025.72 & 1095.32 & 1095.81 &   129 &     8 & 14.44 &  0.03 &      &   \\
 &    -0.2 &  977.02 & 1043.41 & 1043.65 &    74 &     7 & 13.21 &  0.05 &      &   \\
 &    -3.1 & 1031.93 & 1102.00 & 1102.31 &   108 &     7 & 14.13 &  0.03 &      &   \\
 &    -5.1 & 1037.62 & 1108.08 & 1108.39 &    74 &     7 & 14.27 &  0.04 &      &   \\
 &    -3.4 & 1238.82 & 1322.98 & 1323.29 &    48 &     7 & 13.49 &  0.06 &      &   \\
 &     3.0 & 1242.80 & 1327.24 & 1327.55 &    29 &     7 & 13.49 &  0.10 &      &   \\
\hline \\[-1ex]
\multicolumn{11}{c}{MARK132 ($\zem=1.757$)} \\[1ex]
\hline
0.70776 & \nodata & 1548.19 & 2643.55 & 2644.28 &    29 &     9 & 12.88 &  0.13 &  2 & 354 \\
 &     0.4 & 1550.77 & 2647.95 & 2648.68 & $<   17$ & \nodata & $< 12.93$ & \nodata &      &   \\
0.74843 & \nodata & 1548.19 & 2706.10 & 2707.69 &   318 &    13 & 14.13 &  0.03 &  1 & 482 \\
 &    -0.4 & 1550.77 & 2710.60 & 2712.20 &   165 &    12 & 14.08 &  0.03 &      &   \\
0.76352 & \nodata & 1548.19 & 2729.63 & 2730.92 &   110 &    11 & 13.53 &  0.05 &  1 & 482 \\
 &    -1.4 & 1550.77 & 2734.17 & 2735.46 &    75 &    16 & 13.64 &  0.09 &      &   \\
\hline \\[-1ex]
\multicolumn{11}{c}{3C249.1 ($\zem=0.312$)} \\[1ex]
\hline
0.02616 & \nodata & 1548.19 & 1588.55 & 1588.81 &    24 &     7 & 12.81 &  0.14 &  2 & 375 \\
 &    -7.6 & 1550.77 & 1591.19 & 1591.45 & $<   13$ & \nodata & $< 12.81$ & \nodata &      &   \\
 &   -33.4 & 1215.67 & 1246.76 & 1247.75 &   317 &    15 & 13.95 &  0.03 &      &   \\
 &   -31.2 &  977.02 & 1002.15 & 1002.74 &   524 &    78 & $> 13.91$\tablenotemark{a} & \nodata &      &   \\
\hline \\[-1ex]
\multicolumn{11}{c}{PG1206+459 ($\zem=1.155$)} \\[1ex]
\hline
0.60072 & \nodata & 1548.19 & 2477.60 & 2478.89 &   185 &    25 & 13.74 &  0.07 &  2 & 358 \\
 &    -5.4 & 1550.77 & 2481.72 & 2483.01 & $<   45$ & \nodata & $< 13.38$ & \nodata &      &   \\
0.73377 & \nodata & 1548.19 & 2683.80 & 2684.56 &   176 &     7 & 13.80 &  0.02 &  2 & 494 \\
 &    -2.1 & 1550.77 & 2688.26 & 2689.02 &    83 &     9 & 13.68 &  0.05 &      &   \\
0.92677 & \nodata & 1548.19 & 2979.54 & 2985.83 &  2363 &    59 & $> 15.15$ & \nodata &  1 & 477 \\
 &   -75.8 & 1550.77 & 2984.50 & 2990.79 &  2156 &    54 & $> 15.39$ & \nodata &      &   \\
 &   -15.0 & 1215.67 & 2339.49 & 2345.04 &  2384 &    30 & $> 15.07$ & \nodata &      &   \\
 &    66.8 & 1206.50 & 2321.60 & 2326.84 &   983 &    40 & $> 13.97$ & \nodata &      &   \\
 &   -94.8 & 1238.82 & 2384.15 & 2388.91 &   876 &    39 & $> 14.85$ & \nodata &      &   \\
 &   -98.9 & 1242.80 & 2391.82 & 2396.59 &   496 &    34 & 14.82 &  0.03 &      &   \\
 &   100.2 & 1260.42 & 2427.17 & 2430.55 &   431 &    25 & $> 13.73$ & \nodata &      &   \\
 &    21.7 & 1393.76 & 2682.52 & 2687.93 &   778 &    18 & 14.08 &  0.01 &      &   \\
 &    58.3 & 1402.77 & 2699.87 & 2705.31 &   402 &    28 & 14.06 &  0.03 &      &   \\
0.93425 & \nodata & 1548.19 & 2993.93 & 2995.25 &   259 &    22 & $> 14.18$ & \nodata &  1 & 511 \\
 &     0.5 & 1550.77 & 2998.91 & 3000.23 &   200 &    20 & $> 14.34$ & \nodata &      &   \\
 &   -14.4 & 1215.67 & 2350.48 & 2352.09 &   552 &    21 & $> 14.42$ & \nodata &      &   \\
 &     2.6 & 1206.50 & 2333.27 & 2334.11 &   133 &    15 & $> 13.13$ & \nodata &      &   \\
 &     5.9 & 1393.76 & 2695.43 & 2696.40 &   143 &    11 & $> 13.51$ & \nodata &      &   \\
 &     3.6 & 1402.77 & 2712.87 & 2713.84 &    98 &    11 & 13.60 &  0.06 &      &   \\
\hline \\[-1ex]
\multicolumn{11}{c}{PG1211+143 ($\zem=0.081$)} \\[1ex]
\hline
0.05114 & \nodata & 1548.19 & 1626.83 & 1627.67 &   264 &    10 & $> 14.07$ & \nodata &  1 & 509 \\
 &     3.8 & 1550.77 & 1629.54 & 1630.38 &   147 &    10 & 14.01 &  0.03 &      &   \\
 &    36.0 & 1215.67 & 1276.98 & 1279.19 &  1214 &     7 & $> 14.74$ & \nodata &      &   \\
 &    20.8 & 1025.72 & 1077.59 & 1078.89 &   823 &   103 & $> 15.33$ & \nodata &      &   \\
 &   -50.1 &  977.02 & 1026.58 & 1027.04 &   304 &    85 & $> 13.67$\tablenotemark{a} & \nodata &      &   \\
 &    13.5 & 1206.50 & 1268.07 & 1268.37 &   123 &     3 & 12.94 &  0.01 &      &   \\
 &    21.6 & 1260.42 & 1324.88 & 1325.05 &    24 &     2 & 12.29 &  0.04 &      &   \\
 &    12.0 & 1393.76 & 1464.88 & 1465.29 &    66 &     4 & 12.96 &  0.03 &      &   \\
 &    12.3 & 1402.77 & 1474.35 & 1474.76 &    44 &     5 & 13.04 &  0.05 &      &   \\
0.06439 & \nodata & 1548.19 & 1647.56 & 1648.21 &    71 &     9 & 13.29 &  0.05 &  1 & 511 \\
 &    -6.9 & 1550.77 & 1650.30 & 1650.95 &    31 &     9 & 13.24 &  0.12 &      &   \\
 &    36.4 & 1215.67 & 1293.48 & 1295.26 &   861 &     5 & $> 14.55$ & \nodata &      &   \\
 &    30.6 & 1025.72 & 1091.36 & 1092.50 &   448 &    12 & $> 15.14$ & \nodata &      &   \\
 &    -6.4 &  977.02 & 1039.69 & 1040.14 &   144 &     9 & 13.52 &  0.03 &      &   \\
 &    58.3 & 1031.93 & 1098.13 & 1099.07 &   183 &    11 & 14.26 &  0.03 &      &   \\
 &    60.6 & 1037.62 & 1104.19 & 1105.13 &    67 &    11 & 14.09 &  0.07 &      &   \\
 &    -4.4 & 1206.50 & 1283.96 & 1284.43 &    29 &     4 & 12.17 &  0.06 &      &   \\
 &    38.5 & 1260.42 & 1341.58 & 1341.96 &    10 &     3 & 11.85 &  0.12 &      &   \\
\hline \\[-1ex]
\multicolumn{11}{c}{MRK205 ($\zem=0.071$)} \\[1ex]
\hline
0.00427 & \nodata & 1548.19 & 1554.46 & 1555.20 &   292 &    23 & $> 14.14$ & \nodata &  1 & 511 \\
 &     8.6 & 1550.77 & 1557.05 & 1557.78 &   196 &    23 & $> 14.17$ & \nodata &      &   \\
 &    11.3 & 1215.67 & 1220.38 & 1221.47 &   807 &    32 & $> 14.53$ & \nodata &      &   \\
 &    13.3 & 1025.72 & 1029.68 & 1030.64 &   486 &    29 & $> 15.23$ & \nodata &      &   \\
 &    -6.7 &  977.02 &  980.81 &  981.56 &   382 &    46 & $> 14.13$ & \nodata &      &   \\
 &     4.6 &  989.80 &  993.68 &  994.38 &   301 &    41 & $> 14.51$\tablenotemark{a} & \nodata &      &   \\
 &     8.4 & 1206.50 & 1211.42 & 1211.97 &   263 &    32 & $> 13.09$\tablenotemark{a} & \nodata &      &   \\
 &     8.3 & 1260.42 & 1265.57 & 1266.25 &   240 &    11 & $> 13.54$ & \nodata &      &   \\
 &     3.7 & 1393.76 & 1399.43 & 1400.00 &   122 &    11 & 13.25 &  0.04 &      &   \\
 &     1.3 & 1402.77 & 1408.48 & 1409.05 &    75 &    13 & 13.33 &  0.07 &      &   \\
\hline \\[-1ex]
\multicolumn{11}{c}{QSO--123050+011522 ($\zem=0.117$)} \\[1ex]
\hline
0.00574 & \nodata & 1548.19 & 1556.92 & 1557.24 &    65 &    11 & 13.34 &  0.07 &  1 & 511 \\
 &    -0.2 & 1550.77 & 1559.51 & 1559.75 &    36 &     9 & 13.33 &  0.11 &      &   \\
 &    -9.0 & 1215.67 & 1222.14 & 1223.32 &   684 &    21 & $> 14.48$ & \nodata &      &   \\
 &    -9.9 & 1206.50 & 1213.19 & 1213.59 &   175 &    34 & $> 12.91$\tablenotemark{a} & \nodata &      &   \\
 &   -12.6 & 1260.42 & 1267.41 & 1267.77 &    74 &    11 & 12.83 &  0.07 &      &   \\
 &    -4.9 & 1393.76 & 1401.47 & 1401.92 &    66 &     8 & 12.96 &  0.05 &      &   \\
 &    -7.9 & 1402.77 & 1410.54 & 1410.99 &    63 &     9 & 13.21 &  0.06 &      &   \\
\hline \\[-1ex]
\multicolumn{11}{c}{PG1241+176 ($\zem=1.283$)} \\[1ex]
\hline
0.48472 & \nodata & 1548.19 & 2298.17 & 2299.23 &   305 &    46 & $> 13.88$\tablenotemark{a} & \nodata &  2 & 352 \\
 &    -8.9 & 1550.77 & 2302.00 & 2303.05 & $<   93$ & \nodata & $< 14.15$ & \nodata &      &   \\
0.55070 & \nodata & 1548.19 & 2399.44 & 2402.21 &   852 &    44 & $> 14.60$ & \nodata &  1 & 450 \\
 &    12.8 & 1550.77 & 2403.43 & 2406.21 &   580 &    50 & 14.60 &  0.05 &      &   \\
0.55842 & \nodata & 1548.19 & 2412.41 & 2413.21 &   230 &    33 & $> 13.75$\tablenotemark{a} & \nodata &  1 & 482 \\
 &    -4.9 & 1550.77 & 2416.42 & 2417.23 &   204 &    29 & 14.26 &  0.10 &      &   \\
0.75776 & \nodata & 1548.19 & 2721.08 & 2721.63 &   100 &    13 & 13.58 &  0.07 &  1 & 486 \\
 &    -0.1 & 1550.77 & 2725.61 & 2726.06 &    50 &    12 & 13.47 &  0.11 &      &   \\
0.78567 & \nodata & 1548.19 & 2763.64 & 2765.44 &   134 &    14 & 13.58 &  0.05 &  1 & 486 \\
 &     1.4 & 1550.77 & 2768.23 & 2770.04 &    71 &    13 & 13.59 &  0.08 &      &   \\
0.89546 & \nodata & 1548.19 & 2934.10 & 2934.86 &   106 &    25 & $> 13.42$\tablenotemark{a} & \nodata &  1 & 509 \\
 &    -1.5 & 1550.77 & 2938.98 & 2939.74 &   126 &    27 & $> 13.79$\tablenotemark{a} & \nodata &      &   \\
 &    -5.6 & 1215.67 & 2303.47 & 2304.94 &   510 &    42 & $> 13.97$\tablenotemark{a} & \nodata &      &   \\
 &    -3.4 & 1206.50 & 2286.53 & 2287.17 &    79 &    23 & $< 12.76$ & \nodata &      &   \\
\hline \\[-1ex]
\multicolumn{11}{c}{PG1248+401 ($\zem=1.032$)} \\[1ex]
\hline
0.55277 & \nodata & 1548.19 & 2403.65 & 2404.35 &    52 &    15 & 13.16 &  0.13 &  2 & 486 \\
 &     8.5 & 1550.77 & 2407.64 & 2408.35 &    64 &    17 & 13.56 &  0.12 &      &   \\
0.56484 & \nodata & 1548.19 & 2422.25 & 2423.16 &    97 &    14 & 13.51 &  0.06 &  1 & 484 \\
 &     4.9 & 1550.77 & 2426.27 & 2427.19 &    80 &    13 & 13.71 &  0.07 &      &   \\
0.70104 & \nodata & 1548.19 & 2633.11 & 2633.97 &   108 &    24 & 13.64 &  0.13 &  1 & 492 \\
 &    -0.3 & 1550.77 & 2637.49 & 2638.35 &    76 &    22 & 13.72 &  0.13 &      &   \\
0.77291 & \nodata & 1548.19 & 2743.91 & 2745.76 &   619 &    33 & $> 14.56$ & \nodata &  1 & 495 \\
 &     3.7 & 1550.77 & 2748.48 & 2750.33 &   564 &    29 & $> 14.77$ & \nodata &      &   \\
 &     6.2 & 1393.76 & 2470.19 & 2471.84 &   426 &    14 & $> 13.98$ & \nodata &      &   \\
 &     4.7 & 1402.77 & 2486.17 & 2487.83 &   315 &    12 & 14.02 &  0.02 &      &   \\
0.85508 & \nodata & 1548.19 & 2870.57 & 2873.99 &   826 &    39 & $> 14.59$ & \nodata &  1 & 463 \\
 &   -11.5 & 1550.77 & 2875.34 & 2878.77 &   567 &    41 & $> 14.65$ & \nodata &      &   \\
 &   -79.1 & 1260.42 & 2336.91 & 2338.58 &   196 &    29 & 13.34 &  0.11 &      &   \\
 &   -33.2 & 1393.76 & 2584.38 & 2585.96 &   226 &    22 & 13.52 &  0.05 &      &   \\
 &   -45.9 & 1402.77 & 2601.10 & 2602.69 &   137 &    32 & 13.59 &  0.10 &      &   \\
\hline \\[-1ex]
\multicolumn{11}{c}{PG1259+593 ($\zem=0.478$)} \\[1ex]
\hline
0.04615 & \nodata & 1548.19 & 1619.46 & 1619.85 &   103 &    21 & $> 13.41$\tablenotemark{a} & \nodata &  1 & 499 \\
 &     0.9 & 1550.77 & 1622.16 & 1622.54 &    74 &    19 & 13.71 &  0.11 &      &   \\
 &    34.8 & 1215.67 & 1271.17 & 1272.78 &  1043 &    18 & $> 14.65$ & \nodata &      &   \\
 &   -11.3 & 1025.72 & 1072.39 & 1073.64 &   748 &    35 & $> 15.40$ & \nodata &      &   \\
 &    -7.2 &  977.02 & 1021.61 & 1022.67 &   411 &    22 & $> 14.14$ & \nodata &      &   \\
 &    35.0 & 1206.50 & 1261.91 & 1262.83 &   220 &    13 & 13.14 &  0.03 &      &   \\
\hline \\[-1ex]
\multicolumn{11}{c}{PKS1302--102 ($\zem=0.278$)} \\[1ex]
\hline
0.00438 & \nodata & 1548.19 & 1554.81 & 1555.14 &    32 &     8 & 12.94 &  0.11 &  1 & 383 \\
 &    -6.5 & 1550.77 & 1557.40 & 1557.73 & $<   18$ & \nodata & $< 12.98$ & \nodata &      &   \\
 &    -1.3 & 1215.67 & 1220.75 & 1221.23 &   300 &    18 & $> 14.15$ & \nodata &      &   \\
 &    -0.1 & 1025.72 & 1029.83 & 1030.58 &   240 &    14 & $> 14.88$ & \nodata &      &   \\
 &   -17.3 &  977.02 &  980.89 &  981.60 &   234 &    34 & $> 13.91$ & \nodata &      &   \\
\hline \\[-1ex]
\multicolumn{11}{c}{CSO873 ($\zem=1.014$)} \\[1ex]
\hline
0.66089 & \nodata & 1548.19 & 2569.87 & 2572.23 &   424 &    37 & $> 14.24$ & \nodata &  1 & 494 \\
 &     3.7 & 1550.77 & 2574.14 & 2576.51 &   272 &    34 & $> 14.31$ & \nodata &      &   \\
0.73385 & \nodata & 1548.19 & 2683.86 & 2684.74 &    77 &    19 & 13.38 &  0.10 &  1 & 366 \\
 &    -2.2 & 1550.77 & 2688.33 & 2689.21 & $<   39$ & \nodata & $< 13.33$ & \nodata &      &   \\
\hline \\[-1ex]
\multicolumn{11}{c}{PG1630+377 ($\zem=1.476$)} \\[1ex]
\hline
0.75420 & \nodata & 1548.19 & 2715.42 & 2716.22 &    49 &    14 & 13.17 &  0.11 &  2 & 354 \\
 &    -1.2 & 1550.77 & 2719.94 & 2720.74 & $<   27$ & \nodata & $< 13.15$ & \nodata &      &   \\
0.91440 & \nodata & 1548.19 & 2963.34 & 2965.06 &   384 &    16 & $> 14.28$ & \nodata &  1 & 469 \\
 &    10.3 & 1550.77 & 2968.17 & 2969.99 &   317 &    18 & $> 14.43$ & \nodata &      &   \\
 &    -2.0 & 1215.67 & 2325.82 & 2328.60 &  1076 &    33 & $> 14.69$ & \nodata &      &   \\
 &     8.9 & 1206.50 & 2309.25 & 2310.59 &   142 &    19 & 12.99 &  0.07 &      &   \\
 &   -12.6 & 1393.76 & 2667.64 & 2668.60 &   123 &    15 & 13.26 &  0.06 &      &   \\
 &   -12.5 & 1402.77 & 2684.89 & 2685.86 &    98 &    12 & 13.47 &  0.06 &      &   \\
0.95269 & \nodata & 1548.19 & 3022.00 & 3023.84 &   560 &    21 & $> 14.43$ & \nodata &  1 & 477 \\
 &   -24.3 & 1550.77 & 3027.02 & 3028.87 &   276 &    22 & 14.27 &  0.04 &      &   \\
 &    70.6 & 1215.67 & 2372.75 & 2376.16 &  1434 &    25 & $> 14.83$ & \nodata &      &   \\
 &     8.9 & 1206.50 & 2355.41 & 2356.56 &   157 &    19 & 13.00 &  0.06 &      &   \\
 &     3.9 & 1260.42 & 2460.75 & 2461.78 &    43 &    13 & 12.54 &  0.13 &      &   \\
 &    14.7 & 1393.76 & 2720.92 & 2722.40 &   202 &    15 & 13.49 &  0.04 &      &   \\
 &    23.1 & 1402.77 & 2738.52 & 2740.01 &   136 &    21 & 13.59 &  0.07 &      &   \\
\hline \\[-1ex]
\multicolumn{11}{c}{PG1634+706 ($\zem=1.334$)} \\[1ex]
\hline
0.41935 & \nodata & 1548.19 & 2197.10 & 2197.71 &    44 &    11 & 13.07 &  0.11 &  2 & 354 \\
 &     0.4 & 1550.77 & 2200.76 & 2201.37 & $<   22$ & \nodata & $< 13.06$ & \nodata &      &   \\
0.65126 & \nodata & 1548.19 & 2556.26 & 2556.77 &    67 &     6 & 13.33 &  0.04 &  1 & 496 \\
 &    -3.0 & 1550.77 & 2560.51 & 2560.94 &    67 &     6 & 13.61 &  0.04 &      &   \\
 &     0.3 & 1215.67 & 2007.05 & 2007.67 &    78 &    21 & 13.25 &  0.12 &      &   \\
0.65355 & \nodata & 1548.19 & 2559.31 & 2560.62 &   393 &     9 & $> 14.29$ & \nodata &  1 & 499 \\
 &     2.0 & 1550.77 & 2563.57 & 2564.88 &   284 &    10 & 14.34 &  0.02 &      &   \\
 &     0.9 & 1215.67 & 2009.41 & 2010.92 &   630 &    30 & $> 14.45$ & \nodata &      &   \\
 &   -63.3 & 1206.50 & 1993.91 & 1995.48 &   502 &    28 & $> 13.71$ & \nodata &      &   \\
 &    -7.2 & 1393.76 & 2303.95 & 2305.30 &   145 &    11 & 13.30 &  0.04 &      &   \\
 &     0.1 & 1402.77 & 2318.85 & 2320.21 &    73 &    10 & 13.26 &  0.06 &      &   \\
0.81814 & \nodata & 1548.19 & 2814.55 & 2815.30 &    90 &     4 & 13.51 &  0.02 &  1 & 501 \\
 &    -0.1 & 1550.77 & 2819.23 & 2819.98 &    67 &     4 & 13.61 &  0.03 &      &   \\
 &    -0.2 & 1215.67 & 2209.74 & 2210.75 &   201 &    10 & $> 13.92$ & \nodata &      &   \\
 &   -23.7 & 1206.50 & 2193.22 & 2193.83 &   115 &     7 & 12.88 &  0.03 &      &   \\
 &    -2.4 & 1260.42 & 2291.25 & 2291.88 &    21 &     6 & 12.21 &  0.13 &      &   \\
 &    -1.0 & 1393.76 & 2533.54 & 2534.40 & $<   15$ & \nodata & $< 12.25$ & \nodata &      &   \\
 &    -3.4 & 1402.77 & 2549.93 & 2550.79 &    21 &     7 & 12.71 &  0.13 &      &   \\
0.90560 & \nodata & 1548.19 & 2949.76 & 2950.74 &   198 &     7 & 14.04 &  0.04 &  1 & 509 \\
 &    -0.5 & 1550.77 & 2954.66 & 2955.65 &   165 &     7 & 14.20 &  0.03 &      &   \\
 &    -1.6 & 1215.67 & 2316.12 & 2317.05 &   315 &     6 & $> 14.16$ & \nodata &      &   \\
 &    -5.3 & 1025.72 & 1954.23 & 1955.01 &   180 &    20 & $> 14.39$\tablenotemark{a} & \nodata &      &   \\
 &    -4.3 & 1031.93 & 1966.01 & 1966.81 &   104 &    22 & 14.08 &  0.10 &      &   \\
 &    -7.1 & 1037.62 & 1976.85 & 1977.66 &    67 &    18 & 14.11 &  0.12 &      &   \\
 &    -2.7 & 1206.50 & 2298.85 & 2299.38 &    61 &     6 & 12.58 &  0.05 &      &   \\
 &     1.3 & 1238.82 & 2360.43 & 2360.98 &    67 &     6 & 13.57 &  0.04 &      &   \\
 &    -8.8 & 1242.80 & 2368.02 & 2368.57 &    52 &     5 & 13.74 &  0.05 &      &   \\
 &   -21.3 & 1260.42 & 2401.59 & 2402.15 &   107 &     4 & 13.09 &  0.03 &      &   \\
 &     7.0 & 1393.76 & 2655.64 & 2656.26 &   259 &     5 & 13.82 &  0.03 &      &   \\
 &    -3.1 & 1402.77 & 2672.82 & 2673.44 &    40 &     6 & 13.01 &  0.06 &      &   \\
0.91144 & \nodata & 1548.19 & 2958.77 & 2959.74 &    34 &     7 & 12.96 &  0.09 &  2 & 383 \\
 &    -2.8 & 1550.77 & 2963.69 & 2964.66 & $<   17$ & \nodata & $< 12.92$ & \nodata &      &   \\
 &    18.3 & 1215.67 & 2322.48 & 2324.84 &   660 &    11 & $> 14.48$ & \nodata &      &   \\
 &     1.6 & 1025.72 & 1959.76 & 1961.33 &   311 &    25 & $> 14.91$ & \nodata &      &   \\
\hline \\[-1ex]
\multicolumn{11}{c}{HS1700+6416 ($\zem=2.736$)} \\[1ex]
\hline
0.08077 & \nodata & 1548.19 & 1673.11 & 1673.40 &   289 &    90 & $> 13.80$ & \nodata &  2 & 371 \\
 &    -1.5 & 1550.77 & 1675.90 & 1676.18 & $<  154$ & \nodata & $< 14.40$ & \nodata &      &   \\
 &     4.7 & 1215.67 & 1313.28 & 1314.36 &   661 &    28 & $> 14.43$ & \nodata &      &   \\
\hline \\[-1ex]
\multicolumn{11}{c}{PG1718+481 ($\zem=1.084$)} \\[1ex]
\hline
0.45953 & \nodata & 1548.19 & 2259.21 & 2259.97 &   147 &    13 & 13.68 &  0.04 &  1 & 480 \\
 &    -8.6 & 1550.77 & 2262.97 & 2263.72 &   120 &    13 & 13.83 &  0.05 &      &   \\
\hline \\[-1ex]
\multicolumn{11}{c}{H1821+643 ($\zem=0.297$)} \\[1ex]
\hline
0.22503 & \nodata & 1548.19 & 1895.74 & 1897.42 &   137 &    23 & 13.61 &  0.07 &  1 & 381 \\
 &     3.0 & 1550.77 & 1898.90 & 1900.57 & $<   47$ & \nodata & $< 13.38$ & \nodata &      &   \\
 &   -51.7 & 1215.67 & 1488.23 & 1489.74 &   834 &    13 & $> 14.60$ & \nodata &      &   \\
 &   -33.7 & 1025.72 & 1255.90 & 1256.80 &   496 &     9 & $> 15.24$ & \nodata &      &   \\
 &   -17.9 &  977.02 & 1196.50 & 1197.22 &   306 &    13 & $> 13.99$ & \nodata &      &   \\
 &   -11.5 & 1031.93 & 1263.72 & 1264.45 &   172 &     9 & 14.26 &  0.02 &      &   \\
 &   -12.0 & 1037.62 & 1270.68 & 1271.42 &   106 &     9 & 14.30 &  0.04 &      &   \\
 &    -4.0 & 1206.50 & 1477.75 & 1478.19 &    88 &     7 & 12.79 &  0.04 &      &   \\
0.24531 & \nodata & 1548.19 & 1927.27 & 1928.76 &    75 &    22 & 13.32 &  0.12 &  2 & 383 \\
 &    -1.2 & 1550.77 & 1930.47 & 1931.97 & $<   44$ & \nodata & $< 13.35$ & \nodata &      &   \\
 &     3.2 & 1215.67 & 1513.68 & 1514.08 &    47 &     8 & 13.01 &  0.07 &      &   \\
 &   -25.3 & 1025.72 & 1277.01 & 1277.51 &    55 &     8 & 13.95 &  0.06 &      &   \\
 &     1.1 & 1031.93 & 1284.86 & 1285.28 &    50 &     7 & 13.67 &  0.06 &      &   \\
 &     3.8 & 1037.62 & 1291.94 & 1292.37 &    29 &     5 & 13.70 &  0.07 &      &   \\
\hline \\[-1ex]
\multicolumn{11}{c}{PHL1811 ($\zem=0.190$)} \\[1ex]
\hline
0.08091 & \nodata & 1548.19 & 1673.01 & 1673.74 &   189 &    25 & $> 13.93$ & \nodata &  1 & 511 \\
 &    -0.6 & 1550.77 & 1675.79 & 1676.53 &   129 &    22 & 14.03 &  0.10 &      &   \\
 &   -27.8 & 1215.67 & 1313.15 & 1314.77 &   904 &     9 & $> 14.64$ & \nodata &      &   \\
 &   -36.0 & 1025.72 & 1108.19 & 1109.04 &   622 &     8 & $> 15.36$ & \nodata &      &   \\
 &   -19.7 &  977.02 & 1055.61 & 1056.31 &   338 &     9 & $> 14.07$ & \nodata &      &   \\
 &    11.7 &  989.80 & 1069.72 & 1070.02 &   137 &     7 & $> 14.47$ & \nodata &      &   \\
 &    10.9 & 1206.50 & 1303.98 & 1304.27 &   179 &     4 & $> 13.31$ & \nodata &      &   \\
 &     2.2 & 1260.42 & 1362.31 & 1362.60 &   172 &     4 & $> 13.44$ & \nodata &      &   \\
\enddata
\tablenotetext{a}{COG}
\tablecomments{C\,IV systems by target and redshift of C\,IV 1548.
Upper limits are $2\sigma$ limits for both \EWr\ and \logN.
The column densities were measured by the AODM, unless ``COG'' indicated, in which case, the limit is from assuming \EWr\ results from the linear portion of the COG.
The definite C\,IV doublets are labeled group G = 1, while the likely doublets are G = 2.
The binary flag is described in Table \ref{tab.flag} and \S\ \ref{subsec.flag}.
}
\end{deluxetable}

\clearpage 

\subsection{Final \ion{C}{4} Sample}\label{subsec.smpl}

\begin{figure}[!hbt]
  \begin{center}$
    \begin{array}{c}
      \includegraphics[height=0.47\textwidth,angle=90]{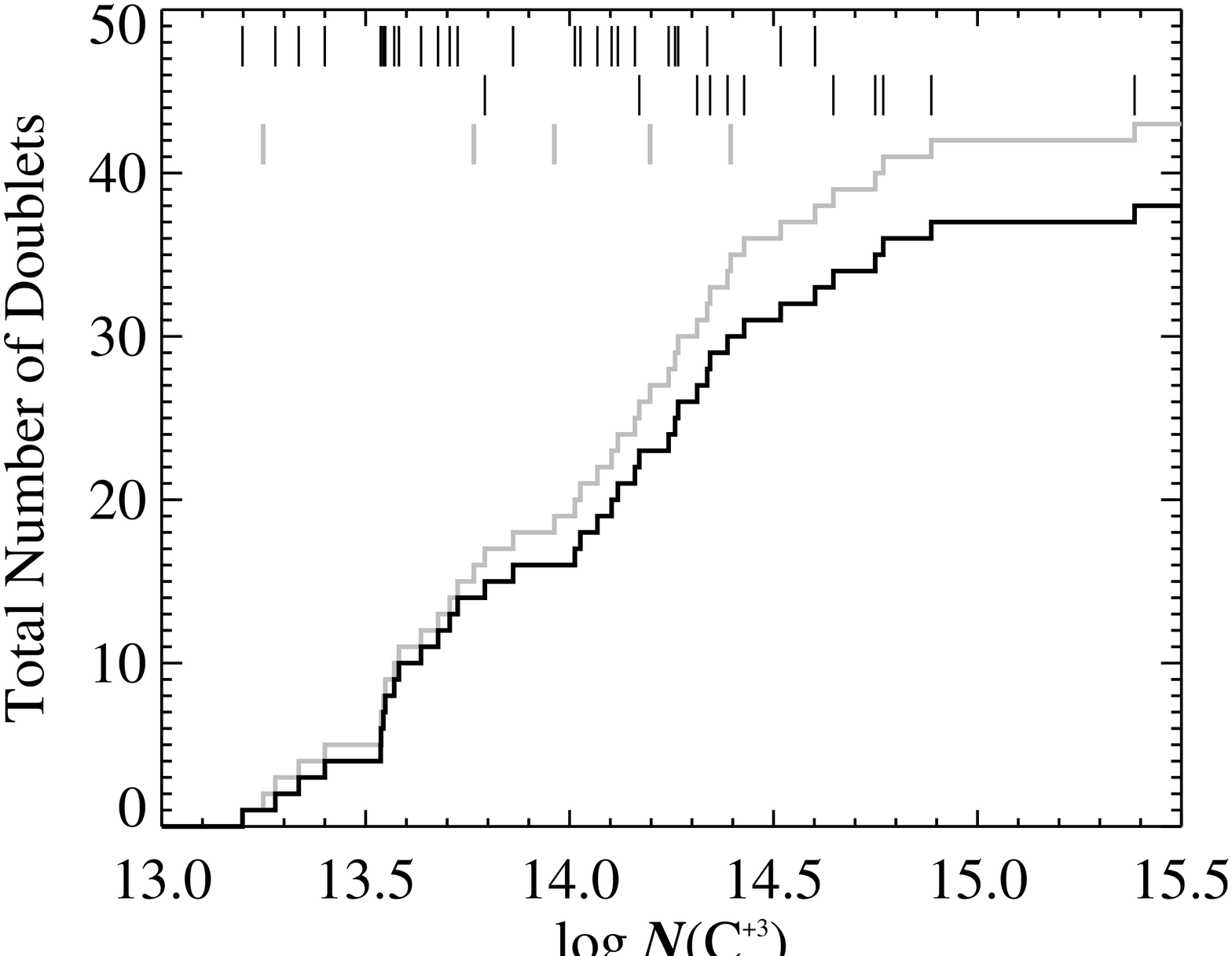} \\
      \includegraphics[height=0.47\textwidth,angle=90]{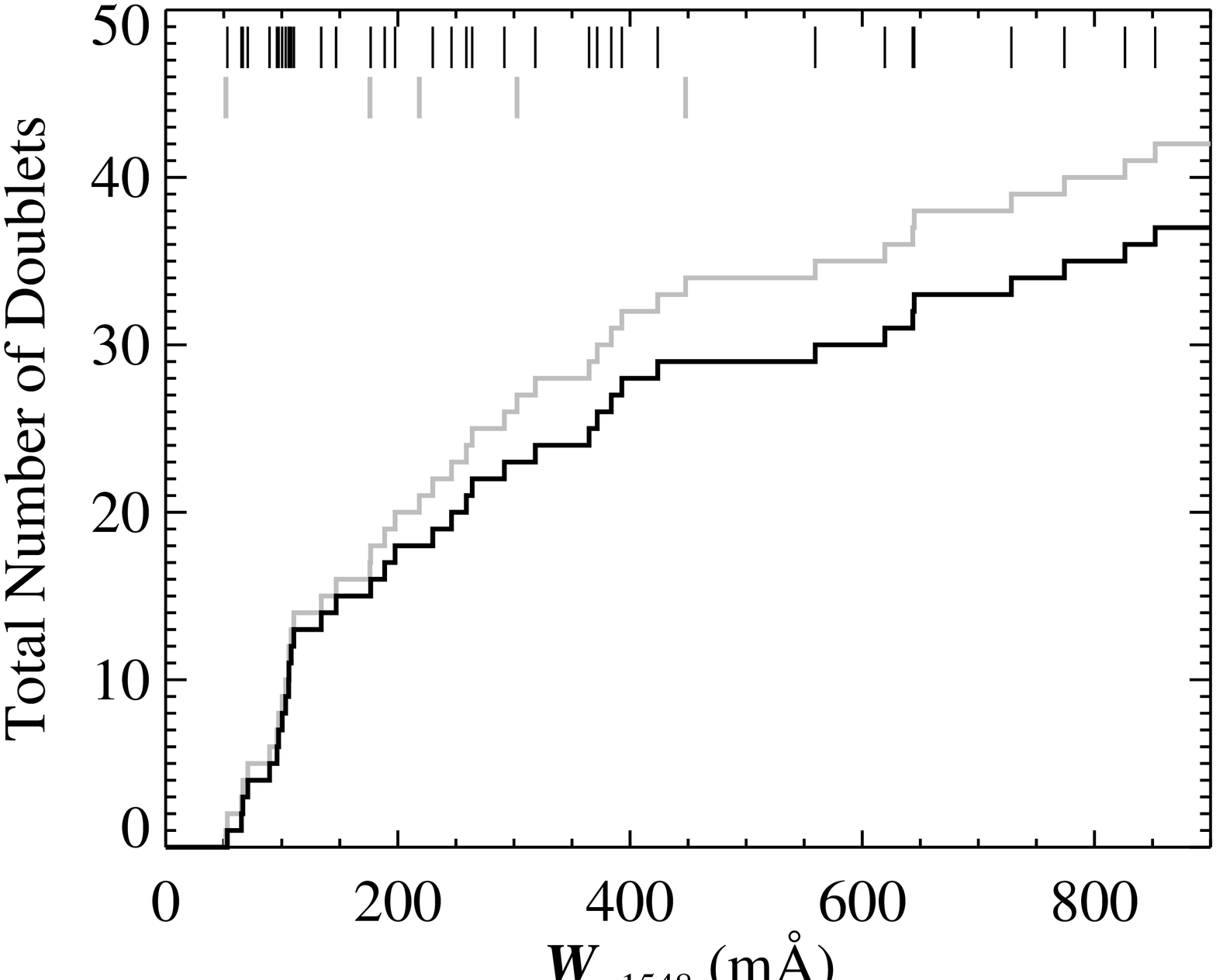}
    \end{array}$
  \end{center} 
  \caption[Cumulative column density and equivalent width distributions for the
    $3\sigEWr$ G = 1 and G = 2 samples.]
    {Cumulative column density and equivalent width distributions for
      the $\EWr\ge3\sigEWr$ G = 1 and G = 2 samples. The black and
      gray curves are the G = 1 and G = 1+2 groups, respectively. For
      each absorber, \logCIV\ and \EWlin{1548}\ are indicated with the
      hashes across the top, with the color indicating the sample.  As
      in Figure \ref{fig.gz}, the saturated doublets are shown in the
      middle row of hashes in the top panel.  The $\zciv=0.92677$
      \ion{C}{4} doublet associated with the DLA in the PG1206+459
      sightline, with $\EWlin{1548}=2364\mA$, is not shown in the
      \EWlin{1548} plot (bottom panel).
    \label{fig.hist}
  }
\end{figure}

The 319 candidate \ion{C}{4} systems with $\EWlin{1548} \ge
3\sigEWlin{1548}$, $R_{W}$ within the expected range, and $\vert
\dvciv \vert \le 10\kms$ were visually evaluated by multiple authors.
The remaining systems were reviewed by at least one author. We
assessed the likelihood of each candidate, utilizing the diagnostics
described in the previous section. Then we agreed upon two groups of
intervening \ion{C}{4} systems that had \zciv\ more than $1000\kms$
redward of the Milky Way Galaxy ($z=0$) and more than $3000\kms$
blueward of the background source ($z=\zem$): 44 definite systems (G =
1) and 19 likely systems (G = 2). The detailed information about all G
= 1+2 systems is listed in Table \ref{tab.sys} and the summary of the
doublets is listed in Table \ref{tab.civ}. An example velocity plot of
a \ion{C}{4} doublet is given in Figure \ref{fig.g1}, and the full
complement of velocity plots are in Appendix \ref{appdx.plots}. The
cumulative distribution of column densities and equivalent widths are
given in Figure \ref{fig.hist}.

The final, visual evaluation of the candidates was subjective. We
required consensus among the authors in defining our final
sample. Typically, the definite (G = 1) systems show other
transitions, usually \Lya\ and/or \ion{Si}{4}. When there was no
coverage of the other transitions, which had lower rest wavelengths,
the G = 1 \ion{C}{4} doublets were multi-component with matching AOD
profiles.

Typically, the G = 2 systems include the \ion{C}{4} doublets detected
in regions where the S/N was low, so the comparison of the profiles
was less conclusive, and there were not enough other positive
diagnostics (\eg detection of \Lya) to boost the confidence of the
identification.

For all \ion{C}{4} systems, we confirmed that the \ion{C}{4} doublet
was not a higher-redshift \ion{O}{1} 1302, \ion{Si}{2} 1304 pair,
which has rest wavelength separation similar to that of the doublet,
by checking whether \ion{Si}{2} 1260 existed and was much stronger
than the 1304 line. We also confirmed that the doublet and the other
transitions were not other, common transitions at different redshifts.

We adjusted the wavelength limits ($\wvlo,\wvhi$) as necessary to
correct for blending. For example, in Figure \ref{fig.civcand}, the
wavelength bounds for the \ion{C}{4} 1548 line was reduced.

For the remaining analyses, we restrict our intergalactic \ion{C}{4}
sample to the systems where both lines of the doublet were detected
with $\EWr \ge 3\,\sigEWr$.\footnote{We list all G = 1+2 \ion{C}{4}
  systems in Table \ref{tab.sys} for the community and any future
  analyses with different criteria.} This left 38 G = 1 systems and
five G = 2 systems.

\begin{deluxetable}{llrr@{}lr@{}lr@{}lr@{}lr@{}l}
\tablecolumns{13}
\tablewidth{0pc}
\tablecaption{\ion{C}{4} DOUBLET SUMMARY \label{tab.civ}}
\tabletypesize{\scriptsize}
\tablehead{ 
\colhead{(1)} & \colhead{(2)} & \colhead{(3)} & 
\multicolumn{2}{c}{(4)} & \multicolumn{2}{c}{(5)} & 
\multicolumn{2}{c}{(6)} & \multicolumn{2}{c}{(7)} & 
\multicolumn{2}{c}{(8)} \\
\colhead{Target} & \colhead{G} & \colhead{$z_{1548}$} & 
\multicolumn{2}{c}{$\EWlin{1548}$} & \multicolumn{2}{c}{$\EWlin{1550}$} & 
\multicolumn{2}{c}{$\logN{}_{1548}$} & \multicolumn{2}{c}{$\logN{}_{1550}$} & 
\multicolumn{2}{c}{$\logCIV$} \\ 
 & & & \multicolumn{2}{c}{(\mA)} & \multicolumn{2}{c}{(\mA)} & 
\multicolumn{2}{c}{} & \multicolumn{2}{c}{} & \multicolumn{2}{c}{} 
}
\startdata
                    PG0117+213 & 2 & 0.51964 & $   94$&$\,\pm\,   24$ & $<   45$ & & $13.47$&$\,\pm\, 0.11$  & $< 13.39$  & & $13.47$&$\,\pm\, 0.11$  \\
 & 1 & 0.57632 & $  728$&$\,\pm\,   28$ & $  442$&$\,\pm\,   25$ & $> 14.56$  & & $14.52$&$\,\pm\, 0.03$  & $14.52$&$\,\pm\, 0.03$  \\
                   PKS0232--04 & 1 & 0.73910 & $  372$&$\,\pm\,   34$ & $  359$&$\,\pm\,   41$ & $14.16$&$\,\pm\, 0.08$  & $> 14.43$  & & $14.16$&$\,\pm\, 0.08$  \\
 & 2 & 0.86818 & $   69$&$\,\pm\,   22$ & $<   37$ & & $< 13.13$  & & $< 13.32$  & & $< 13.13$ &  \\
                  PKS0405--123 & 1 & 0.36071 & $   86$&$\,\pm\,   17$ & $<   35$ & & $13.37$&$\,\pm\, 0.09$  & $< 13.27$  & & $13.37$&$\,\pm\, 0.09$  \\
                   PKS0454--22 & 2 & 0.20645 & $  142$&$\,\pm\,   24$ & $   81$&$\,\pm\,   26$ & $13.64$&$\,\pm\, 0.08$  & $13.70$&$\,\pm\, 0.14$  & $13.65$&$\,\pm\, 0.07$  \\
 & 1 & 0.24010 & $  644$&$\,\pm\,   57$ & $  495$&$\,\pm\,   54$ & $> 14.20$\tablenotemark{a} & & $> 14.39$\tablenotemark{a} & & $> 14.39$ &  \\
 & 1 & 0.27797 & $  274$&$\,\pm\,   52$ & $<  104$ & & $> 13.83$\tablenotemark{a} & & $< 14.20$  & & $[13.83$&$,14.14]$  \\
 & 1 & 0.38152 & $   96$&$\,\pm\,   32$ & $  110$&$\,\pm\,   27$ & $< 13.55$  & & $> 13.73$\tablenotemark{a} & & $[13.43$&$,13.77]$\tablenotemark{b} \\
 & 1 & 0.40227 & $  177$&$\,\pm\,   29$ & $  140$&$\,\pm\,   31$ & $< 13.72$  & & $> 13.84$\tablenotemark{a} & & $[13.84$&$,13.88]$  \\
 & 2 & 0.42955 & $   81$&$\,\pm\,   25$ & $<   50$ & & $< 13.20$  & & $< 13.44$  & & $< 13.44$ &  \\
 & 1 & 0.47436 & $  645$&$\,\pm\,   37$ & $  524$&$\,\pm\,   33$ & $> 14.55$  & & $> 14.75$  & & $> 14.75$ &  \\
 & 1 & 0.48328 & $  246$&$\,\pm\,   27$ & $  183$&$\,\pm\,   29$ & $14.04$&$\,\pm\, 0.08$  & $14.12$&$\,\pm\, 0.09$  & $14.07$&$\,\pm\, 0.06$  \\
                  HE0515--4414 & 2 & 0.50601 & $  448$&$\,\pm\,   28$ & $  333$&$\,\pm\,   21$ & $> 14.36$  & & $14.39$&$\,\pm\, 0.04$  & $14.39$&$\,\pm\, 0.04$  \\
 & 2 & 0.73082 & $   27$&$\,\pm\,    9$ & $<   17$ & & $12.86$&$\,\pm\, 0.14$  & $< 12.96$  & & $12.86$&$\,\pm\, 0.14$  \\
 & 1 & 0.94042 & $  365$&$\,\pm\,   20$ & $  242$&$\,\pm\,   23$ & $> 14.27$  & & $14.24$&$\,\pm\, 0.05$  & $14.24$&$\,\pm\, 0.05$  \\
                   HS0624+6907 & 1 & 0.06351 & $  106$&$\,\pm\,   10$ & $   61$&$\,\pm\,   10$ & $13.54$&$\,\pm\, 0.04$  & $13.55$&$\,\pm\, 0.07$  & $13.54$&$\,\pm\, 0.03$  \\
 & 1 & 0.07574 & $   53$&$\,\pm\,    6$ & $   32$&$\,\pm\,    8$ & $13.19$&$\,\pm\, 0.05$  & $13.23$&$\,\pm\, 0.11$  & $13.20$&$\,\pm\, 0.04$  \\
                   HS0747+4259 & 2 & 0.83662 & $  303$&$\,\pm\,   25$ & $  232$&$\,\pm\,   30$ & $> 14.11$  & & $14.20$&$\,\pm\, 0.06$  & $14.20$&$\,\pm\, 0.06$  \\
                   HS0810+2554 & 1 & 0.83135 & $  774$&$\,\pm\,   49$ & $  724$&$\,\pm\,   53$ & $> 14.63$  & & $> 14.89$  & & $> 14.89$ &  \\
 & 2 & 0.87687 & $  219$&$\,\pm\,   55$ & $  162$&$\,\pm\,   49$ & $> 13.73$\tablenotemark{a} & & $< 13.94$  & & $[13.73$&$,14.11]$  \\
                    PG0953+415 & 1 & 0.06807 & $  136$&$\,\pm\,   26$ & $<   50$ & & $> 13.53$\tablenotemark{a} & & $< 13.63$  & & $[13.53$&$,13.77]$  \\
                       MARK132 & 2 & 0.70776 & $   29$&$\,\pm\,    9$ & $<   17$ & & $12.88$&$\,\pm\, 0.13$  & $< 12.93$  & & $12.88$&$\,\pm\, 0.13$  \\
 & 1 & 0.74843 & $  318$&$\,\pm\,   13$ & $  165$&$\,\pm\,   12$ & $14.13$&$\,\pm\, 0.03$  & $14.08$&$\,\pm\, 0.03$  & $14.10$&$\,\pm\, 0.02$  \\
 & 1 & 0.76352 & $  110$&$\,\pm\,   11$ & $   75$&$\,\pm\,   16$ & $13.53$&$\,\pm\, 0.05$  & $13.64$&$\,\pm\, 0.09$  & $13.55$&$\,\pm\, 0.04$  \\
                       3C249.1 & 2 & 0.02616 & $   24$&$\,\pm\,    7$ & $<   13$ & & $12.81$&$\,\pm\, 0.14$  & $< 12.81$  & & $12.81$&$\,\pm\, 0.14$  \\
                    PG1206+459 & 2 & 0.60072 & $  185$&$\,\pm\,   25$ & $<   45$ & & $13.74$&$\,\pm\, 0.07$  & $< 13.38$  & & $13.74$&$\,\pm\, 0.07$  \\
 & 2 & 0.73377 & $  176$&$\,\pm\,    7$ & $   83$&$\,\pm\,    9$ & $13.80$&$\,\pm\, 0.02$  & $13.68$&$\,\pm\, 0.05$  & $13.77$&$\,\pm\, 0.02$  \\
 & 1 & 0.92677 & $ 2363$&$\,\pm\,   59$ & $ 2156$&$\,\pm\,   54$ & $> 15.15$  & & $> 15.39$  & & $> 15.39$ &  \\
 & 1 & 0.93425 & $  259$&$\,\pm\,   22$ & $  200$&$\,\pm\,   20$ & $> 14.18$  & & $> 14.34$  & & $> 14.34$ &  \\
                    PG1211+143 & 1 & 0.05114 & $  264$&$\,\pm\,   10$ & $  147$&$\,\pm\,   10$ & $> 14.07$  & & $14.01$&$\,\pm\, 0.03$  & $14.01$&$\,\pm\, 0.03$  \\
 & 1 & 0.06439 & $   71$&$\,\pm\,    9$ & $   31$&$\,\pm\,    9$ & $13.29$&$\,\pm\, 0.05$  & $13.24$&$\,\pm\, 0.12$  & $13.28$&$\,\pm\, 0.05$  \\
                        MRK205 & 1 & 0.00427 & $  292$&$\,\pm\,   23$ & $  196$&$\,\pm\,   23$ & $> 14.14$  & & $> 14.17$  & & $> 14.17$ &  \\
            QSO--123050+011522 & 1 & 0.00574 & $   65$&$\,\pm\,   11$ & $   36$&$\,\pm\,    9$ & $13.34$&$\,\pm\, 0.07$  & $13.33$&$\,\pm\, 0.11$  & $13.34$&$\,\pm\, 0.06$  \\
                    PG1241+176 & 2 & 0.48472 & $  305$&$\,\pm\,   46$ & $<   93$ & & $> 13.88$\tablenotemark{a} & & $< 14.15$  & & $[13.88$&$,14.01]$  \\
 & 1 & 0.55070 & $  852$&$\,\pm\,   44$ & $  580$&$\,\pm\,   50$ & $> 14.60$  & & $14.60$&$\,\pm\, 0.05$  & $14.60$&$\,\pm\, 0.05$  \\
 & 1 & 0.55842 & $  230$&$\,\pm\,   33$ & $  204$&$\,\pm\,   29$ & $> 13.75$\tablenotemark{a} & & $14.26$&$\,\pm\, 0.10$  & $14.26$&$\,\pm\, 0.10$  \\
 & 1 & 0.75776 & $  100$&$\,\pm\,   13$ & $   50$&$\,\pm\,   12$ & $13.58$&$\,\pm\, 0.07$  & $13.47$&$\,\pm\, 0.11$  & $13.54$&$\,\pm\, 0.06$  \\
 & 1 & 0.78567 & $  134$&$\,\pm\,   14$ & $   71$&$\,\pm\,   13$ & $13.58$&$\,\pm\, 0.05$  & $13.59$&$\,\pm\, 0.08$  & $13.58$&$\,\pm\, 0.04$  \\
 & 1 & 0.89546 & $  106$&$\,\pm\,   25$ & $  126$&$\,\pm\,   27$ & $> 13.42$\tablenotemark{a} & & $> 13.79$\tablenotemark{a} & & $> 13.79$ &  \\
                    PG1248+401 & 2 & 0.55277 & $   52$&$\,\pm\,   15$ & $   64$&$\,\pm\,   17$ & $13.16$&$\,\pm\, 0.13$  & $13.56$&$\,\pm\, 0.12$  & $13.25$&$\,\pm\, 0.09$  \\
 & 1 & 0.56484 & $   97$&$\,\pm\,   14$ & $   80$&$\,\pm\,   13$ & $13.51$&$\,\pm\, 0.06$  & $13.71$&$\,\pm\, 0.07$  & $13.57$&$\,\pm\, 0.05$  \\
 & 1 & 0.70104 & $  108$&$\,\pm\,   24$ & $   76$&$\,\pm\,   22$ & $13.64$&$\,\pm\, 0.13$  & $13.72$&$\,\pm\, 0.13$  & $13.68$&$\,\pm\, 0.09$  \\
 & 1 & 0.77291 & $  619$&$\,\pm\,   33$ & $  564$&$\,\pm\,   29$ & $> 14.56$  & & $> 14.77$  & & $> 14.77$ &  \\
 & 1 & 0.85508 & $  826$&$\,\pm\,   39$ & $  567$&$\,\pm\,   41$ & $> 14.59$  & & $> 14.65$  & & $> 14.65$ &  \\
                    PG1259+593 & 1 & 0.04615 & $  103$&$\,\pm\,   21$ & $   74$&$\,\pm\,   19$ & $> 13.41$\tablenotemark{a} & & $13.71$&$\,\pm\, 0.11$  & $13.71$&$\,\pm\, 0.11$  \\
                  PKS1302--102 & 1 & 0.00438 & $   32$&$\,\pm\,    8$ & $<   18$ & & $12.94$&$\,\pm\, 0.11$  & $< 12.98$  & & $12.94$&$\,\pm\, 0.11$  \\
                        CSO873 & 1 & 0.66089 & $  424$&$\,\pm\,   37$ & $  272$&$\,\pm\,   34$ & $> 14.24$  & & $> 14.31$  & & $> 14.31$ &  \\
 & 1 & 0.73385 & $   77$&$\,\pm\,   19$ & $<   39$ & & $13.38$&$\,\pm\, 0.10$  & $< 13.33$  & & $13.38$&$\,\pm\, 0.10$  \\
                    PG1630+377 & 2 & 0.75420 & $   49$&$\,\pm\,   14$ & $<   27$ & & $13.17$&$\,\pm\, 0.11$  & $< 13.15$  & & $13.17$&$\,\pm\, 0.11$  \\
 & 1 & 0.91440 & $  384$&$\,\pm\,   16$ & $  317$&$\,\pm\,   18$ & $> 14.28$  & & $> 14.43$  & & $> 14.43$ &  \\
 & 1 & 0.95269 & $  560$&$\,\pm\,   21$ & $  276$&$\,\pm\,   22$ & $> 14.43$  & & $14.27$&$\,\pm\, 0.04$  & $14.27$&$\,\pm\, 0.04$  \\
                    PG1634+706 & 2 & 0.41935 & $   44$&$\,\pm\,   11$ & $<   22$ & & $13.07$&$\,\pm\, 0.11$  & $< 13.06$  & & $13.07$&$\,\pm\, 0.11$  \\
 & 1 & 0.65126 & $   67$&$\,\pm\,    6$ & $   67$&$\,\pm\,    6$ & $13.33$&$\,\pm\, 0.04$  & $13.61$&$\,\pm\, 0.04$  & $13.40$&$\,\pm\, 0.03$  \\
 & 1 & 0.65355 & $  393$&$\,\pm\,    9$ & $  284$&$\,\pm\,   10$ & $> 14.29$  & & $14.34$&$\,\pm\, 0.02$  & $14.34$&$\,\pm\, 0.02$  \\
 & 1 & 0.81814 & $   90$&$\,\pm\,    4$ & $   67$&$\,\pm\,    4$ & $13.51$&$\,\pm\, 0.02$  & $13.61$&$\,\pm\, 0.03$  & $13.54$&$\,\pm\, 0.02$  \\
 & 1 & 0.90560 & $  198$&$\,\pm\,    7$ & $  165$&$\,\pm\,    7$ & $14.04$&$\,\pm\, 0.04$  & $14.20$&$\,\pm\, 0.03$  & $14.12$&$\,\pm\, 0.03$  \\
 & 2 & 0.91144 & $   34$&$\,\pm\,    7$ & $<   17$ & & $12.96$&$\,\pm\, 0.09$  & $< 12.92$  & & $12.96$&$\,\pm\, 0.09$  \\
                   HS1700+6416 & 2 & 0.08077 & $  289$&$\,\pm\,   90$ & $<  154$ & & $> 13.80$  & & $< 14.40$  & & $[13.80$&$,14.06]$  \\
                    PG1718+481 & 1 & 0.45953 & $  147$&$\,\pm\,   13$ & $  120$&$\,\pm\,   13$ & $13.68$&$\,\pm\, 0.04$  & $13.83$&$\,\pm\, 0.05$  & $13.73$&$\,\pm\, 0.03$  \\
                     H1821+643 & 1 & 0.22503 & $  137$&$\,\pm\,   23$ & $<   47$ & & $13.61$&$\,\pm\, 0.07$  & $< 13.38$  & & $13.61$&$\,\pm\, 0.07$  \\
 & 2 & 0.24531 & $   75$&$\,\pm\,   22$ & $<   44$ & & $13.32$&$\,\pm\, 0.12$  & $< 13.35$  & & $13.32$&$\,\pm\, 0.12$  \\
                       PHL1811 & 1 & 0.08091 & $  189$&$\,\pm\,   25$ & $  129$&$\,\pm\,   22$ & $> 13.93$  & & $14.03$&$\,\pm\, 0.10$  & $14.03$&$\,\pm\, 0.10$  \\
\enddata
\tablenotetext{a}{\logN\ measured by assuming \EWr\ results from the linear portion of the COG.}
\tablenotetext{b}{Limits set by doublet lines increased/decreased by $1\sigma$ to set \logCIV\ range.}
\tablecomments{
Summary of C\,IV doublets by target and redshift of C\,IV 1548.
The definite C\,IV doublets are labeled group G = 1, while the likely doublets are G = 2.
Upper limits are $2\sigma$ limits for both \EWr\ and \logN.
The adopted column density for the C\,IV doublets are listed in the last column (see \S\ \ref{subsec.meas}).
}
\end{deluxetable}


\section{Survey Sensitivity}\label{sec.cmplt}

\subsection{Monte-Carlo Completeness Tests}\label{subsec.mc}

We used Monte-Carlo tests to determine the column density and
equivalent width limit, as a function of the redshift, where we
recovered 95\% of the simulated \ion{C}{4} doublets for each spectrum.
From these sensitivity functions, we estimated the unblocked co-moving
absorption pathlengths $\DX{\NCIV}$ and $\DX{\EWlin{1548}}$ for our
survey, as defined below.

The simulated \ion{C}{4} doublets were single- or multi-component
Voigt profiles. Each Voigt profile was defined by a randomly selected
redshift, column density, and Doppler parameter. For the
multi-component simulated doublets, each component was assigned a
decreasing fraction of the total column density and a velocity offset
from the assigned redshift. For each column density bin, $\Delta
\logCIV = 0.1$, in the range $12.7 \le \logCIV \le 14.3$, one-hundred
randomly-generated Voigt profiles were added to the spectrum for each
redshift bin, $\delta z=0.005$, in the range covered by the
spectrum. The one-hundred simulated doublets were distributed in a
manner that avoided blending. The simulated doublets had Doppler
parameters $5\kms \le b \le 25\kms$ and from one to five components in
the range $-100\kms \le \dvabs \le 100\kms$. We visually inspected the
synthetic doublets to verify our model profiles resembled observations
of $z>2$ \ion{C}{4} doublets.

The profiles were added to the ``cleaned'' spectrum with the
appropriate noise.  The cleaned spectrum was the original spectrum
with all automatically-detected absorption features replaced with
continuum and noise. The noise was randomly sampled from the
neighboring pixels, not in the absorption features. We elected to
modify the original spectra (as opposed to generating completely
synthetic spectra) to keep the pixel-by-pixel signal-to-noise
properties of the spectra realistic.  The minimum flux level (\ie the
flux at line black) for the optically-thick Voigt profiles was
measured empirically from the troughs of the strongest absorption
lines, typically Galactic, in the original spectrum.

\begin{figure*}[!hbt]
  \includegraphics[width=0.75\textwidth,angle=90]{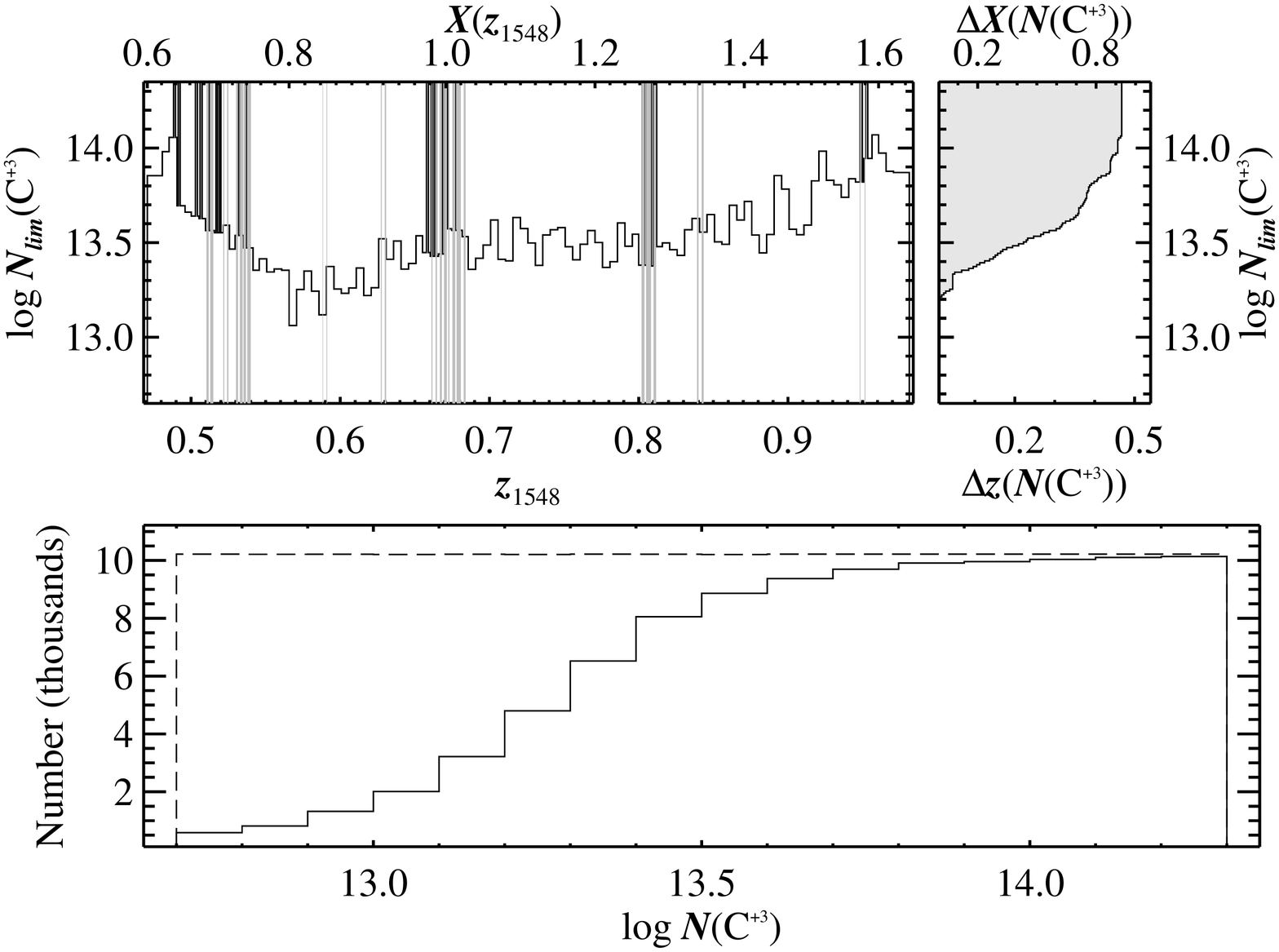}
  \caption[Example Monte-Carlo completeness test for \logCIVlim.]
{Example Monte-Carlo completeness test for \logCIVlim. The
    column density limit where our survey was 95\% complete is shown
    for the E230M spectrum of CSO873 (top, left panel). The bottom axis is
    the redshift of the 1548 line, and the top axis is the corresponding co-moving
    absorption pathlength from Equation \ref{eqn.x}. 
The general trend follows the S/N profile of the spectrum whose peak
sensitivity lies at $\lambda \approx 2600\Ang$, corresponding to 
$z_{1548} \approx 0.68$.
The redshift
    ranges excluded due to Galactic lines are shaded (gray); each
    Galactic line is counted twice, as a contaminant to both \ion{C}{4}
    lines. The abrupt, narrow spikes to infinity are the excluded
    saturated pixels in the \Lya\ forest. The total pathlength as a
    function of \NCIV\ to which the survey is 
    95\% complete, \DX{\NCIV}, is shown for this single sightline
    (top, right panel). For clarity, the gray shaded region lays {\it under} the
    \DX{\NCIV} curve (also see the total distribution shown in Figure
    \ref{fig.x}).  Also  
    shown  is the total  number of  simulated doublets tested (dashed
    line) compared to the number of recovered doublets (solid line) as a
    function of \logCIV\ (bottom panel).
    \label{fig.cmpltn}
  }
\end{figure*}

\begin{figure*}[!hbt]
  \includegraphics[width=0.75\textwidth,angle=90]{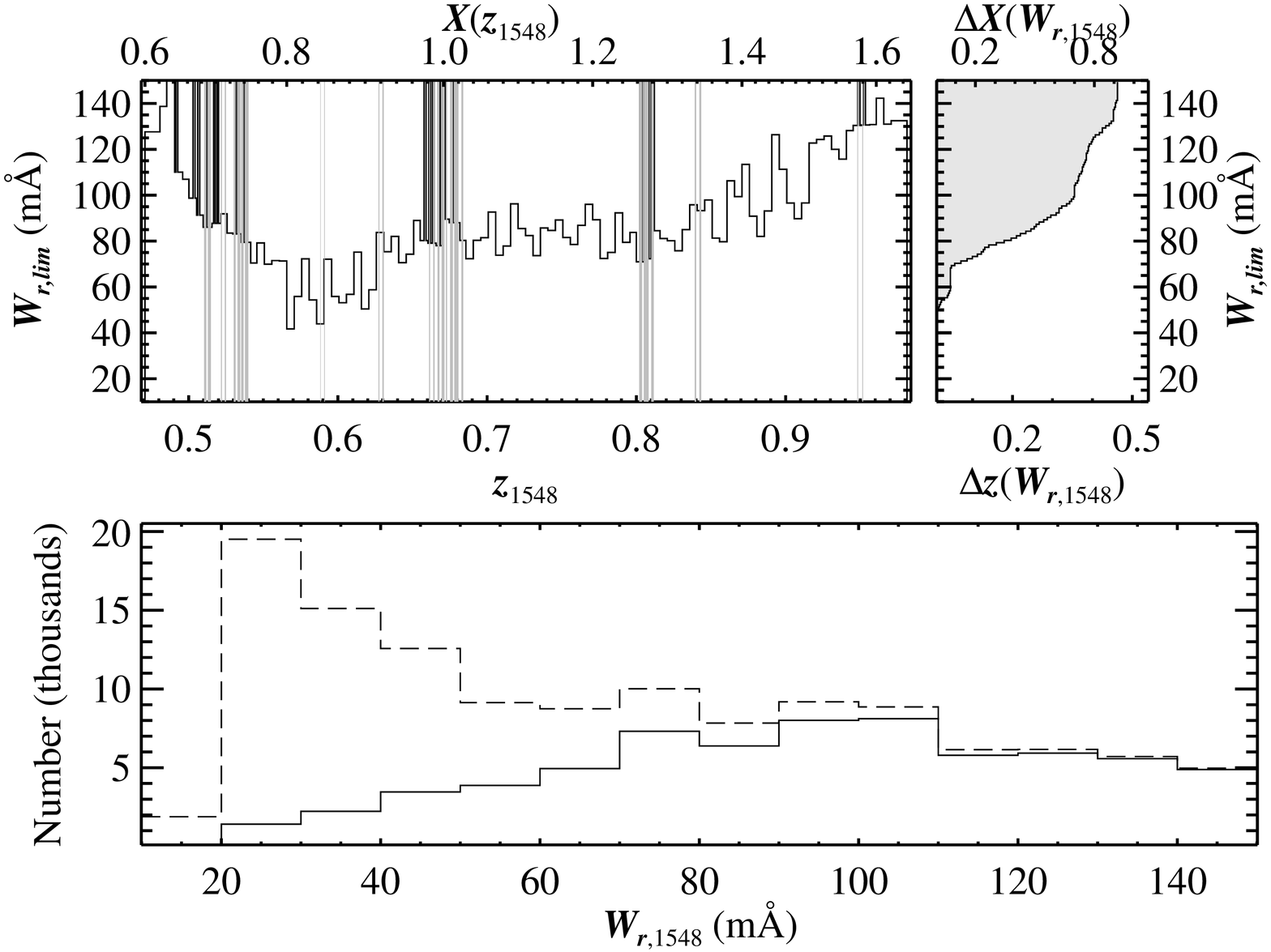}
  \caption[Example Monte-Carlo completeness test for
    \EWlin{lim}, similar to Figure \ref{fig.cmpltn}.]
{Example Monte-Carlo completeness test for
    \EWlin{lim}, similar to Figure \ref{fig.cmpltn}. The rest
    equivalent width limit for the \ion{C}{4} 
    1548 line where our survey was 95\% complete is shown for the E230M
    spectrum of CSO873 (top, left panel). The total number of simulated
    \ion{C}{4} doublets per \EWlin{1548}\ bin was not constant because the
    simulated doublets were assigned randomly-drawn \NCIV\ and
    $b$ (bottom panel). 
    \label{fig.cmpltew}
  }
\end{figure*}

The blind doublet search described in \S\ \ref{subsec.detec} was
conducted on the synthetic spectra. The recovered doublets were
matched to the input simulated doublets. The 95\% completeness limits
\logCIVlim\ and \EWlin{lim}\ (for the 1548 line) were measured for
each redshift bin. For each redshift bin, we fit a fourth-order
polynomial to the percentage of simulated \ion{C}{4} doublets
recovered and solved for \logCIVlim\ and \EWlin{lim}\ (see Figures
\ref{fig.cmpltn} and \ref{fig.cmpltew}, respectively).

\subsection{Search Pathlength}

\begin{figure}[!hbt]
  \begin{center}$
    \begin{array}{c}
      \includegraphics[height=0.47\textwidth,angle=90]{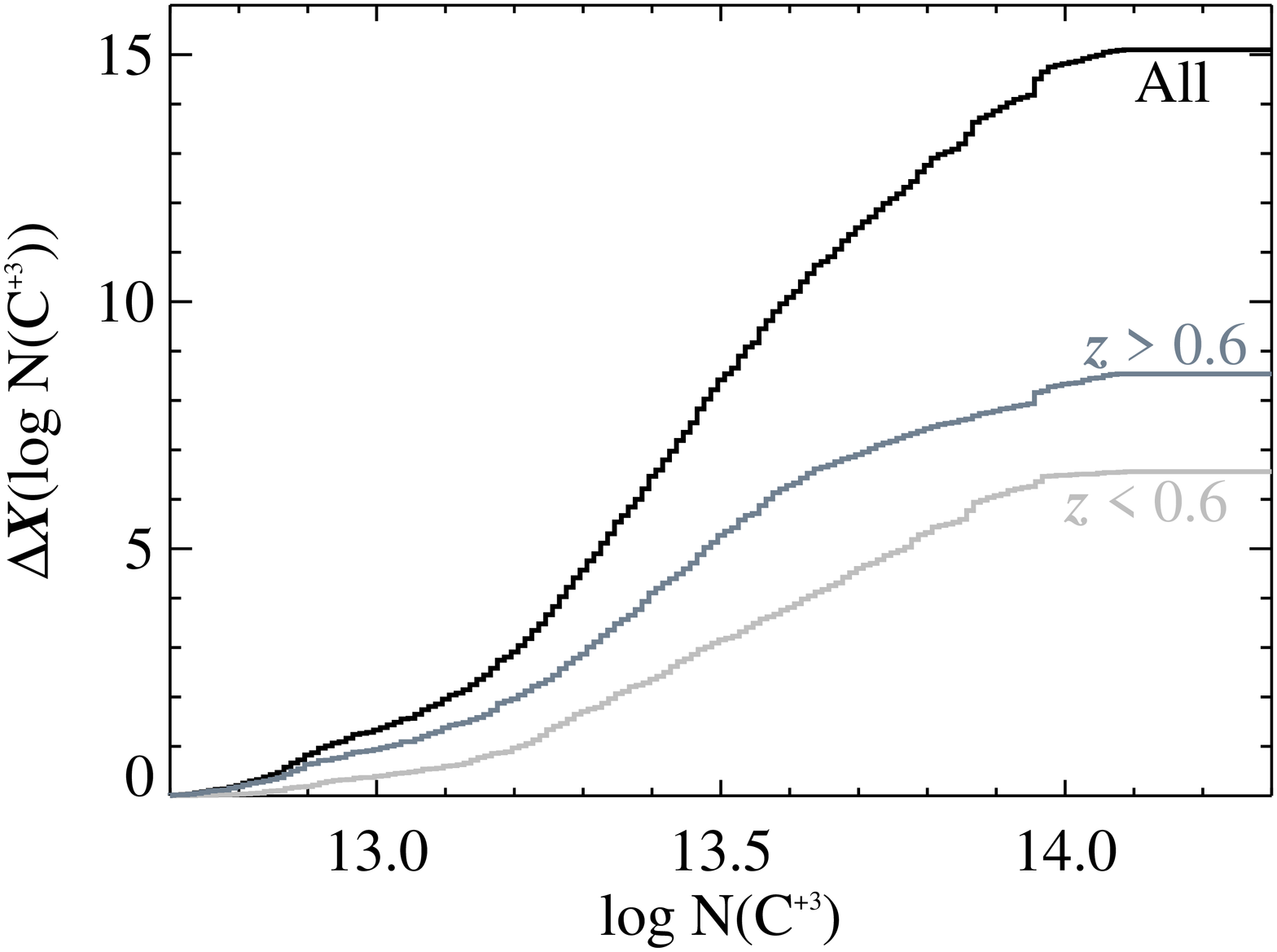} \\
      \includegraphics[height=0.47\textwidth,angle=90]{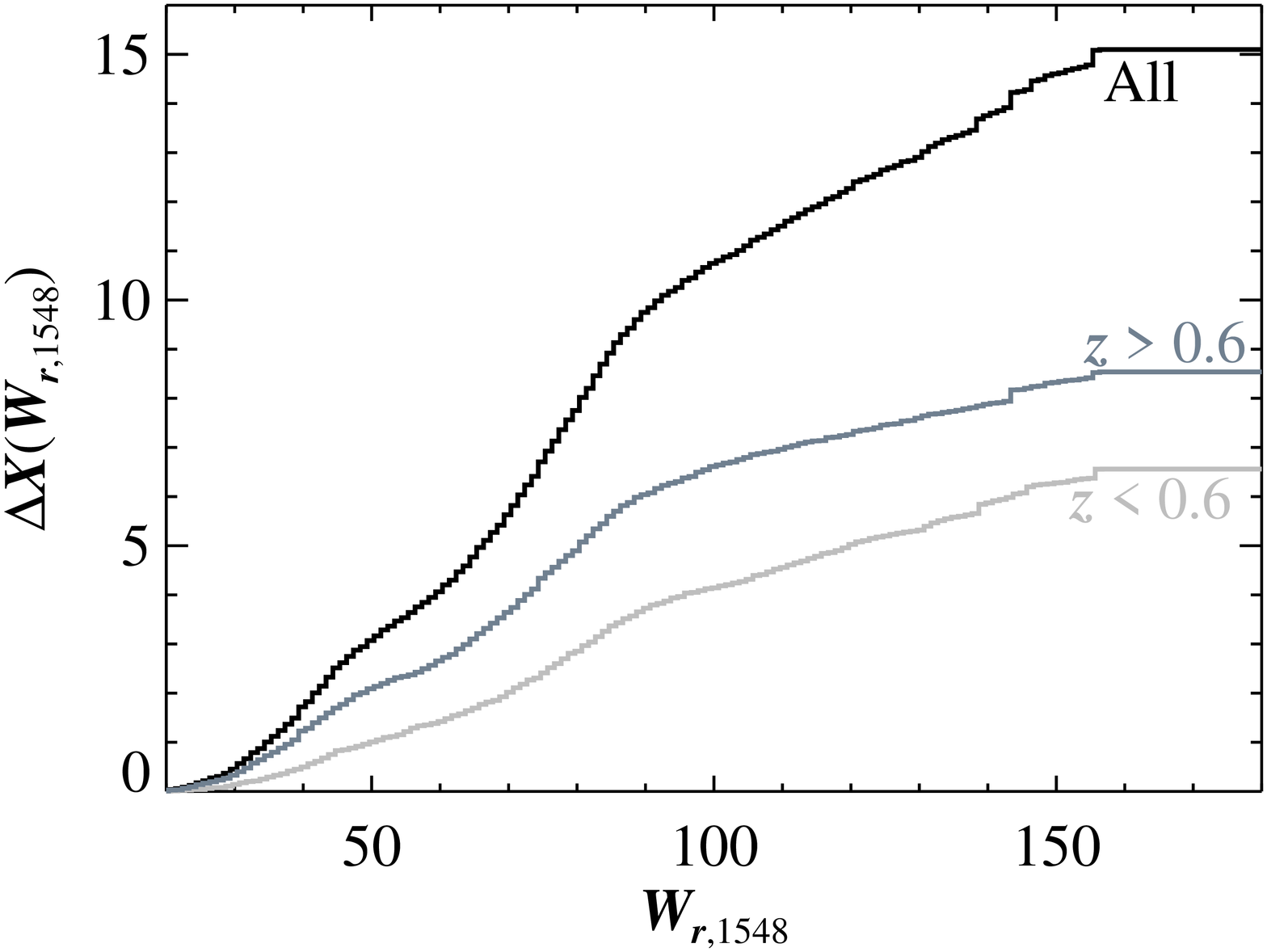}
    \end{array}$
  \end{center} 
  \caption[Redshift pathlength \DX{\logCIV} and \DX{\EWlin{1548}}.]
  {Redshift pathlength \DX{\logCIV} and \DX{\EWlin{1548}}\
    as a function of \ion{C}{4} 1548 column density (top) and
    equivalent width respectively (bottom). The black curve is for the
    full redshift ($z<1$ sample, the light gray for $z<0.6$, and the
    dark gray for $z\ge0.6$. These estimates are based on Monte-Carlo
    analysis and correspond to 95\% completeness limits.
    \label{fig.x}
  }
\end{figure}

The sensitivity functions were constructed for every sightline by the
previously described Monte-Carlo method. From these, we measured the
co-moving absorption pathlength \DX{\NCIV}\ and \DX{\EWlin{1548}}\
(see Figure \ref{fig.x}). We converted the sensitivity functions from
redshift space ($\delta z = 0.005$ from the Monte-Carlo tests) to
pathlength space as follows:
\begin{eqnarray}
  X(z) & = & \frac{2}{3\Omega_{\rm M}} \sqrt{\Omega_{\rm M}(1+z)^{3} +
 \Omega_{\Lambda}} \label{eqn.x} \\
\delta X(z) & = & X(z + 0.5\delta z) - X(z - 0.5\delta z) {\rm ,}\nonumber
\end{eqnarray}
where $\Omega_{\rm M}=0.3$ and $\Omega_{\Lambda}=0.7$ for our adopted
cosmology. The total pathlength as a function of column density
\DX{\NCIV}\ is the sum of the bins $\delta X$ where the 95\%
completeness limit is $\le \NCIV$:
\begin{eqnarray}
\DX{\NCIV} & = & \sum_{{\rm N}_{lim} \le {\rm N}_{i}} \delta X({\rm
  N}_{i}) \label{eqn.dx} \\
\sigDX^{2}  & = & \sigma_{\sigNCIV}^{2} + \sigma_{\delta {\rm N}}^{2}  {\rm .}\nonumber
\end{eqnarray}
The error $\sigma_{\Delta X}$ reflects the uncertainty induced by the
uncertainty in \NCIV\ and the interpolation of \DXp\ due to the column
density bins $\delta {\rm N}$. The aforementioned also applies to the
pathlength as a function of equivalent width \DX{\EWlin{1548}}. For
example, in Figure \ref{fig.cmpltn}, the bins where $\logCIVlim \le
13.5$ totaled $\DX{\NCIV} = 0.44$ for CSO873
($\zem=1.014$). Similarly, we found $\DX{\EWlin{1548}}=0.44$ for
$\EWlin{lim} \le 83\mA$. The total absorption pathlength,  \DX{\NCIV}\ (or
\DX{\EWlin{1548}}) is the sum (Equation \ref{eqn.dx}) over all sightlines.

We masked out common Galactic lines, which we determined from stacked
\stis\ E140M, E230M, and G230M spectra. We measured the wavelength
bounds ($\wvlo,\wvhi$) of the common Galactic lines in the stacked
spectra. We buffered the redshift range excluded due to Galactic lines
by $\pm20\%$ of the width of the lines (see Figures \ref{fig.cmpltn}
and \ref{fig.cmpltew}). Each Galactic line affected the search path
twice, as a contaminant to both \ion{C}{4} lines. We excluded the
regions $1000\kms$ redward of the Galaxy and $3000\kms$ blueward of
the the background source.

We also masked out the saturated regions in the \Lya\ forest for each
spectrum. We excluded pixels with $\fx{}_{i} < \sigfx{}_{i}$, and $\pm
3$ neighboring pixels, in all features detected by the automatic
feature-finding algorithm (see \S\ \ref{subsec.detec}). The excluded
pixels included ones from the strongest \ion{C}{4} doublets, but this
amounted to a small fraction ($<1\%$) of the total \DXp.

\subsection{\Lya\ Forest Contamination}\label{subsec.lya}

The \Lya\ forest was the largest contamination in the \ion{C}{4}
doublet survey. Approximately 60\% of the total pathlength resides in
the \Lya\ forest. For most of these systems, the detection of
associated transitions such as \Lya\ absorption or the \ion{Si}{4} doublet lent
credibility to the \ion{C}{4} identification. For several of our
systems, however, we did not have coverage of other transitions. An
example of this scenario is the G = 1 doublet at $\zciv=0.74843$ in
the MARK132 sightline (which also had a nondescript profile; see
Appendix \ref{appdx.plots}). 
For cases such as this, one must consider
the possibility that a pair of \Lya\ lines mimicked a \ion{C}{4}
doublet by chance. Instead of excluding systems where we did not have
the wavelength coverage or data quality to detect associated
transitions, we accounted for the contamination by considering the
rate at which \Lya\ transitions masquerade as \ion{C}{4}.

In principle, the contamination rate could be addressed in a variety
of ways. While we could have considered random lines of sight through
cosmological simulations, our purpose was first to identify if
contamination by \Lya\ was significant. We thus adopted the simpler
approach of generating synthetic spectra from the known properties of
\Lya\ statistics and searched this pure \Lya\ spectrum for putative
\ion{C}{4} doublets. This approach had the advantages that artificial
spectra were generated very quickly and we could assess our
contamination in a Monte-Carlo sense. Since the higher redshift E230M
data were far more susceptible to contamination (the \Lya\ forest lies
in the range $0.8 \lesssim z \lesssim 1.5$ in these data), we took the
E230M wavelength range as our template.

In order to generate the spectra, we took the statistics of the \Lya\
forest from \citet{janknechtetal06}. In the redshift range of
interest, there was only very weak evolution of the \Lya\ line
density, so for simplicity, we adopted a fixed line
density\footnote{Above $z \sim 1.5$, the \Lya\ line density evolves
  strongly ({\it c.f.} \citealt{kimetal01})} \dNLyadz\ = 150 for $13
\le \logN_{\rm H\,I} \le 16.5$. The column density and Doppler
parameter distributions were taken to be a power law and truncated
Gaussian, respectively, with the parameters from
\citet{janknechtetal06}. The placement of the \Lya\ lines was somewhat
less clear. At high redshift, strong \Lya\ absorbers, identified in
high-resolution data by the presence of \ion{C}{4} absorption, have
been shown to be highly clustered \citep{fernandezsotoetal96}. By
using heavier carbon atoms to trace the underlying gas component
structure, those authors showed \Lya\ absorption to be highly
clustered on small velocity scales. In the E230M data, however, we did
not have the resolution to resolve such small-scale structure, and
this was reflected in the lack of clustering detected in the UV data
of \citet{janknechtetal06}. In order to reproduce the observed trend
as closely as possible, we thus distributed our \Lya\ lines in a
uniform fashion. Further, since the observed statistics did not
account for the small-scale blending mentioned above, we imposed a
minimum separation for two lines $|\Dz| > 7\times10^{-5}$. Finally, we
added to our spectra the corresponding higher-order Lyman lines
stronger than a nominal detection level of $15\mA$, since they too
could contribute to the contamination rate.

The line lists generated in the above fashion, which could be
generated very quickly, constituted the full information available in
a ``spectrum.'' We thus adopted a Monte-Carlo approach, which was
appropriate for both low- and high-contamination rates. An automated
set of criteria was used to select pairs of lines that could be
misidentified as \ion{C}{4} doublets. We selected pairs of lines
that lay within $30\kms$ of the expected position, had $b$-values
within $10\kms$ of each other, and had an equivalent-width ratio
between $2:1$ and $1:1$. These fake doublets were saved for further
examination, and the simulation was iterated 1000 times.

\begin{figure}[!hbt]
  \begin{center}
  \includegraphics[height=0.47\textwidth,angle=90]{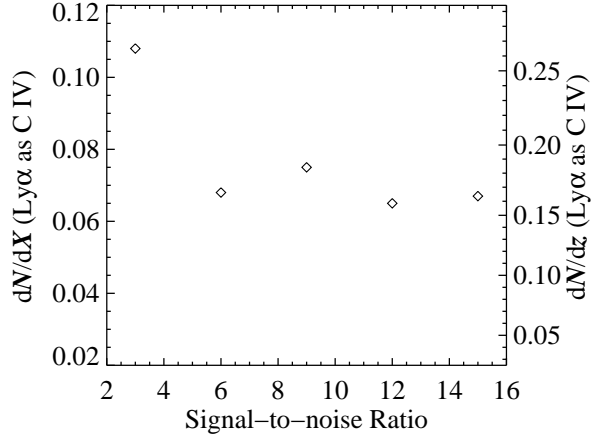}
  \end{center}
  \caption[Rate of coincident \Lya\ forest lines misidentified as
    \ion{C}{4} doublets.]
    {Rate of coincident \Lya\ forest lines misidentified as \ion{C}{4}
      doublets as function of S/N. The \Lya\ forest at $0.8\le z \le
      1.5$ was simulated in a Monte-Carlo fashion, and the incidence
      of \Lya\ forest lines that were ``observed'' to have similar
      line profiles and wavelength separation as a \ion{C}{4} doublet
      was estimated for a range of S/N.
    \label{fig.dnlyadx}
  }
\end{figure}

For each misidentified doublet, an actual spectrum was generated, and
noise added at the given S/N level. All the candidates were visually
inspected to exclude clearly mismatched line pairs. The final number
of accepted doublets was averaged over the 1000 simulation iterations,
yielding \dNCIVdX\ for misidentified doublets. Since lines could be
quickly generated, and noise easily added, we assessed the
contamination rate at several S/N levels that spanned the range of
typical values in our survey, with the results shown in
Figure~\ref{fig.dnlyadx}. In all cases, the contamination rate was not
a significant effect, but in \S\ \ref{subsec.wolya}, we discuss
several G = 2 \ion{C}{4} doublets that might be the result of \Lya\
forest contamination.

\section{Analysis}\label{sec.analysis}

\subsection{Frequency Distributions}\label{subsec.freqdistr}


\begin{deluxetable}{cccccccccccc}
\rotate
\tablewidth{0pc}
\tablecaption{\ff{\NCIV}, \dNCIVdX, AND \OmCIV\ SUMMARY \label{tab.fn}}
\tabletypesize{\scriptsize}
\tablehead{ 
\colhead{(1)} & \colhead{(2)} & \colhead{(3)} & \colhead{(4)} & \colhead{(5)} & 
\colhead{(6)} & \colhead{(7)} & \colhead{(8)} & \colhead{(9)} & \colhead{(10)} & 
\colhead{(11)} & \colhead{(12)} \\
\colhead{G} & \colhead{$\langle z \rangle$} & \colhead{$z_{l}$} & \colhead{$z_{h}$} & 
\colhead{\Num} & \colhead{\logCIV} & 
\colhead{\dNCIVdz} & \colhead{\dNCIVdX} & \colhead{\OmCIV} & 
\colhead{\kff{14}} & \colhead{\aff{N}} & \colhead{P$_{\rm KS}$} \\ 
 & & & & & & & & 
\colhead{($10^{-8}$)} & \colhead{(\!\cm{2})} & & 
}
\startdata
\hline \\[-1ex]
\multicolumn{12}{c}{Samples by Redshift} \\[1ex]
\hline
1 &   0.65355 &   0.00574 &   0.95269 & 36 & (13.15,\,16.00) & $ 8.0^{+ 2.5}_{- 2.0}$ & $ 4.2^{+ 1.3}_{- 1.0}$ & $ 6.20^{+ 1.82}_{- 1.52}$ & $ 0.67^{+ 0.18}_{- 0.16}$ & $-1.50^{+ 0.17}_{- 0.19}$ & 0.364 \\
 &   0.65355 &   0.00427 &   0.95269 & 27 & (13.20,\,15.39) & $ 6.0^{+ 1.2}_{- 1.0}$ & $ 3.5^{+ 0.7}_{- 0.6}$ & $> 3.43$ & \nodata &  \nodata &  \nodata \\
1+2 &   0.65355 &   0.00574 &   0.95269 & 41 & (13.15,\,16.00) & $ 9.3^{+ 2.7}_{- 2.2}$ & $ 4.9^{+ 1.4}_{- 1.1}$ & $ 6.99^{+ 1.91}_{- 1.61}$ & $ 0.76^{+ 0.20}_{- 0.17}$ & $-1.52^{+ 0.16}_{- 0.18}$ & 0.484 \\
 &   0.65355 &   0.00427 &   0.95269 & 32 & (13.20,\,15.39) & $ 7.0^{+ 1.3}_{- 1.1}$ & $ 4.1^{+ 0.7}_{- 0.6}$ & $> 4.13$ & \nodata &  \nodata &  \nodata \\
1 &   0.38152 &   0.00574 &   0.57632 & 17 & (13.15,\,16.00) & $ 9.3^{+ 5.3}_{- 3.4}$ & $ 5.9^{+ 3.4}_{- 2.2}$ & $ 6.24^{+ 2.88}_{- 2.14}$ & $ 0.79^{+ 0.32}_{- 0.25}$ & $-1.75^{+ 0.28}_{- 0.33}$ & 0.234 \\
 &   0.38152 &   0.00427 &   0.57632 & 15 & (13.20,\,14.75) & $ 6.8^{+ 2.1}_{- 1.6}$ & $ 5.0^{+ 1.5}_{- 1.2}$ & $> 4.81$ & \nodata &  \nodata &  \nodata \\
1+2 &   0.40227 &   0.00574 &   0.57632 & 19 & (13.15,\,16.00) & $10.5^{+ 5.6}_{- 3.7}$ & $ 6.6^{+ 3.5}_{- 2.3}$ & $ 7.01^{+ 3.05}_{- 2.30}$ & $ 0.88^{+ 0.33}_{- 0.27}$ & $-1.74^{+ 0.28}_{- 0.31}$ & 0.049 \\
 &   0.40227 &   0.00427 &   0.57632 & 17 & (13.20,\,14.75) & $ 8.0^{+ 2.3}_{- 1.9}$ & $ 6.0^{+ 1.7}_{- 1.4}$ & $> 5.64$ & \nodata &  \nodata &  \nodata \\
1 &   0.78567 &   0.65126 &   0.95269 & 19 & (13.37,\,16.00) & $ 8.1^{+ 3.7}_{- 2.5}$ & $ 3.9^{+ 1.8}_{- 1.2}$ & $ 6.35^{+ 2.52}_{- 1.99}$ & $ 0.62^{+ 0.27}_{- 0.21}$ & $-1.39^{+ 0.24}_{- 0.27}$ & 0.268 \\
 &   0.81814 &   0.65126 &   0.95269 & 12 & (13.40,\,15.39) & $ 5.6^{+ 1.6}_{- 1.2}$ & $ 2.7^{+ 0.8}_{- 0.6}$ & $> 2.48$ & \nodata &  \nodata &  \nodata \\
1+2 &   0.81814 &   0.65126 &   0.95269 & 22 & (13.37,\,16.00) & $ 9.8^{+ 4.4}_{- 3.0}$ & $ 4.7^{+ 2.1}_{- 1.4}$ & $ 7.23^{+ 2.65}_{- 2.14}$ & $ 0.75^{+ 0.28}_{- 0.23}$ & $-1.45^{+ 0.23}_{- 0.25}$ & 0.263 \\
 &   0.81814 &   0.65126 &   0.95269 & 15 & (13.40,\,15.39) & $ 6.4^{+ 1.6}_{- 1.3}$ & $ 3.1^{+ 0.8}_{- 0.6}$ & $> 3.10$ & \nodata &  \nodata &  \nodata \\
\hline \\[-1ex]
\multicolumn{12}{c}{Samples by Instrument} \\[1ex]
\hline
E140M &   0.06351 &   0.00574 &   0.08091 & 7 & (13.15,\,14.25) & $10.5^{+11.8}_{- 5.4}$ & $ 9.6^{+10.7}_{- 4.9}$ & $ 5.65^{+11.79}_{- 2.91}$ & $ 0.73^{+ 1.05}_{- 0.51}$ & $-2.19^{+ 0.88}_{- 0.95}$ & 0.281 \\
 &   0.06351 &   0.00427 &   0.08091 & 7 & (13.20,\,14.17) & $ 7.7^{+ 3.9}_{- 2.7}$ & $ 7.2^{+ 3.6}_{- 2.6}$ & $> 3.74$ & \nodata &  \nodata &  \nodata \\
E230M &   0.73910 &   0.24010 &   0.95269 & 34 & (13.14,\,16.00) & $ 9.2^{+ 2.7}_{- 2.2}$ & $ 4.6^{+ 1.4}_{- 1.1}$ & $ 7.51^{+ 2.18}_{- 1.83}$ & $ 0.73^{+ 0.22}_{- 0.18}$ & $-1.38^{+ 0.17}_{- 0.19}$ & 0.160 \\
 &   0.73910 &   0.24010 &   0.95269 & 25 & (13.25,\,15.39) & $ 6.9^{+ 1.4}_{- 1.2}$ & $ 3.6^{+ 0.7}_{- 0.6}$ & $> 4.32$ & \nodata &  \nodata &  \nodata \\
\enddata
\tablecomments{
Parameters from the maximum-likelihood analysis for $\ff{\NCIV} = \kff{14}\,(\NCIV/\mathrm{N}_{0})^{\aff{N}}$, where $\log \mathrm{N}_{0} = 14$.
The C\,IV doublets were divided into several sub-samples by the group G and the redshift range.
For each sub-sample, the first line refers to the maximum-likelihood analysis and the second line, to the observed quantities.
\dNCIVdX, listed in the first sub-sample row, is the integral of \ff{\NCIV}\ from $\logCIV=13$ to infinity with the best-fit \kff{14} and \aff{N} (see Equation \ref{eqn.dndx_int}).
Also in the first sub-sample row, the integrated $\dNCIVdz \equiv \dNCIVdX \cdot \ud X/\ud z$, where the latter term is the derivative of Equation \ref{eqn.x}, evaluated at $\langle z \rangle$.
The observed \dNCIVdz\ and \dNCIVdX\ are from the sum of the {\it total} number of doublets, weighted by the pathlength available to detect the doublet (based on its \NCIV; see Equation \ref{eqn.dndx_sum}).
\OmCIV, listed in the first sub-sample row, is the integral of $\ff{\NCIV}\cdot\NCIV$ from $13\le \logCIV \le15$ with the best-fit \kff{14} and \aff{N} (see Equation \ref{eqn.omciv_int}).
The observed \OmCIV\ were from the sum of the {\it unsaturated} doublets, as given by \Num\ (see Equation \ref{eqn.omciv_sum}).
P$_{KS}$ is the significance of the one-sided K-S statistic of the best-fit power law.
}
\end{deluxetable}


\begin{deluxetable}{ccccccccccc}
\tablewidth{0pc}
\tablecaption{\ff{\EWlin{1548}}\ AND \dNCIVdX\ SUMMARY \label{tab.few}}
\tabletypesize{\scriptsize}
\tablehead{ 
\colhead{(1)} & \colhead{(2)} & \colhead{(3)} & \colhead{(4)} & \colhead{(5)} & 
\colhead{(6)} & \colhead{(7)} & \colhead{(8)} & 
\colhead{(9)} & \colhead{(10)} & \colhead{(11)} \\ 
\colhead{G} & \colhead{$\langle z \rangle$} & 
\colhead{$z_{l}$} & \colhead{$z_{h}$} & 
\colhead{\Num} & \colhead{\EWlin{1548}} & 
\colhead{\dNCIVdz} & \colhead{\dNCIVdX} & 
\colhead{\kff{3}} & \colhead{\aff{W}} & \colhead{P$_{\rm KS}$} \\ 
 & & & & & \colhead{($\mA$)} & & & \colhead{($\!\mA^{-1}$)} & & 
}
\startdata
\hline \\[-1ex]
\multicolumn{11}{c}{Samples by Redshift} \\[1ex]
\hline
1 &   0.65355 &   0.00427 &   0.95269 & 38 & $(  47,\,2423)$ & $ 7.0^{+   1.9}_{-   1.6}$ & $ 3.7^{+   1.0}_{-   0.8}$ & $ 1.59^{+ 0.47}_{- 0.40}$ & $-1.60^{+ 0.25}_{- 0.26}$ & 0.714 \\
 &  & & &  & $(  53,\,2363)$ & $ 5.7^{+ 1.1}_{- 0.9}$ & $ 3.4^{+ 0.7}_{- 0.6}$ &  \nodata &  \nodata &  \nodata \\
1+2 &   0.65355 &   0.00427 &   0.95269 & 43 & $(  37,\,2423)$ & $ 7.8^{+   2.3}_{-   1.7}$ & $ 4.1^{+   1.2}_{-   0.9}$ & $ 1.78^{+ 0.53}_{- 0.42}$ & $-1.57^{+ 0.22}_{- 0.23}$ & 0.374 \\
 &  & & &  & $(  52,\,2363)$ & $ 6.7^{+ 1.2}_{- 1.1}$ & $ 3.9^{+ 0.7}_{- 0.6}$ &  \nodata &  \nodata &  \nodata \\
1 &   0.38152 &   0.00427 &   0.57632 & 18 & $(  47,\, 896)$ & $ 7.5^{+  11.5}_{-   2.4}$ & $ 4.8^{+   7.3}_{-   1.6}$ & $ 2.07^{+ 1.21}_{- 0.88}$ & $-1.60^{+ 0.48}_{- 0.49}$ & 0.200 \\
 &  & & &  & $(  53,\, 852)$ & $ 6.2^{+ 1.9}_{- 1.5}$ & $ 4.6^{+ 1.4}_{- 1.1}$ &  \nodata &  \nodata &  \nodata \\
1+2 &   0.40227 &   0.00427 &   0.57632 & 20 & $(  37,\, 896)$ & $ 8.4^{+  11.0}_{-   2.6}$ & $ 5.3^{+   6.9}_{-   1.7}$ & $ 2.29^{+ 1.26}_{- 0.91}$ & $-1.58^{+ 0.44}_{- 0.45}$ & 0.038 \\
 &  & & &  & $(  52,\, 852)$ & $ 7.7^{+ 2.3}_{- 1.9}$ & $ 5.7^{+ 1.7}_{- 1.4}$ &  \nodata &  \nodata &  \nodata \\
1 &   0.81814 &   0.65126 &   0.95269 & 20 & $(  61,\,2423)$ & $ 7.6^{+   9.0}_{-   2.4}$ & $ 3.6^{+   4.3}_{-   1.1}$ & $ 1.61^{+ 0.65}_{- 0.52}$ & $-1.46^{+ 0.34}_{- 0.36}$ & 0.501 \\
 &  & & &  & $(  67,\,2363)$ & $ 5.5^{+ 1.5}_{- 1.2}$ & $ 2.7^{+ 0.8}_{- 0.6}$ &  \nodata &  \nodata &  \nodata \\
1+2 &   0.81814 &   0.65126 &   0.95269 & 23 & $(  61,\,2423)$ & $ 8.7^{+   6.2}_{-   2.5}$ & $ 4.1^{+   2.9}_{-   1.2}$ & $ 1.83^{+ 0.69}_{- 0.55}$ & $-1.49^{+ 0.32}_{- 0.34}$ & 0.624 \\
 &  & & &  & $(  67,\,2363)$ & $ 6.3^{+ 1.6}_{- 1.3}$ & $ 3.1^{+ 0.8}_{- 0.6}$ &  \nodata &  \nodata &  \nodata \\
\hline \\[-1ex]
\multicolumn{11}{c}{Samples by Instrument} \\[1ex]
\hline
E140M &   0.06351 &   0.00427 &   0.08091 & 8 & $(  47,\, 315)$ & $\approx 7.4$ & $\approx 6.7$ & $ 2.36^{+ 6.29}_{- 1.87}$ & $-1.89^{+ 1.09}_{- 1.11}$ & 0.248 \\
 &  & & &  & $(  53,\, 264)$ & $ 6.6^{+ 3.6}_{- 2.5}$ & $ 6.1^{+ 3.4}_{- 2.3}$ &  \nodata &  \nodata &  \nodata \\
E230M &   0.73910 &   0.24010 &   0.95269 & 35 & $(  37,\,2423)$ & $ 8.7^{+   5.1}_{-   2.2}$ & $ 4.3^{+   2.5}_{-   1.1}$ & $ 1.90^{+ 0.56}_{- 0.48}$ & $-1.42^{+ 0.24}_{- 0.25}$ & 0.355 \\
 &  & & &  & $(  52,\, 852)$ & $ 5.2^{+ 1.3}_{- 1.1}$ & $ 2.7^{+ 0.7}_{- 0.6}$ &  \nodata &  \nodata &  \nodata \\
\enddata
\tablecomments{
Parameters from the maximum-likelihood analysis for $\ff{\EWr} = \kff{3}\,(\EWr/\EWlin{0})^{\aff{W}}$, where $\EWlin{0} = 400\mA$.
The tabulated information is similar to that presented in Table \ref{tab.fn}, except that the integrated line density limit is $\EWr = 50\mA$.
}
\end{deluxetable}


\begin{deluxetable}{ccccccccc}
\tablecolumns{10}
\tablewidth{0pc}
\tablecaption{\ff{\NCIV}\ AND \ff{\EWlin{1548}}\ CORRELATION MATRICES BY ABSORBER \label{tab.corrm}}
\tabletypesize{\scriptsize}
\tablehead{ 
 & & \multicolumn{3}{c}{\ff{\NCIV}} & & \multicolumn{3}{c}{\ff{\EWlin{1548}}} \\ 
 \cline{3-5} \cline{7-9} \\ 
\colhead{G} & \colhead{$z$} & 
\colhead{$\langle k_{14} \rangle$} & \colhead{$\langle \alpha_{N} \rangle$} & 
\colhead{$r_{k_{i},\alpha_{i}}$} & & 
\colhead{$\langle k_{3} \rangle$} & \colhead{$\langle \alpha_{W} \rangle$} & 
\colhead{$r_{k_{i},\alpha_{i}}$} \\ 
 & & 
\colhead{$(\!\cm{2})$} & & & &
\colhead{$(\!\mA^{-1})$} & &
}
\startdata
1 & $< 1$ & $ 0.65\pm 0.04$ & $-1.50\pm 0.12$ & -0.63 &  & $ 1.57\pm 0.62$ & $-1.59\pm 0.33$ &  0.97 \\ 
1+2 & $< 1$ & $ 0.75\pm 0.04$ & $-1.52\pm 0.12$ & -0.69 &  & $ 1.76\pm 0.72$ & $-1.56\pm 0.33$ &  0.97 \\ 
1 & $< 0.6$ & $ 0.75\pm 0.05$ & $-1.76\pm 0.21$ &  0.43 &  & $ 1.96\pm 0.49$ & $-1.60\pm 0.34$ &  0.99 \\ 
1+2 & $< 0.6$ & $ 0.84\pm 0.05$ & $-1.74\pm 0.22$ &  0.33 &  & $ 2.18\pm 0.51$ & $-1.58\pm 0.31$ &  0.99 \\ 
1 & $> 0.6$ & $ 0.60\pm 0.11$ & $-1.40\pm 0.17$ & -0.78 &  & $ 1.57\pm 0.66$ & $-1.44\pm 0.38$ &  0.87 \\ 
1+2 & $> 0.6$ & $ 0.72\pm 0.13$ & $-1.46\pm 0.16$ & -0.76 &  & $ 1.79\pm 0.81$ & $-1.47\pm 0.41$ &  0.88 \\ 
\enddata
\tablecomments{
Mean values and correlation coefficient from jack-knife analysis of power-law fit to \ff{\NCIV} and \ff{\EWr}.
Each absorber was iteratively excluded from the maximum likelihood analysis of the frequency distributions.
}
\end{deluxetable}


\begin{deluxetable}{ccccccccc}
\tablecolumns{10}
\tablewidth{0pc}
\tablecaption{\ff{\NCIV}\ AND \ff{\EWlin{1548}}\ CORRELATION MATRICES BY SIGHTLINE \label{tab.loscorrm}}
\tabletypesize{\scriptsize}
\tablehead{ 
 & & \multicolumn{3}{c}{\ff{\NCIV}} & & \multicolumn{3}{c}{\ff{\EWlin{1548}}} \\ 
 \cline{3-5} \cline{7-9} \\ 
\colhead{G} & \colhead{$z$} & 
\colhead{$\langle k_{14} \rangle$} & \colhead{$\langle \alpha_{N} \rangle$} & 
\colhead{$r_{k_{i},\alpha_{i}}$} & & 
\colhead{$\langle k_{3} \rangle$} & \colhead{$\langle \alpha_{W} \rangle$} & 
\colhead{$r_{k_{i},\alpha_{i}}$} \\ 
 & & 
\colhead{$(\!\cm{2})$} & & & &
\colhead{$(\!\mA^{-1})$} & &
}
\startdata
1 & $< 1$ & $ 0.63\pm 0.11$ & $-1.50\pm 0.10$ & -0.12 &  & $ 1.54\pm 0.62$ & $-1.58\pm 0.31$ &  0.88 \\ 
1+2 & $< 1$ & $ 0.73\pm 0.11$ & $-1.52\pm 0.10$ & -0.14 &  & $ 1.73\pm 0.67$ & $-1.55\pm 0.32$ &  0.92 \\ 
1 & $< 0.6$ & $ 0.72\pm 0.20$ & $-1.76\pm 0.27$ &  0.70 &  & $ 1.89\pm 0.80$ & $-1.61\pm 0.39$ &  0.94 \\ 
1+2 & $< 0.6$ & $ 0.81\pm 0.20$ & $-1.75\pm 0.27$ &  0.45 &  & $ 2.11\pm 0.79$ & $-1.58\pm 0.37$ &  0.91 \\ 
1 & $> 0.6$ & $ 0.56\pm 0.11$ & $-1.40\pm 0.18$ & -0.72 &  & $ 1.52\pm 0.60$ & $-1.43\pm 0.37$ &  0.82 \\ 
1+2 & $> 0.6$ & $ 0.69\pm 0.09$ & $-1.46\pm 0.14$ & -0.39 &  & $ 1.75\pm 0.65$ & $-1.46\pm 0.41$ &  0.90 \\ 
\enddata
\tablecomments{
Mean values and correlation coefficient from jack-knife analysis of power-law fit to \ff{\NCIV} and \ff{\EWr}.
All absorbers from each sightline were iteratively excluded from the maximum likelihood analysis of the frequency distributions.
}
\end{deluxetable}

The frequency distribution is the number of absorbers $\Num$ per
column density \NCIV\ or equivalent width $\EWlin{1548}$ bin per
absorption pathlength sensitive to those absorbers (see \S\
\ref{subsec.mc}):
\begin{equation}
  \ff{\NCIV} \equiv \frac{\Delta \Num}{\Delta \NCIV \,\DX{\NCIV}} \label{eqn.fndef}
\end{equation}
and
\begin{equation}
  \ff{\EWr} \equiv \frac{\Delta \Num}{\Delta \EWr \,\DX{\EWr}} \label{eqn.fewdef} {\rm .}
\end{equation}
This quantity represents a nearly full description of an absorption
line survey and is akin to the luminosity function as used in galaxy
studies.

\begin{figure}[!hbt]
  \begin{center}$
    \begin{array}{c}
      \includegraphics[height=0.47\textwidth,angle=90]{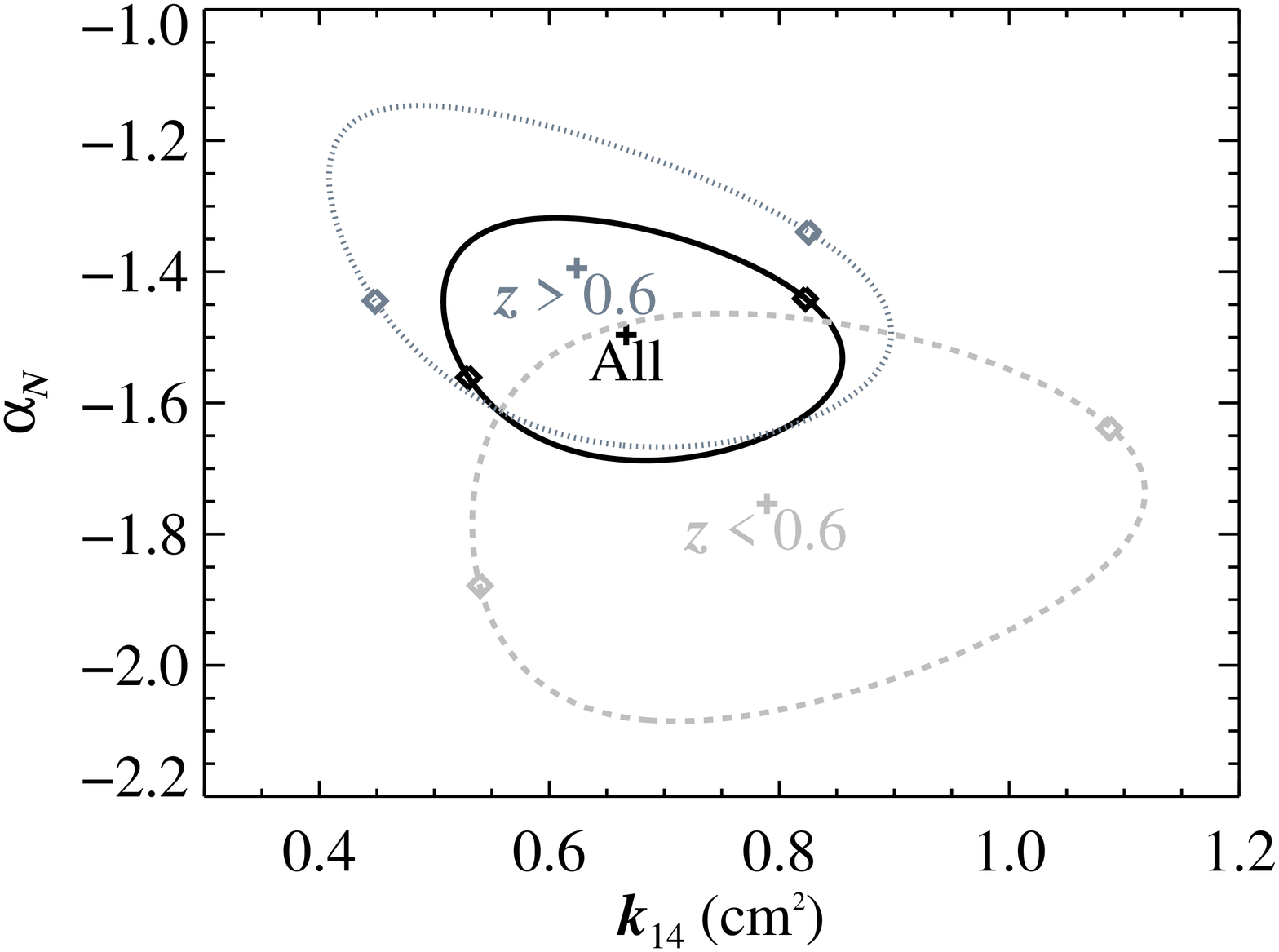} \\
      \includegraphics[height=0.47\textwidth,angle=90]{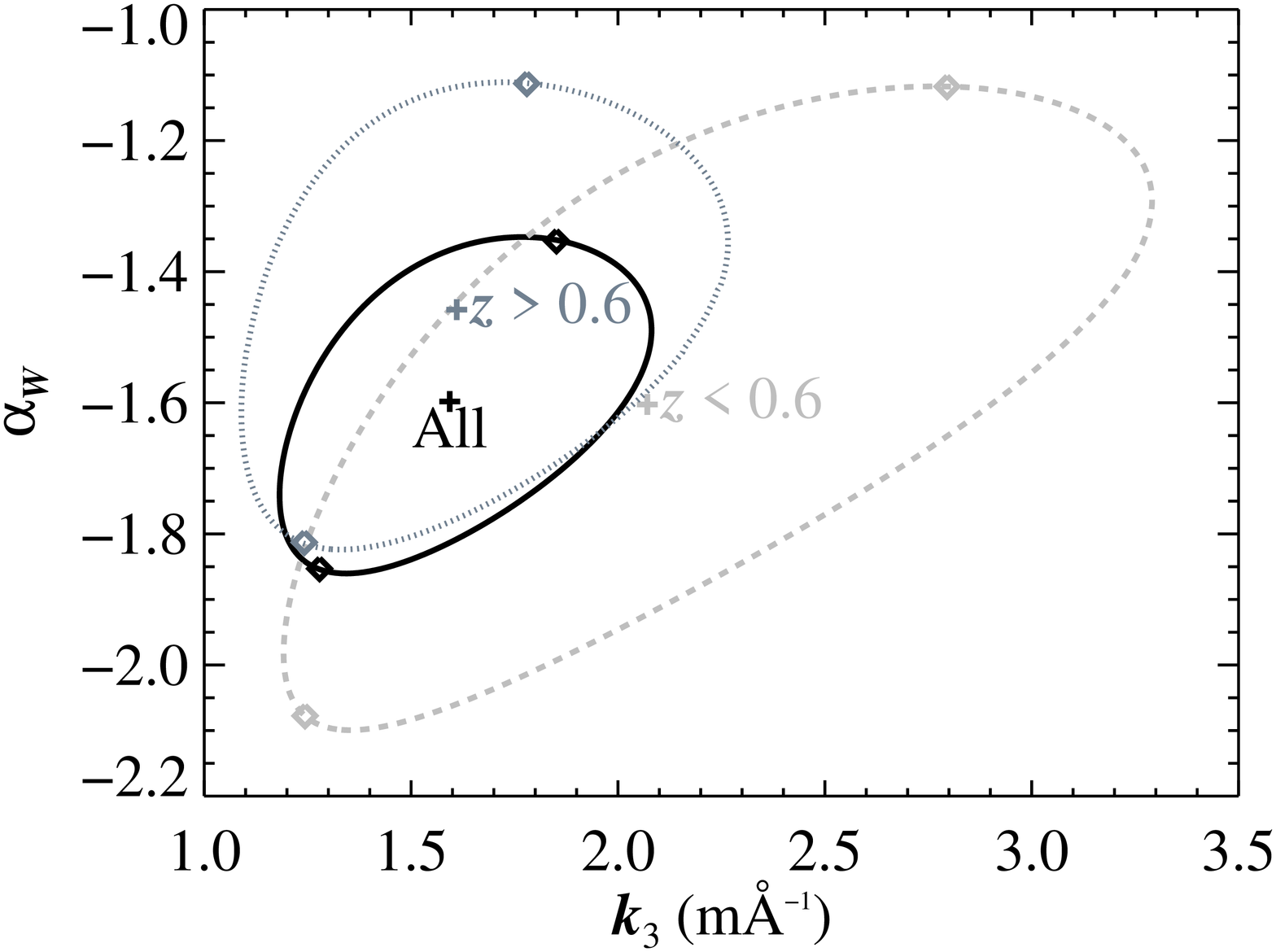}
    \end{array}$
  \end{center} 
  \caption[\ff{\NCIV}\ and \ff{\EWlin{1548}}\ 1-$\sigma$ ellipses.]
  {\ff{\NCIV}\ and \ff{\EWlin{1548}}\ 1-$\sigma$ ellipses from the
    maximum- likelihood analysis of the G = 1 sample. The sample was
    divided by redshift: $z<1$ (black ellipse and points), $z<0.6$
    (light gray, dashed), and $0.6\le z < 1$ (gray, dotted). The plus
    signs indicate the best-fit values (see Tables \ref{tab.fn} and
    \ref{tab.few}). The diamonds indicate the coefficient and exponent
    that define the 1-$\sigma$ error on the integrated \OmCIV\ and/or
    \dNCIVdX. The coefficient \kff{14}\ and the exponent \aff{N}\ were
    not highly correlated for \ff{\NCIV} (top panel) due to the
    treatment of saturated absorbers in the maximum-likelihood
    analysis (see Appendix \ref{appdx.maxL}). The values were highly
    correlated for \ff{\EWlin{1548}} (bottom panel).
    \label{fig.fnerr}
  }
\end{figure}

We chose to model the frequency distributions with power-law functions
because it approximated the frequency distributions well. The
power-law functions are of the form:
\begin{equation}
\ff{\NCIV}= \kff{14}\,\bigg(\frac{\NCIV}{\N{0}}\bigg)^{\aff{N}} \label{eqn.fn}
\end{equation}
and
\begin{equation}
\ff{\EWr}= \kff{3}\,\bigg(\frac{\EWr}{\EWlin{0}}\bigg)^{\aff{W}} {\rm ,} \label{eqn.few}
\end{equation}
where $\N{0} = 10^{14}\cm{-2}$ and $\EWlin{0}=400\mA$ and the
subscripts on the coefficients $k$ indicate the normalization (\eg
$\kff{14} = k/10^{14}\cm{2}$). We used the conjugate gradient method
to maximize the likelihood function $\mathcal{L}$ and to simultaneously
fit for the coefficient $k$ and exponent $\alpha$. From the 1-$\sigma$
error ellipse where $\ln \mathcal{L} - \ln \mathcal{L}_{\rm max} \ge
-1.15$,\footnote{We empirically determined the $\ln \mathcal{L} - \ln
  \mathcal{L}_{\rm max} \ge -1.15$ constraint. This limit defines a
  contour that contains 68.3\% of the likelihood surface. For a
  Gaussian distribution, the 1-$\sigma$ contour would be defined by
  $\ln \mathcal{L} - \ln \mathcal{L}_{\rm max} \ge -0.5$} we
estimated the errors in $k$ and $\alpha$ (see Figure \ref{fig.fnerr}).

We derived $\mathcal{L}$ in a similar manner to that outlined in
\citet{storrielombardietal96} and detailed in our Appendix
\ref{appdx.maxL}. The main difference was to include the observed
number of strong, saturated absorbers, where we only had a lower-limit
estimate of \NCIV, as a constraint in $\mathcal{L}$. The best-fit
power law must allow for a number of strong (saturated) absorbers
consistent with the observed number. We set the saturation limit to
$\logCIV = 14.3$, which we determined empirically (see Figure
\ref{fig.hist}). In the maximum-likelihood analysis, the unsaturated
doublets with $\logCIV \ge 14.3$ were counted as saturated absorber,
and the two saturated (G = 1+2) doublets with $\logCIV < 14.3$ were
excluded. Without the saturation term in $\mathcal{L}$, the best-fit
$\alpha$ would have been significantly steeper, depending on the
integration limits, since it would have been a strong statistical
statement to not detect {\it any} strong absorbers.

For the same reason, the choice of integration limits in the
maximum-likelihood analysis influenced the result. We set the lower
integration limit to the smallest observed value less one sigma for
the sample analyzed, \eg $\N{min}-\sigma_{{\rm N},min}$. For
\ff{\NCIV}, the saturation limit was $10^{14.3}\cm{-2}$ and the upper
limit ``infinity'' was $10^{16}\cm{-2}$. For \ff{\EWlin{1548}}, the
upper limit was the largest observed value plus one sigma
$\EWr{}_{,max} + \sigEWr{}_{,max}$.

\begin{figure}[!hbt]
  \begin{center}
  \includegraphics[height=0.47\textwidth,angle=90]{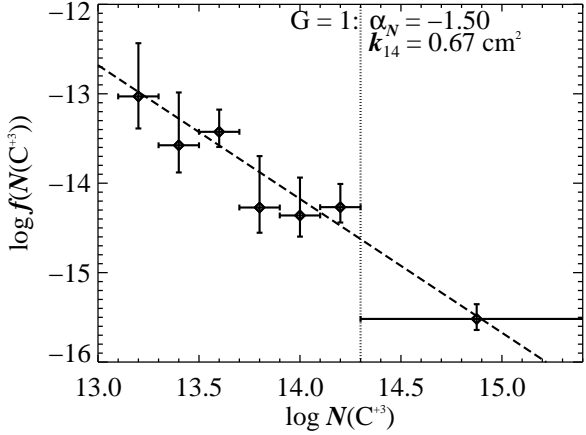}
  \end{center}
  \caption[Column density frequency distribution for the full $z<1$ sample.]
  {Column density frequency distribution for the full $z<1$
    sample. The G = 1 observations are the black diamonds, and the
    long-dash line indicates the best power-law fit. The observed and
    fitted G = 1+2 \ff{\NCIV}\ agrees very well. The AODM column
    densities are lower limits for saturated systems; our adopted
    saturation limit $\logCIV = 14.3$ is indicated (vertical, dotted
    line).
    \label{fig.fN}
  }
\end{figure}

\begin{figure}[!hbt]
  \begin{center}
  \includegraphics[height=0.47\textwidth,angle=90]{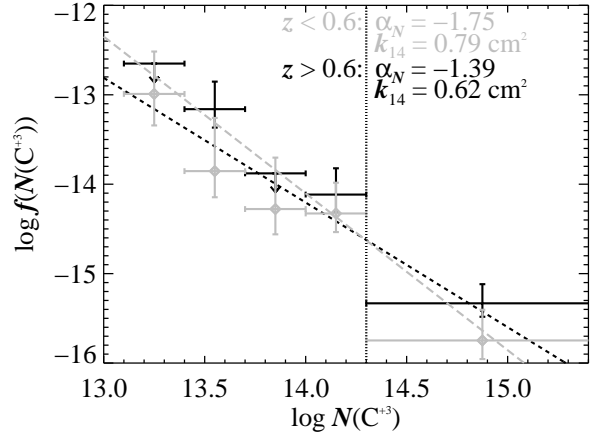}
  \end{center}
  \caption[Column density frequency distribution for two
    redshift bins of the G = 1 sample.]
    {Column density frequency distribution for two redshift bins of
      the G = 1 sample. The best-fit \aff{N}\ for the $\zciv < 0.6$
      \ff{\NCIV}\ (black, long-dash line) is steeper than that for the
      $0.6 \le \zciv < 1$ bin (gray, dashed line).  As in Figure
      \ref{fig.fN}, the best-fit \aff{N} and \kff{14} from the
      maximum-likelihood analysis are shown, as is the saturation
      limit $\logCIV = 14.3$ (vertical, dotted line).
    \label{fig.fN_byz}
  }
\end{figure}

The best-fit power-law \ff{\NCIV}\ for the G = 1 and the G = 1+2
samples are shown in Figure \ref{fig.fN} and enumerated in Table
\ref{tab.fn}. For the results of the maximum likelihood analysis of
\ff{\EWlin{1548}}, see Table \ref{tab.few}. They were consistent
within the errors. To examine the temporal evolution of \ion{C}{4}
absorbers in our survey, we divided the G = 1 sample into two redshift
bins, defined by, approximately, the median redshift: $\zciv<0.6$ and
$0.6 \le z < 1$ (see Figure \ref{fig.fN_byz}).

We performed a jackknife re-sampling analysis to estimate the errors of
and correlation between $k$ and $\alpha$ from our maximum-likelihood
analyses. For the jackknife, each \ion{C}{4} doublet $i$ was excluded
from the full sample of $\Num$ absorbers. Then, the frequency
distributions \ff{\NCIV}\ and \ff{\EWlin{1548}} were re-fit with
integration limits set as described previously. This was done for the
G = 1 and G = 1+2 samples for all three redshift cuts ($\zciv < 1$,
$<0.6$, and $\ge0.6$; see Table \ref{tab.corrm}). The variance in the
\eg coefficient distribution $k_{i}$ from the jackknife is:
\begin{equation} \sigma_{k_{i}}^{2} = \frac{1}{\Num-1}
  \sum_{i}^{\Num} (k_{i} - \langle k \rangle )^{2} {\rm ,}
\end{equation}
where $\langle k \rangle$ is the mean of the best-fit values from the
re-sampling analysis. The same applies to the exponent distribution
$\alpha_{i}$. The mean coefficients and exponents from the
maximum-likelihood analyses to the various redshift samples agreed
with the results from the analyses of the full samples (\ie compare
Tables \ref{tab.fn} and \ref{tab.few} with Table \ref{tab.corrm}).

The correlation between the parameters $k_{i}$ and $\alpha_{i}$ is the
ratio of the covariance to the product of $\sigma_{k_{i}}$ and
$\sigma_{\alpha_{i}}$: \begin{equation}
  r_{k_{i},\alpha_{i}} = \frac{\Num-1}{\Num}\frac{\displaystyle \sum_{i}^{\Num}
   \Big(k_{i} - \langle k \rangle \Big) \Big(\alpha_{i} - \langle \alpha_{i}
  \rangle \Big) } {\displaystyle \sigma_{k_{i}}\sigma_{\alpha_{i}} }  {\rm .}
\end{equation}
The coefficient and exponent were strongly correlated for the
\ff{\EWlin{1548}} power-law fits (\ie $r_{k_{i},\alpha_{i}} \approx 1$). By
accounting for the saturated doublets in the maximum-likelihood
function, \kff{14}\ and \aff{N}\ were less correlated for the
\ff{\NCIV}\ fits.

Since \ion{C}{4} absorbers cluster \citep{fernandezsotoetal96}, we
also performed a jackknife of the sightline and iteratively fit the
frequency distributions after excluding all absorbers from 
each sightline. The results are compiled in Table \ref{tab.loscorrm};
they agree well with results for the full sample (Tables \ref{tab.fn}
and \ref{tab.few}). We did not re-estimate \DXp\ without the excluded
sightline since each sightline contributes $\lesssim10\%$ to \DXp\ of
any redshift range.


\subsection{\ion{C}{4} Absorber Line Density}\label{subsec.dndx}

\begin{figure}[!hbt]
  \begin{center}$
    \begin{array}{c}
      \includegraphics[height=0.47\textwidth,angle=90]{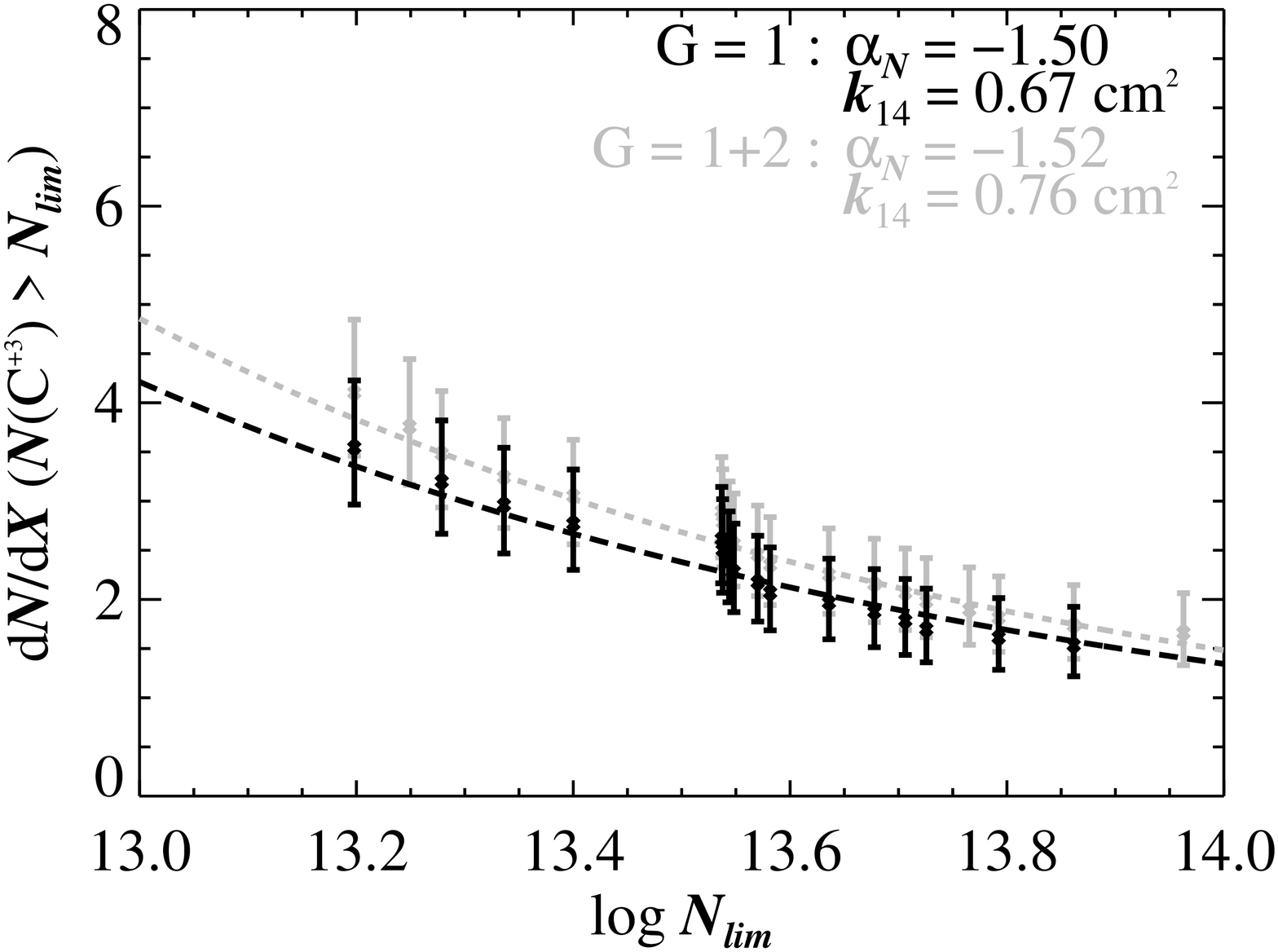} \\
      \includegraphics[height=0.47\textwidth,angle=90]{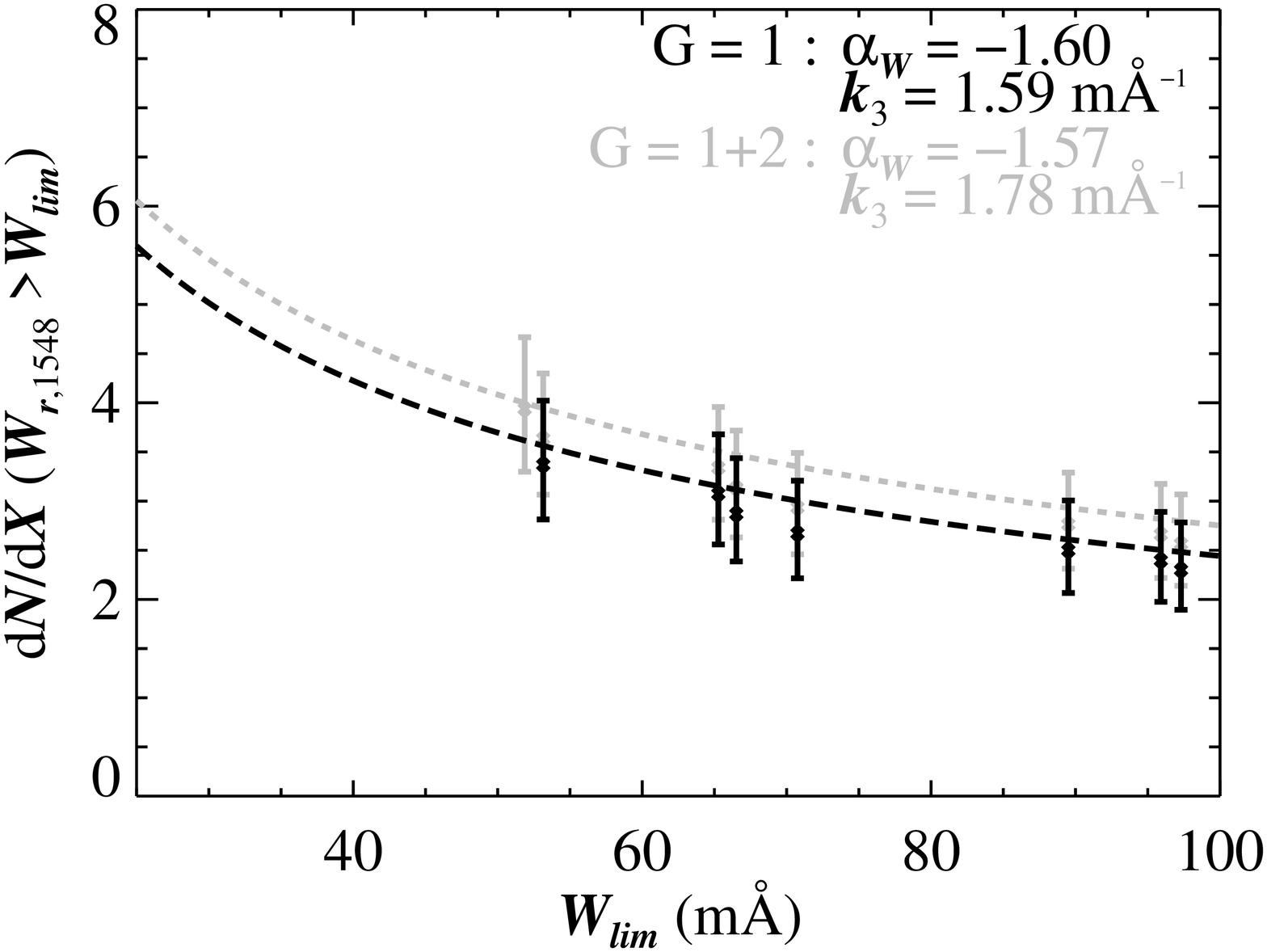}
      \end{array}$
    \end{center}
  \caption[Absorber line density as function of \NCIV\ and
    \EWlin{1548}.]
    {Absorber line density as function of \NCIV\ and \EWlin{1548}. The
      summed \dNCIVdX\ from the G = 1 and 1+2 groups are shown with
      the solid black and gray crosses, respectively (see Equation
      \ref{eqn.dndx_sum}). The dashed lines show the integrated
      \dNCIVdX\ from the fits to \ff{\NCIV} and \ff{\EWlin{1548}} (see
      Equation \ref{eqn.dndx_int}).
    \label{fig.dndx}
  }
\end{figure}

We measured the density of \ion{C}{4} doublets with a minimum
column density as follows:
\begin{eqnarray}
\frac{\displaystyle \ud \Num_{\mathrm{C\,IV}}}
{\displaystyle \ud X} (\NCIV \ge \N{lim}) = \sum_{\N{i} \ge
  \N{lim}} 
\frac{\displaystyle 1}{\displaystyle \DX{\N{i}}}
\label{eqn.dndx_sum} \\
\sigma_{\ud \Num/\ud X}^{2} = \sum_{\N{i} \ge \N{lim}}
\Big( \frac{\displaystyle \sigDX}{\displaystyle \DX{\N{i}}^2}
\Big)^{2} {\rm .}  \nonumber 
\end{eqnarray}
The absorber line density can be measured for any equivalent width limit in a
similar fashion. We estimated the contribution of the low-number statistics in
\Num\ by adding the Poisson counting variance to $\sigma_{\ud \Num/\ud
  X}^{2}$. In Figure \ref{fig.dndx}, we show \dNCIVdX\ for the full G
= 1 and G = 1+2 samples as functions of column density (top panel) and
equivalent width (bottom panel). The weakest system we found had
$\logCIV = 13.2$ and $\EWlin{1548} = 52\mA$. At this limit, $\dNCIVdX
= 4.1^{+0.7}_{-0.6}$ for the G = 1+2 sample.

In Figure \ref{fig.dndx}, we
extrapolated our \dNCIVdX\ to lower \NCIV\ and \EWr\ by integrating the
power-law fits of the frequency distribution over \eg $\N{lim} \le \NCIV < \infty$:
\begin{equation} 
\frac{\displaystyle \ud \Num_{\mathrm{C\,IV}}}{\displaystyle \ud X}
(\NCIV \ge \N{lim}) 
= -\frac{\kff{14}}{1 + \aff{N}}
\frac{\N{lim}^{1+\aff{N}}}{\N{0}^{\aff{N}}}  {\rm .}
\label{eqn.dndx_int}
\end{equation}
The summed and integrated \ion{C}{4} absorber line density by \NCIV\ and
\EWlin{1548} are listed in Tables \ref{tab.fn} and \ref{tab.few},
respectively. The error in \dNCIVdX\ from the power-law model of
\ff{\NCIV}\ was defined by the error ellipse in the
maximum-likelihood analysis. However, the extrema in \dNCIVdX\ allowed
by the 1-$\sigma$ ellipse did not occur at the extrema of \kff{14}\ and
\aff{N}. As shown in Figure \ref{fig.fnerr} (bottom panel), the
extrema in the 1-$\sigma$-allowed values of \dNCIVdX\ occurred on the
sides of the ellipse where \kff{14}\ and \aff{N} were maximized (or
minimized). Again, the equations are similar for line density as a
function of equivalent width.

\subsection{\Cthr\ Mass Density}\label{subsec.omciv}

\begin{figure}[!hbt]
  \begin{center}
  \includegraphics[height=0.47\textwidth,angle=90]{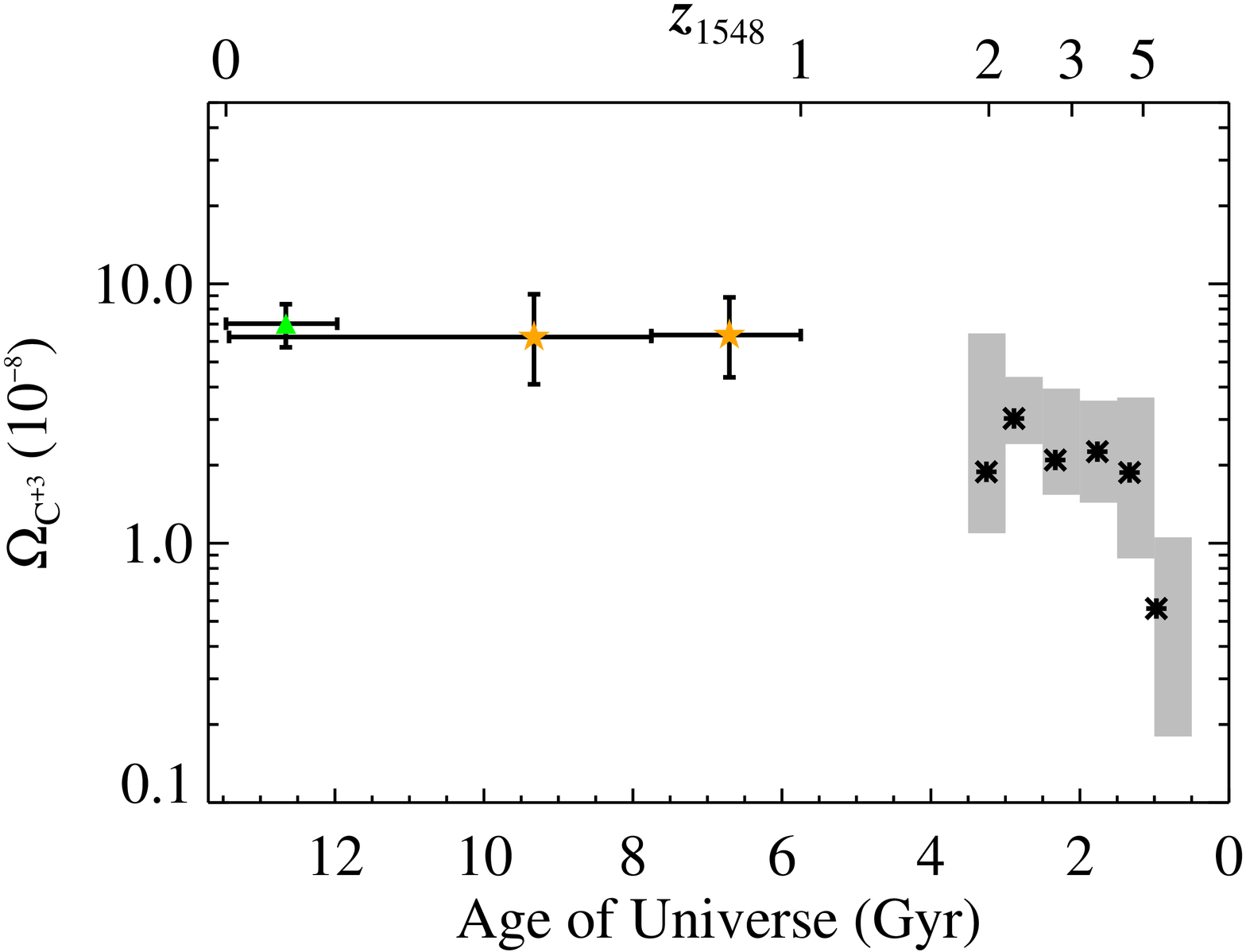}
  \end{center}
  \caption[\Cthr\ mass density relative to the critical density over
    age of Universe.]
    {\Cthr\ mass density relative to the critical density over age of
      Universe. The integrated \OmCIV\ (see Equation
      \ref{eqn.omciv_int}) for the G = 1 sample, divided by redshift,
      are the (orange) stars. The value from
      \citet{danforthandshull08} is the (green) triangle and is not an
      independent measurement of \OmCIV. The grey boxes indicate the
      binned, cosmology-adjusted \OmCIV\ from several high-redshift
      studies: \citet{songaila01, pettinietal03, boksenbergetal03ph,
        scannapiecoetal06, beckeretal09}; and \citet{ryanweberetal09}
      (see Appendix \ref{appdx.adjOmCIV}). The horizontal extent of
      the boxes indicate the bin size. The vertical extent represents
      the maximum range spanned by $\OmCIV \pm \sigOmCIV$ (the
      published errors were adjusted to be $1\sigma$, as
      necessary). The average time and error-weighted \OmCIV\ per bin
      are shown with black asterisks.
    \label{fig.omciv} 
  }
\end{figure}

In principle, \OmCIV\ is the ratio of the mass in \Cthr\ ions relative
to the critical density $\rho_{c,0}$. In practice, the observations
limit \OmCIV\ to include only \ion{C}{4} absorbers within a range of
column densities. The lower \NCIV\ bound reflects the limit where the
observations can confidently detect and identify \ion{C}{4} doublets,
typically $\N{min} \approx 10^{13}\cm{-2}$. The upper bound is usually
$\N{max} = 10^{15}\cm{-2}$. Doublets with $\logCIV > 15$ are rare and
often associated with galaxies (for example, the DLA \ion{C}{4}
doublet towards PG1206+459 with $\logCIV > 15.4$). In addition, there
has been no observed break in \ff{\NCIV}, so the integrated \OmCIV\
is infinite.

The metallicity and ionizing 
background of the intergalactic medium affect the \Cthr\ mass density,
and the observed evolution of \OmCIV\ over time (or redshift) add a
constraint to the changes in the cosmic metallicity and ionizing
background. While one expects that the former increases monotonically with
time, the latter is believed to be decreasing since $z \approx 1$.
\OmCIV\ is the integrated column density-``weighted'' 
frequency distribution \ff{\NCIV}: 
\begin{equation} \OmCIV =
  \frac{H_{0}\,\mathrm{m}_{\rm C}} {c\,\rho_{c,0}}
  \int_{\N{min}}^{\N{max}} \ff{\NCIV}\, \NCIV\, \ud\, \NCIV {\rm
    ,} \label{eqn.omciv}
\end{equation}
where $H_{0} = 70\kms\,\mathrm{Mpc}^{-1}$ is the Hubble constant
today; $\mathrm{m}_{\rm C} =2\times10^{-23}\,$g is the mass of the
carbon atom; $c$ is the speed of light; and
$\rho_{c,0}=3H_{0}^2(8\pi G)^{-1} =9.26\times10^{-30}\,{\rm
    g}\cm{-3}$ for our assumed Hubble constant.

The observed \OmCIV\ can be approximated by the sum of the detected
\ion{C}{4} absorbers:
\begin{eqnarray}
  \OmCIV & = & \frac{H_{0}\,\mathrm{m}_{\rm C}} {c\,\rho_{c,0}}
  \sum_{\Num} \frac{\NCIV}{\DX{\NCIV}}  \label{eqn.omciv_sum}
  \\ 
  \sigOmCIV^{2} & = & \bigg(\frac{H_{0}\, \mathrm{m}_{\rm
      C}}{c\,\rho_{c,0}} \bigg)^{2} \Bigg( \sum_{\Num}
  \bigg(\frac{\sigNCIV}{\DX{\NCIV}}\bigg)^{2} \\
  & & + \sum_{\Num} \bigg(\frac{\NCIV \sigDX}{\DX{\NCIV}} 
  \bigg)^{2}\Bigg) \nonumber 
\end{eqnarray}
\citep{lanzettaetal91}. We list the summed \OmCIV\ values for various
redshift samples in Table \ref{tab.fn}. We only include absorbers that
are unsaturated in at least one line, typically with $\logCIV < 14.3$,
because we only have a lower limit to \NCIV\ for saturated absorbers.
Therefore, the {\it summed} \OmCIV\ is a lower limit, since some
saturated absorbers may have $\logCIV \le 15$ and could have been
included in the \OmCIV\ sum. 

To measure a value for \OmCIV, we use our best-fit \ff{\NCIV} power
law (Equation \ref{eqn.fn}) and integrate Equation \ref{eqn.omciv}
analytically: 
\begin{equation}
  \OmCIV = \frac{H_{0}\,\mathrm{m}_{\rm C}} {c\,\rho_{c,0}} 
  \frac{10^{-14}\kff{14}}{2+\aff{N}}
  \bigg(\frac{\N{max}^{2+\aff{N}} -
    \N{min}^{2+\aff{N}}}{\N{0}^{\aff{N}}} \bigg) 
  {\rm .}  \label{eqn.omciv_int}
\end{equation}
With this, we can measure \OmCIV\ over the column density range of
unsaturated and saturated absorbers or even extrapolate the model to
different ranges. Like the errors for \dNCIVdX\ (see \S\
\ref{subsec.dndx}, the errors for the integrated \OmCIV\ were taken as
the extrema of the 1-$\sigma$ \ff{\NCIV} error ellipses (see top
panel, Figure \ref{fig.fnerr}).

The choice for the limits of integration $\N{min}$ to $\N{max}$ was
critical, since there has been no observed downturn in \ff{\NCIV},
measured at any redshift. We used $13 \le \logCIV \le 15$ in our
analysis (see Table \ref{tab.fn} and Figures \ref{fig.omciv} and
\ref{fig.omcivfit}) because this range reflects that of the observed
column densities and overlapped with the majority of studies used for
comparison \citep[\eg][]{songaila01}. The integrated \OmCIV\ is most
influenced by the strongest absorbers, because they contain
substantially more mass. For example, according to Equation
\ref{eqn.omciv_int} and with $\aff{N}=-1.5$, the $14.3 \le \logCIV
\le 15$ component contributes $\approx 60\%$ of \OmCIV, integrated over the
range $13 \le \logCIV \le 15$. In comparison, the $13 \le \logCIV \le
13.7$ component accounts for $\approx 14\%$. The proportions change as the
power-law exponent changes, increasing the lower column density
contribution as \aff{N} decreases. A detailed discussion of the
effects of the column density range on the summed \OmCIV\ is given in
Appendix \ref{appdx.adjOmCIV}.

\section{Discussion} \label{sec.disc}

\subsection{Comparisons with Previous Results}\label{subsec.comp}

In \citet{danforthandshull08}, $\dNCIVdz = 10^{+4}_{-2}$ for $\EWr \ge
30\mA$. They measured the absorber line density by summing their
observed number of doublets (24) and dividing by the unblocked
redshift pathlength ($\Dz = 2.42$). They estimated \Dz\ by identifying
regions of the spectra where the \ion{C}{4} 1548 line could be
detected at $\EWr \ge 4\sigEWr$ and 1550 at $\ge 2\sigEWr$. We measure
$\dNCIVdz = 6.2^{+1.9}_{-1.6}$ for the $z < 0.6$, G = 1 doublets with
$\EWlin{1548} \ge 53\mA$ (see Table \ref{tab.fn}). The $>1\sigma$
difference between our \dNCIVdz\ values is largely due to the
equivalent width limit. For $\EWr \ge 50\mA$, $\dNCIVdz =
7^{+3}_{-2}$ (C. Danforth, private communication), which is within
$0.3\sigma$ of our value. Another source for the discrepancy is the
detection criterion used (\eg both lines detected with
$\EWr\ge3\sigEWr$), which would affect the number of doublets included
and the unblocked redshift pathlength.\footnote{We
 measure $\Dz=2.2$ for the E140M spectra.}

\begin{figure}[!hbt]
  \begin{center}
  \includegraphics[height=0.47\textwidth,angle=90]{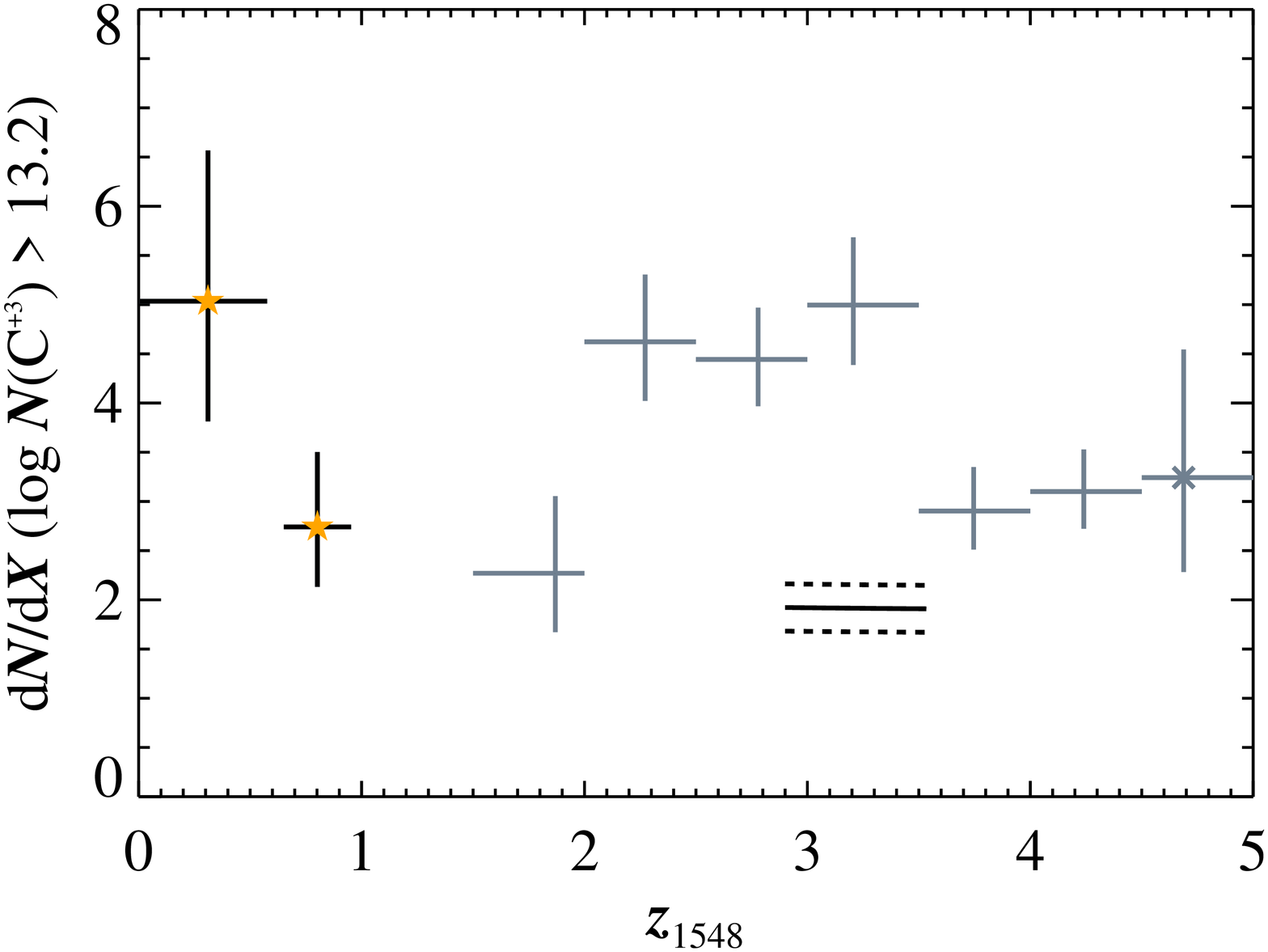} \\
  \end{center}
  \caption[\dNCIVdX\ as a function of redshift.]
 {\dNCIVdX\ as a function of redshift. The number density of \ion{C}{4} absorbers
    has not changed significantly since $z=5$.  The (orange) stars are
    our summed $z<1$ values for $\logCIV \ge 13.2$ (see Equation
    \ref{eqn.dndx_sum}). The solid and dashed horizontal lines are the
    integrated \dNCIVdX\ and 1-$\sigma$ error, respectively, from the
    best-fit \ff{N} from \citet{songaila01}. She only fit her $2.9\le
    z \le 3.54$ observations, but the result agreed with the full
    $1.5\le z \le 5$ sample.  For reference, $\dNCIVdX \equiv
    \Num/\DXp$ from the tabulated \Num\ and \DXp\ in \citet[][$12 \le
    \logCIV \le 14.9$]{songaila01} and \citet[][$12.5 \le \logCIV \le
    14$]{pettinietal03} are shown with the gray pluses and cross,
    respectively; the error bars assume Poisson counting statistics on
    \Num. We have adjusted \DXp\ for differences in cosmology (see
    Appendix \ref{appdx.adjOmCIV}).
    \label{fig.dndx_evo}
  }
\end{figure}

The \ion{C}{4} absorber line density has not changed significantly
since $z=5$ for $\logCIV \ge 13$ absorbers. \citet{songaila01} and
\citet{boksenbergetal03ph} did not detect any redshift evolution for
$\logCIV \ge 13$ doublets over $1.5 \lesssim z \lesssim 4.5$; they
measured $\dNCIVdX \approx 3$. In Figure \ref{fig.dndx_evo}, we
compare the \dNCIVdX\ measurements from \citet{songaila01} with the
current study. In order to compare \dNCIVdX\ for $\logCIV \ge 13.2$,
we use the best-fit coefficient ($k_{14} = 0.63\cm{2}$) and exponent
($\aff{N}=-1.8\pm0.1$) from \citet{songaila01} in Equation
\ref{eqn.dndx_int}: $\dNCIVdX = 1.9\pm0.2$ (adjusted for cosmology).
For the current survey, the summed absorber line density is $\dNCIVdX
= 3.4^{+0.7}_{-0.6}$ at $\langle z \rangle = 0.654$. This is a
$1.8\pm0.4$ increase over the integrated \dNCIVdX\ value from
\citet{songaila01}. However, since she did not provide an error for
the coefficient, we likely underestimate her error on \dNCIVdX\ above.

The best-fit $\aff{N}=-1.75^{+0.28}_{-0.331}$ for
the lower-redshift bin was consistent with $\aff{N}=-1.79\pm0.17$
from \citet{danforthandshull08}, which covered $\zciv < 0.12$.

There has been no consensus on the best power-law exponent for
\ff{\NCIV}\ at $z > 1$. \citet{songaila01} and
\citet{boksenbergetal03ph} conducted surveys similar to the current
study, and they measured $\aff{N}=-1.8\pm0.1$ ($2.9\le z \le 3.54$)
and $\aff{N} = -1.6$ ($1.6 \le z \le 4.4$), respectively. For our $0.6
\le \zciv < 1$ sample, the best-fit $\aff{N}=-1.39^{+0.24}_{-0.27}$
agreed better with \citet{boksenbergetal03ph} but was consistent
with both studies at the $2\sigma$ level.

More detailed comparison with other studies of \ion{C}{4} absorbers at
$z<1$ are given in Appendix \ref{appdx.comp}.

\subsection{\OmCIV\ Evolution}\label{subsec.evo}

We have measured a statistically significant increase in \OmCIV\ at
$z<1$ compared to the roughly constant value observed at $1 < z <
5$. The error-weighted average of the $1 < z < 5$ measurements in
Figure \ref{fig.omciv} is $\overline \Omega_{\Cthr} = (2.2\pm0.2)
\times 10^{-8}$ at $\overline z = 3.240$. Our integrated $\OmCIV =
(6.20^{+1.82}_{-1.52}) \times10^{-8}$ at $\langle z \rangle = 0.654$
is a $2.8\pm0.7$ increase. We recognize the significance of the
increase in \OmCIV\ at $z<1$ because we have carefully adjusted the $1
< z < 5$ values of other authors (see Appendix \ref{appdx.adjOmCIV}).
\citet{danforthandshull08} stated that their \OmCIV\ value agreed with
that of \citet{scannapiecoetal06} without considering the effect of
the column density limits. \citet{fryeetal03} interpreted their
$z<0.1$ \OmCIV\ value as an increase but without comment on its
significance.

\begin{figure}[!hbt]
  \begin{center}$
    \begin{array}{c}
      \includegraphics[height=0.47\textwidth,angle=90]{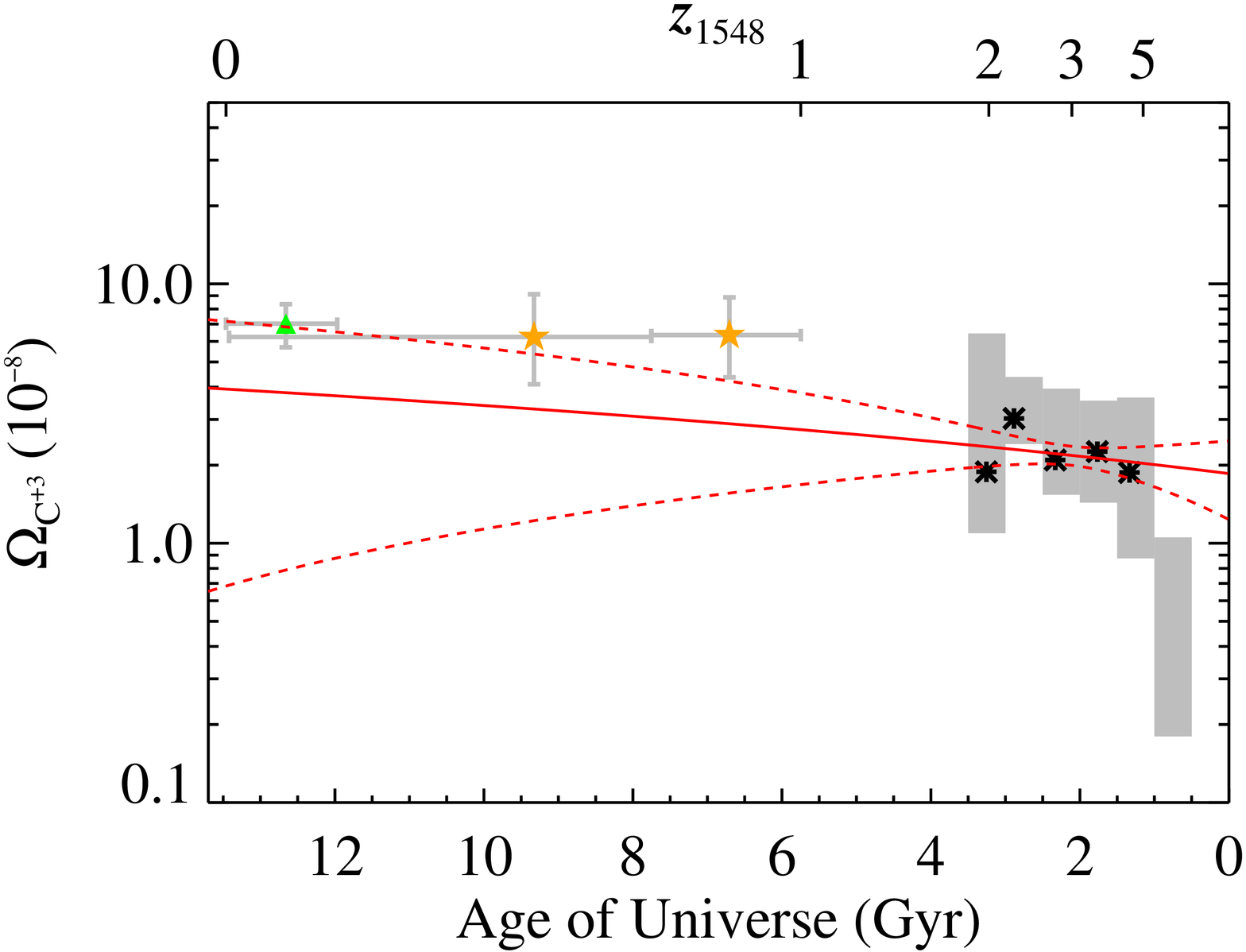} \\
      \includegraphics[height=0.47\textwidth,angle=90]{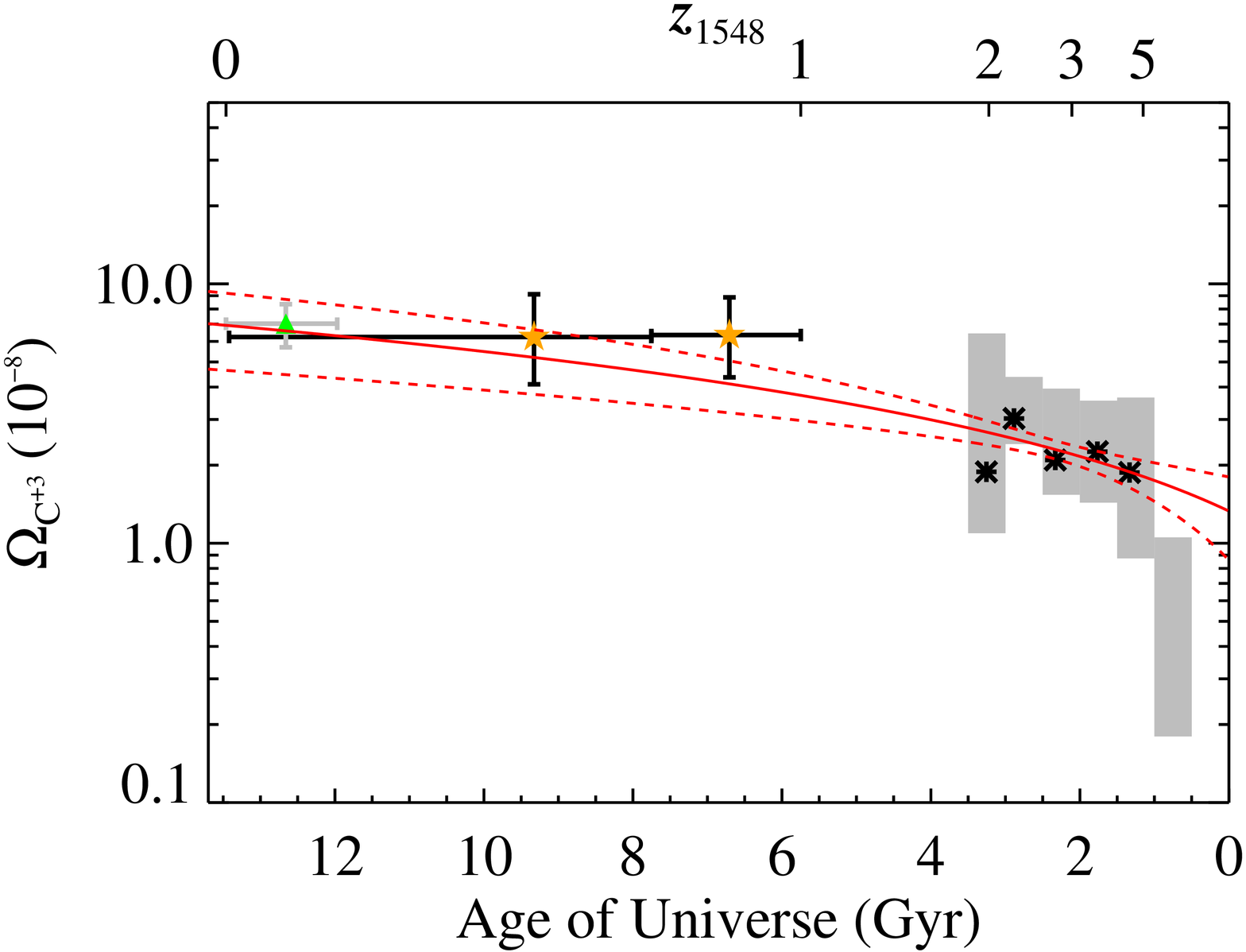}
    \end{array}$
  \end{center}
  \caption[Linear regressions of \OmCIV\ over age of Universe.]
  {Linear regressions of \OmCIV\ over age of Universe. The (red)
    dashed lines are the $\chi^{2}$-minimization fit to subsets of the
    data described in Figure \ref{fig.omciv}, and the (red) dotted
    lines indicate the 1-$\sigma$ range of the model. In the upper
    panel, the slope for the $1 < z < 5$ fit is consistent with no
    evolution $a_{1} = (0.15\pm0.3)\times10^{-8}\,{\rm Gyr}^{-1}$. For
    $z < 5$, in the lower panel, the slope is $a_{1} =
    (0.42\pm0.2)\times10^{-8}\,{\rm Gyr}^{-1}$. The evidence that the
    \Cthr\ mass density has been slowly increasing since $\approx
    1\,$Gyr requires the measurements at low redshift.
    \label{fig.omcivfit}
  }
\end{figure}

To estimate the rate of evolution of the \Cthr\ mass density, we adopt
a simple linear model:
\begin{eqnarray}
\OmCIV & = & a_{0} + a_{1}t_{age}\label{eqn.omciv_fit} \\
\sigOmCIV^{2} & = & \sigma_{a_{0}}^{2} + t_{age}^{2}
\sigma_{a_{1}}^{2} + 2t_{age}COV_{01}  {\rm ,} \label{eqn.omcvi_evo}\nonumber
\end{eqnarray}
where $t_{age}$ is the age of the Universe and $COV_{01}$ is the
covariance of the intercept $a_{0}$ and the slope $a_{1}$. The
observed 1-$\sigma$ uncertainties in \OmCIV were used in the
$\chi^2$-minimization algorithm. The results of the linear regression
for the $1 < z < 5$ ($1\,{\rm Gyr} \le t_{age} \le 6\,{\rm Gyr}$) and
$z < 5$ ($t_{age} > 1\,{\rm Gyr}$) samples are shown in Figure
\ref{fig.omcivfit}. We included the following \OmCIV\ measurements,
adjusted for cosmology and/or column density range (see Appendix
\ref{appdx.adjOmCIV}): \citet[][$1.5 < z <
4.5$]{songaila01};\footnote{The results from \citet{pettinietal03} are
  considered a revision to the $4.5 \le z < 5$ bin since
  \citet{pettinietal03} used higher S/N data. The $5 \le z < 5.5$ bin
  from \citet{songaila01}, which only included one detected absorber,
  was excluded.} \citet[][as revision of \citet{songaila01} $z \approx
4.7$ value]{pettinietal03}; \citet{boksenbergetal03ph};
\citet{scannapiecoetal06}; and the current study.\footnote{We excluded
  the \citet{danforthandshull08} measurement since our $z<0.6$
  measurement includes the E140M data.}

Whether or how \OmCIV\ evolves from $z=5 \rightarrow 1$ is not
statistically constrained, assuming the simple linear model.  The
best-fit slope for this range was $a_{1} =
(0.15\pm0.3)\times10^{-8}\,{\rm Gyr}^{-1}$ with the intercept $a_{0} =
(1.9\pm0.6)\times10^{-8}$ (at $t_{age} = 0\,$Gyr) and $\chi^{2}$
probability ${\rm P_{\chi^{2}}} = 34\%$.

The linear regression for the $z < 5$ data indicates a statistically
significant trend in \OmCIV\ evolution. The fitted parameters are
consistent with \OmCIV\ evolving slightly: $a_{1} =
(0.42\pm0.2)\times10^{-8}\,{\rm Gyr}^{-1}$ and intercept $a_{0} =
(1.33\pm0.5)\times10^{-8}$ (${\rm P_{\chi^{2}}} = 36\%$). Several
high-redshift studies \citep{songaila01, pettinietal03,
  boksenbergetal03ph} concluded that \OmCIV\ evolved very little to
not at all for $z \approx 5 \rightarrow 1$, which is consistent with
the previous linear regression of the $1 < z < 5$ data. Incorporating
the new $z < 1$ measurements, there is evidence that \OmCIV\ has
slowly but steadily increased since $z \approx 5$, at the 97\%
confidence level.

We acknowledge that our accounting for differences in cosmology and
column density range were imperfect.  However, whether or not we
adjust for differences in the \NCIV\ range does not significantly affect
the previous results: our \OmCIV\ value is still a statistically
significant increase over $\overline \Omega_{\Cthr}$, and the rate of
evolution ($a_{1}$) from $z = 5 \rightarrow 0$ is also significant,
assuming the simple linear model.

We emphasize that Equation \ref{eqn.omciv_fit} was {\it not}
physically motivated. The temporal evolution of \OmCIV\ is influenced
by multiple, complex physical processes (\eg star formation, UV
background). In addition, only \ion{C}{4} absorbers with $13 \le
\logCIV \le 15$ are included in the measurements of \OmCIV. However,
as we discuss in \S\ \ref{subsec.whatisciv}, the physical nature of
doublets in this column density range likely changes over the 12\,Gyr
from $z=5 \rightarrow 0$. The linear regression analysis was simply a
secondary way to gauge whether our observed \OmCIV\ values suggest a
significant increase compared to the $z > 1$ values. In the context of
this simple model, we rule out null evolution at high confidence.

\subsection{Changing Nature of \ion{C}{4}
  Absorbers}\label{subsec.whatisciv}

In \S\ \ref{subsec.comp}, we noted that \dNCIVdX\ for
\linebreak[4]$\logCIV \ge 13$ has not evolved with redshift (see
Figure \ref{fig.dndx_evo}). However, there is evidence for evolution
of \dNCIVdX\ when only the strong systems ($\EWlin{1548} \ge 150\mA$ and
$\logCIV \gtrsim 13.7$) are included \citep{steidel90, misawaetal02,
  boksenbergetal03ph}. The weak systems dominate the number counts,
resulting in no evolution of \dNCIVdX\ for the full sample. Since the
stronger doublets dominate the \Cthr\ mass density, we observe the
increased number of strong \ion{C}{4} doublets in the significant
increase of \OmCIV\ at $z < 1$.

Even without limiting redshift-evolution analysis to the
strongest absorbers, \citet{boksenbergetal03ph} noted that complex
\ion{C}{4} systems have higher mean column densities at lower
redshift, where ``complex'' indicates that the number of Voigt profile
components is $\ge 7$. \citet{misawaetal02} predicted this result by
tying their observed increase of high-\EWr\ doublets at low redshift
with \citet{petitjeanandbergeron94} observations that higher \EWr\
systems have more components. From visual inspection of the
$\EWlin{1548} \ge 150\mA$ doublets, we see that more than half have
multiple, prominent components. 

Perhaps the trend to higher column density, more multi-component
\ion{C}{4} absorbers at low redshift supports the claim that
$\approx50\%$ of $\logCIV > 13.3$ doublets are associated with
galactic outflows \citep{songaila06}. Alternatively, the strong
absorbers might trace infall or high-velocity cloud-like halo gas. In
either case, the complex profiles might be a result of the
heterogeneous nature of outflowing or infalling gas. Observations at
low-redshift have indicated that \ion{C}{4} absorption often reside in
galaxy halos, on $\approx100\kpc$ scales \citep{chenetal01}.

What \ion{C}{4} absorption traces at all redshifts affects the
interpretation of the evolution of \OmCIV. In \S\ \ref{subsec.evo}, we
discussed the time evolution of \OmCIV\ for $13 \le \logCIV \le 15$
doublets at $z<5$, and we showed that there has been a significant
increase at $z<1$ and that null evolution is ruled out. However, if
the majority of $13 \le \logCIV \le 15$ \ion{C}{4} doublets are
(low-density) intergalactic at $z>1$ while the majority are
circum-galactic at $z<1$, then the \OmCIV\ observations at high and
low redshift are not directly comparable. In a future paper, we will
explore the changing nature of \ion{C}{4} doublets by comparing the
properties and environments of \ion{C}{4} absorption in cosmological
hydrodynamic simulations\footnote{The simulations are from the
  OverWhelmingly Large Simulations project \citep{schayeetal09ph}.}
with observations \citep[also see][]{cooksey09}.

\subsection{Sightline Selection Bias}\label{subsec.losbias}

Several sightlines in our sample were originally observed for
reasons that increase the likelihood that \ion{C}{4} absorption would
be detected. For example, there were quasars targeted because
intergalactic \ion{Mg}{2} absorption was observed in optical spectra.
Studies have shown that \ion{Mg}{2} absorbers frequently exhibit
\ion{C}{4} absorption \citep{churchilletal99b}. Including such
sightlines in our survey might have increased our detected doublets
and biased our results to have more \ion{C}{4} doublets than would be
observed in an {\it un}biased survey of the IGM. On the other hand,
the sensitivity limits of high-resolution spectrometers on \hst\
(prior to the installation of the Cosmic Origins Spectrograph) imply
only a small number of quasars could have been observed, independent
of known absorbers.

To explore the bias in our survey due to the sightline
selection criteria, we compared our \ion{C}{4} absorber line density
\dNCIVdz\ with \citet{barlowandtytler98}, who performed a survey for
strong \ion{C}{4} absorbers in 15 sightlines from the \hst\ Faint
Object Spectrograph (FOS) archives at the redshifts of known \Lya\
absorbers. The FOS quasar key line project targeted bright quasars,
without prior knowledge of \eg \ion{Mg}{2}, damped \Lya, or Lyman
limit systems \citep{bahcalletal93}. \citet{barlowandtytler98}
measured $\dNCIVdz=2.3\pm0.9$ for $\EWr > 400\mA$ and $0.2<\zciv<0.8$
(six doublets).

We detected six G = 1 \ion{C}{4} doublets that met the
\citet{barlowandtytler98} criteria and find for this sub-sample, $\dNCIVdz
= 1.3^{+0.8}_{-0.5}$. Thus, our measured absorber line density agreed with
the unbiased result within $1\sigma$. Despite including sightlines
observed with prior knowledge about the IGM, we agreed with an
unbiased survey and, in fact, record a smaller incidence of strong
\ion{C}{4} absorbers.

As a second check for sightline bias, we reanalyzed our survey after
excluding 11 G = 1 \ion{C}{4} doublets that were possibly associated
with targeted \ion{Mg}{2} absorbers, DLAs, or LLS (see Table
\ref{tab.selec}).\footnote{Several sightlines were targeted for a
  specific absorption system but the archival spectra did not cover
  the \ion{C}{4} doublet. The following are the sightlines and bias:
  PKS0454--22, \ion{Mg}{2} at $\zabs=0.6248$ and $0.9315$
  \citep{churchillandlebrun98}; HE0515--4414, DLA at $\zabs=1.15$;
  MARK132, LLS at $\zabs=1.7306$; and PKS1127--145, DLA at
  $\zabs=0.312$. The PG1634+706 sightline was also targeted for a D/H
  study and \ion{Mg}{2} absorber at $\zabs=0.9902$, but we did not
  include that echelle order in our co-added spectra.} For this
sub-sample, $\dNCIVdz = 0.4^{+0.6}_{-0.3}$ for the two G = 1
\ion{C}{4} doublets with $\EWr > 400\mA$ and $0.2<\zciv<0.8$, which
underestimates the \citet{barlowandtytler98} results by
$1.8\sigma$. The best-fit coefficient and exponent from the
maximum-likelihood analysis of \ff{\EWlin{1548}} were $k_{3} =
1.40^{+0.67}_{-0.51}\mA^{-1}$ and $\aff{W} = -1.51^{+0.39}_{-0.41}$,
which lie within $0.5\sigma$ of the values in Table \ref{tab.few}. For
\ff{\NCIV}, the best-fit values were $k_{14} =
0.49^{+0.16}_{-0.13}\cm{2}$ and $\aff{N} = -1.71^{+0.22}_{-0.26}$,
which were within $1\sigma$ of the values in Table \ref{tab.fn},
respectively. The change in coefficient and exponent change the
integrated $\OmCIV = (3.96^{+1.46}_{-1.15})\times10^{-8}$, lower than
that from the full sample but a $1.8\pm0.5$ increase over the
error-weighted average of the $1<z<5$ measurements.

Excluding the 11 \ion{C}{4} doublets possibly associated with known
absorption features yields an integrated \OmCIV\ within $3\sigma$ of
the value extrapolated from the model to $1 < z < 5$ (see top panel,
Figure \ref{fig.omcivfit}). However, by excluding the 11 absorbers,
most with $\logCIV > 14$, we under-sampled the strong doublets as
expected from the \citet{barlowandtytler98} study, and the high-\NCIV\
systems dominate the \Cthr\ mass density. In the attempt to eliminate
the effects of the sightline selection bias, we introduced a possible bias
against strong \ion{C}{4} doublets. We have proceeded, therefore, with
the full sample of archival spectra. 

\begin{deluxetable}{llllcll}
  \tablewidth{0pt}
  \tabletypesize{\scriptsize}
  \tablecaption{SIGHTLINE SELECTION BIAS\label{tab.selec} }
  \tablehead{ \colhead{Target} & \colhead{$z_{abs}$} &
  \colhead{$\vert\delta v_{excl}\vert$} & \colhead{Bias} & \colhead{$\mathcal{N}_{excl}$} &
  \colhead{\dvabs} & \colhead{\logCIV} \\
  & & \colhead{(\!\kms)} & & & \colhead{(\!\kms)} & } 
  \startdata
  PG0117+213   &  0.5763\tablenotemark{a}  &  600 & Mg\,II & 1 &
  3 & $14.52\pm0.03$ \\
     &  0.72907\tablenotemark{b} &  600 & Mg\,II & 0 & \nodata & \nodata \\
  PG1206+459   &  0.93\tablenotemark{c} & 1500 & Mg\,II & 2 & $-503$ & $>15.39$ \\
     &      &  &  &  & $661$ & $14.25\pm0.09$ \\
  PG1211+143   &  0.051  &  400 & D/H & 1 & $39$ & $14.01\pm0.03$ \\
     &  0.0652  &  400 & D/H & 1 & $-228$ & $13.28\pm0.05$ \\
  PG1241+176   &  0.5504  &  400 & Mg\,II & 1 & $57$ & $14.60\pm0.05$ \\ 
     &  0.5584  &  400 & Mg\,II & 1 & $3$ & $14.26\pm0.10$ \\
     &  0.8954  &  400 & Mg\,II & 1 & $9$ & $< 13.81$ \\
  PG1248+401   &  0.7729  &  400 & Mg\,II & 1 & $2$ & $14.66\pm0.06$ \\ 
     &  0.8545  &  400 & Mg\,II & 1 & $94$ & $14.61\pm0.04$ \\ 
  PKS1302--102 &  0.0940   &  400 & D/H & 0 & \nodata & \nodata \\
  CSO873       &  0.66    &  400 & Mg\,II & 1 & $162$ & $14.27\pm0.02$ \\
  PG1634+706   &  0.701   &  400 & D/H, LLS & 0 & \nodata & \nodata \\
  \enddata
  \tablenotetext{a}{\citet{churchilletal00}} 
  \tablenotetext{b}{\citet{churchilletal99a}} 
  \tablenotetext{c}{$z_{abs} = 0.9254, 0.9276,$ and 0.9342}
  \tablecomments{Sightlines targeted for specific absorption
    systems. We quote the target absorption system redshift \zabs\
    from the proposal unless noted otherwise. By default, we excluded
    absorbers with $\vert \dvabs \vert \le 400\kms$, the clustering
    scale measured by \citet{churchillandvogt01}. The exceptions are
    taken from the literature.}
\end{deluxetable}

\subsection{\ion{C}{4} Doublets without \Lya\ Absorption}\label{subsec.wolya}

\begin{figure}[!hbt]
  \begin{center}
  \includegraphics[height=0.47\textwidth,angle=90]{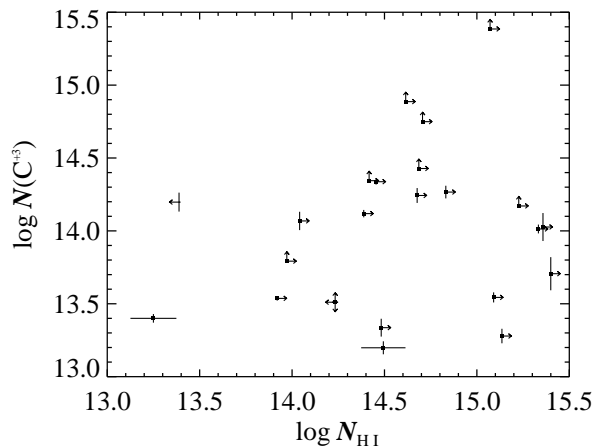}
  \end{center}
  \caption[\Cthr\ and \ion{H}{1} column densities.]  {\Cthr\ and
    \ion{H}{1} column densities. The symbols indicate the location or
    limit of all \ion{C}{4} doublets with both lines detected at $\EWr
    \ge 3\sigEWr$ and with \Lya\ coverage. When possible, the
    \ion{H}{1} column densities included the constraint given by \Lyb\
    detection.  The G = 1 sample are highlighted by the filled black
    square.  The upper limits are $2\sigma$.
    \label{fig.abund}
  }
\end{figure}

We conducted a blind survey for \ion{C}{4} doublets, relying initially
on the characteristic wavelength separation of the doublet. From the
resulting list of candidates, we drew on other known characteristics
of \ion{C}{4} doublets to distinguish the final sample (see \S\
\ref{sec.selec} and Table \ref{tab.flag}).  Although the detection of
associated \Lya\ absorption was a diagnostic and based on previous
observations \citep{ellisonetal99,simcoeetal04}, it was not
required.  \ion{C}{4} doublets without \Lya\ absorption have been
observed at $z>1$ \citep{schayeetal07}. Here we discuss the likelihood
that some \ion{C}{4} doublets in our sample were actually due to \Lya\
forest contamination (see \S\ \ref{subsec.lya}) and if we indeed
detect any ``naked'' \ion{C}{4} absorption.

Our sample includes two G = 2 \ion{C}{4} doublets without \Lya\
absorption detected ($\N{\rm H\,I} \le 10^{13.4}\cm{-2}$) that were in
the \Lya\ forest: $\zciv=0.83662$, $\logCIV = 14.2$ doublet toward
HS0747+4259 and $\zciv=0.87687$, $\logCIV \approx 14$ doublet in toward
HS0810+2554. In the scatter plot of \NCIV\ versus $N_{\rm H\,I}$, the
HS0747+4259 system has the highest \Cthr\ column density for any $\log
\N{\rm H\,I} < 14$ system (see Figure \ref{fig.abund}). The
HS0810+2554 system does not have a useful constraint on $\N{\rm H\,I}$
(see Table \ref{tab.civ}) and could not be plotted. Though these two
systems could be high-metallicity systems akin to those in
\citet{schayeetal07}, it is more likely that their atypical column
densities are due to their low signal-to-noise ratios, which increases
the uncertainties in continuum fitting and column densities. The
HS0747+4259 and HS0810+2554 E230M spectra have ${\rm S/N} = 6\,{\rm
  pix}^{-1}$ and $3\,{\rm pix}^{-1}$, respectively (see Table
\ref{tab.obssumm}).

The low S/N also increases the likelihood that \Lya\ forest
lines will be (incorrectly) identified as \ion{C}{4} doublets (see \S\
\ref{subsec.lya}).  In the most extreme scenario, the maximum
absorption pathlength in the \Lya\ forest is $\DXp=8.9$ for the strongest
Monte-Carlo absorbers ($\logCIV \ge 14.1$ or $\EWlin{1548} \ge
157\mA$). The estimated rate of \Lya\ forest lines masquerading as
\ion{C}{4} doublets was $\ud \Num/\ud X \le 0.11$ for $0.8 \le z \le
1.5$ (see Figure \ref{fig.dnlyadx}). Therefore, the maximum expected
number of \ion{C}{4}-like \Lya\ pairs would be $1^{+2}_{-1}$, which is
consistent with \Lya\ forest lines coinciding to be the two G = 2
doublets in question.

Next, we evaluate how many of the 25 G = 1+2 \ion{C}{4} doublets
detected in the \Lya\ forest might be false-positive
identifications. Two were discussed previously. There were 19 from the
``definitely \ion{C}{4}'' (G = 1) sample. In 15 of the forest
doublets, the detection of other associated transitions\footnote{The G
  = 1 $\zciv=0.81814$ system towards PG1634+706 is an example of a
  \ion{C}{4} doublet with associated transitions such as \Lya\ and the
  \ion{Si}{4} doublet.}  and/or multi-component profiles\footnote{For
  example, the G = 1 $\zciv=0.57632$ doublet towards PG0117+213
  clearly has a multi-component profile, with \ion{C}{4} absorption
  correlated for almost 200\kms.} lent credibility to the
identification as \ion{C}{4} absorption.

There were eight doublets in the \Lya\ forest that had no associated
transitions, no \Lya\ coverage, and nondescript profiles.\footnote{The
  G = 1 $\zciv = 0.74843$ doublet towards MARK132 has a nondescript
  profile.}  Therefore, there were fewer diagnostics available for
evaluating the doublet identification. We include these doublets in
our sample because we estimate the \Lya\ contamination with
Monte-Carlo simulations (\S\ \ref{subsec.lya}). There is a total of
ten systems in the \Lya\ forest that satisfy few of the \ion{C}{4}
characteristics (see Table \ref{tab.flag}): the eight doublets with
nondescript profiles and the two that lack significant \Lya\
absorption, discussed previously. The simulations of the \Lya\ forest
contamination rate excludes all ten from being \Lya\ forest
contamination at the 99.9\% confidence level.

Ultimately, there were no definite detections of \ion{C}{4} doublets
without associated \Lya\ absorption.

\section{Summary}\label{sec.summ}

We conducted the largest survey for $\zciv < 1$ \ion{C}{4} absorbers
to date. We surveyed 49 sightlines with the \hst\ \stis\ and/or \ghrs\
archival spectra with moderate signal-to-noise ratios and
resolution. All absorption-line features and actual absorption lines
were identified by an automated feature-finding algorithm. From this
list, candidate \ion{C}{4} systems were assembled, based solely on
the doublet's characteristic wavelength separation (see Table
\ref{tab.cand}). 

We visually inspected all candidates with rest equivalent widths
$\EWlin{1548} \ge 3\sigEWlin{1548}$.  After considering the various
diagnostics (\eg profile; see Table \ref{tab.flag}), we identified 44
definite (G = 1) and 19 likely (G = 2) \ion{C}{4} systems. Of these,
we only analyzed the 38 G = 1 and five G = 2 doublets where both lines
were detected with $\EWr \ge 3\sigEWr$. All subsequent analyses
considered the full sample and two redshift bins, divided at,
approximately, the median redshift: $\zciv < 0.6$ (20 G = 1+2
doublets) and $0.6 \le \zciv < 1$ (23).

From synthetic spectra, we estimated the unblocked, co-moving
pathlength \DXp\ to which our survey was $95\%$ complete as a function
of \NCIV\ and \EWlin{1548} (see Figure \ref{fig.x}). For the strongest
absorbers ($\logCIV \ge 14.1$ and $\EWlin{1548} \ge 157\mA$), $\DXp =
15.1$. 

There were ten G = 1+2 doublets detected in the \Lya\ forest that had
nondescript profiles, no associated metal lines, and either no \Lya\
absorption or \Lya\ coverage did not exist. We estimated the
contamination rate that \Lya\ forest lines would be mistaken for \ion{C}{4}
doublets, with Monte-Carlo simulations. The rate is small: $\ud N/\ud X
\le 0.11$.  Therefore, it was ruled out at the $99.9\%$ confidence level
that all of the ten doublets were \Lya-as-\ion{C}{4} pairs. However,
the two G = 2 doublets without \Lya\ absorption ($\EWlin{\Lya} <
3\sigEWr$) were consistent with being \Lya\ forest lines masquerading
as \ion{C}{4} doublets.

With a maximum-likelihood analysis, we modeled the column density
frequency distribution with a power law: $\ff{\NCIV} =
\kff{14}(\N{\Cthr}/\N{0})^{\aff{N}}$. The best-fit exponent was
$\aff{N}=-1.75^{+0.28}_{-0.33}$ for the $\zciv < 0.6$ sample. For the $0.6
\le \zciv < 1$ sample, the best-fit value was
$\aff{N}=-1.39^{+0.24}_{-0.27}$. There has been no consensus on the
power-law exponent in the various $\zciv > 1.5$ surveys
\citep{ellisonetal00, songaila01, boksenbergetal03ph, songaila05}.

We measured the \ion{C}{4} absorber line density to be $\dNCIVdX =
4.1^{+0.7}_{-0.6}$ for $\logCIV \ge 13.2$ and $\EWlin{1548} \ge 52\mA$
from the full G = 1+2 sample. \dNCIVdX\ has not evolved significantly
since $z=5$ (see Figure \ref{fig.dndx_evo}).

The sightlines analyzed in the current study were observed for a
variety of reasons: \eg the targets were UV bright or a known damped
\Lya\ system (DLA) lay along the sightline. The latter reason, and
similar ones, might have introduced a bias into our survey, since
sightlines with DLAs, Lyman-limit systems, and \ion{Mg}{2} absorbers
typically correlate with \ion{C}{4} doublets. Assuming that
\citet{barlowandtytler98} provide an unbiased survey for \ion{C}{4}
absorbers, we first compared our measured redshift densities for $0.2
< \zciv < 0.8$ and $\EWr > 400\mA$. We agreed within $1\sigma$ and
actually measured a lower incidence of strong absorbers.

We also excluded the 11 doublets close to the redshift of the targeted
systems (see Table \ref{tab.selec}), then re-measured the absorber
line density and re-fit the frequency distributions. The ``unbiased''
absorber line density agreed less well with
\citet[][$1.7\sigma$]{barlowandtytler98}. The best-fit \ff{\NCIV}\
exponent and coefficient differed by $<1\sigma$, respectively,
compared to the values for the full sample. Categorically excluding
the \ion{C}{4} doublets possibly associated with \eg \ion{Mg}{2}
absorbers biased our survey against strong $\logCIV > 14$
doublets. Therefore, we concluded that the results for the full
\ion{C}{4} sample represented the results of a truly unbiased survey,
akin to \citet{barlowandtytler98}.

Measuring the \Cthr\ mass density relative to the critical density at
$\zciv < 1$ was a principle aim of this study.  High-redshift studies
have agreed that \OmCIV\ changed little or not at all from $z = 5
\rightarrow 1$, once cosmology was properly taken into account
\citep{songaila01, boksenbergetal03ph, schayeetal03, pettinietal03,
  songaila05, scannapiecoetal06}. For the full G = 1 sample with
$\langle z \rangle = 0.654$ ($t_{age}=7.5\,{\rm Gyr}$), the integrated
$\OmCIV = (6.20^{+1.82}_{-1.52})\times10^{-8}$ for $13 \le \logCIV \le
15$. This was a $2.8\pm0.7$ increase over the $1 < z < 5$ values.

We assumed a simple linear model for the temporal evolution of \OmCIV\
in order to estimate the rate of change in the mass density.  The
linear regression for the $z < 5$ data indicated a slowly increasing
\OmCIV\ with increasing age of the Universe, at the
$>3$-$\sigma$ level: $\ud \OmCIV/\ud t_{age} =
(0.42+0.2)\times10^{-8}\,{\rm Gyr}^{-1}$. This result relied on the
measurements at $z<1$; without which, the slope was unconstrained ($\ud
\OmCIV/\ud t_{age} = (0.15+0.3)\times10^{-8}\,{\rm
  Gyr}^{-1}$). 

Intuitively, it might not appear surprising that \OmCIV\
continually increases from $z = 5 \rightarrow 0$, since the overall
metallicity of the Universe increases. However, increasing metallicity
would not translate directly to increasing \Cthr\ mass density. The
\ion{C}{4} mass density is also subject to the changing ionizing background
in the Universe, which favors the \ion{C}{4} transition {\it less} at
low redshift \citep[see \eg][]{madauetal09}.
\citet{oppenheimeranddave08} concluded that metallicity and ionizing
background balanced one another from $z \approx 5 \rightarrow 1$,
resulting in a nearly constant \OmCIV.

We emphasized that the simple linear model for \OmCIV\ over $t_{age}$
was not physically motivated. The {\it observed} \Cthr\ mass density
only accounted for doublets with $13 \le \logCIV \le 15$, and
absorbers in this range likely arise in different physical
environments at high and low redshift. We discussed the changing
nature of what \ion{C}{4} absorption traces; perhaps it probed the
low-density IGM at $z>1$ and galaxy halos at $z<1$. We will explore,
in a future paper, the physical conditions and environments of
\ion{C}{4} absorbers in cosmological simulations with various feedback
prescriptions.


\acknowledgements We thank P. Jonsson for programming advice on the Monte-Carlo simulations and the maximum-likelihood analysis. We also thank B. Oppenheimer for useful discussions about the evolution of \OmCIV\ and the effects of cosmology.  Based on observations made with the NASA-CNES-CSA \emph{Far Ultraviolet Spectroscopic Explorer}. \fuse\ is operated for NASA by the Johns Hopkins University under NASA contract NAS5-32985.  Based on observations made with the \linebreak[4]NASA/ESA \emph{Hubble Space
  Telescope} Space Telescope Imaging Spectrograph and Goddard
High-Resolution Spectrograph, obtained from the
data archive at the Space Telescope Institute. STScI is operated by
the association of Universities for Research in Astronomy, Inc. under
the NASA contract NAS 5-26555.
The current study was funded by the HST archival grant 10679 and the
NSF CAREER grant AST 05\_48180.

{\it Facilities:} \facility{FUSE}, \facility{HST (STIS)},
\facility{HST (GHRS)}

\bibliographystyle{apj} \bibliography{../civabsorbers}

\appendix 
\section{Velocity Plots}\label{appdx.plots}

\begin{figure}[h!]
  \begin{center}$
    \begin{array}{cc}
      \includegraphics[width=0.45\textwidth]{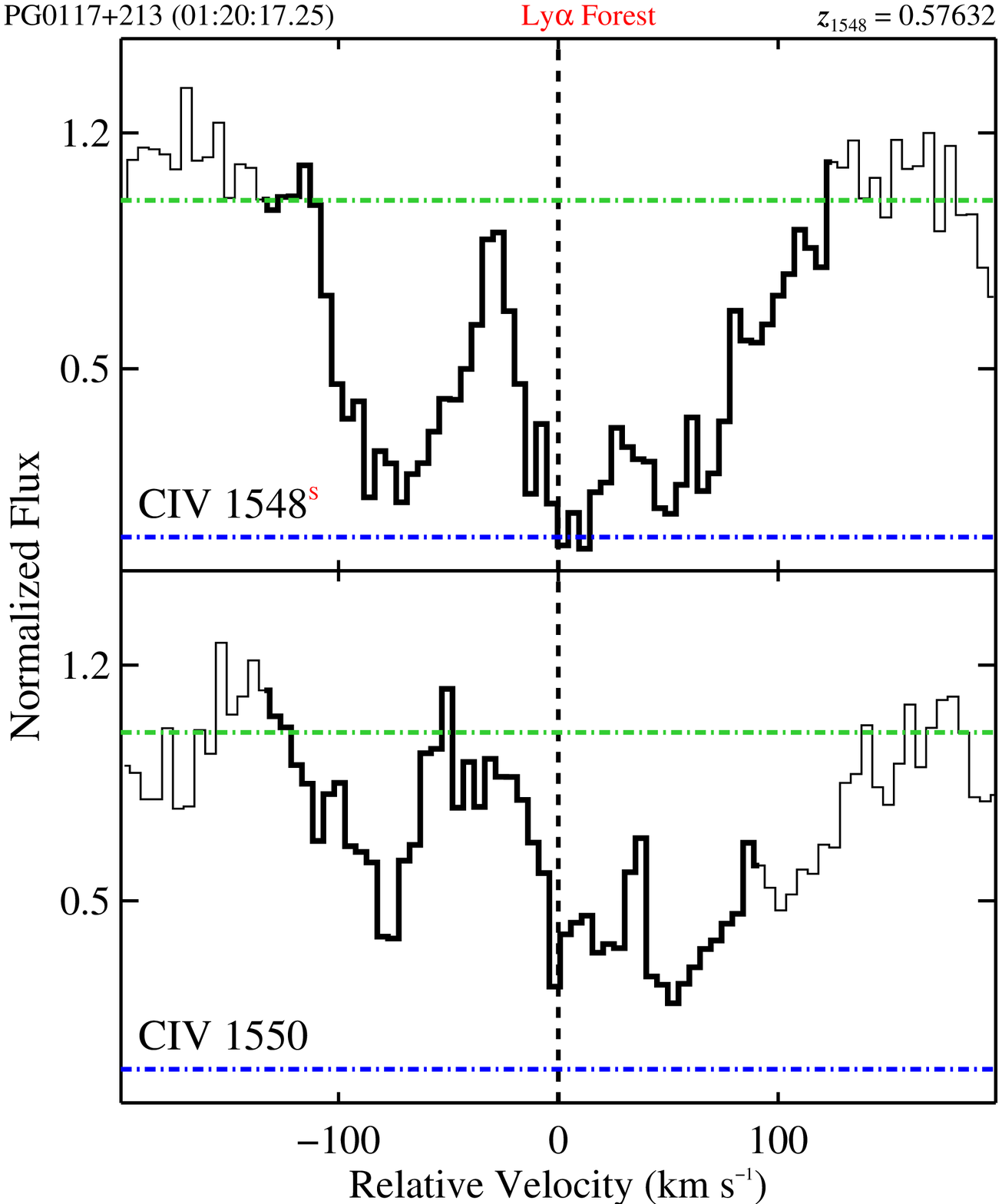} & 
      \includegraphics[width=0.45\textwidth]{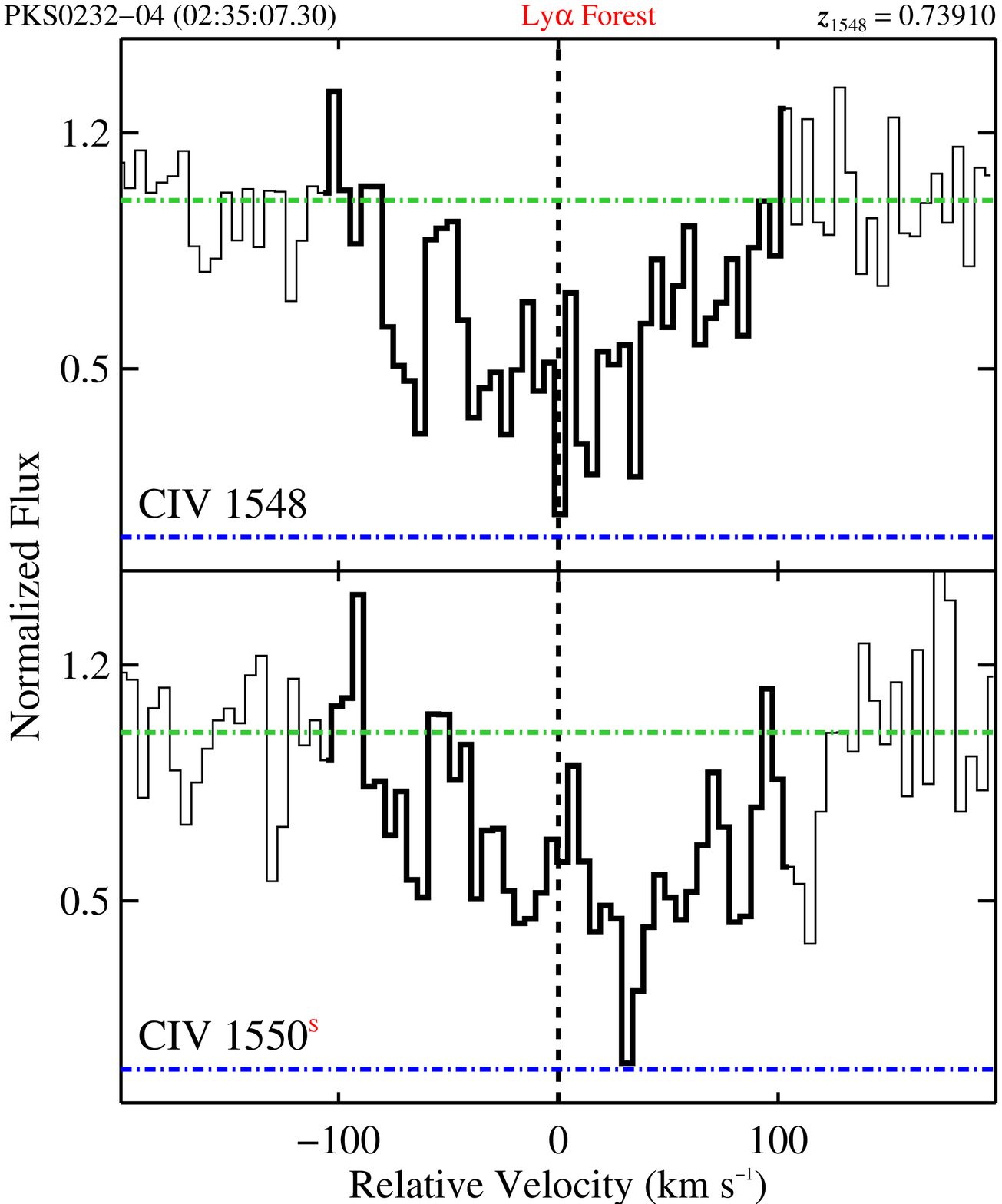} 
    \end{array}$                                      
  \end{center}
  \caption[Velocity plots of G = 1 \ion{C}{4} systems.]  {Velocity
    plots of G = 1 \ion{C}{4} systems.  The regions of spectra around
    each absorption line are aligned in velocity space with respect to
    the rest wavelength of the transition and $\zciv$. Saturated
    transitions are indicated with the (red) `S'; transitions with
    $\NCIV < 3\sigNCIV$ are indicated with the (red) `W.'  The regions
    used to measure \EWr\ and \logN\ are shown by the dark
    outline. The flux at zero and unity are shown with the dash-dot
    lines (blue and green, respectively); the (black) vertical dashed
    line indicates $v=0\kms$.
    \label{fig.g1appdx}
  }
\end{figure}
\addtocounter{figure}{-1}

\begin{figure}[!hbt]
  \begin{center}$
    \begin{array}{cc}
      \includegraphics[width=0.45\textwidth]{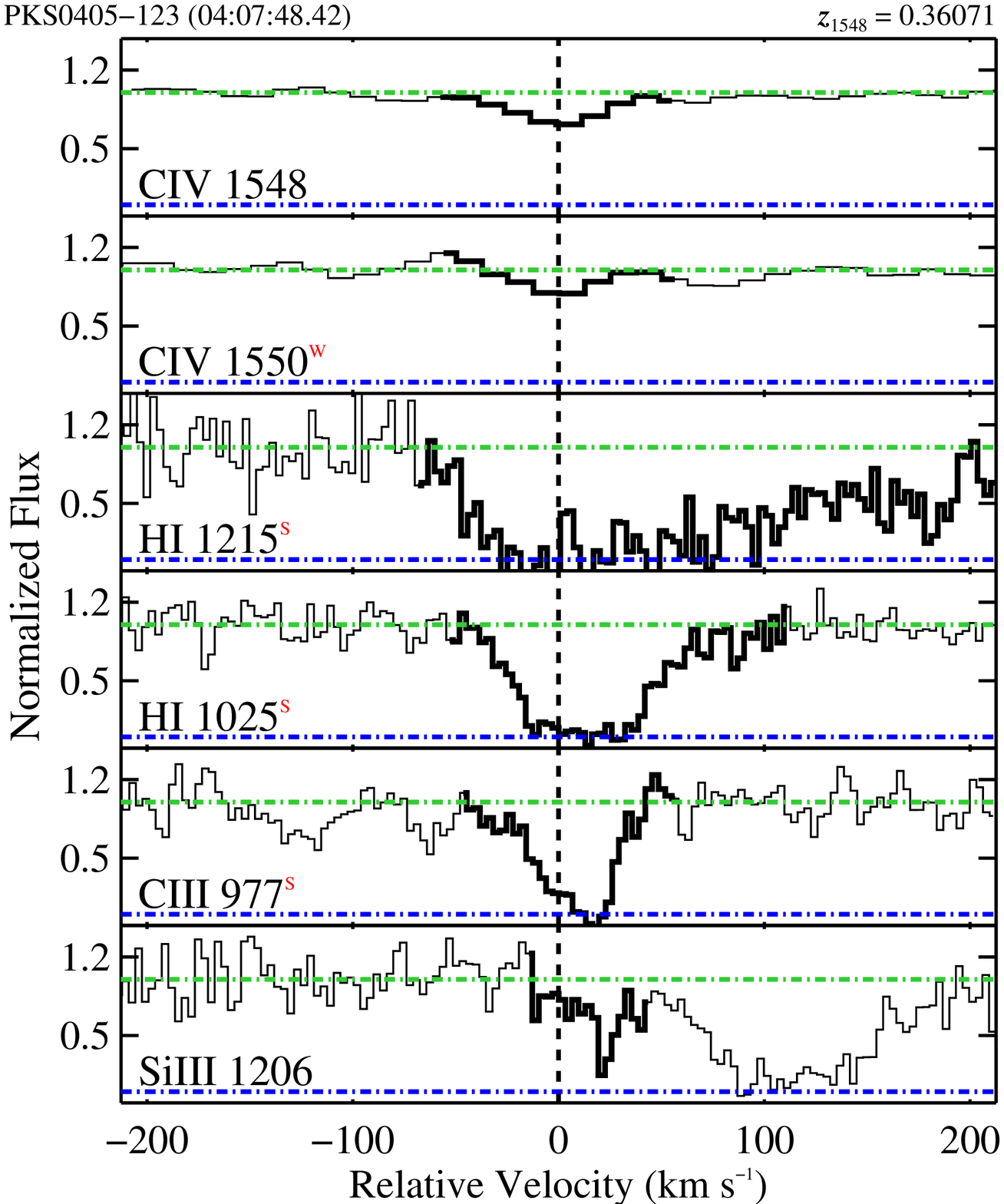} & 
      \includegraphics[width=0.45\textwidth]{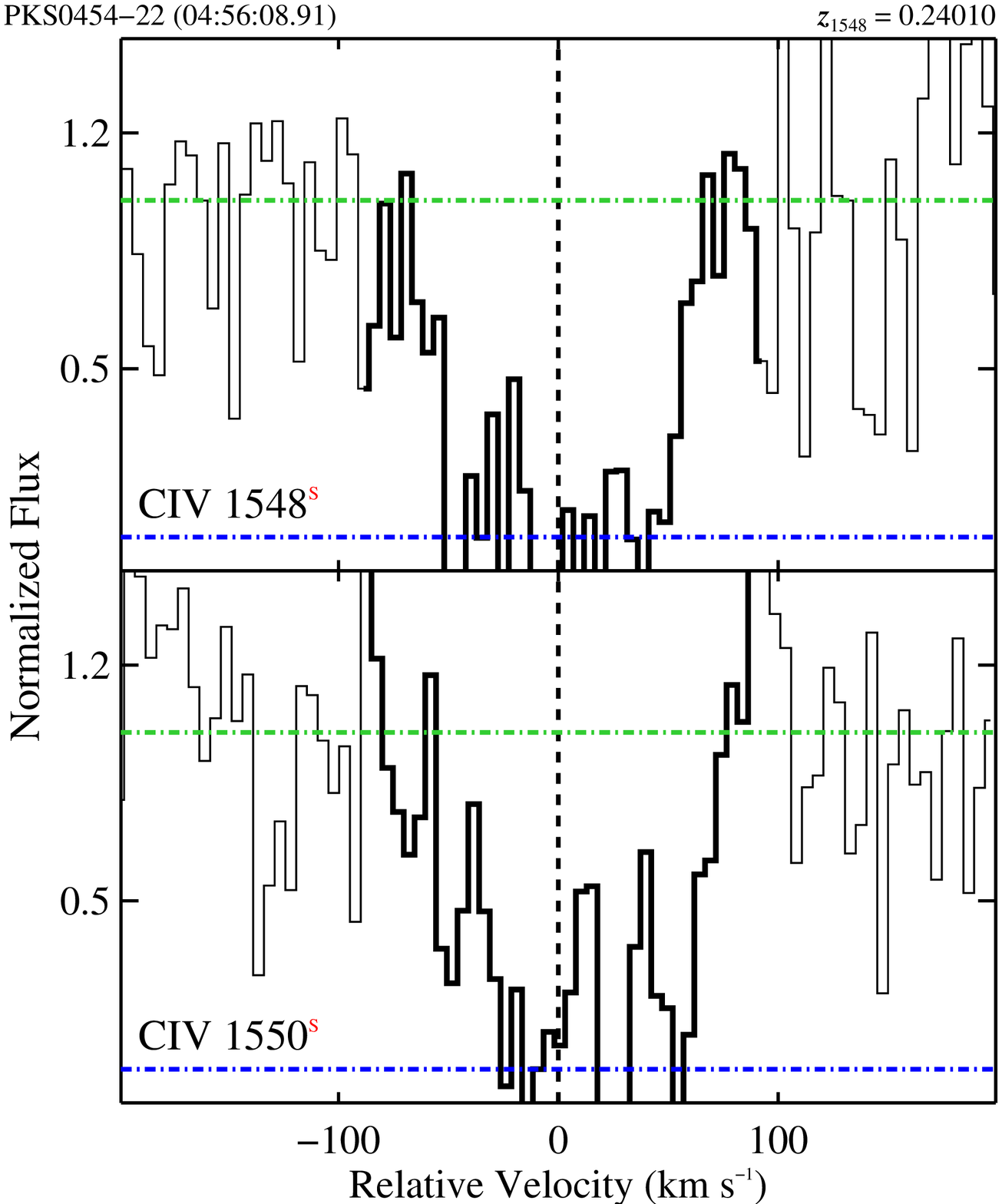} \\
      \includegraphics[width=0.45\textwidth]{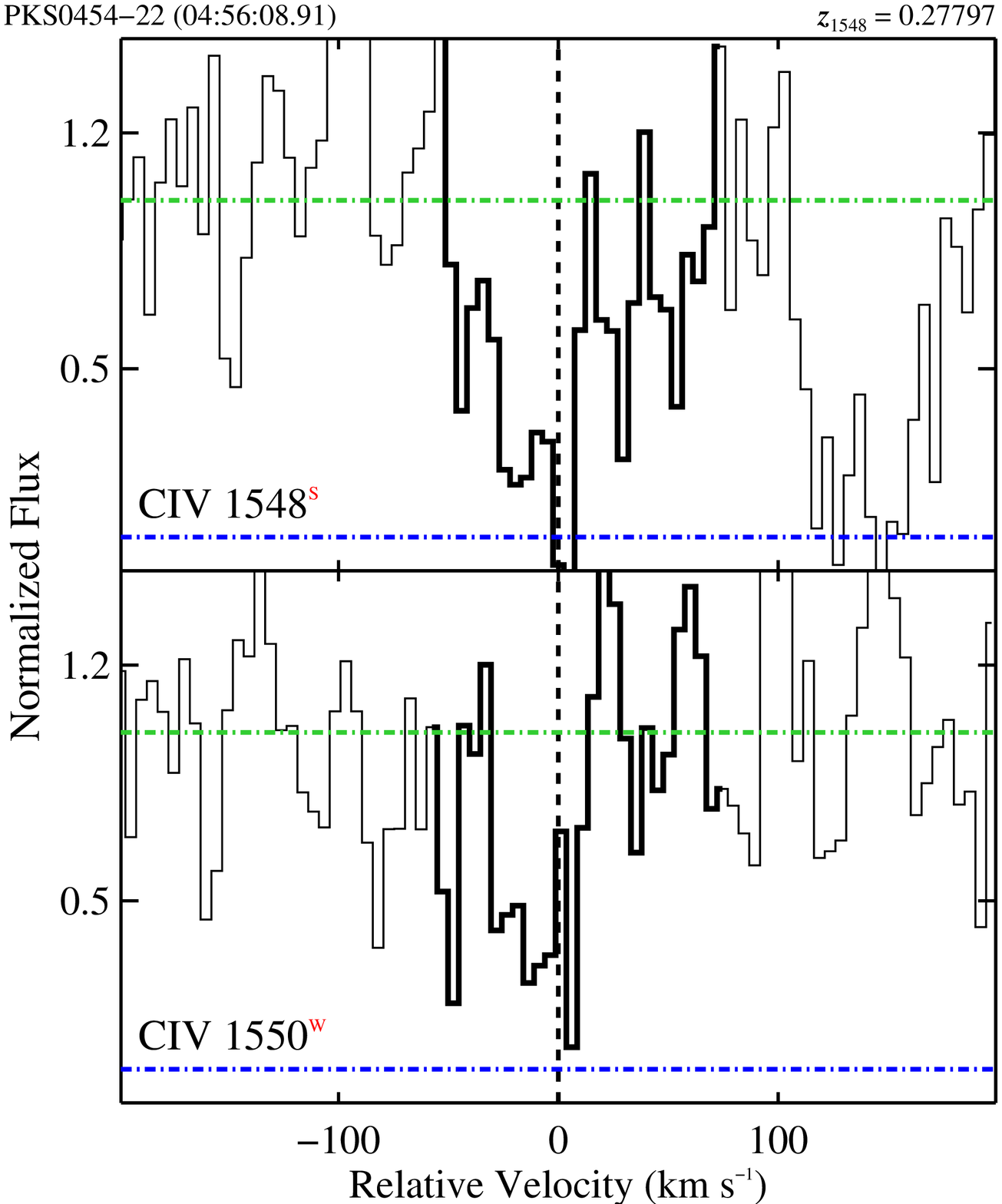} & 
      \includegraphics[width=0.45\textwidth]{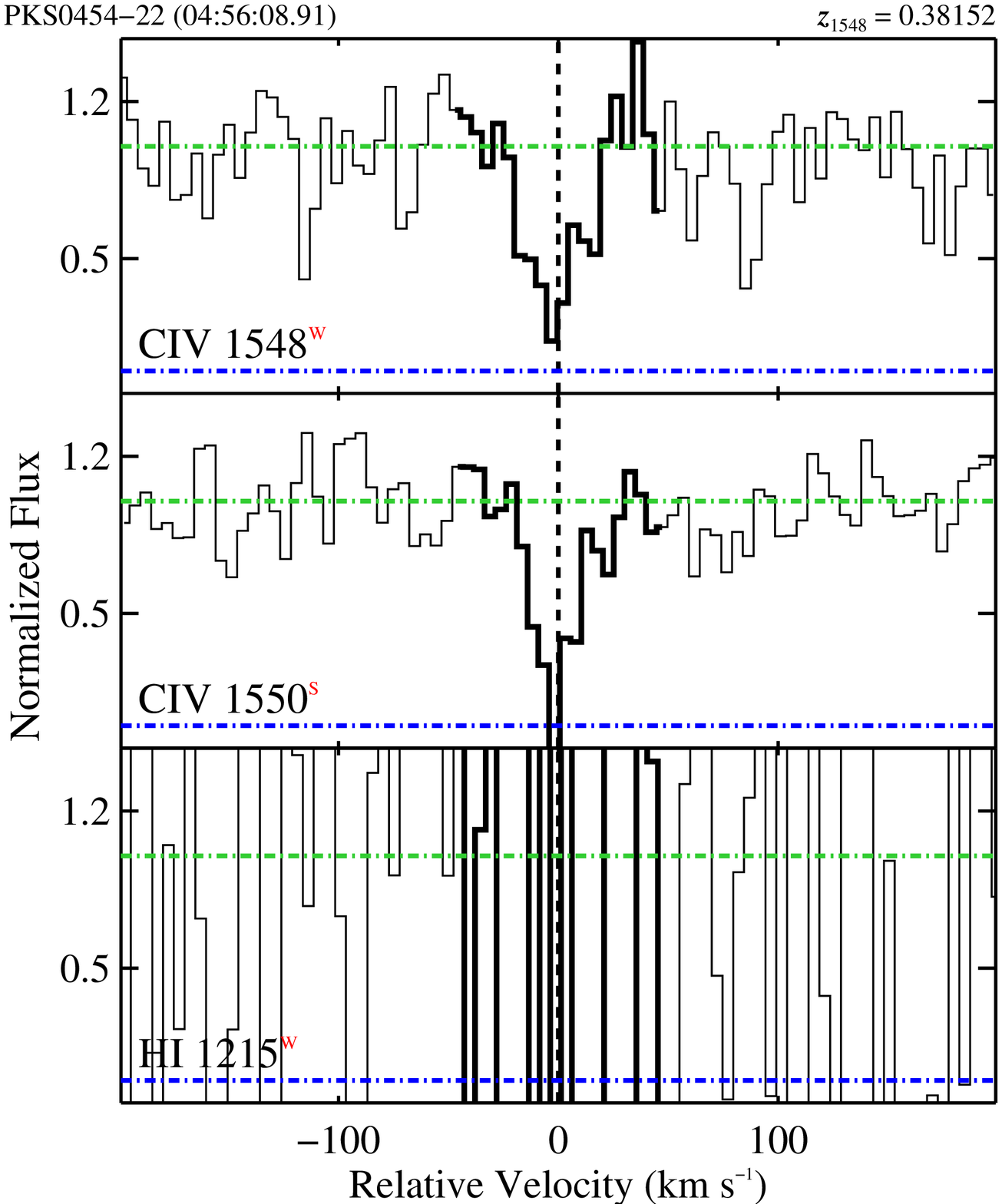} 
    \end{array}$                                      
  \end{center}
  \caption{G = 1 velocity plots (continued) }
\end{figure}
\addtocounter{figure}{-1}

\begin{figure}[!hbt]
  \begin{center}$
    \begin{array}{cc}
      \includegraphics[width=0.45\textwidth]{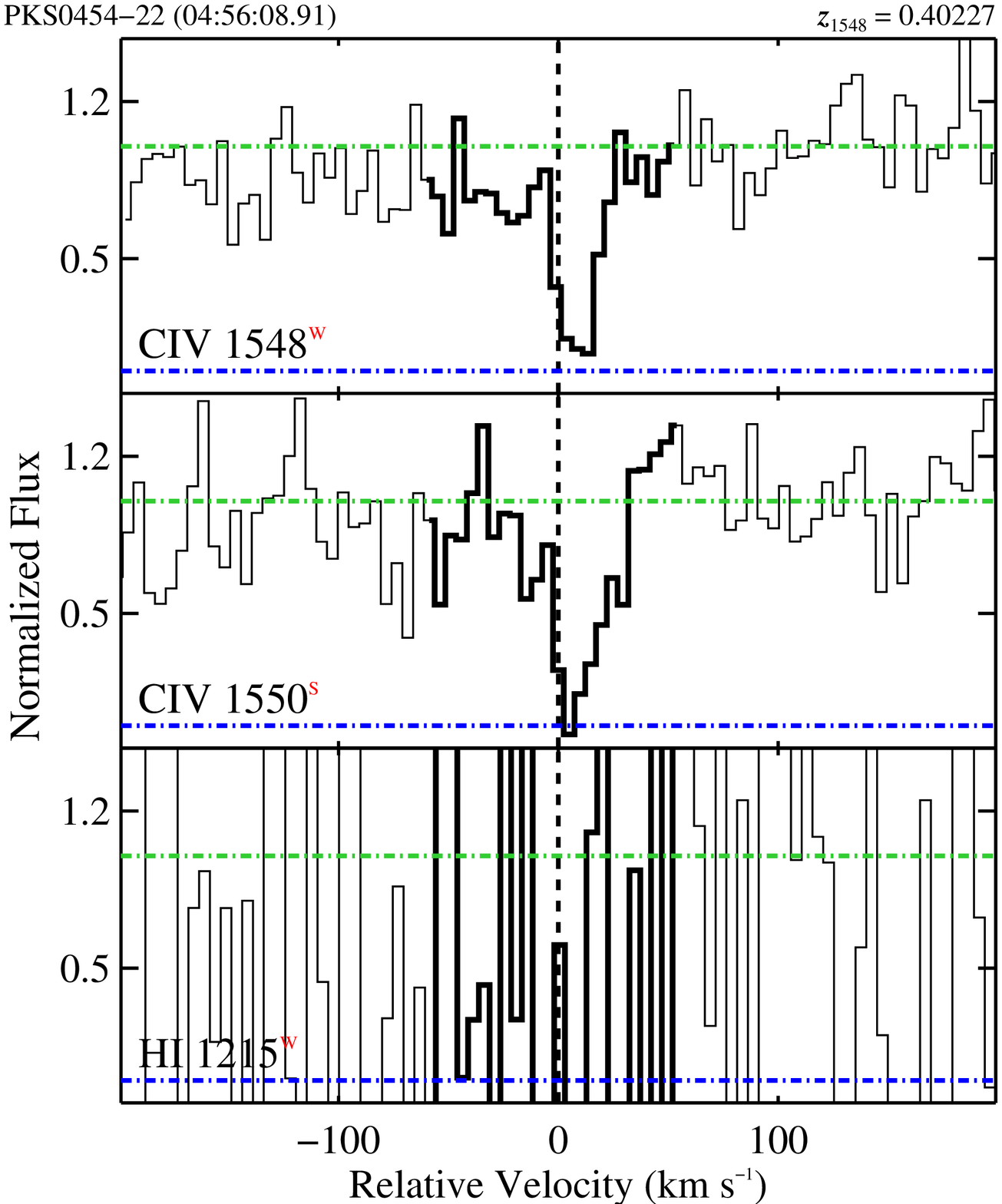} & 
      \includegraphics[width=0.45\textwidth]{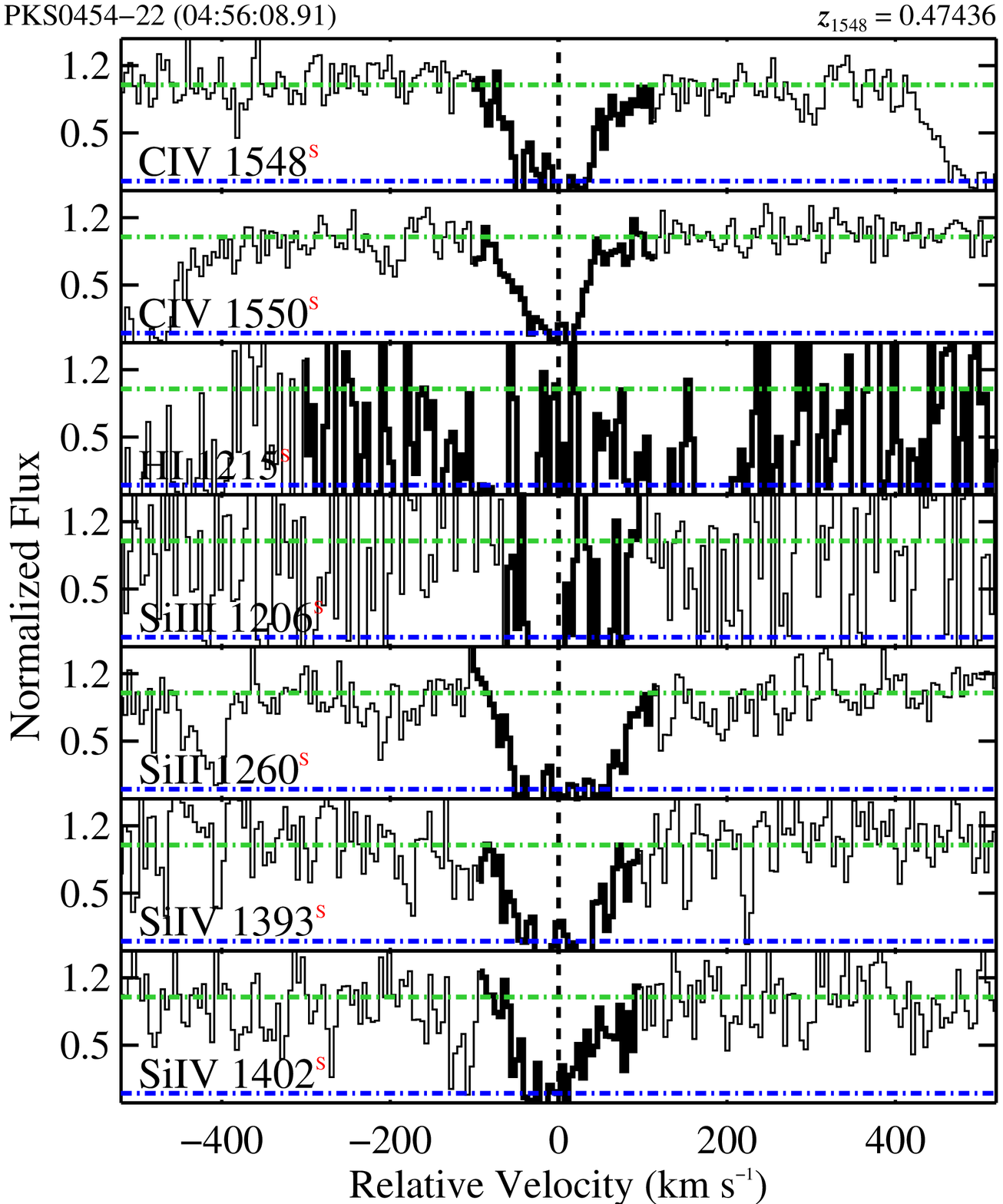} \\
      \includegraphics[width=0.45\textwidth]{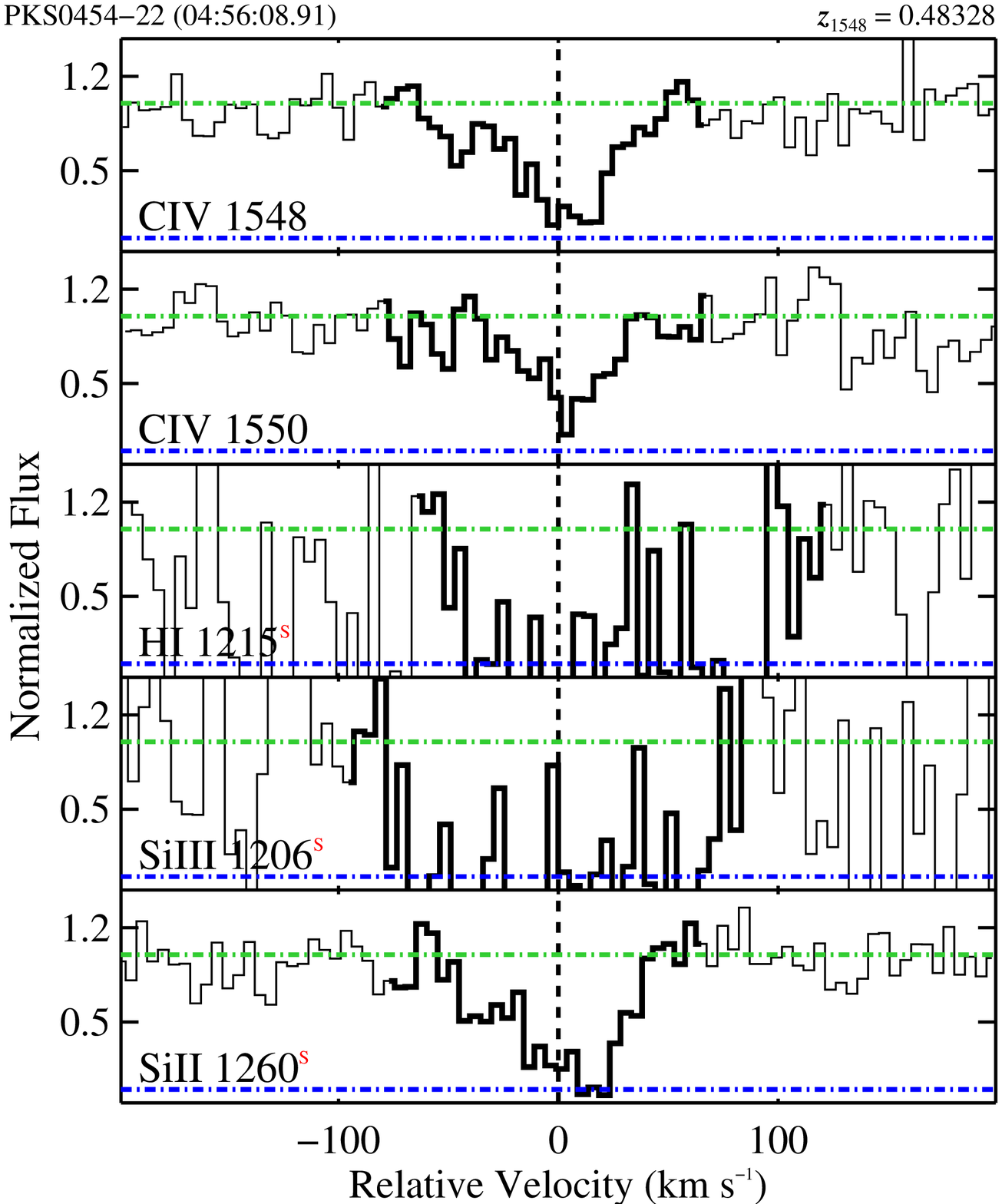} & 
      \includegraphics[width=0.45\textwidth]{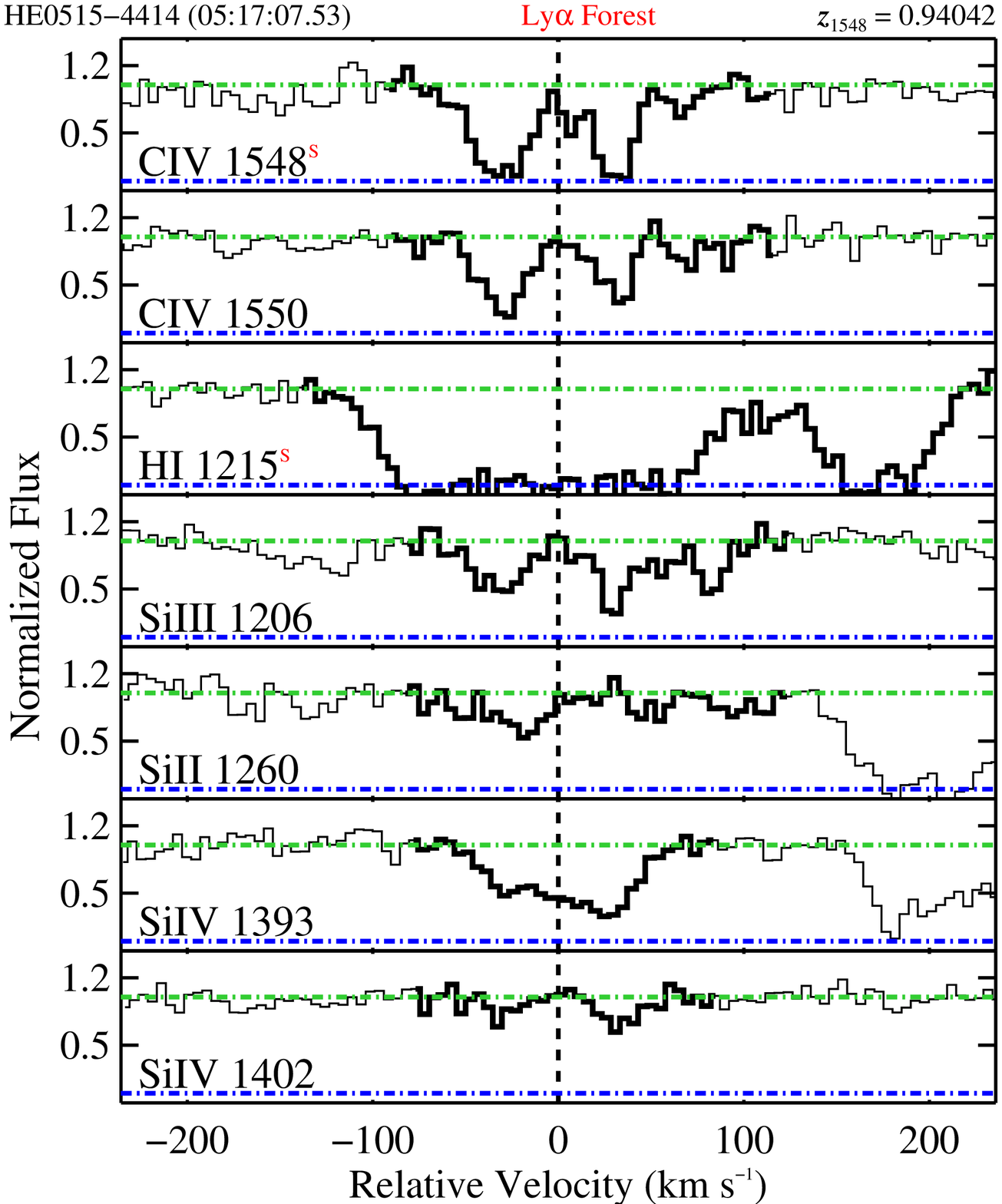} 
    \end{array}$                                      
  \end{center}
  \caption{G = 1 velocity plots (continued) }
\end{figure}
\addtocounter{figure}{-1}

\begin{figure}[!hbt]
  \begin{center}$
    \begin{array}{cc}
      \includegraphics[width=0.45\textwidth]{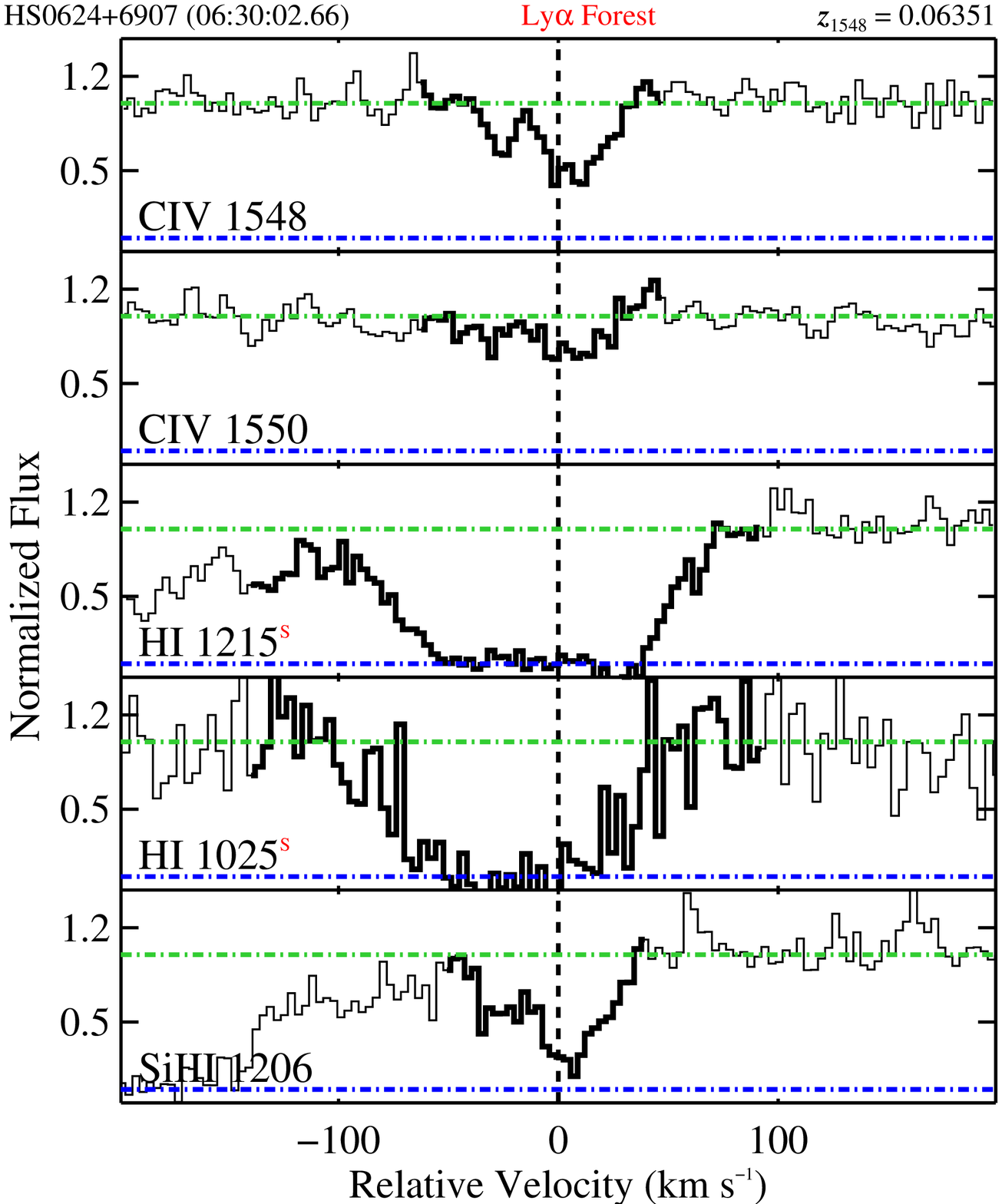} & 
      \includegraphics[width=0.45\textwidth]{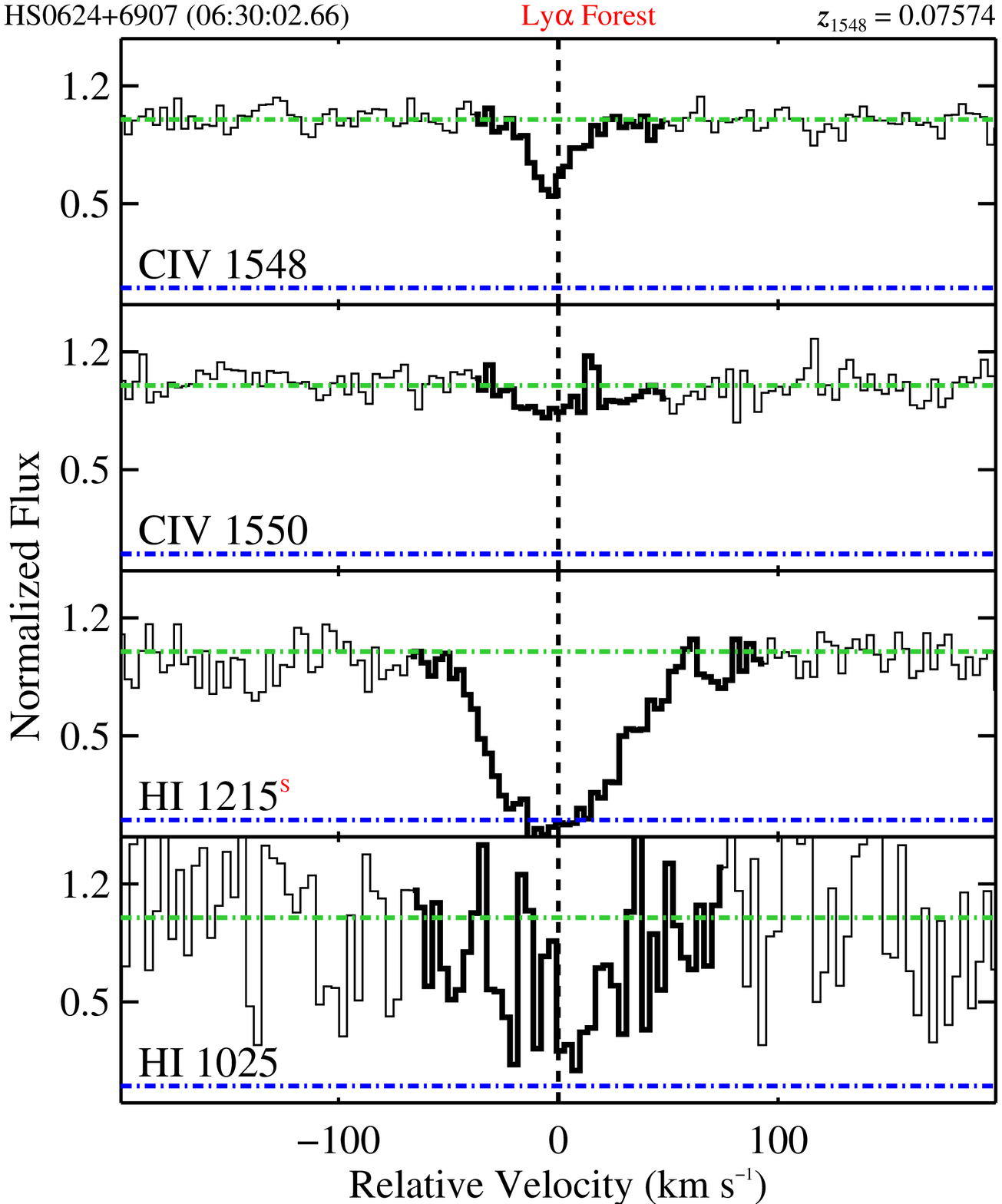} \\
      \includegraphics[width=0.45\textwidth]{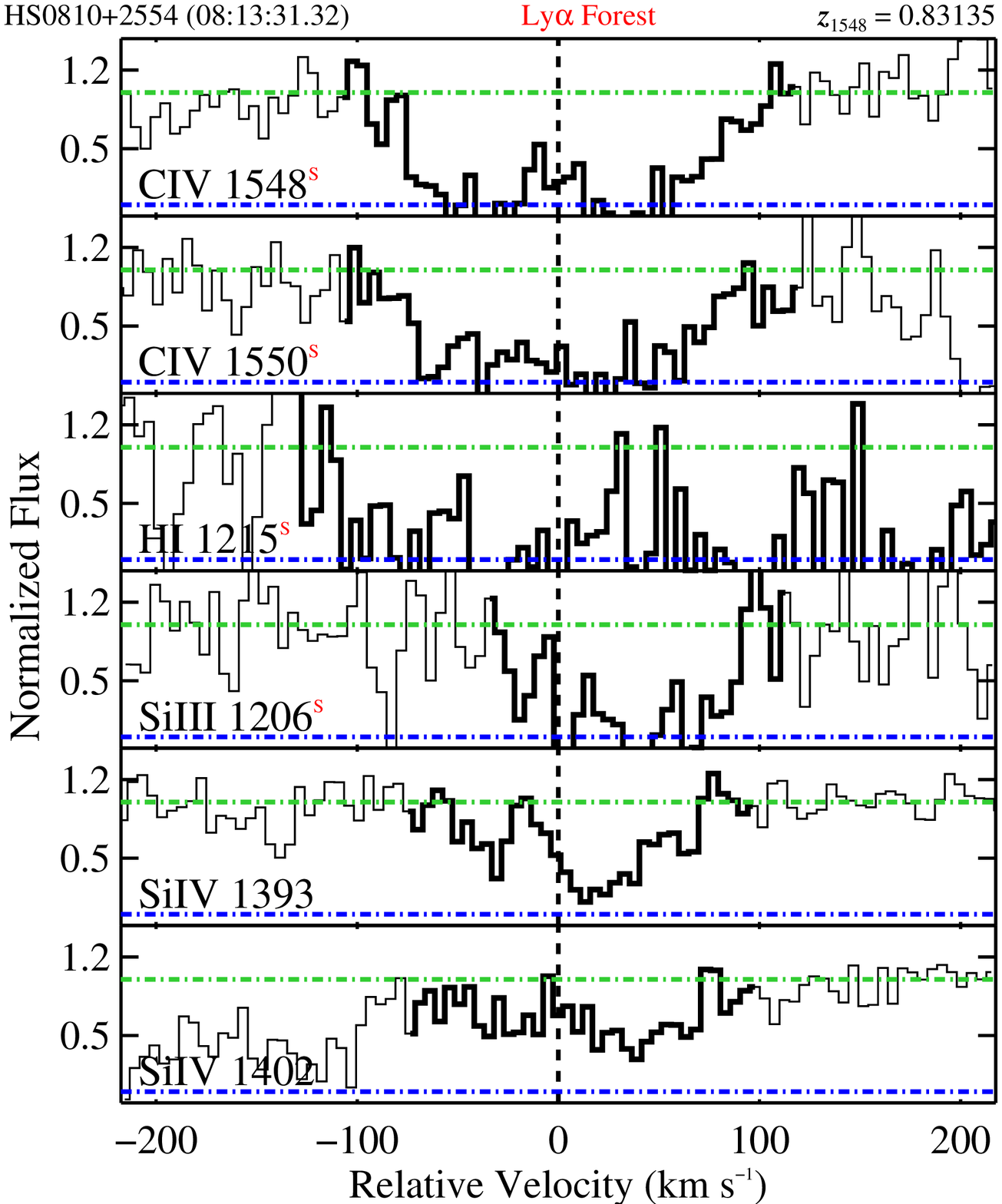} & 
      \includegraphics[width=0.45\textwidth]{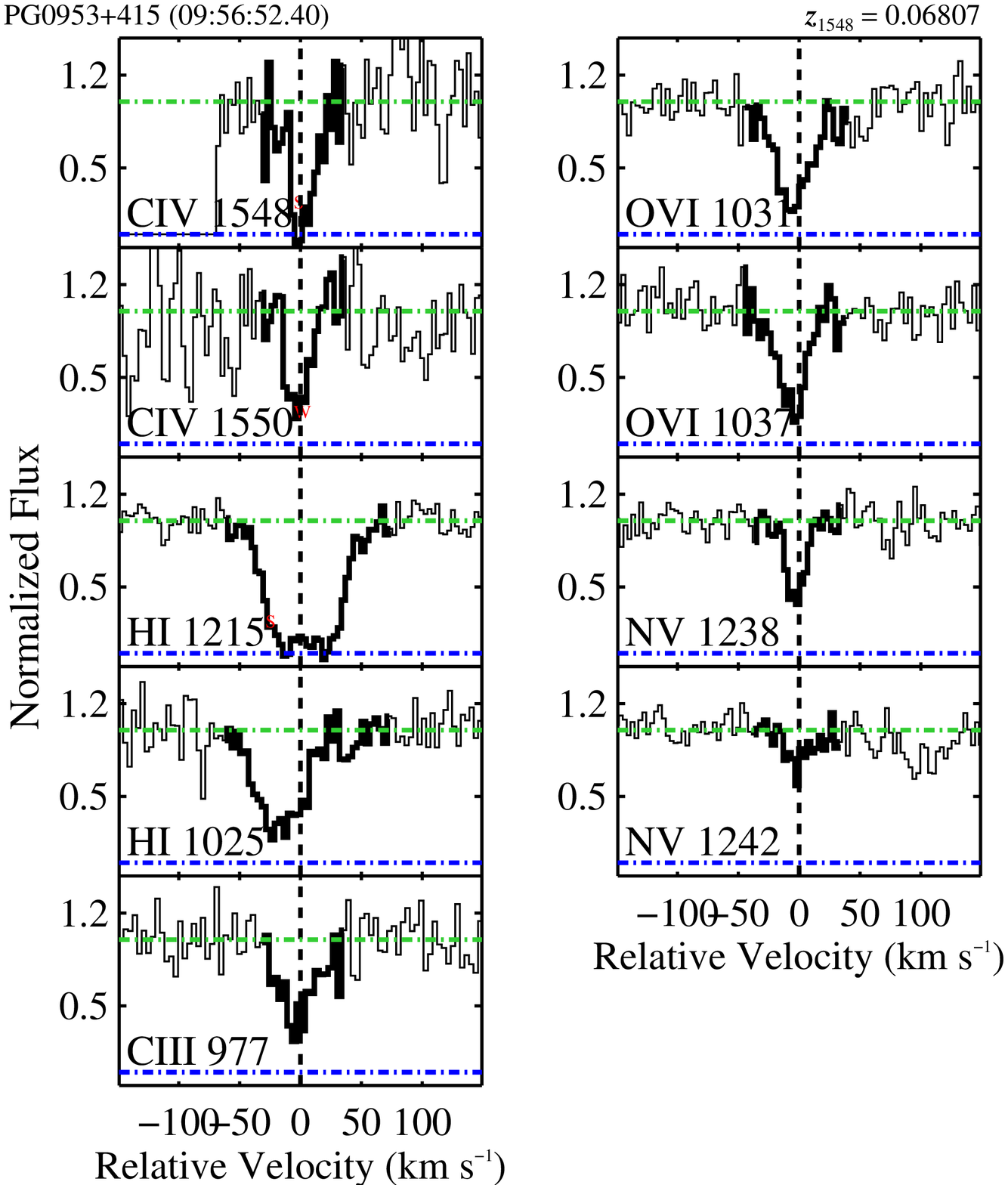} 
    \end{array}$                                      
  \end{center}
  \caption{G = 1 velocity plots (continued) }
\end{figure}
\addtocounter{figure}{-1}

\begin{figure}[!hbt]
  \begin{center}$
    \begin{array}{cc}
      \includegraphics[width=0.45\textwidth]{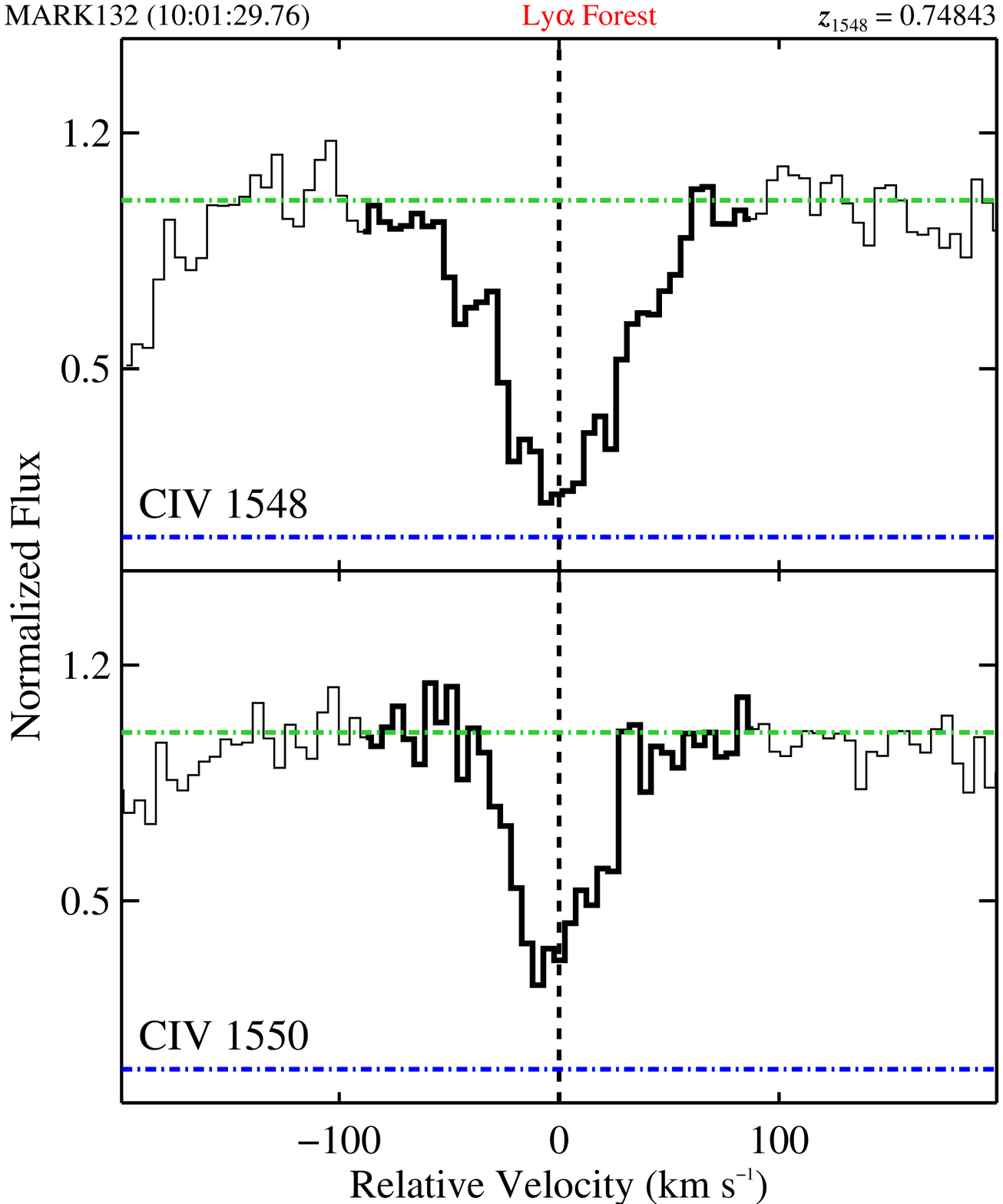} & 
      \includegraphics[width=0.45\textwidth]{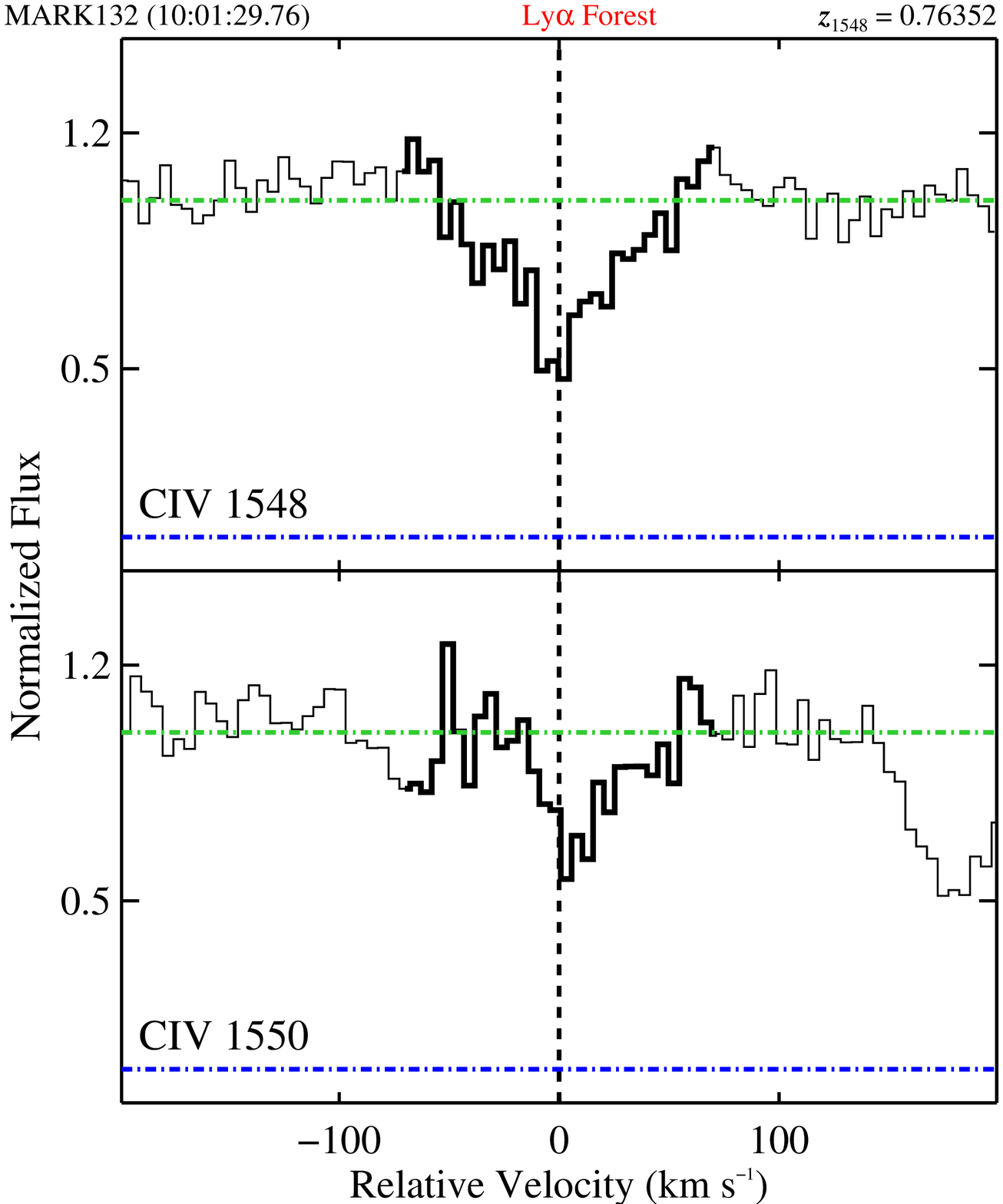} \\
      \includegraphics[width=0.45\textwidth]{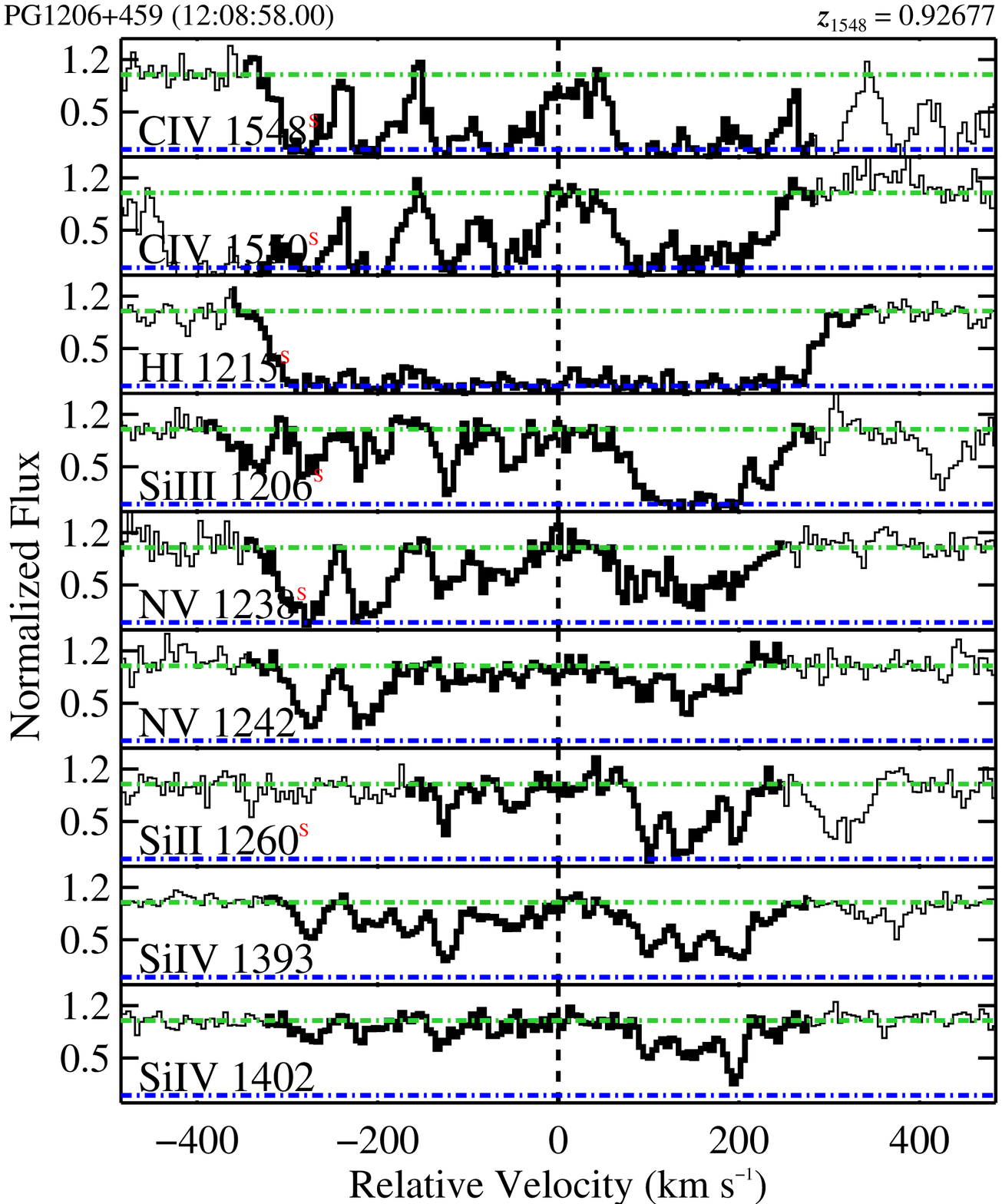} & 
      \includegraphics[width=0.45\textwidth]{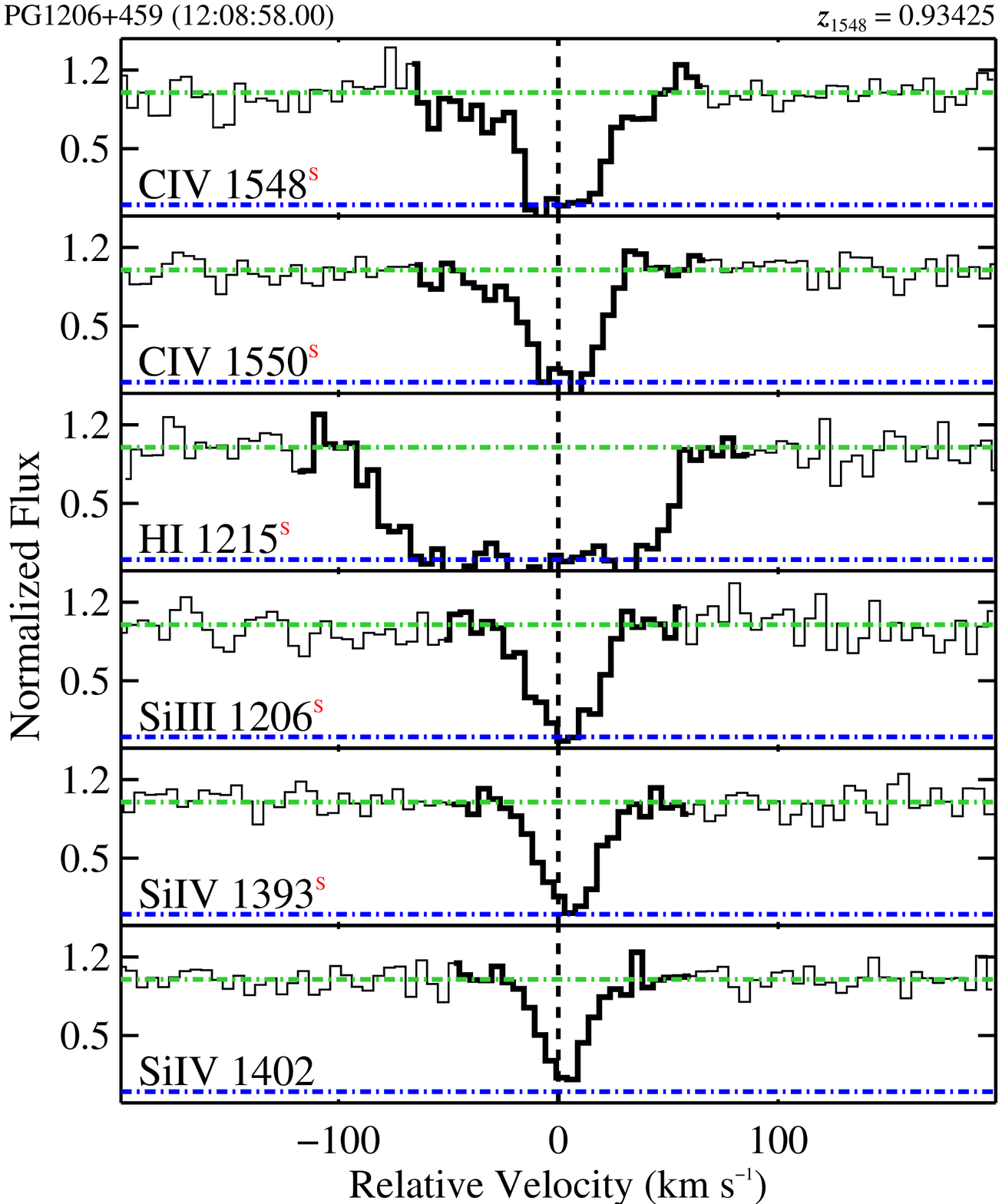} 
    \end{array}$                                      
  \end{center}
  \caption{G = 1 velocity plots (continued) }
\end{figure}
\addtocounter{figure}{-1}

\begin{figure}[!hbt]
  \begin{center}$
    \begin{array}{cc}
      \includegraphics[width=0.45\textwidth]{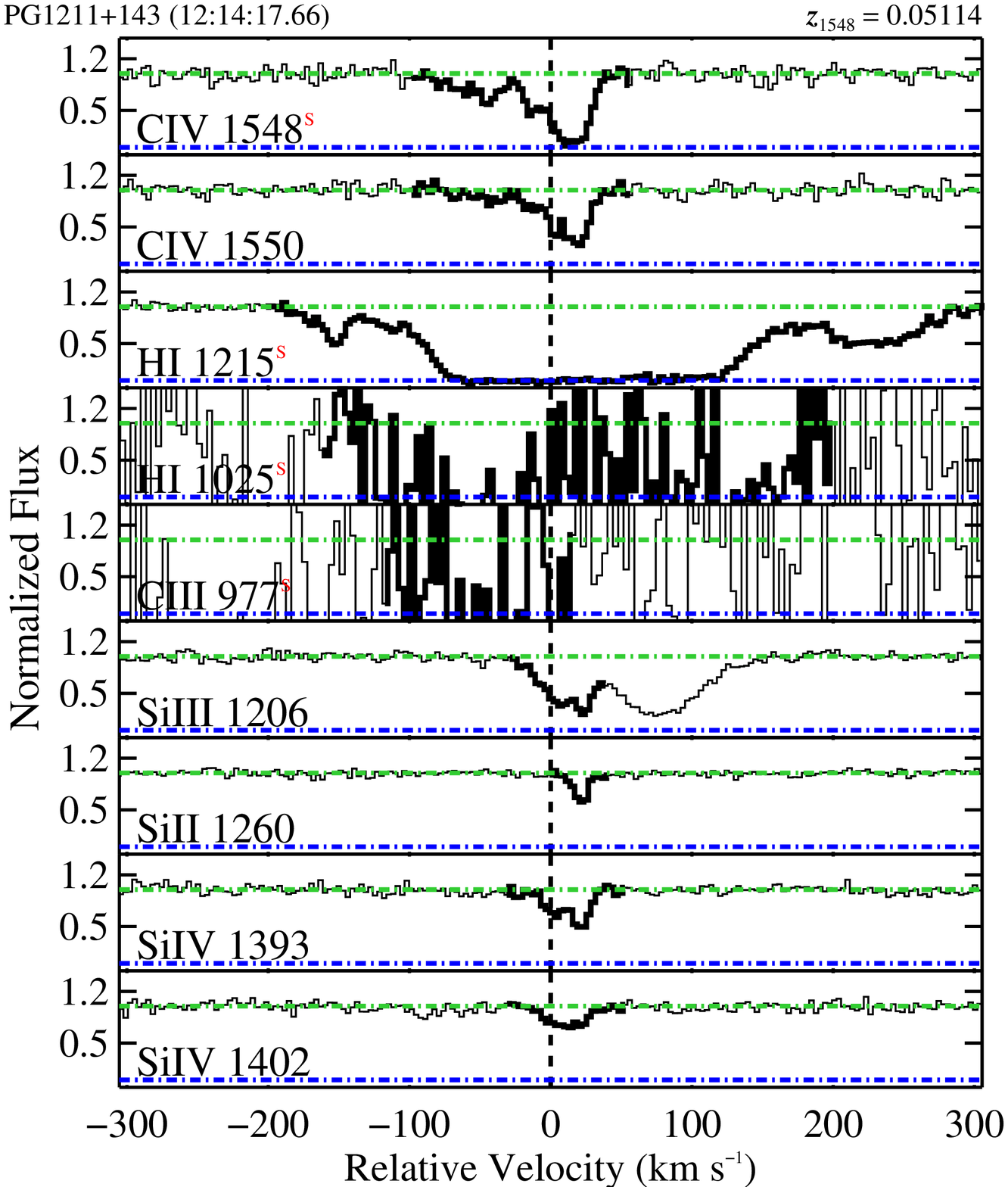} & 
      \includegraphics[width=0.45\textwidth]{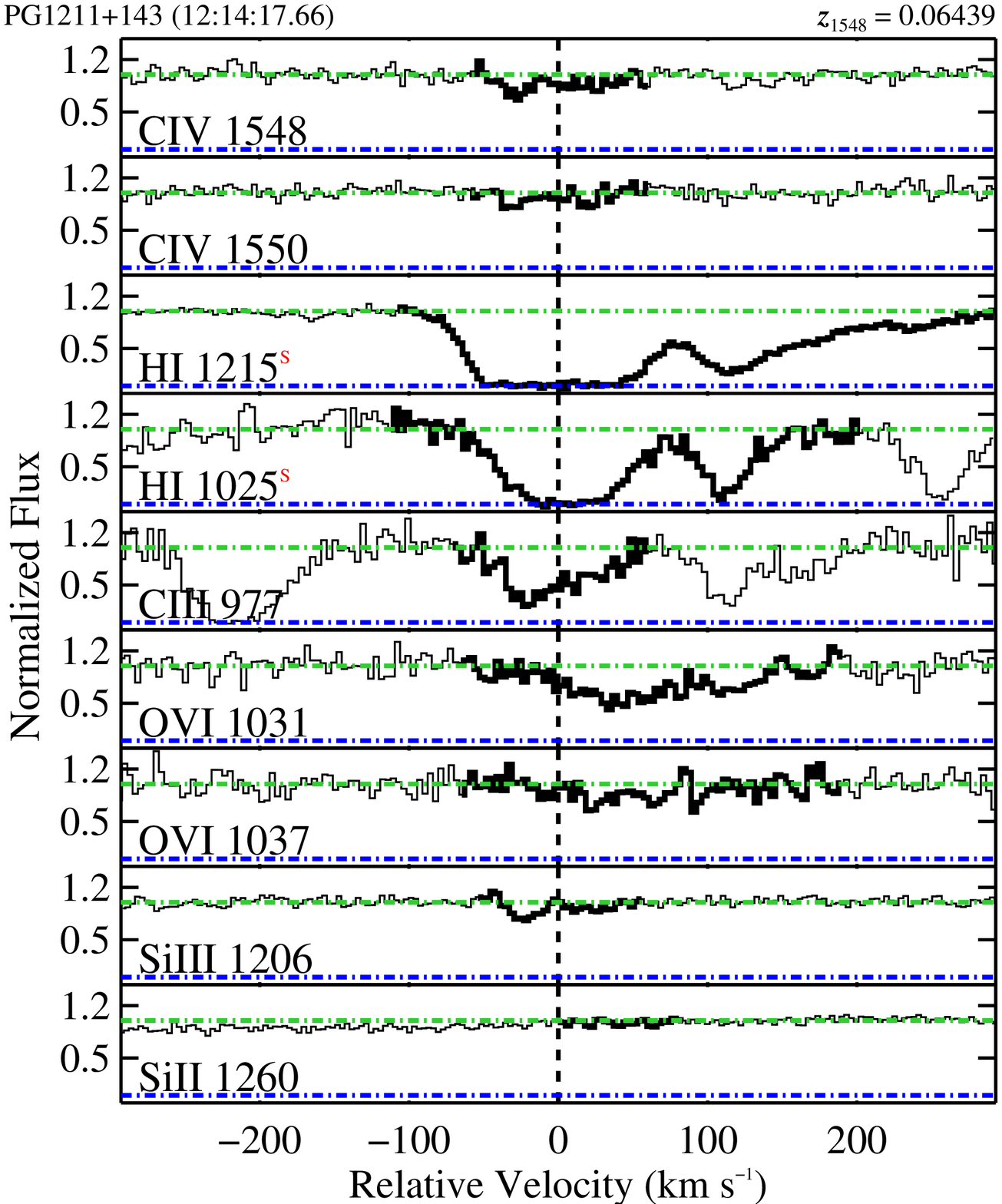} \\
      \includegraphics[width=0.45\textwidth]{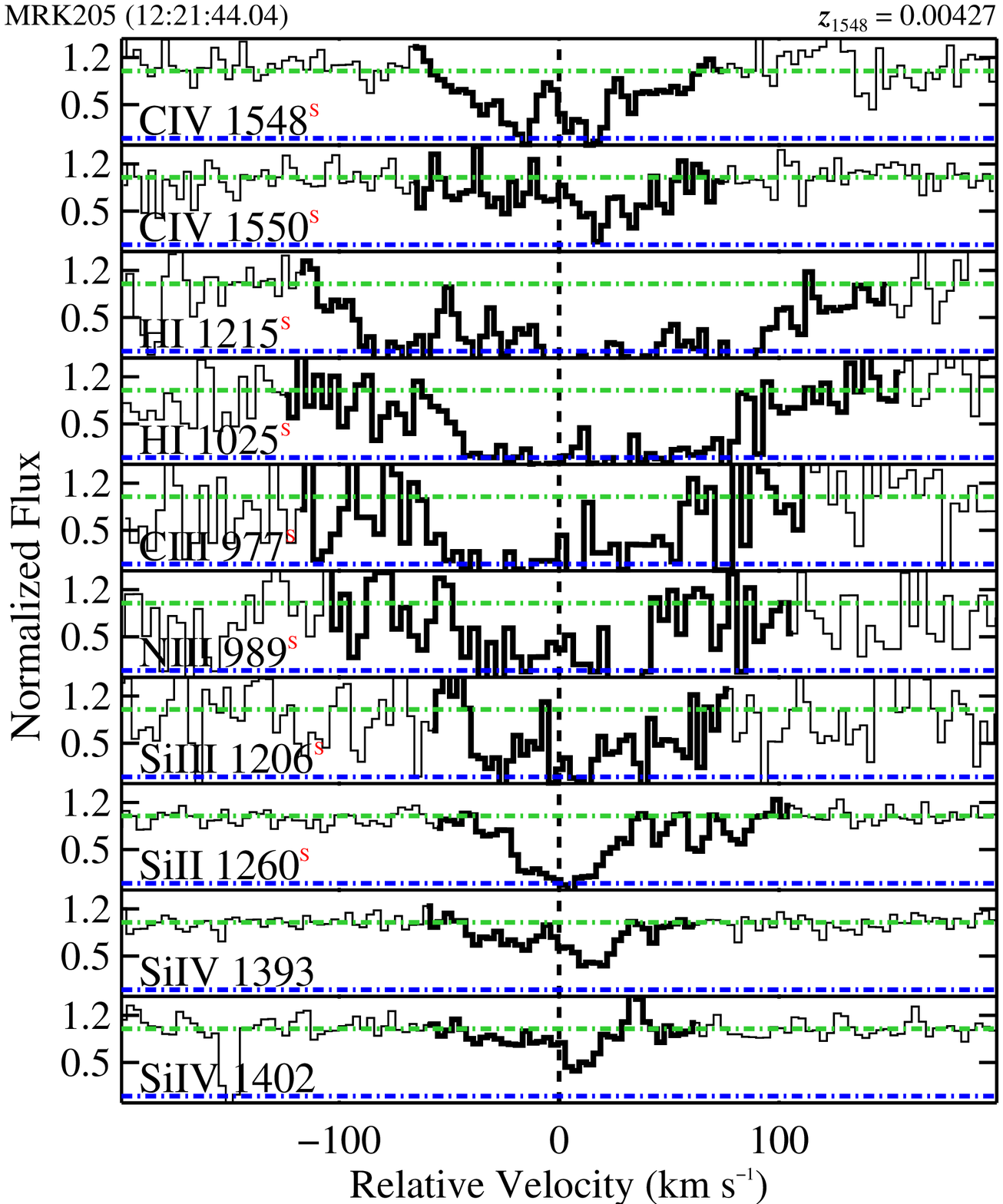} & 
      \includegraphics[width=0.45\textwidth]{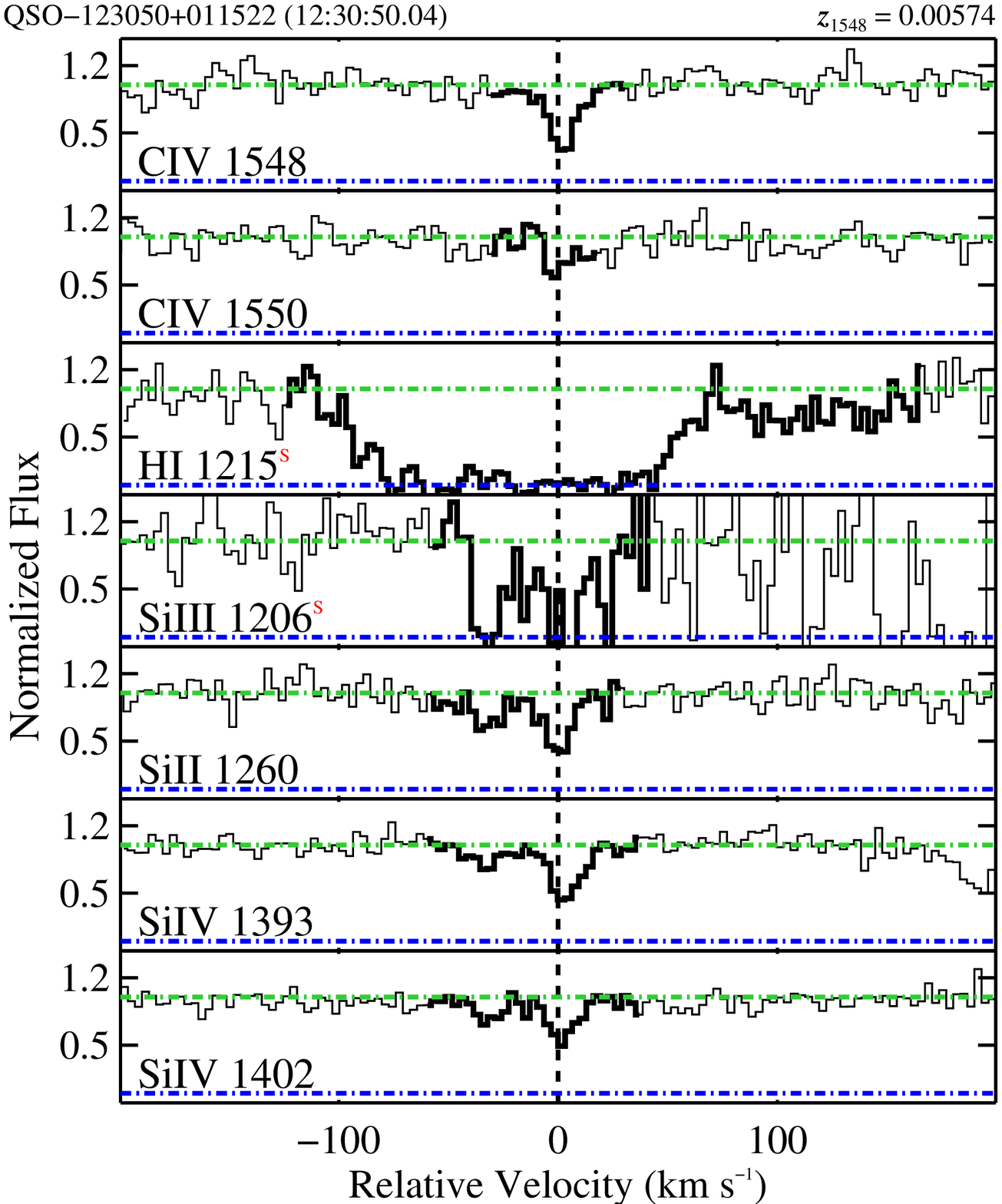} 
    \end{array}$                                      
  \end{center}
  \caption{G = 1 velocity plots (continued) }
\end{figure}
\addtocounter{figure}{-1}

\begin{figure}[!hbt]
  \begin{center}$
    \begin{array}{cc}
      \includegraphics[width=0.45\textwidth]{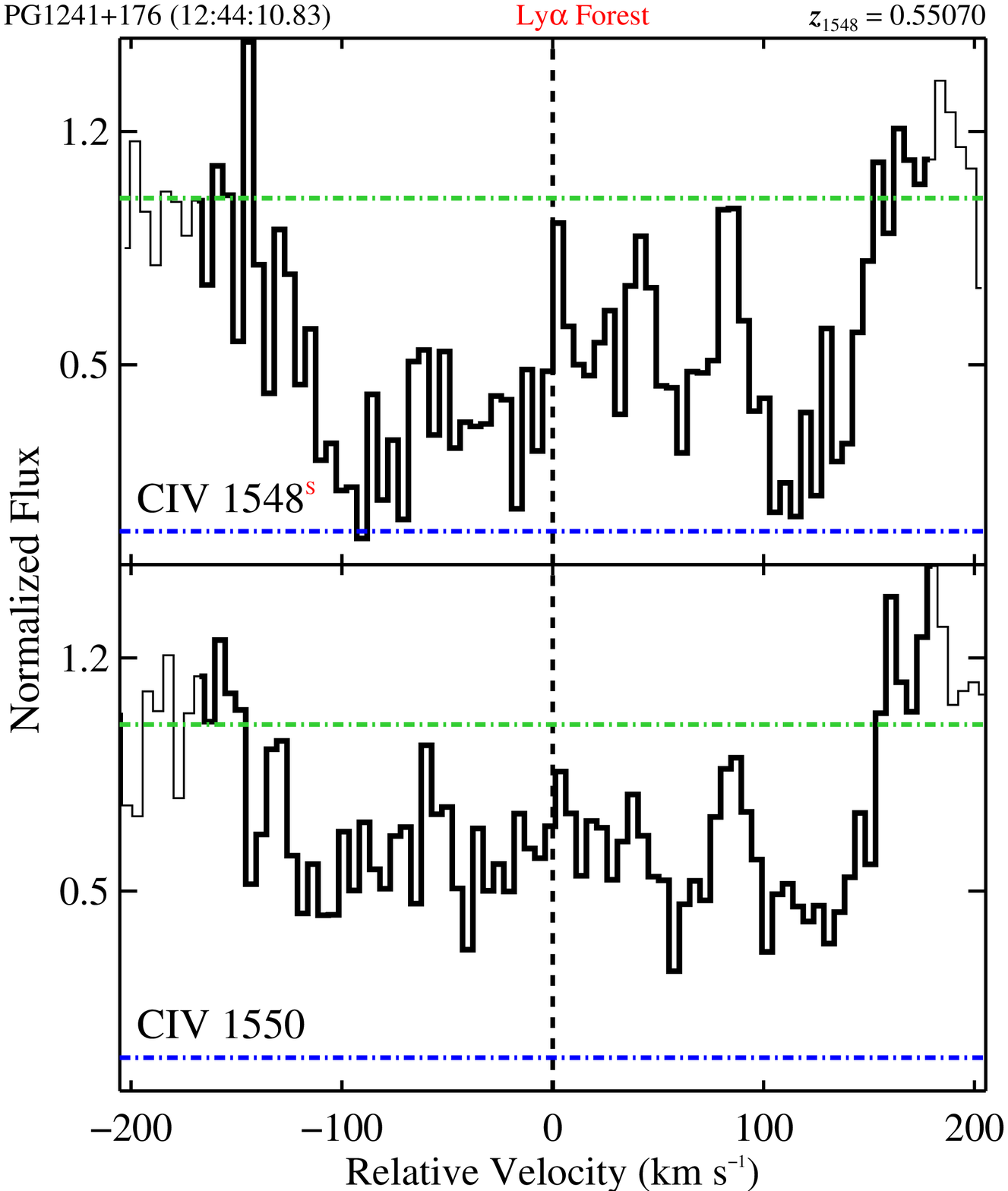} & 
      \includegraphics[width=0.45\textwidth]{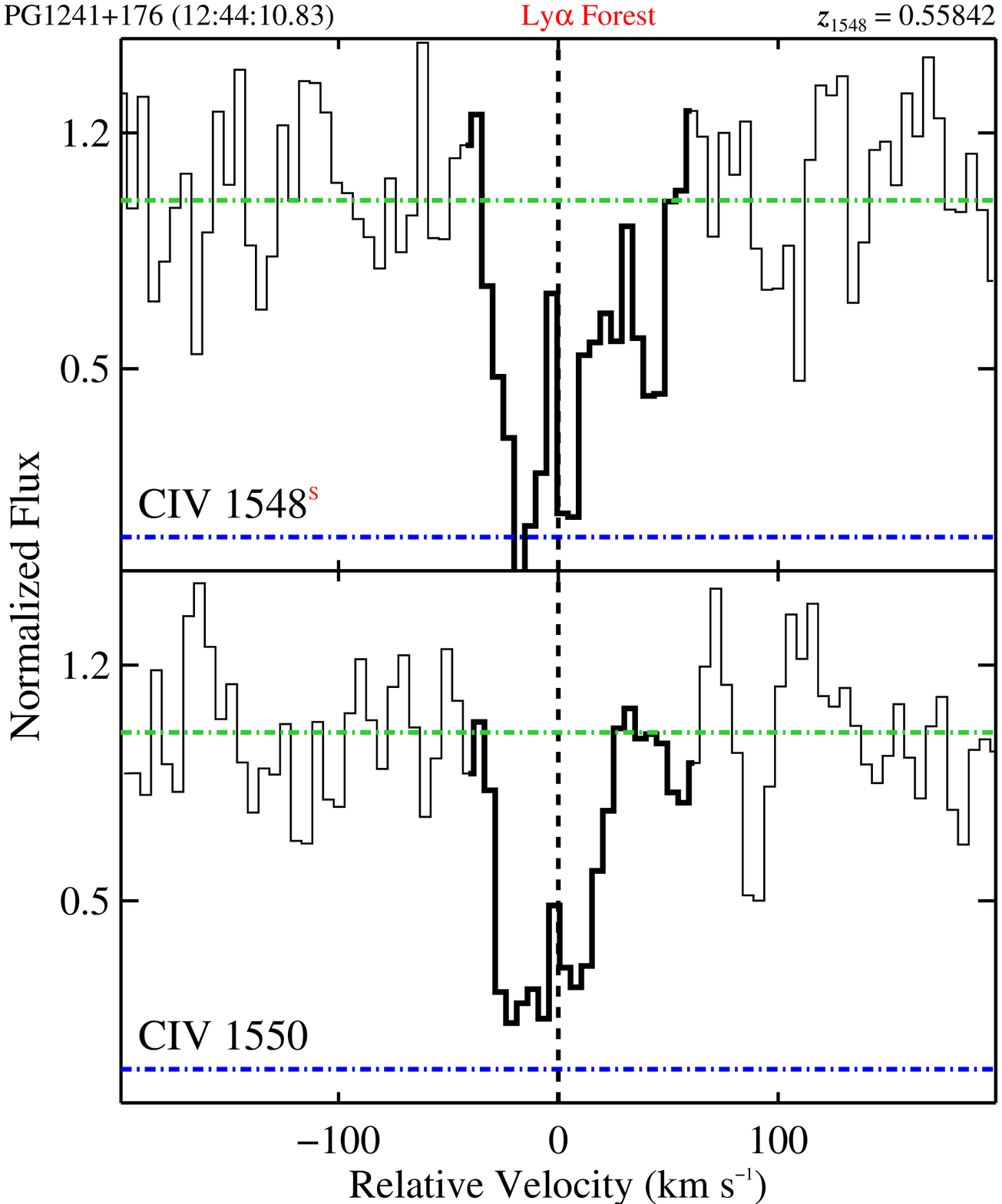} \\
      \includegraphics[width=0.45\textwidth]{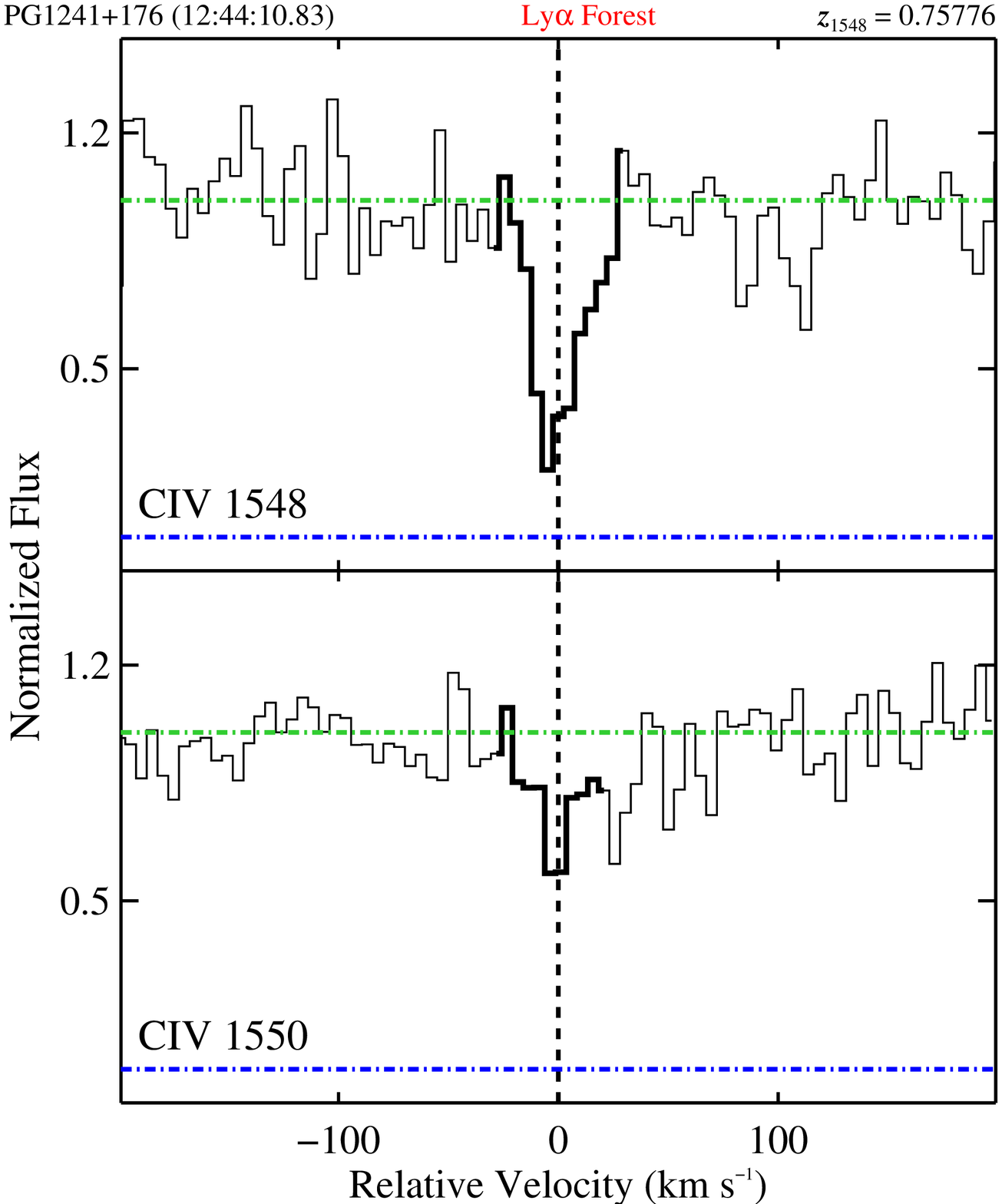} & 
      \includegraphics[width=0.45\textwidth]{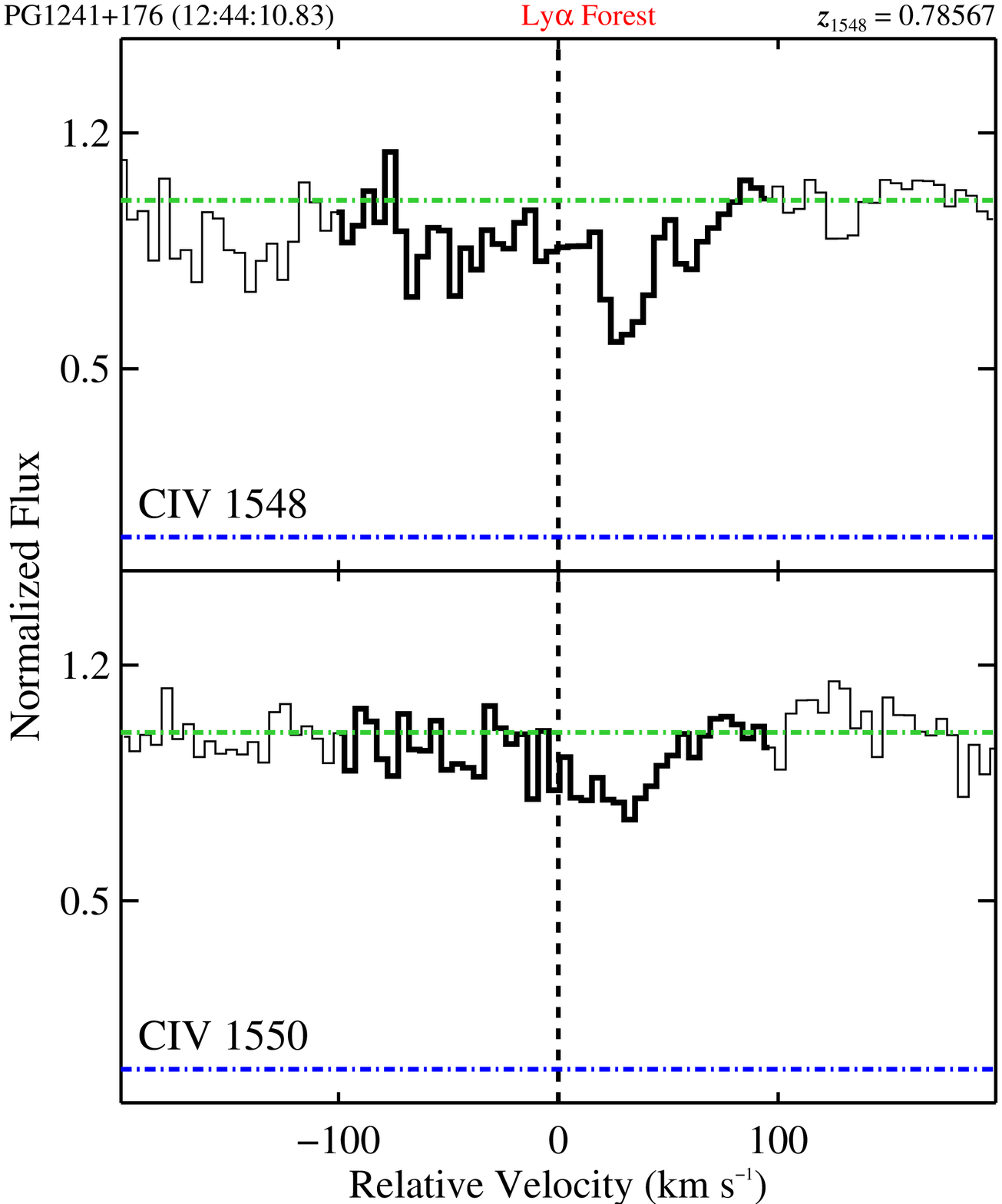} 
    \end{array}$                                      
  \end{center}
  \caption{G = 1 velocity plots (continued) }
\end{figure}
\addtocounter{figure}{-1}

\begin{figure}[!hbt]
  \begin{center}$
    \begin{array}{cc}
      \includegraphics[width=0.45\textwidth]{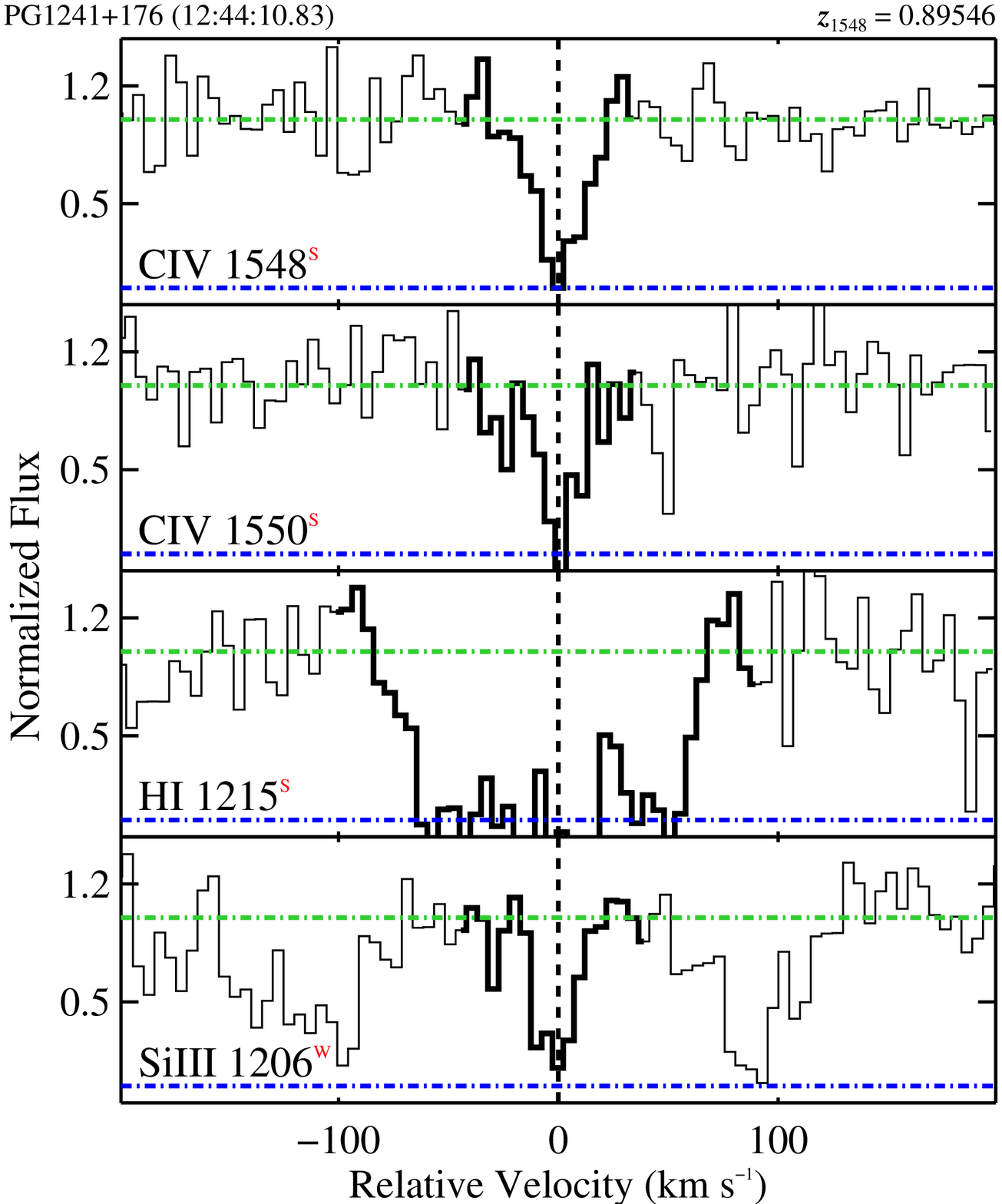} & 
      \includegraphics[width=0.45\textwidth]{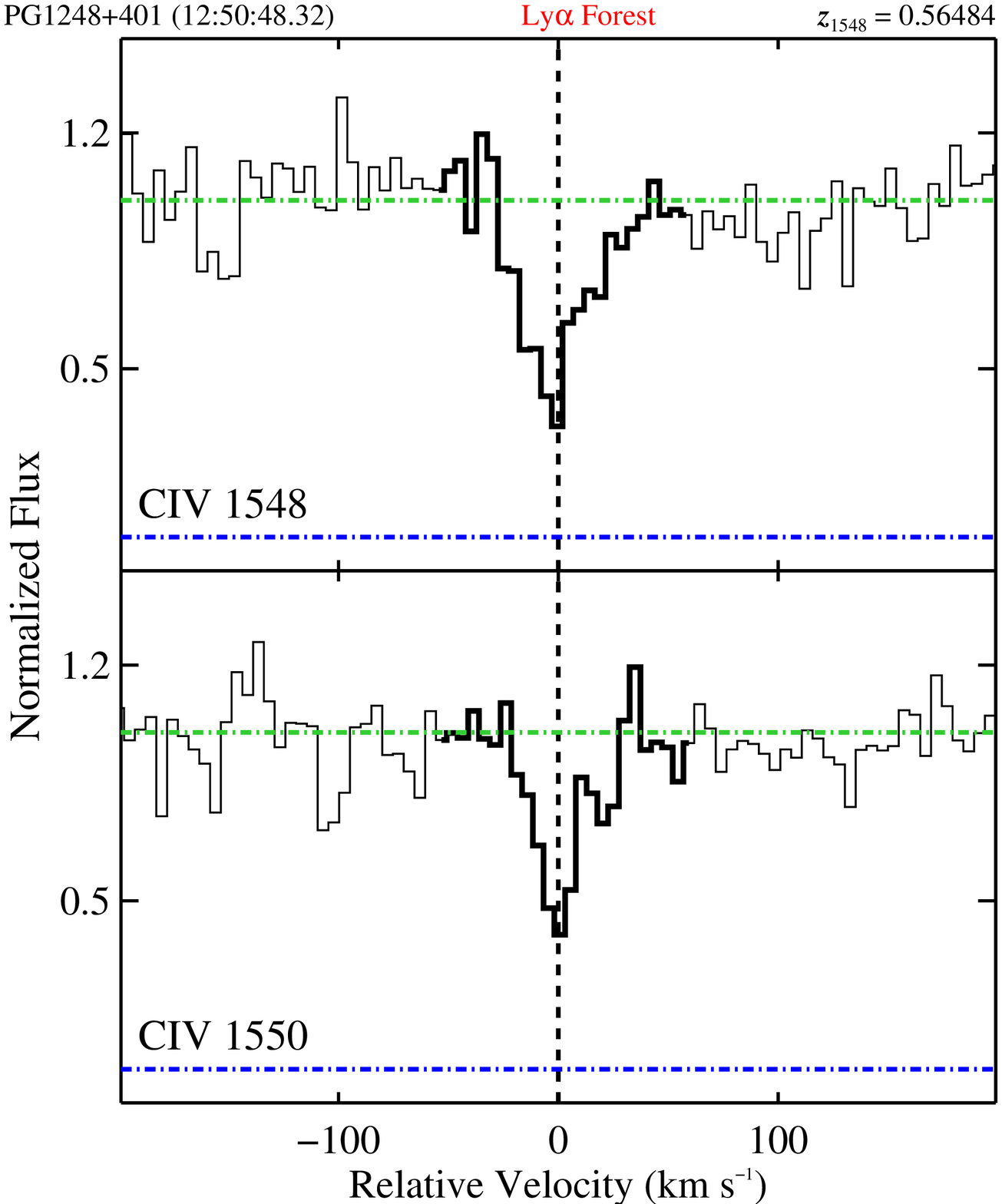} \\
      \includegraphics[width=0.45\textwidth]{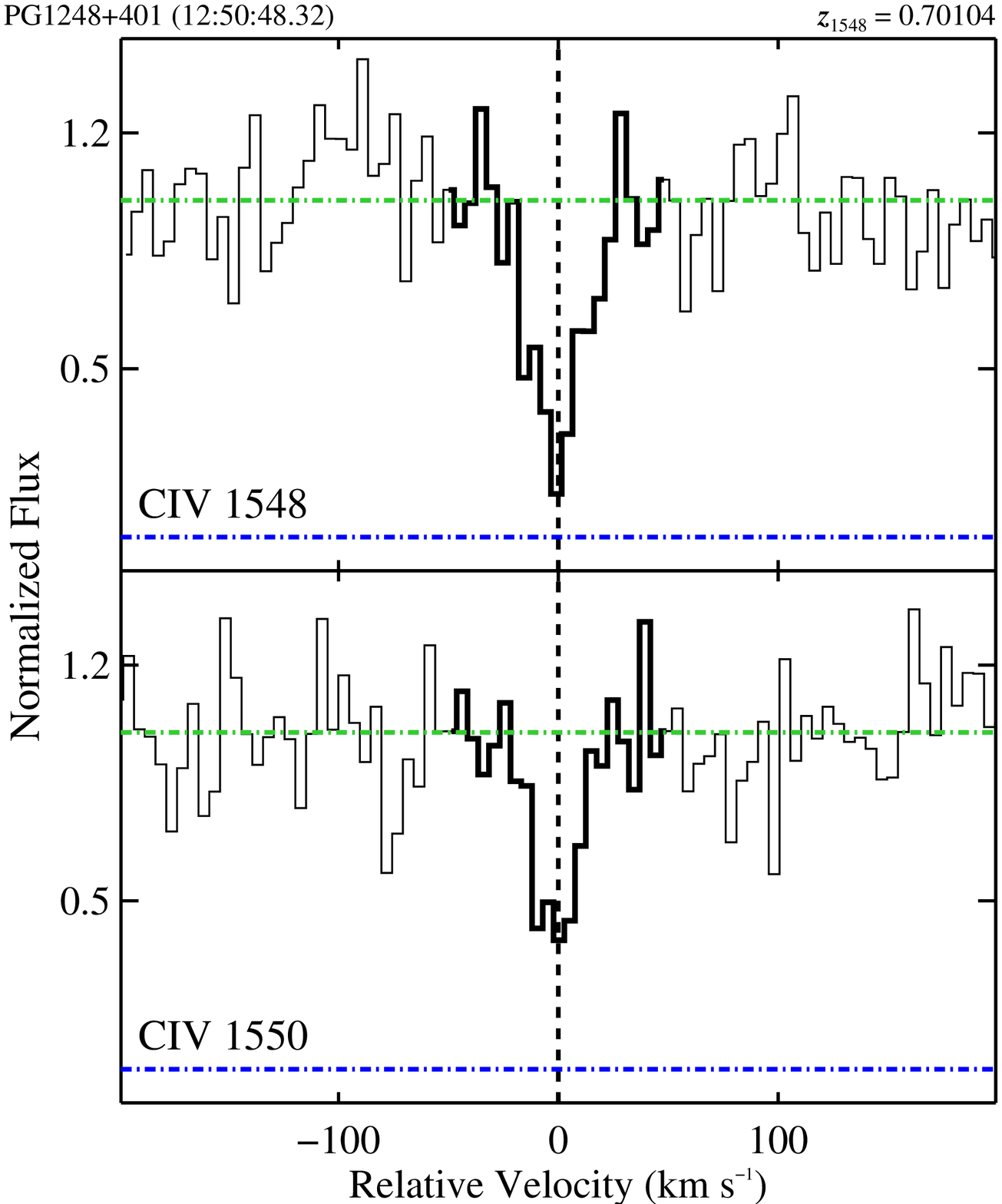}  & 
      \includegraphics[width=0.45\textwidth]{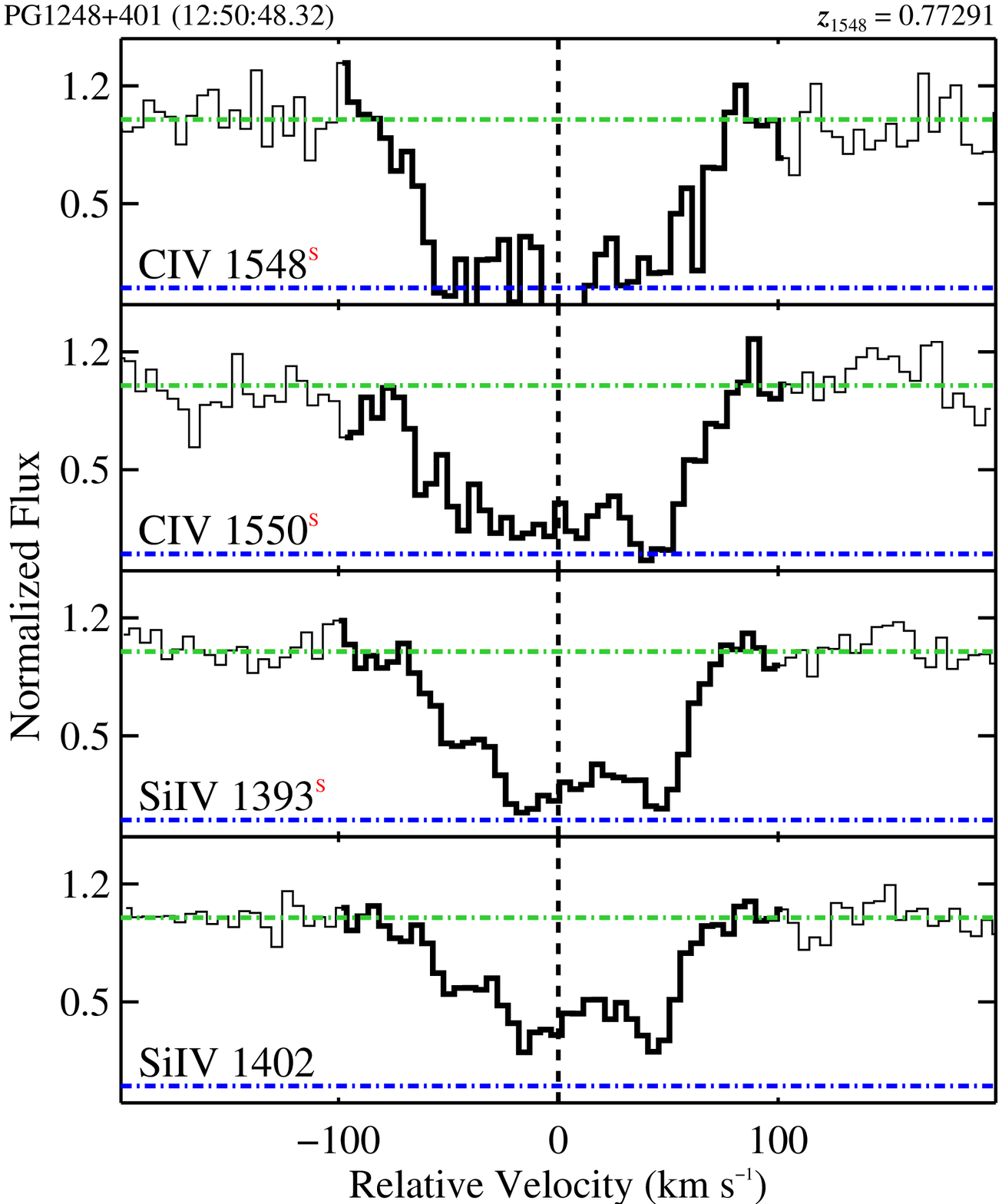} 
    \end{array}$                                      
  \end{center}
  \caption{G = 1 velocity plots (continued) }
\end{figure}
\addtocounter{figure}{-1}

\begin{figure}[!hbt]
  \begin{center}$
    \begin{array}{cc}
      \includegraphics[width=0.45\textwidth]{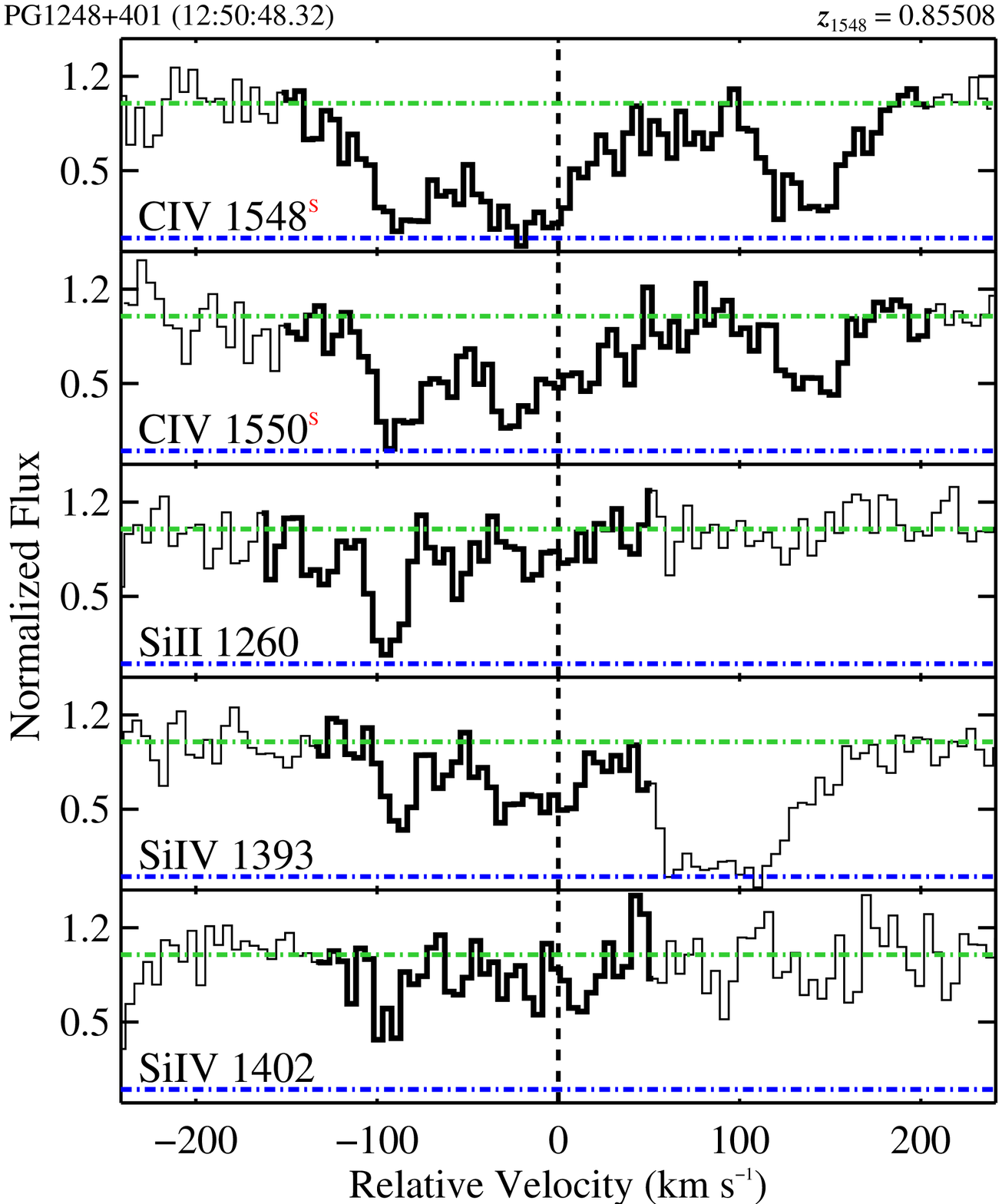} & 
      \includegraphics[width=0.45\textwidth]{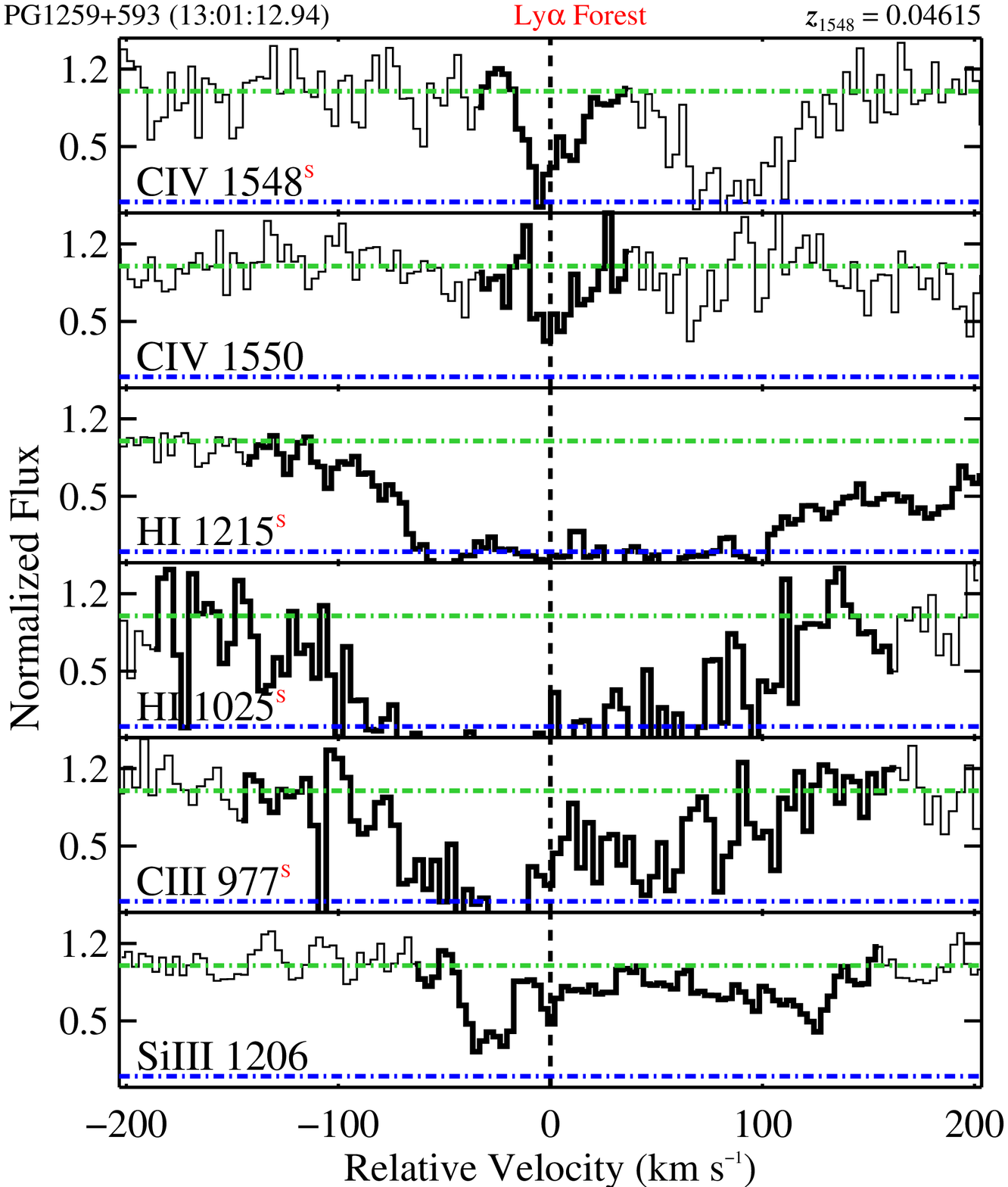} \\
      \includegraphics[width=0.45\textwidth]{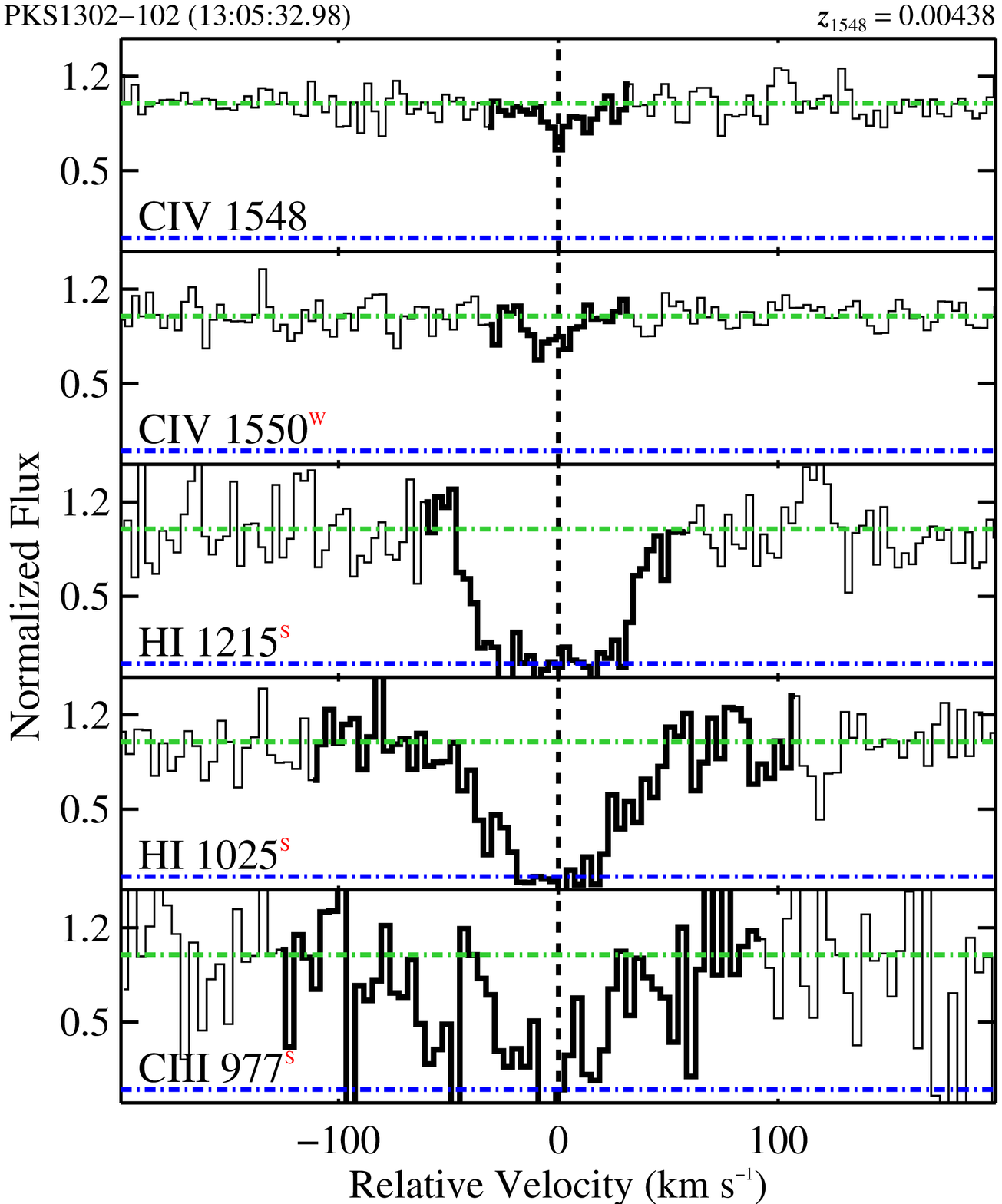} & 
      \includegraphics[width=0.45\textwidth]{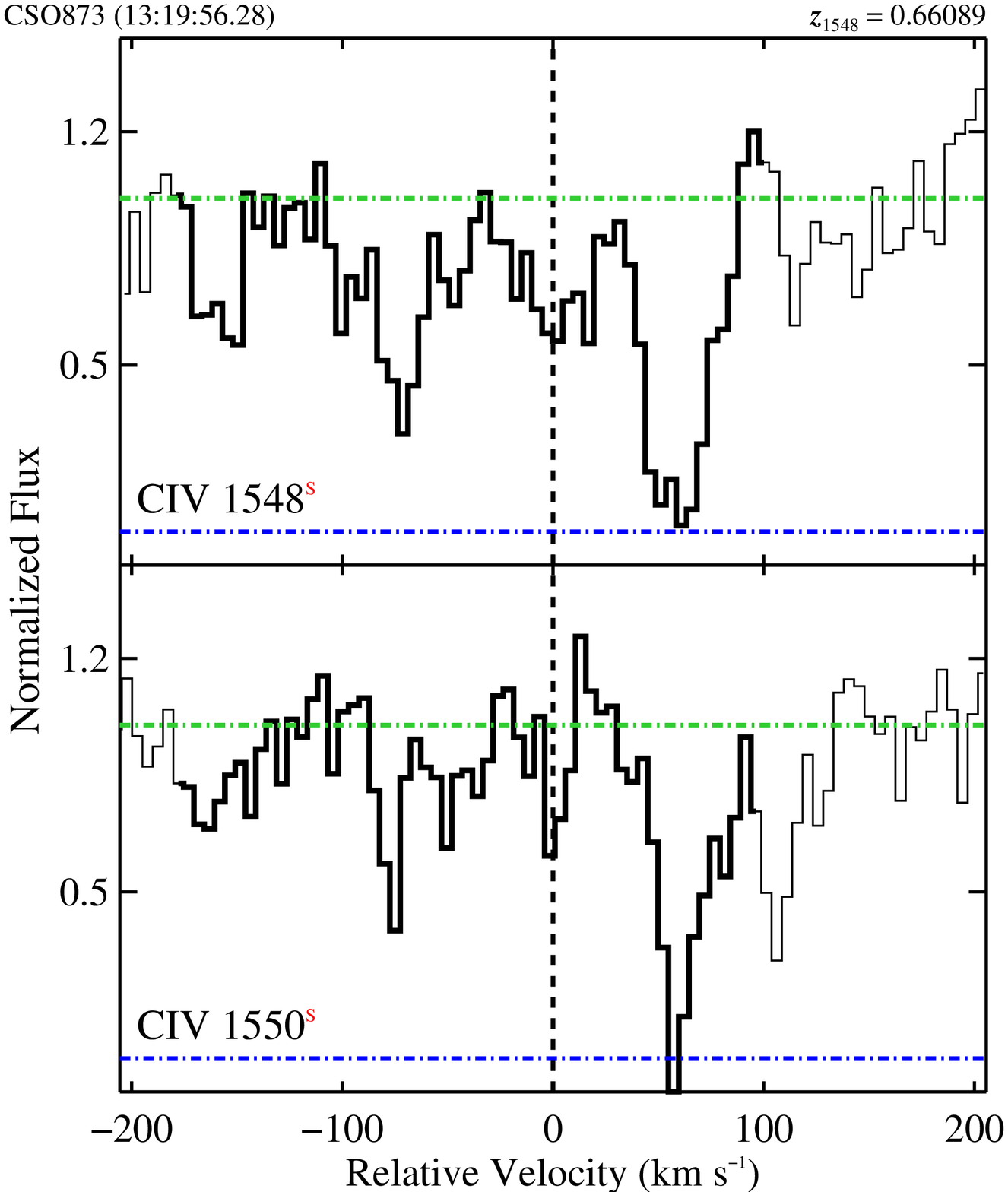} 
    \end{array}$                                      
  \end{center}
  \caption{G = 1 velocity plots (continued) }
\end{figure}
\addtocounter{figure}{-1}

\begin{figure}[!hbt]
  \begin{center}$
    \begin{array}{cc}
      \includegraphics[width=0.45\textwidth]{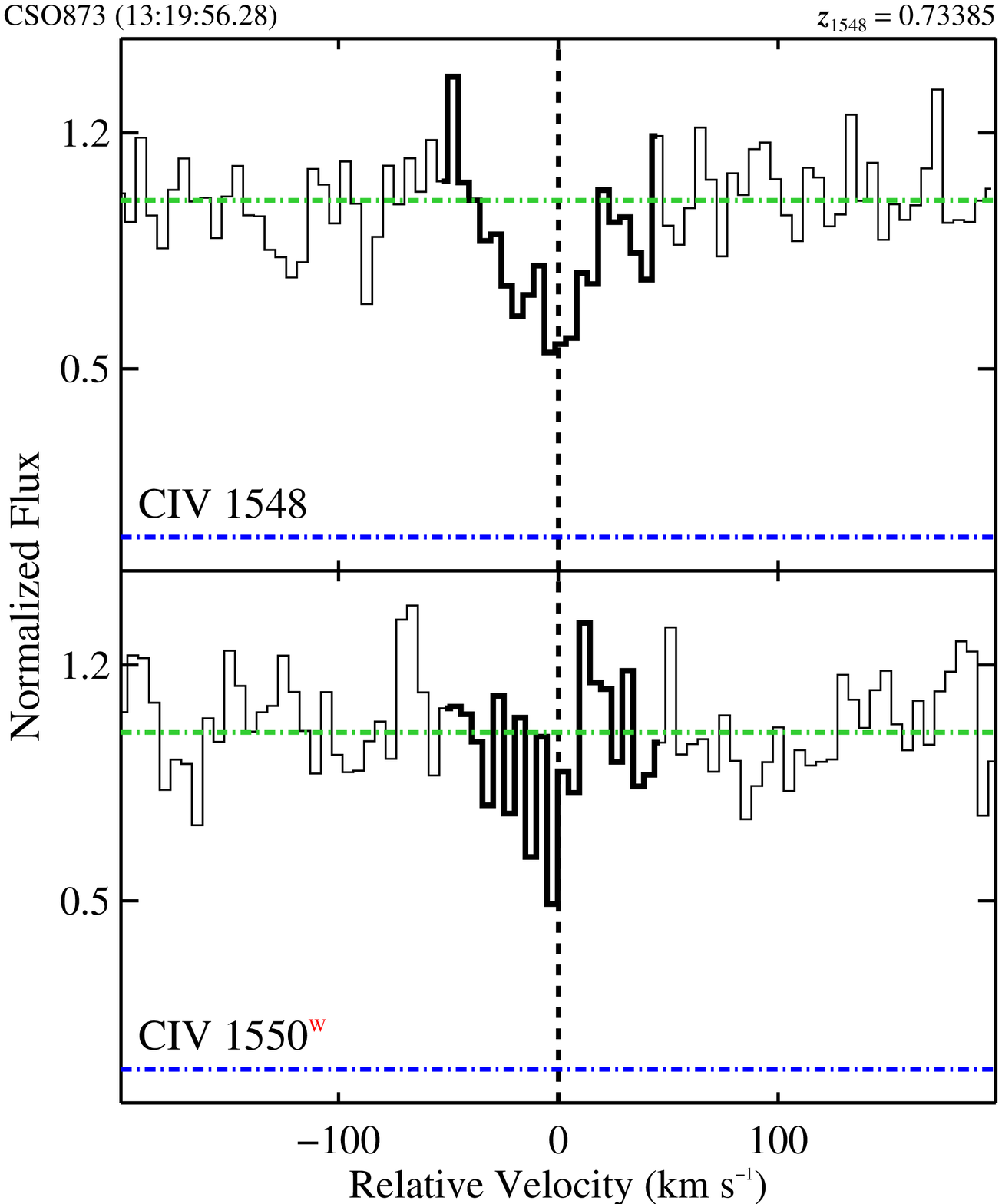} & 
      \includegraphics[width=0.45\textwidth]{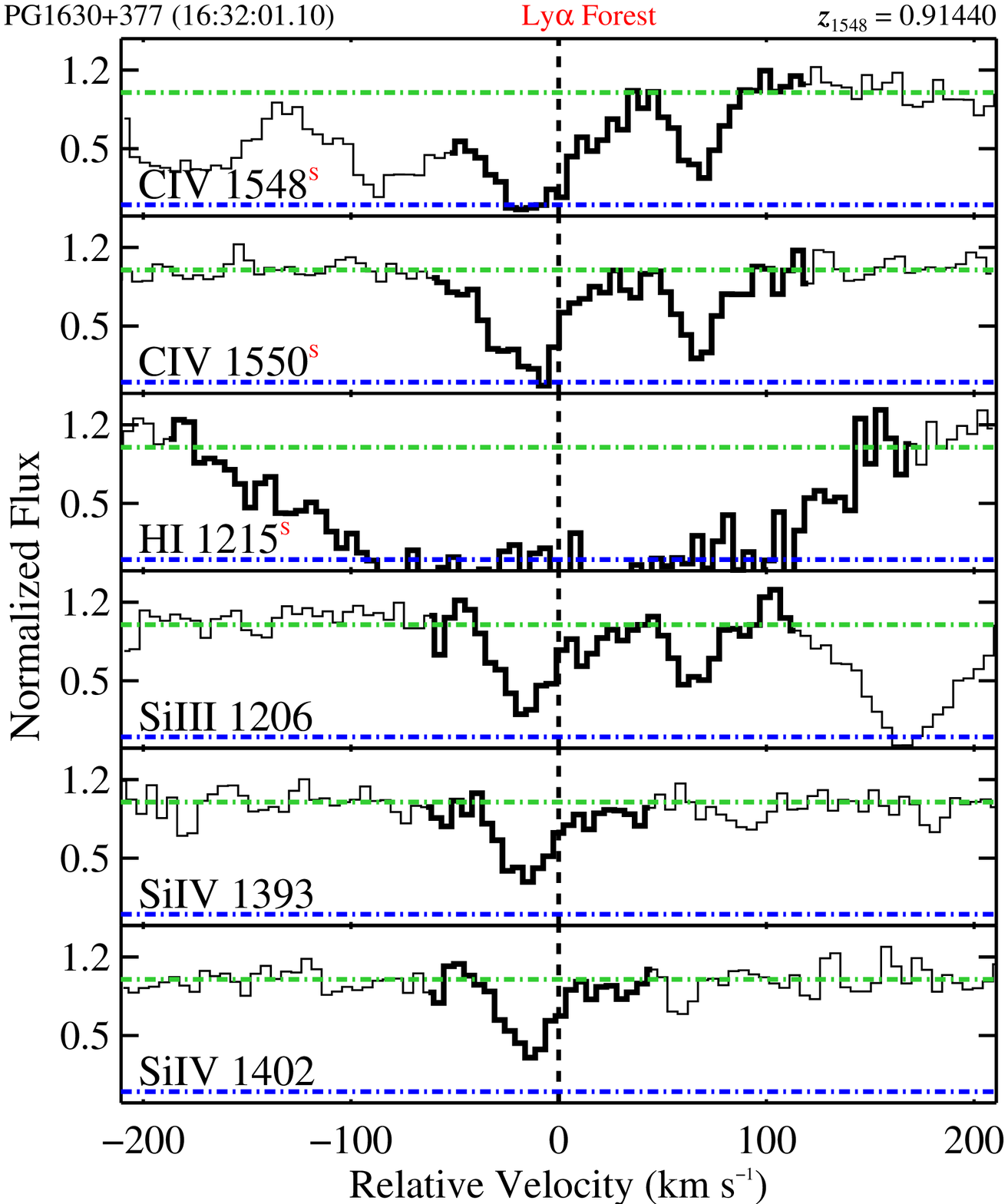} \\
      \includegraphics[width=0.45\textwidth]{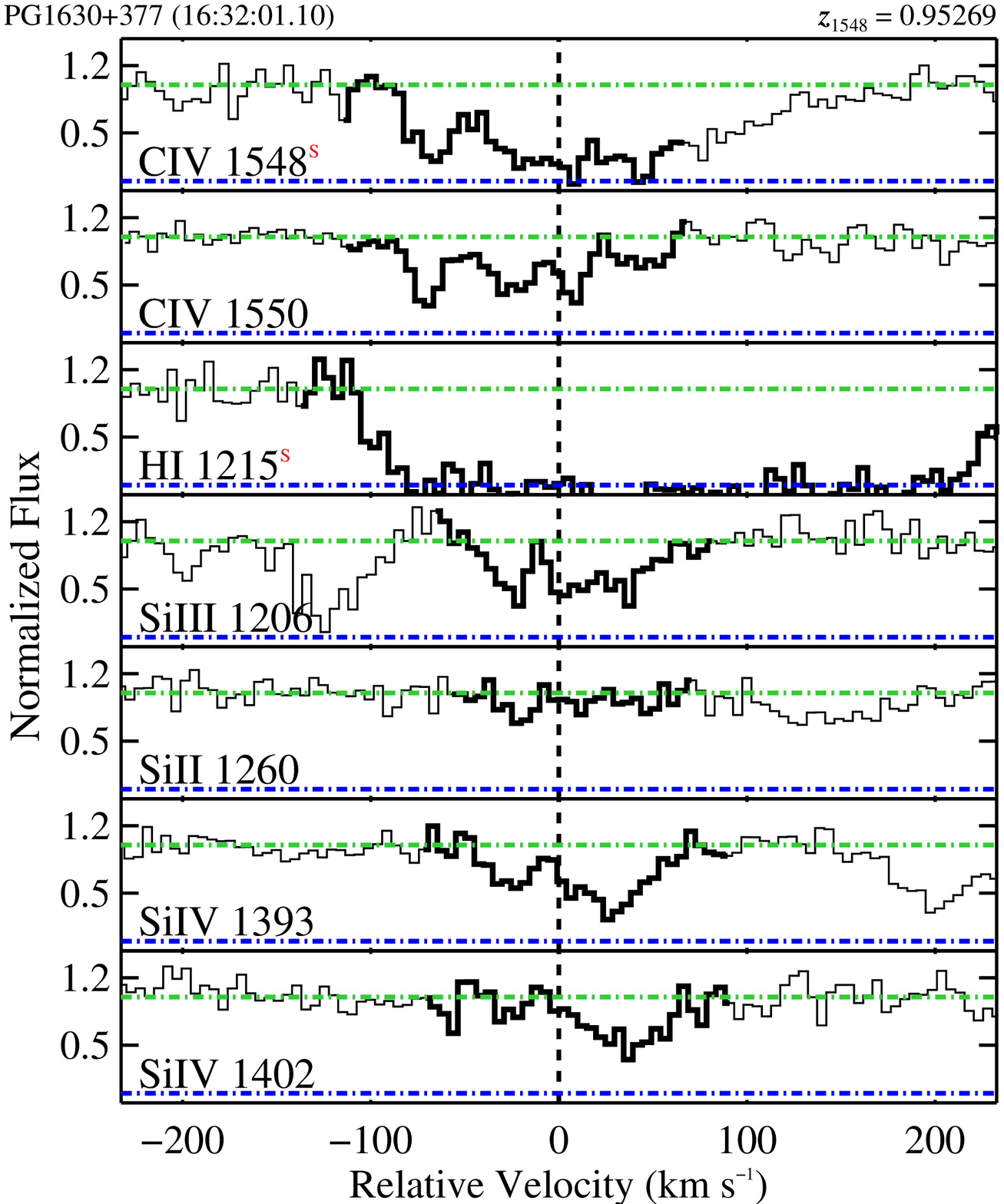} & 
      \includegraphics[width=0.45\textwidth]{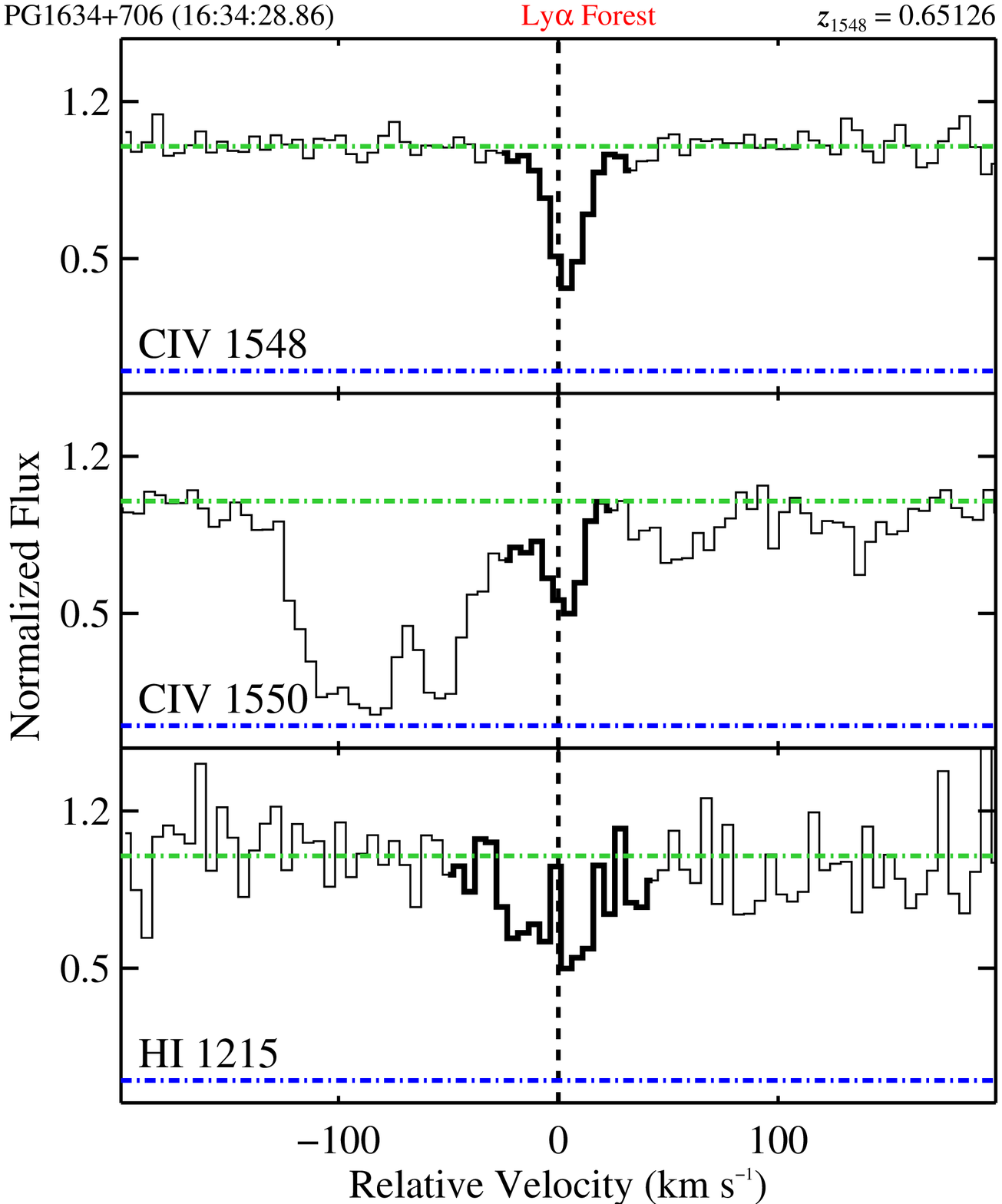} 
    \end{array}$                                      
  \end{center}
  \caption{G = 1 velocity plots (continued) }
\end{figure}
\addtocounter{figure}{-1}

\begin{figure}[!hbt]
  \begin{center}$
    \begin{array}{cc}
      \includegraphics[width=0.45\textwidth]{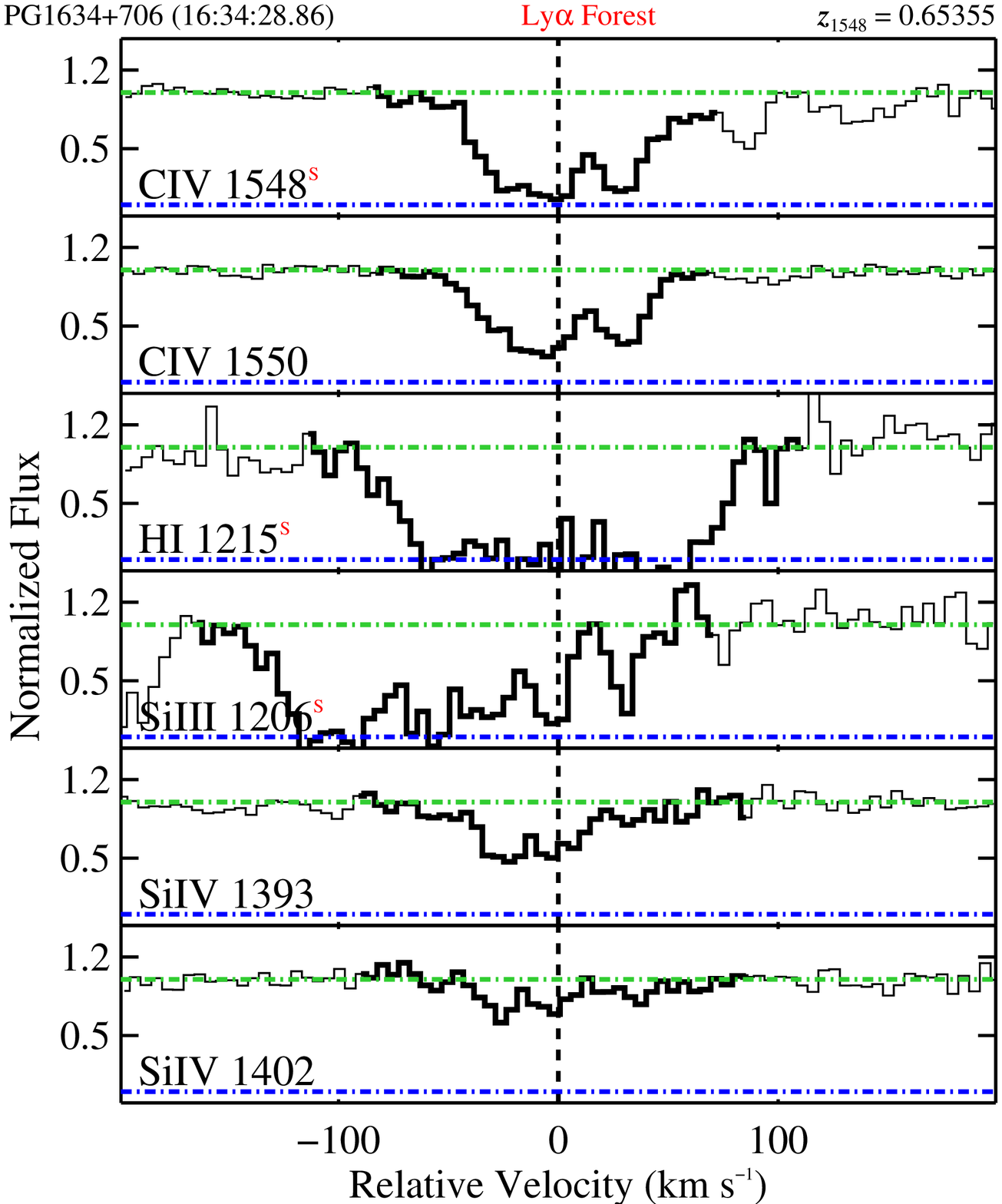} & 
      \includegraphics[width=0.45\textwidth]{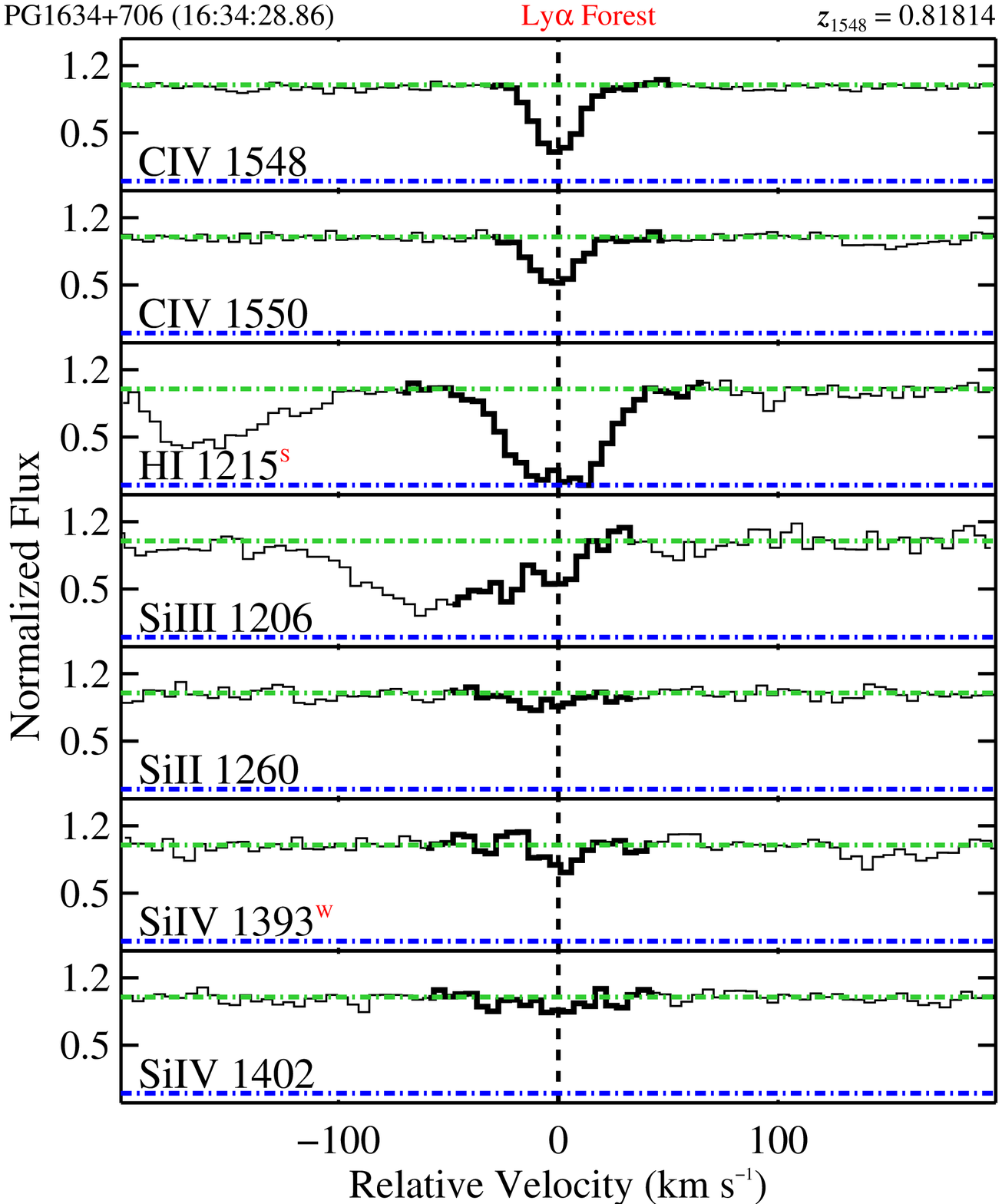} \\
      \includegraphics[width=0.45\textwidth]{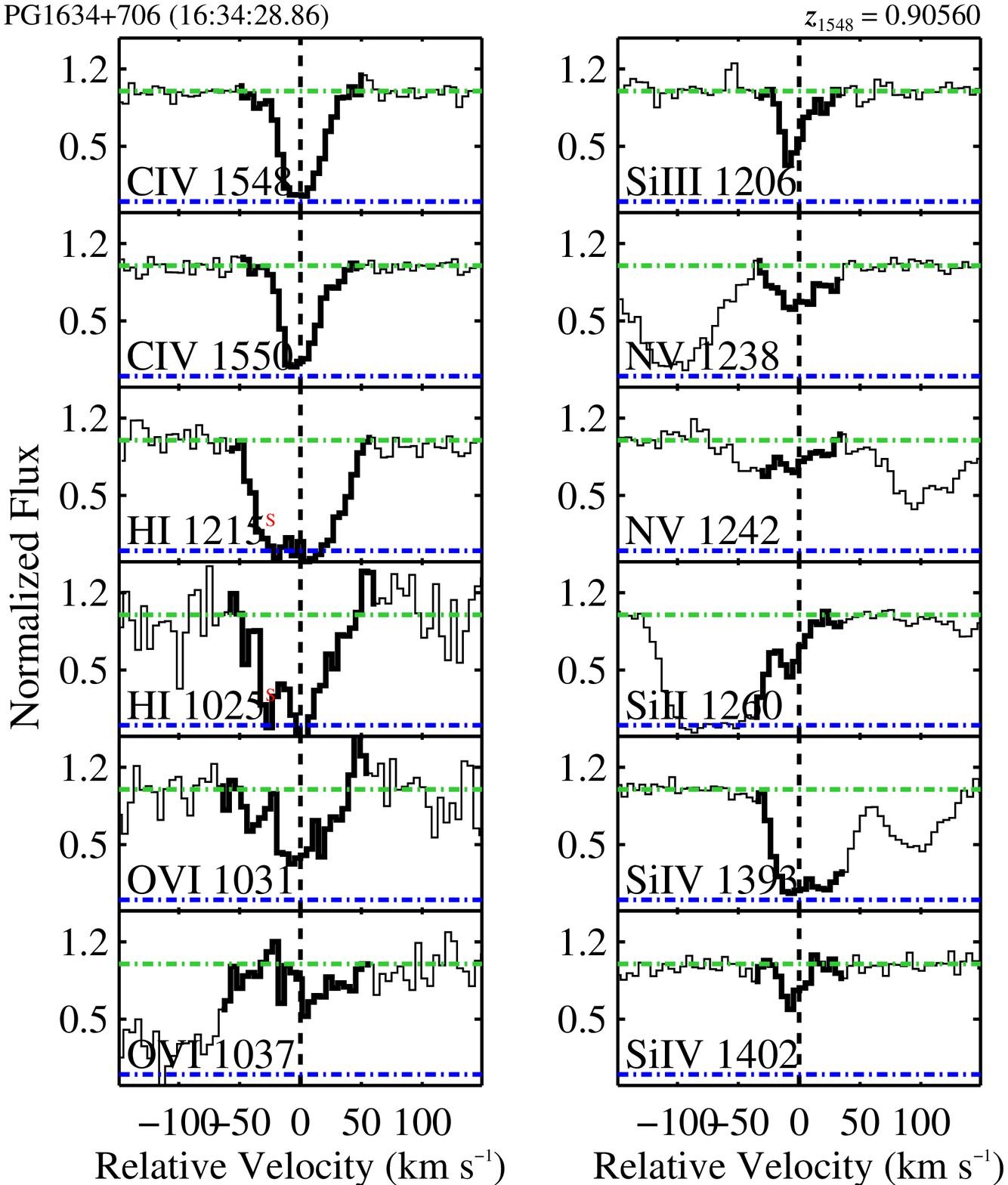} & 
      \includegraphics[width=0.45\textwidth]{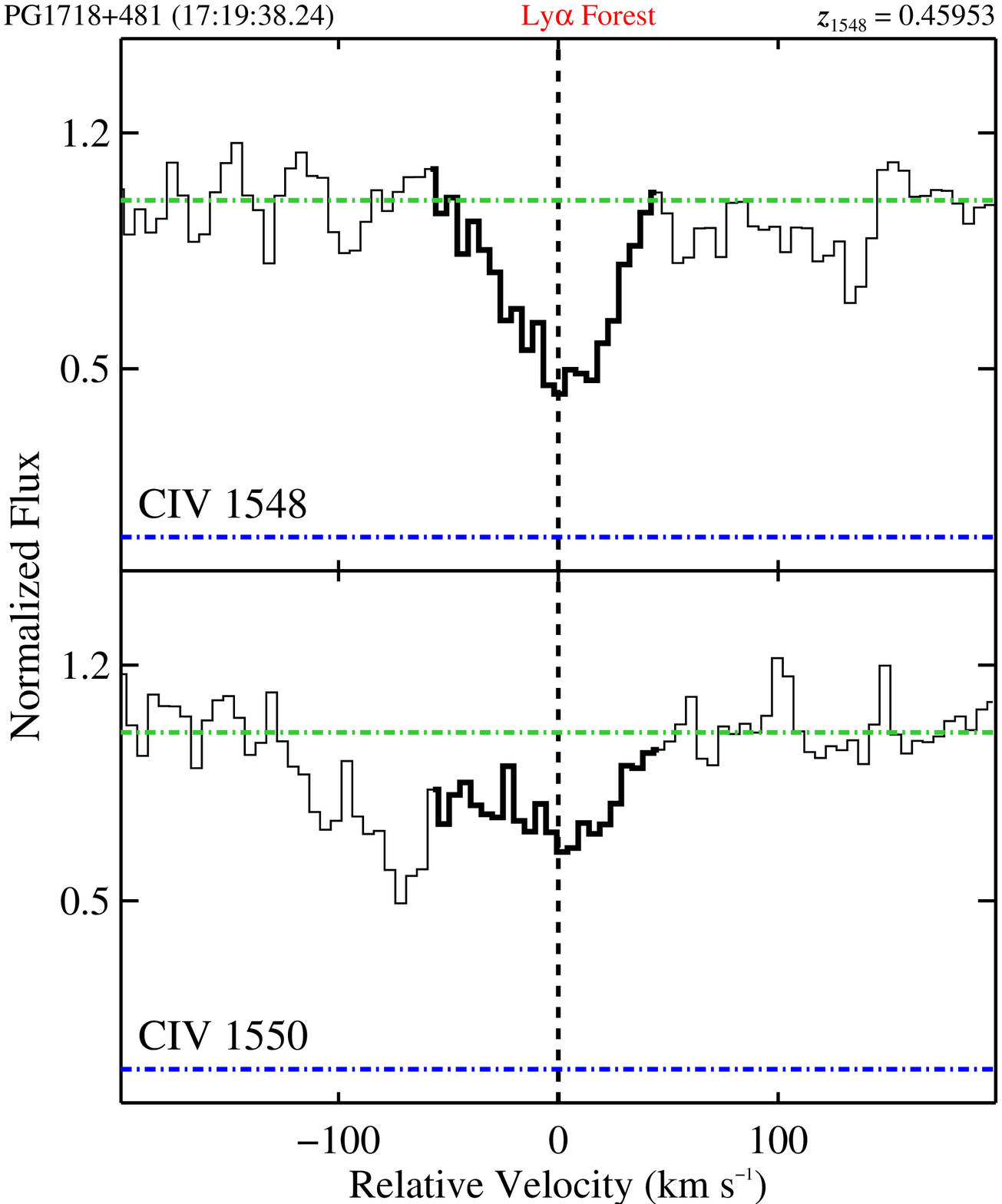} 
    \end{array}$                                      
  \end{center}
  \caption{G = 1 velocity plots (continued) }
\end{figure}
\addtocounter{figure}{-1}

\begin{figure}[!hbt]
  \begin{center}$
    \begin{array}{cc}
      \includegraphics[width=0.45\textwidth]{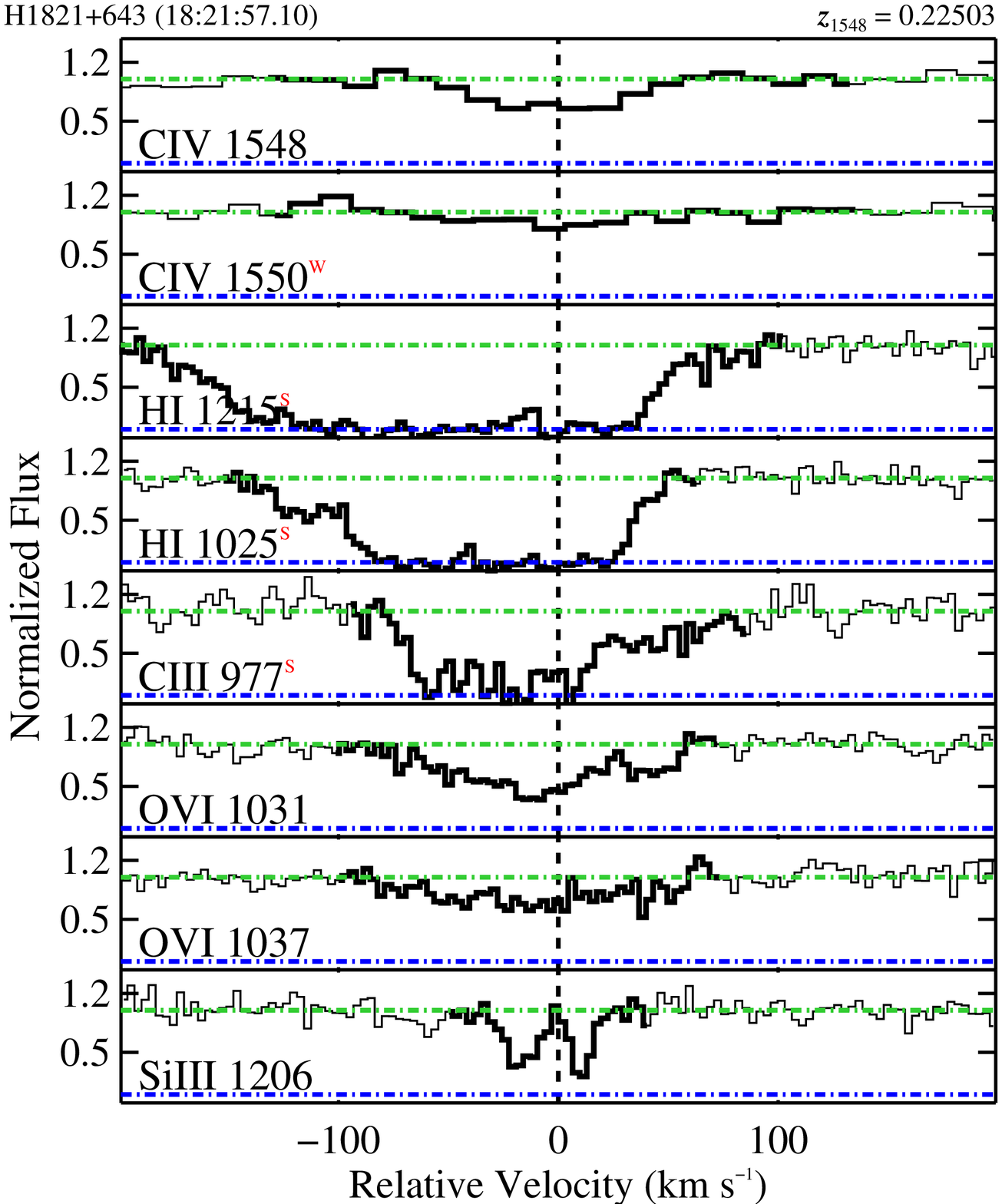} & 
      \includegraphics[width=0.45\textwidth]{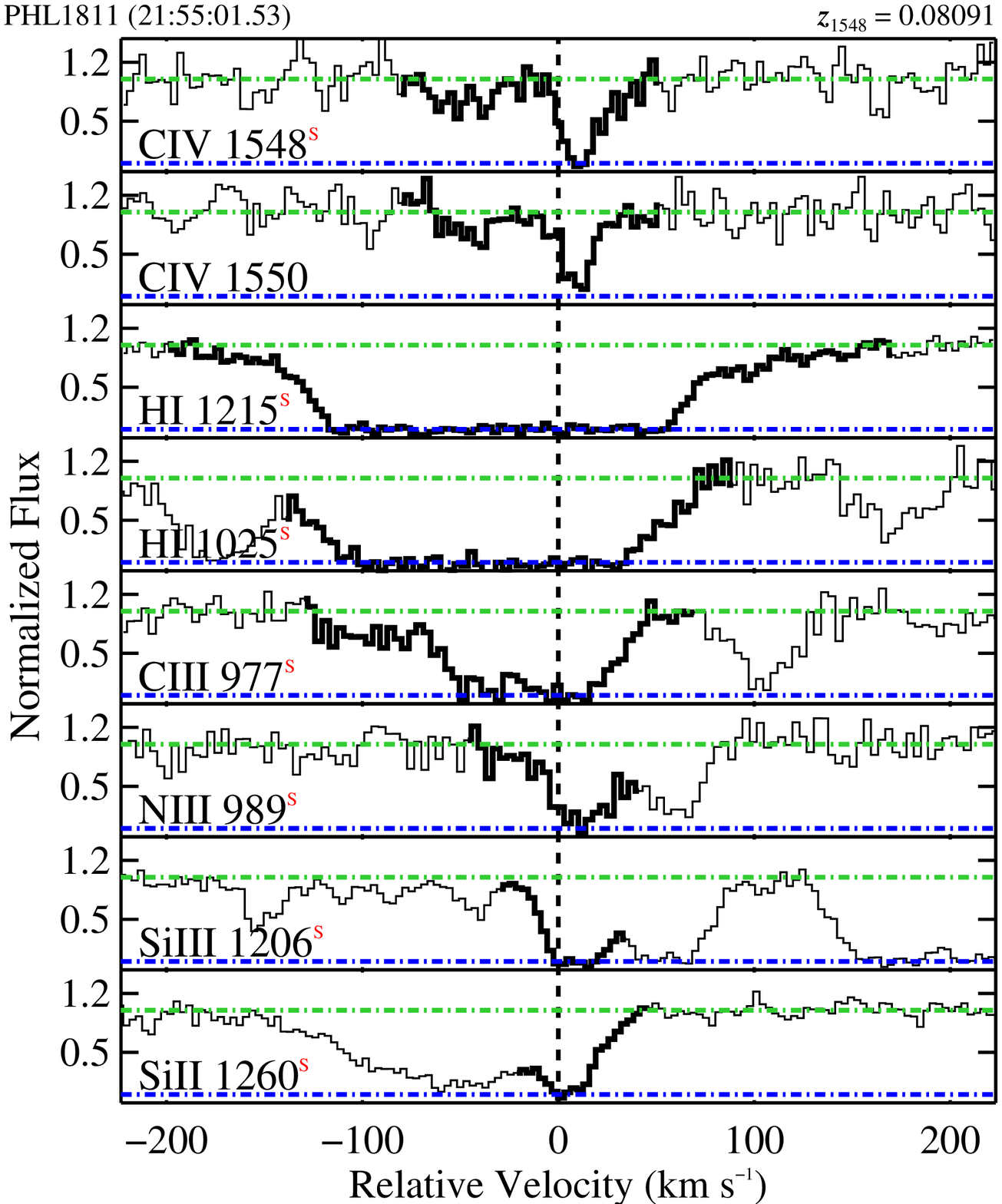} 
    \end{array}$                                      
  \end{center}
  \caption{G = 1 velocity plots (continued) }
\end{figure}

\begin{figure}[!hbt]
  \begin{center}$
    \begin{array}{cc}
      \includegraphics[width=0.45\textwidth]{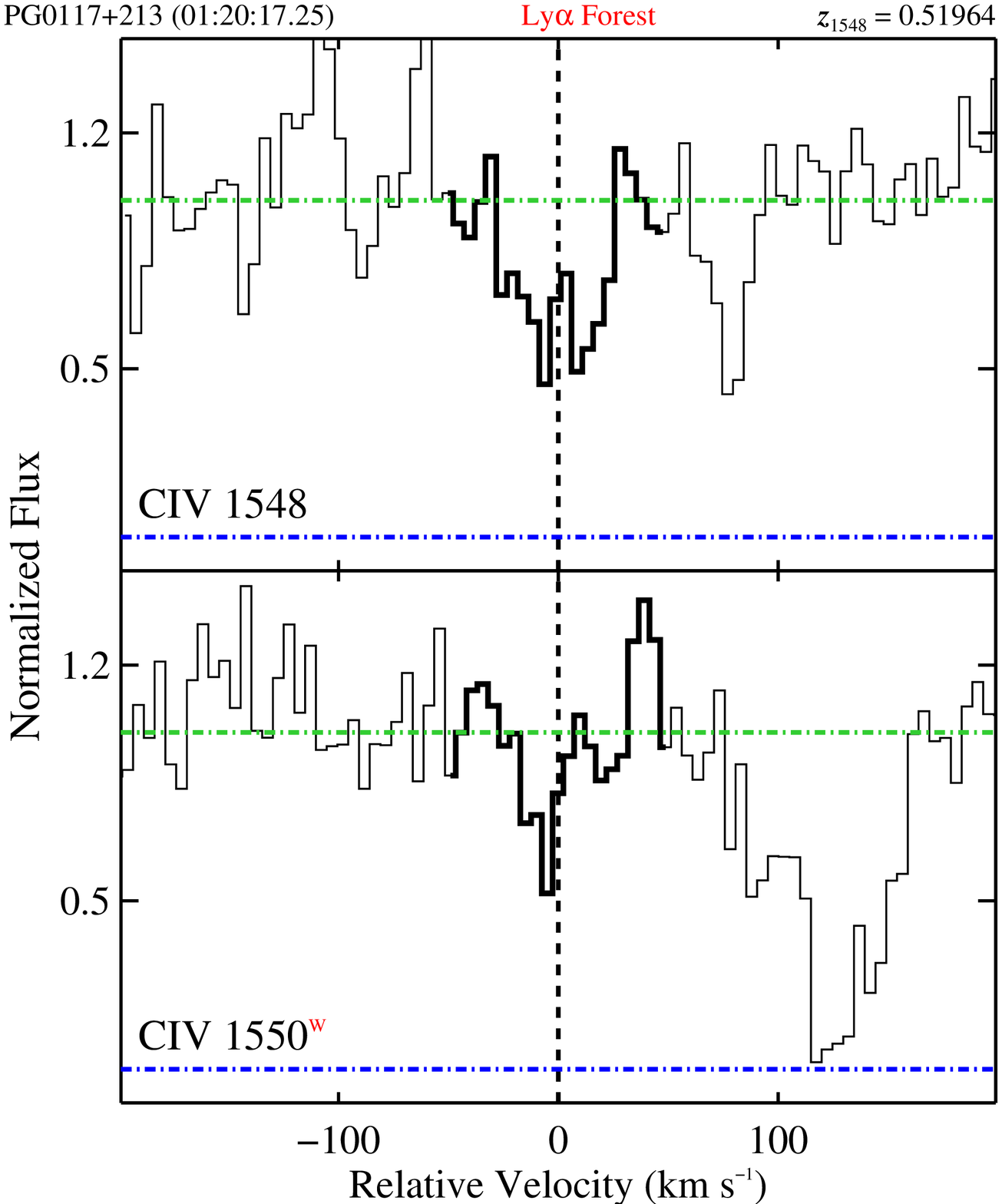} & 
      \includegraphics[width=0.45\textwidth]{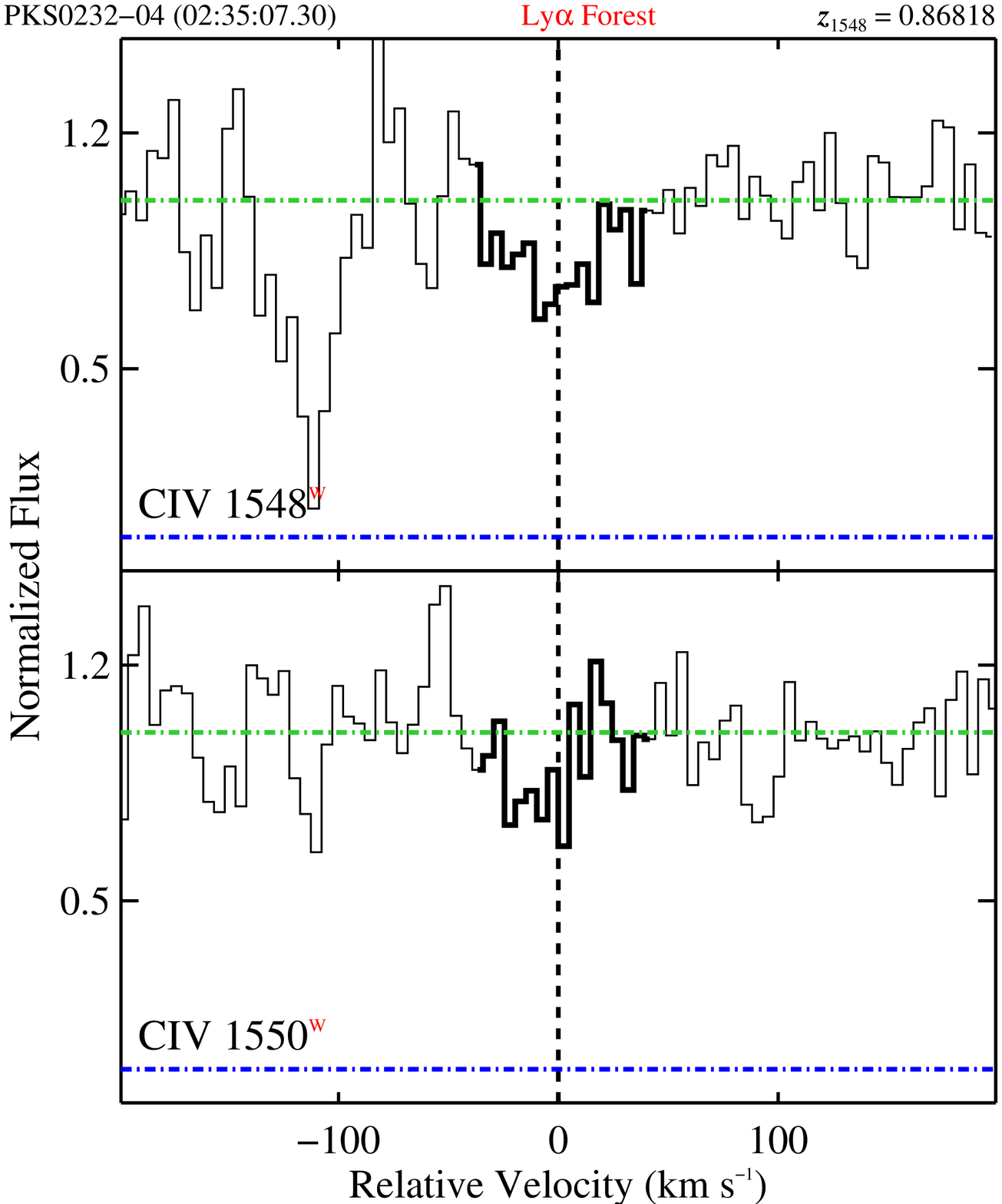} \\
      \includegraphics[width=0.45\textwidth]{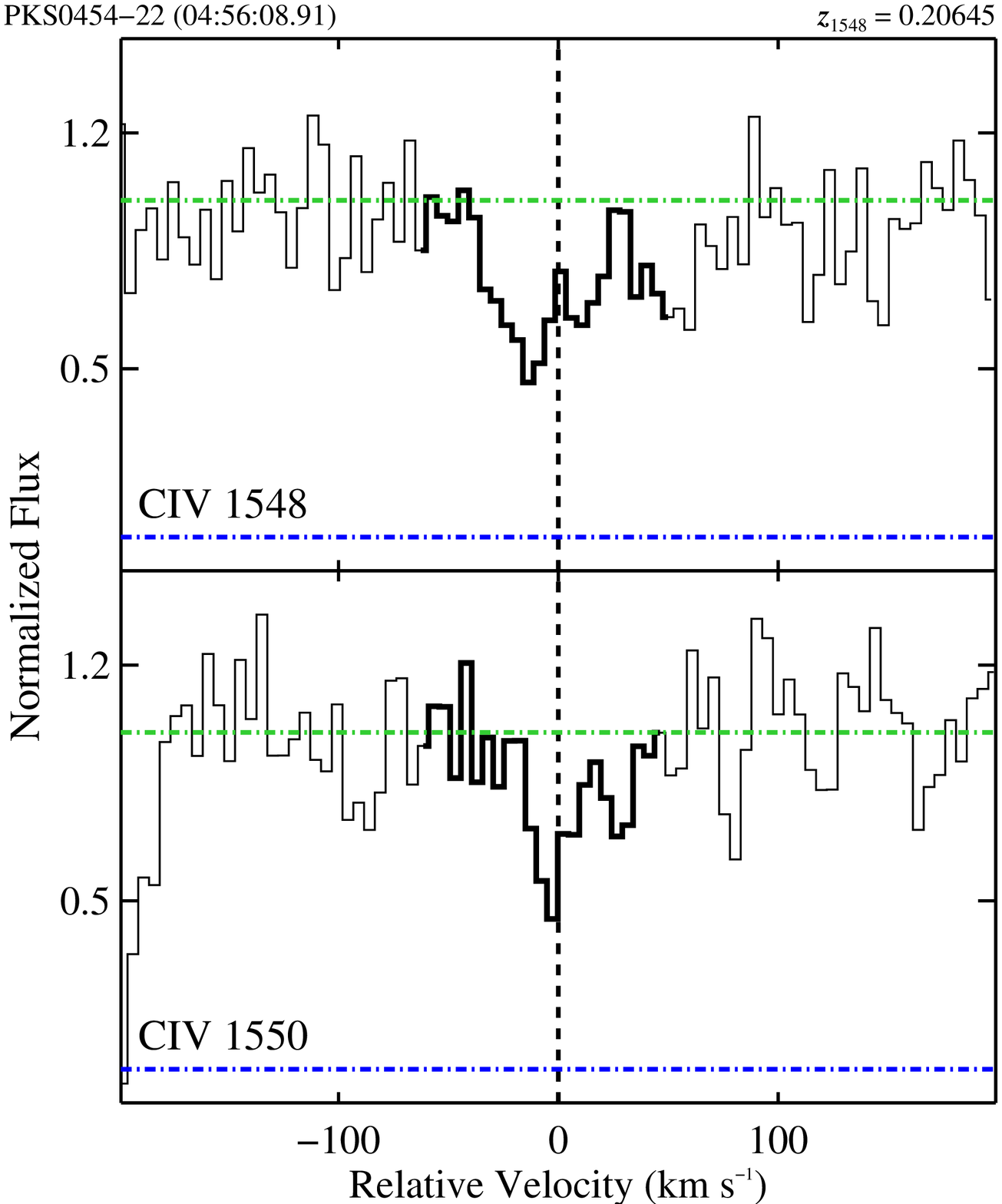} & 
      \includegraphics[width=0.45\textwidth]{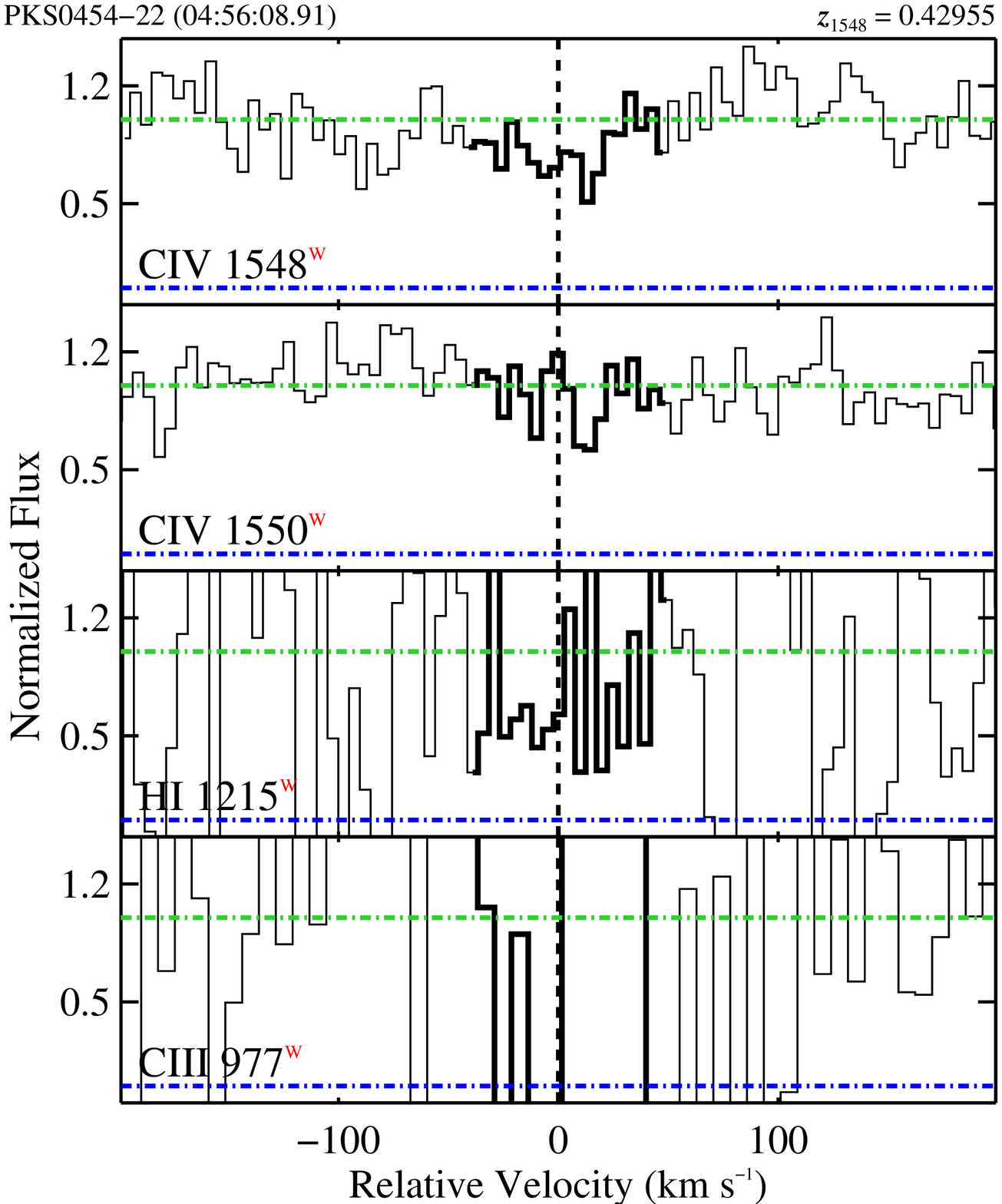} 
    \end{array}$                                      
  \end{center}
  \caption{Velocity plots of G = 2 \ion{C}{4} systems. (See
    Figure \ref{fig.g1appdx} for description of velocity plot.)
    \label{fig.g2}
  }
\end{figure}
\addtocounter{figure}{-1}

\begin{figure}[!hbt]
  \begin{center}$
    \begin{array}{cc}
      \includegraphics[width=0.45\textwidth]{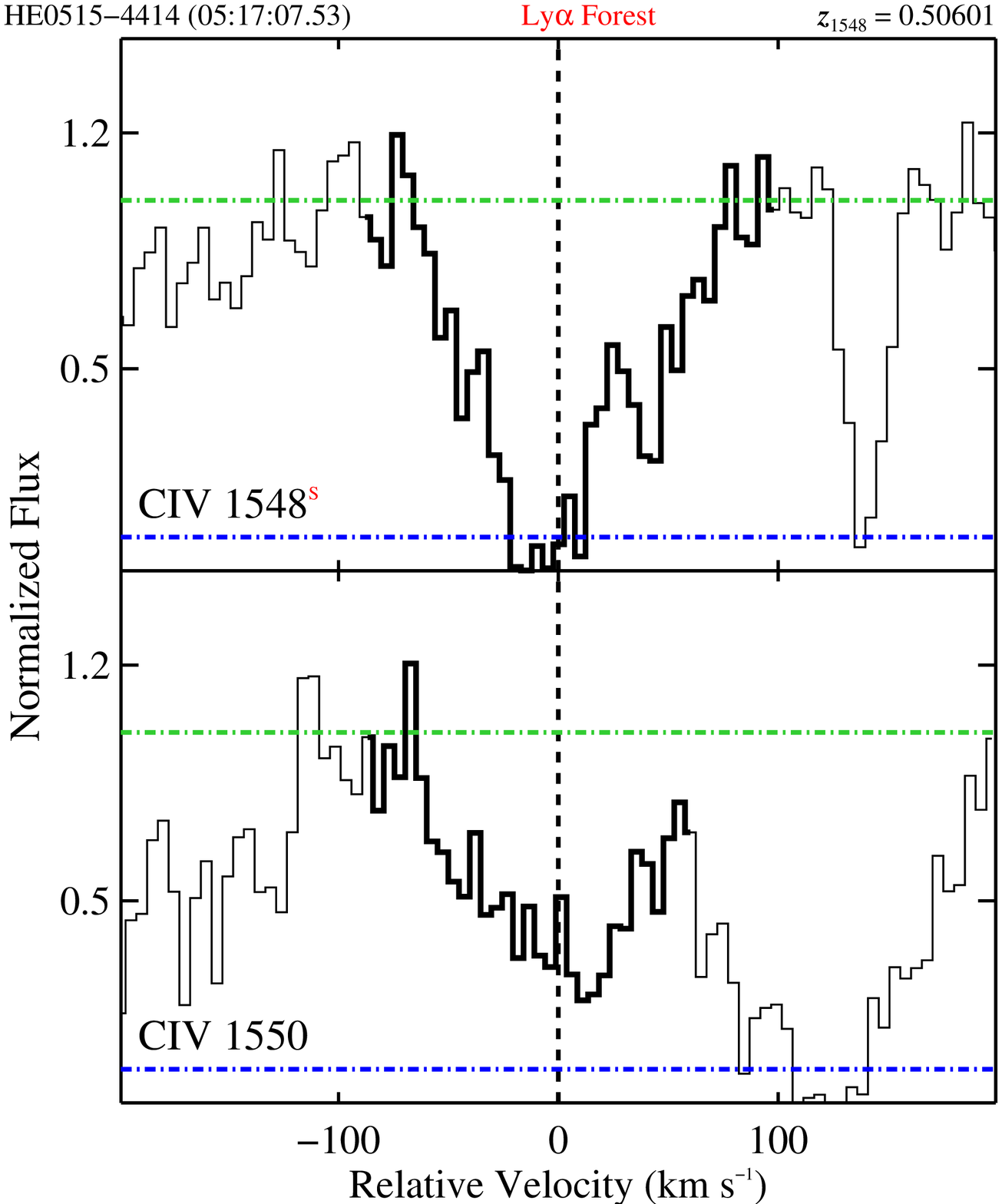} & 
      \includegraphics[width=0.45\textwidth]{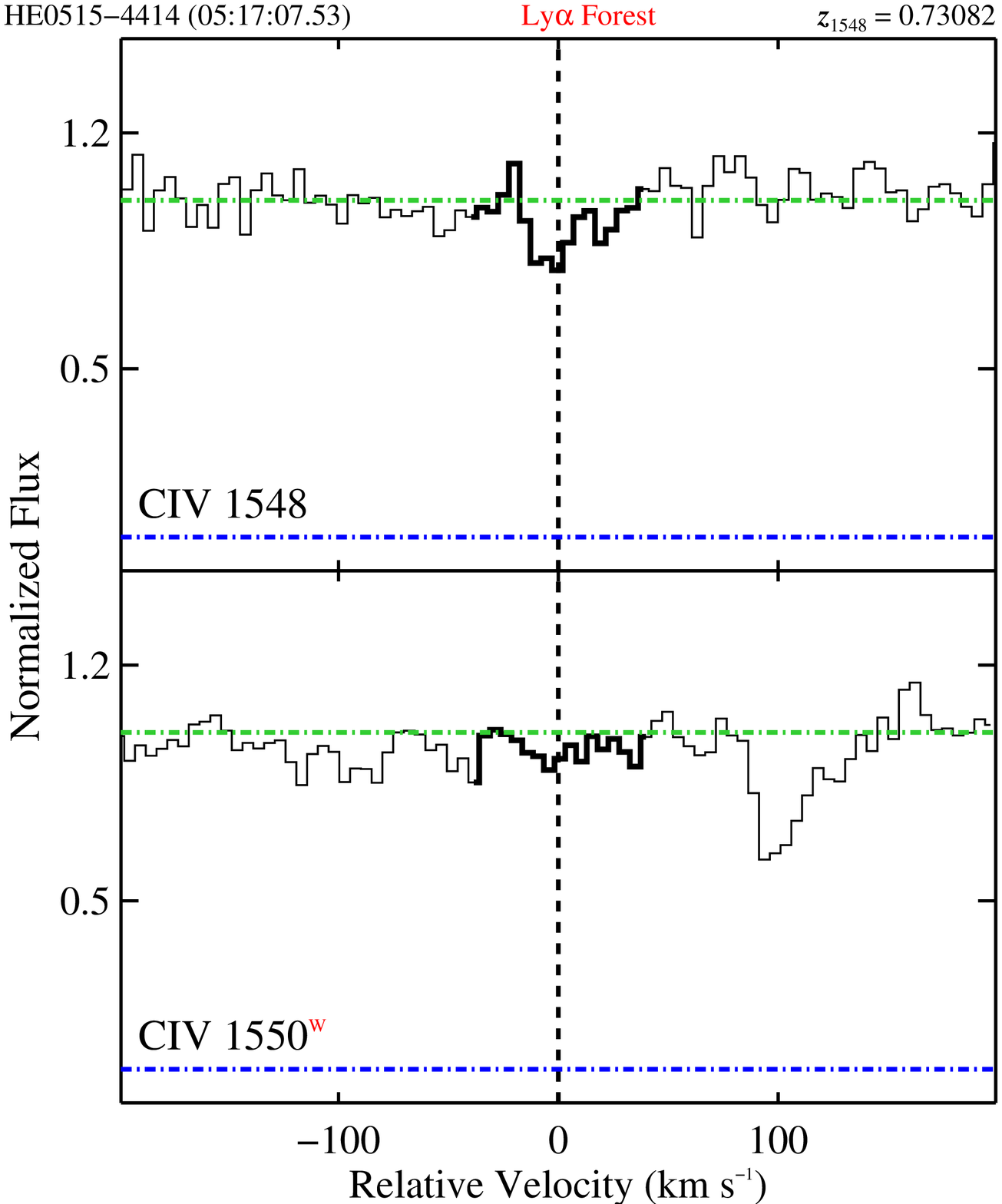} \\
      \includegraphics[width=0.45\textwidth]{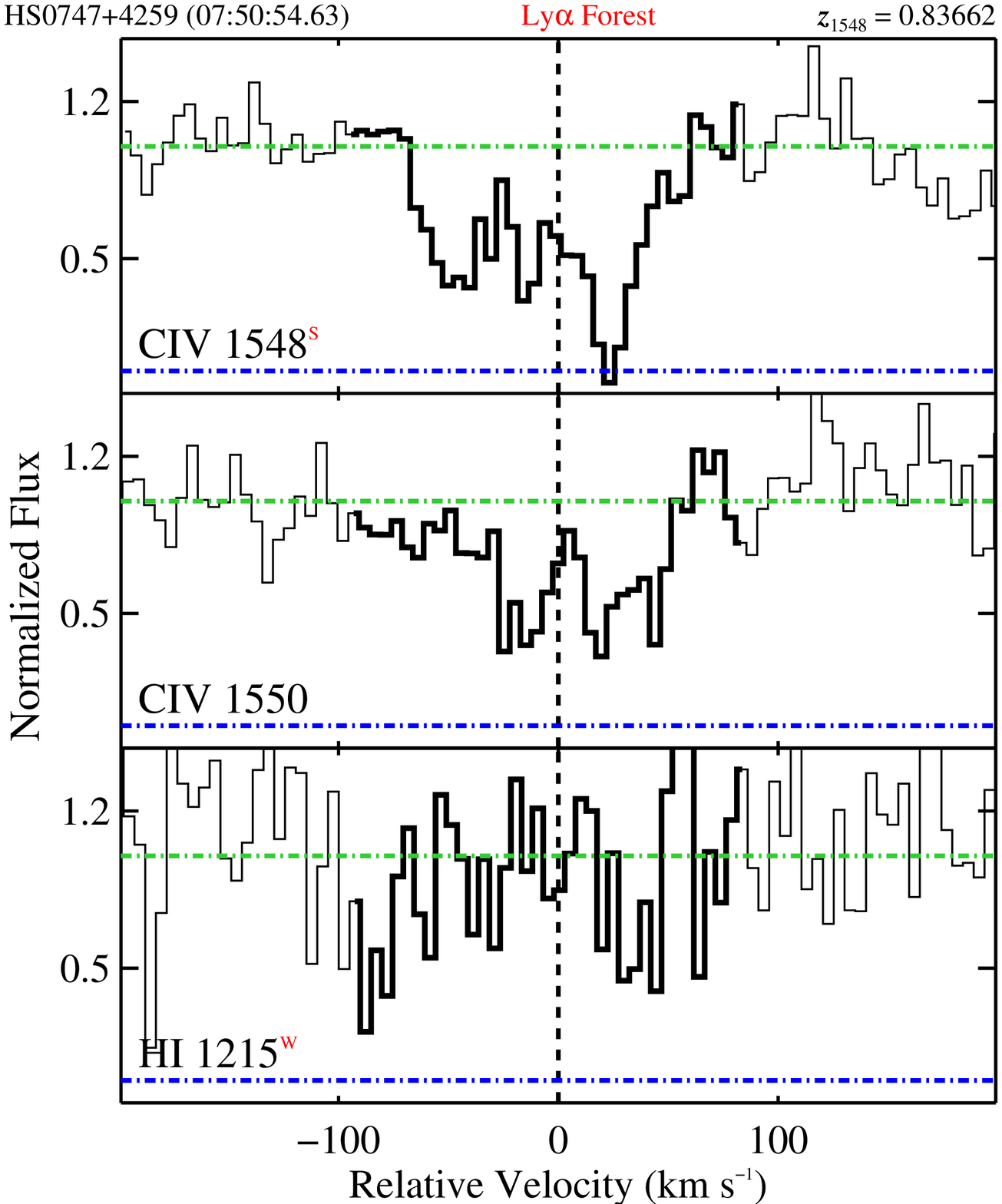} & 
      \includegraphics[width=0.45\textwidth]{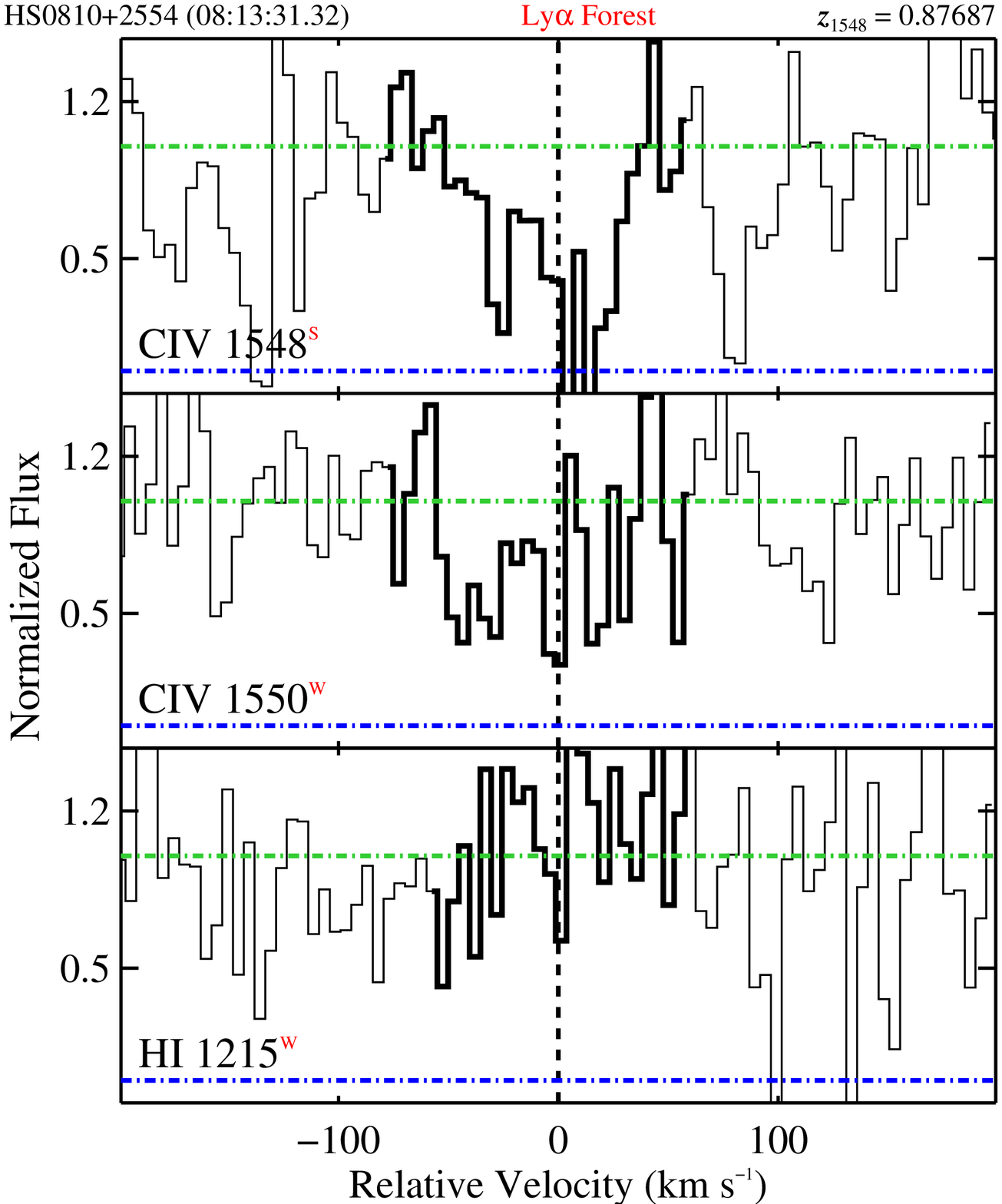} 
    \end{array}$                                      
  \end{center}
  \caption{G = 2 velocity plots (continued) }
\end{figure}
\addtocounter{figure}{-1}

\begin{figure}[!hbt]
  \begin{center}$
    \begin{array}{cc}
      \includegraphics[width=0.45\textwidth]{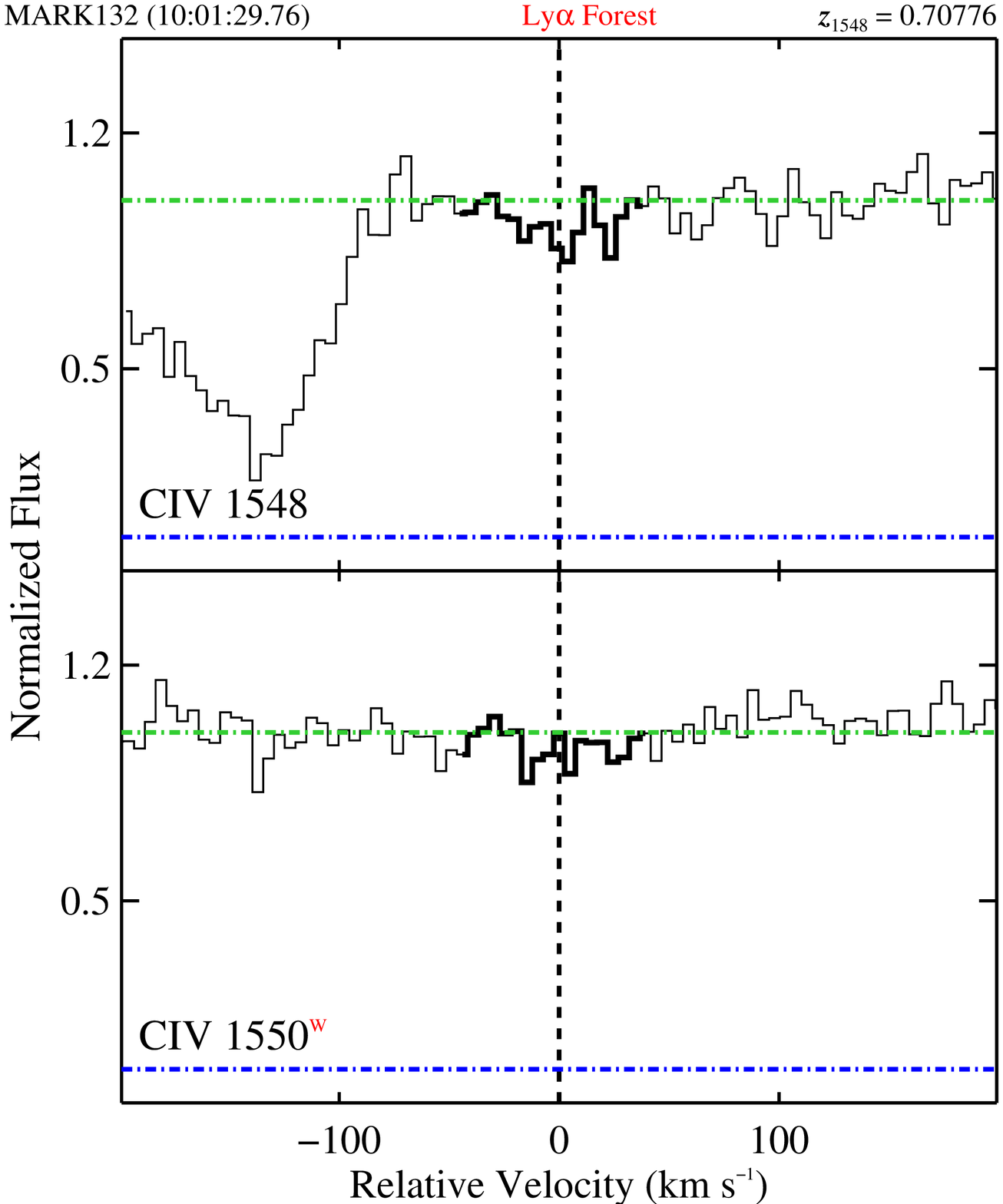} & 
      \includegraphics[width=0.45\textwidth]{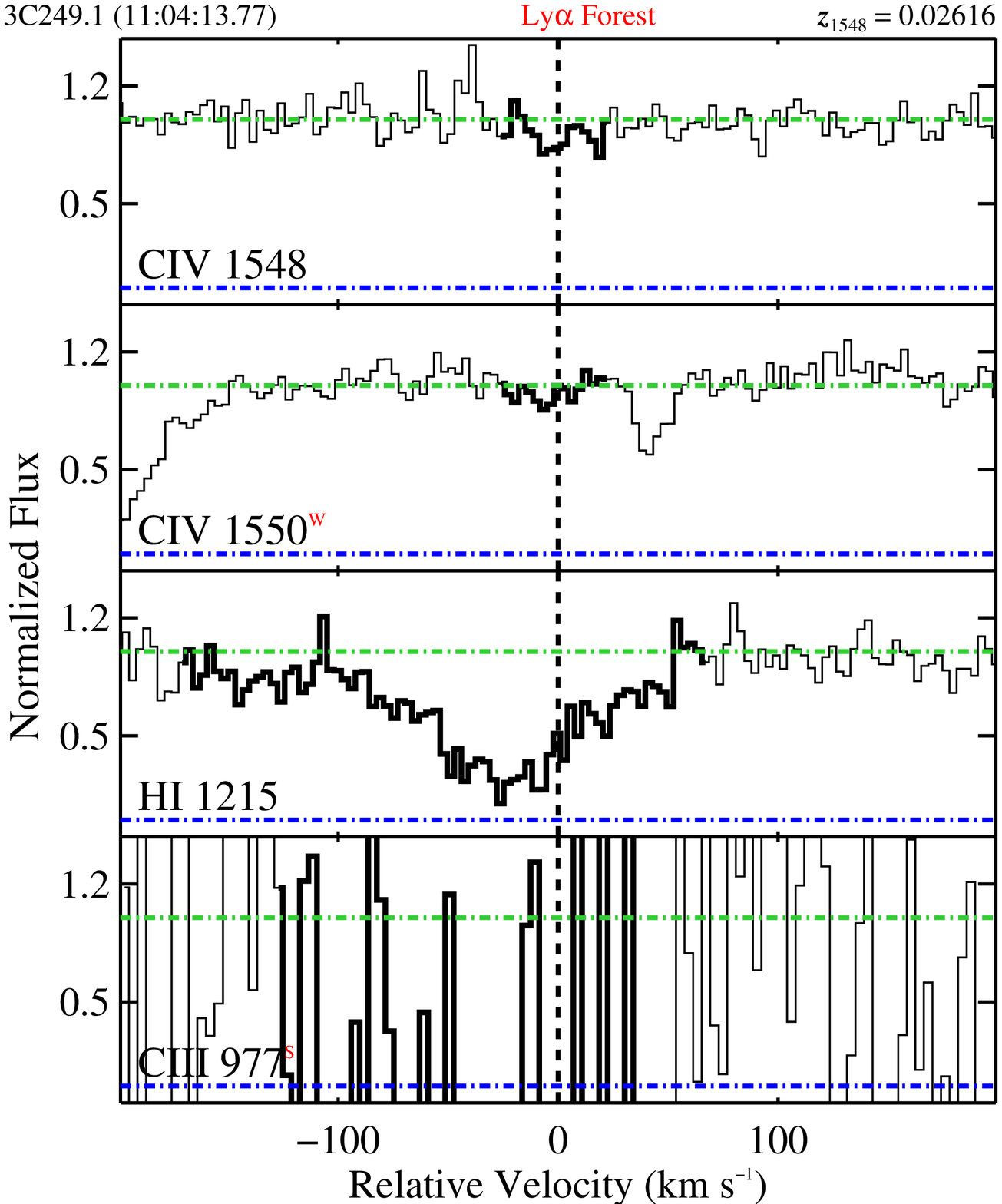} \\
      \includegraphics[width=0.45\textwidth]{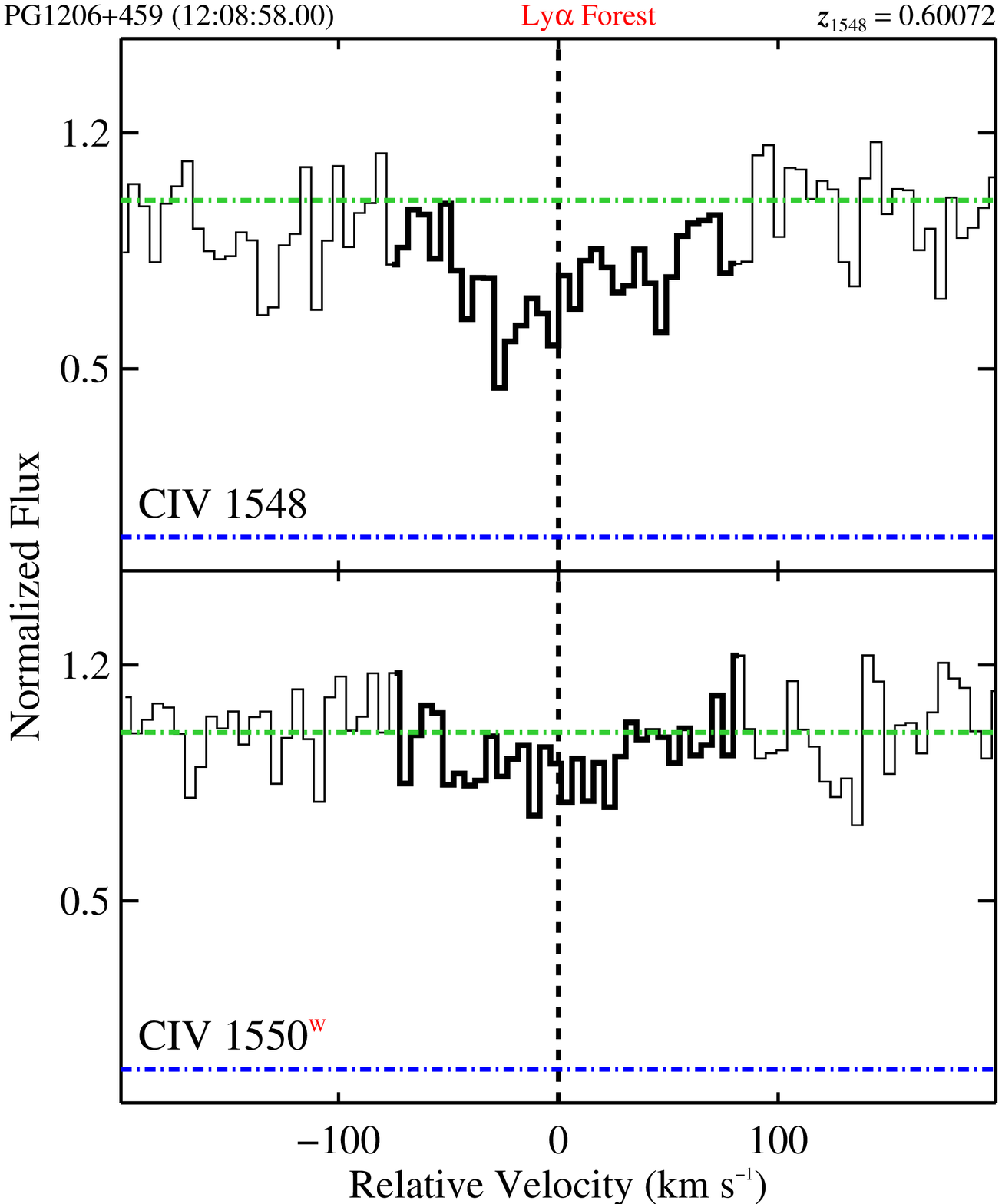} & 
      \includegraphics[width=0.45\textwidth]{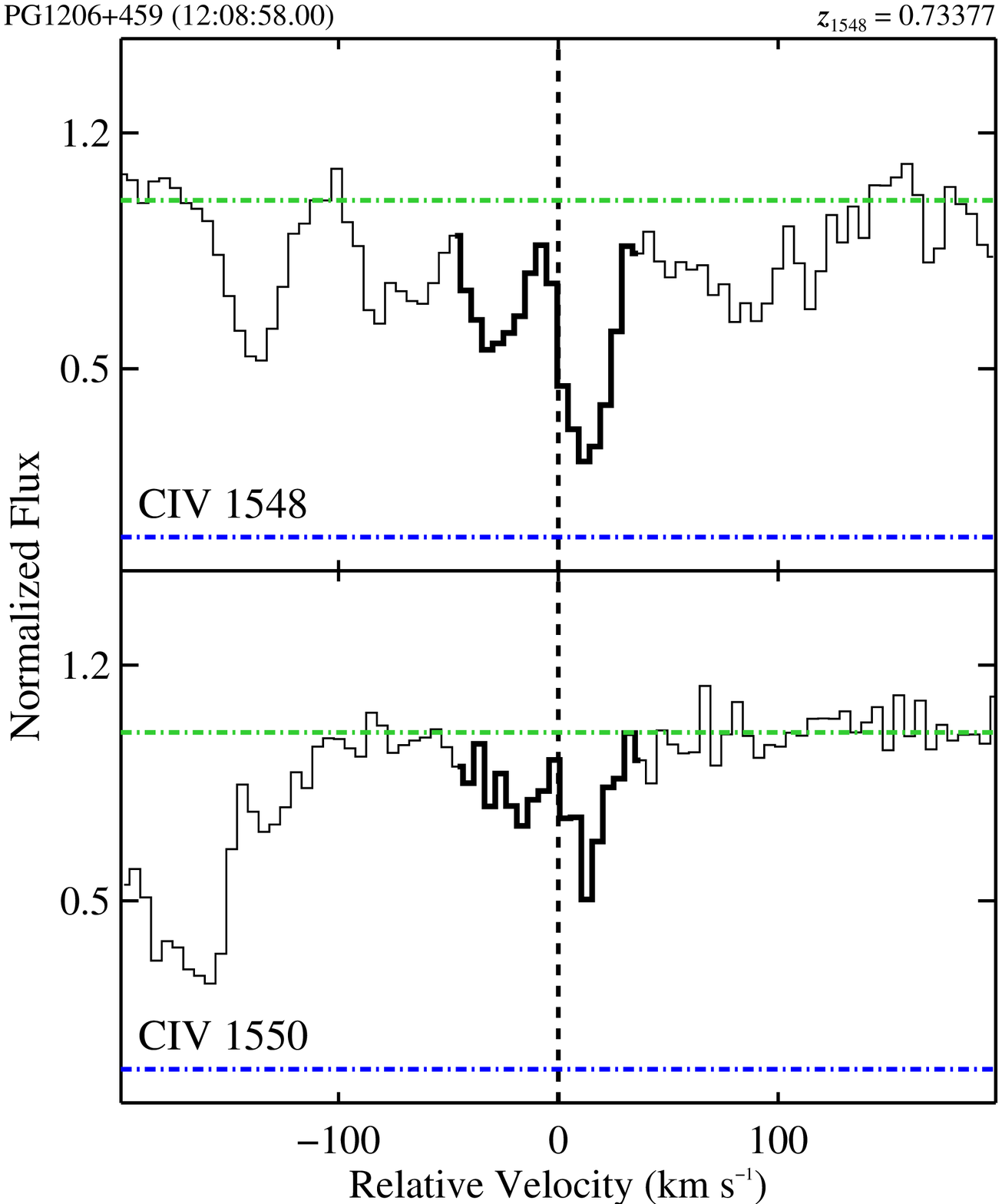} 
    \end{array}$                                      
  \end{center}
  \caption{G = 2 velocity plots (continued) }
\end{figure}
\addtocounter{figure}{-1}

\begin{figure}[!hbt]
  \begin{center}$
    \begin{array}{cc}
      \includegraphics[width=0.45\textwidth]{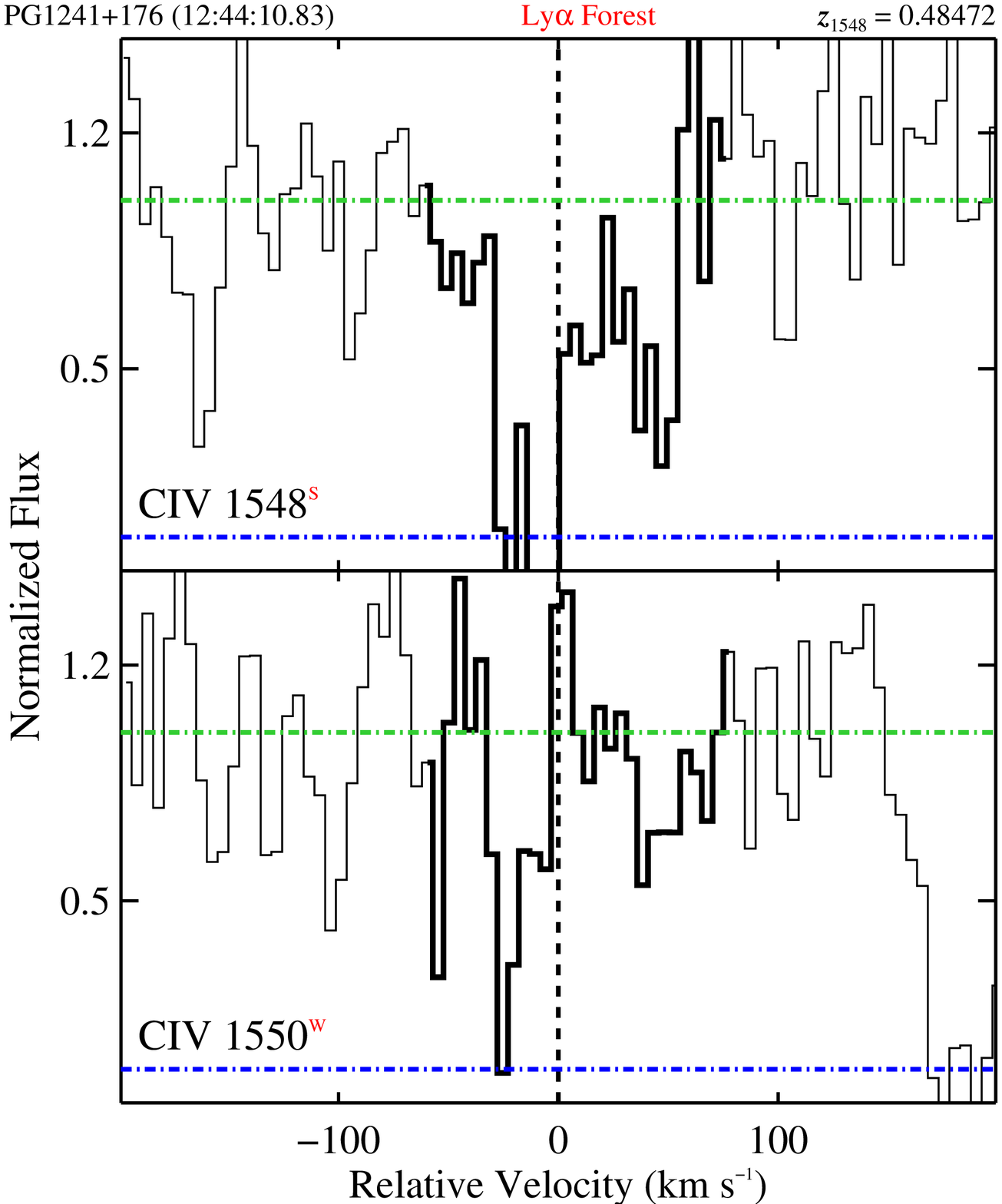} & 
      \includegraphics[width=0.45\textwidth]{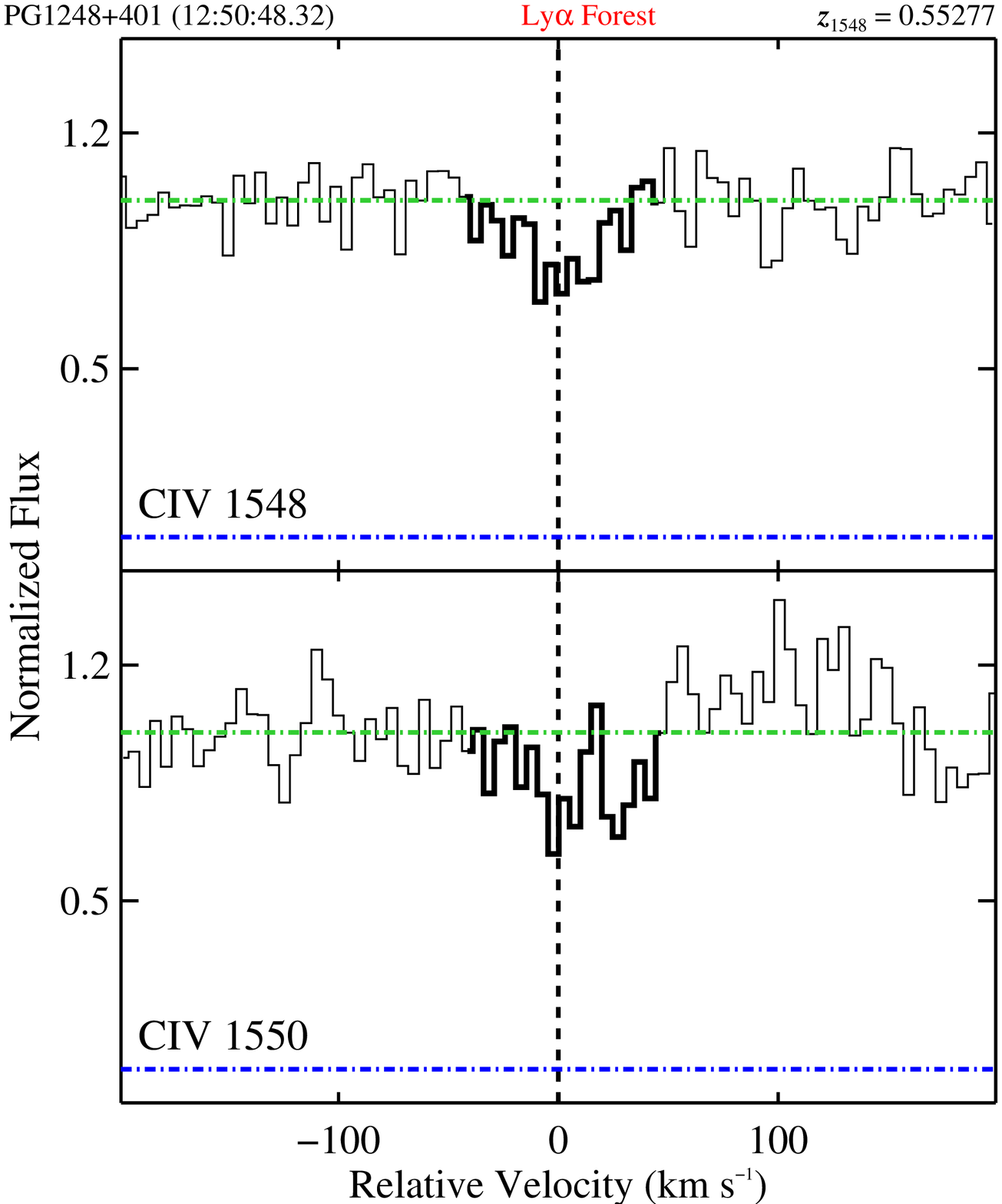} \\
      \includegraphics[width=0.45\textwidth]{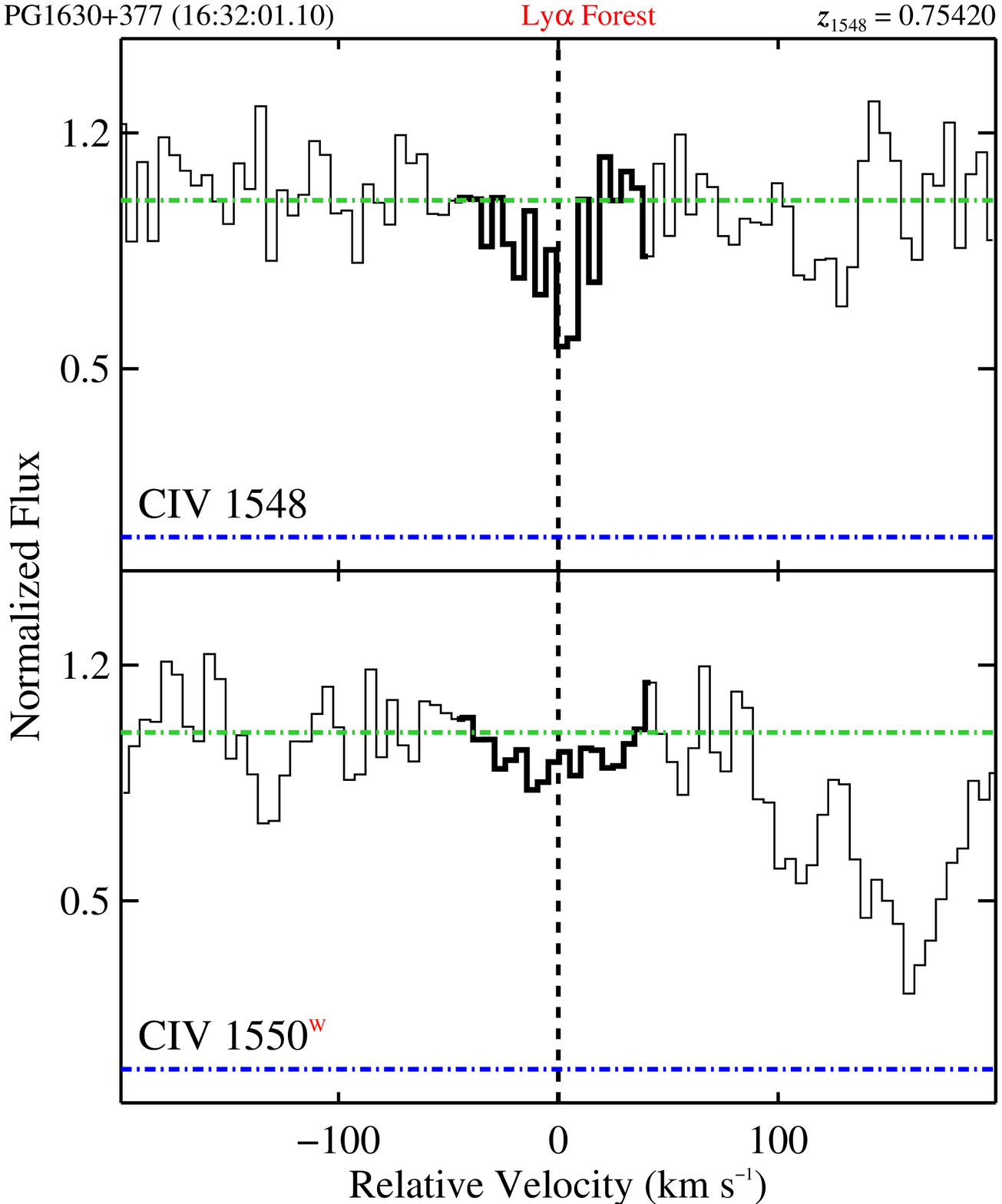} & 
      \includegraphics[width=0.45\textwidth]{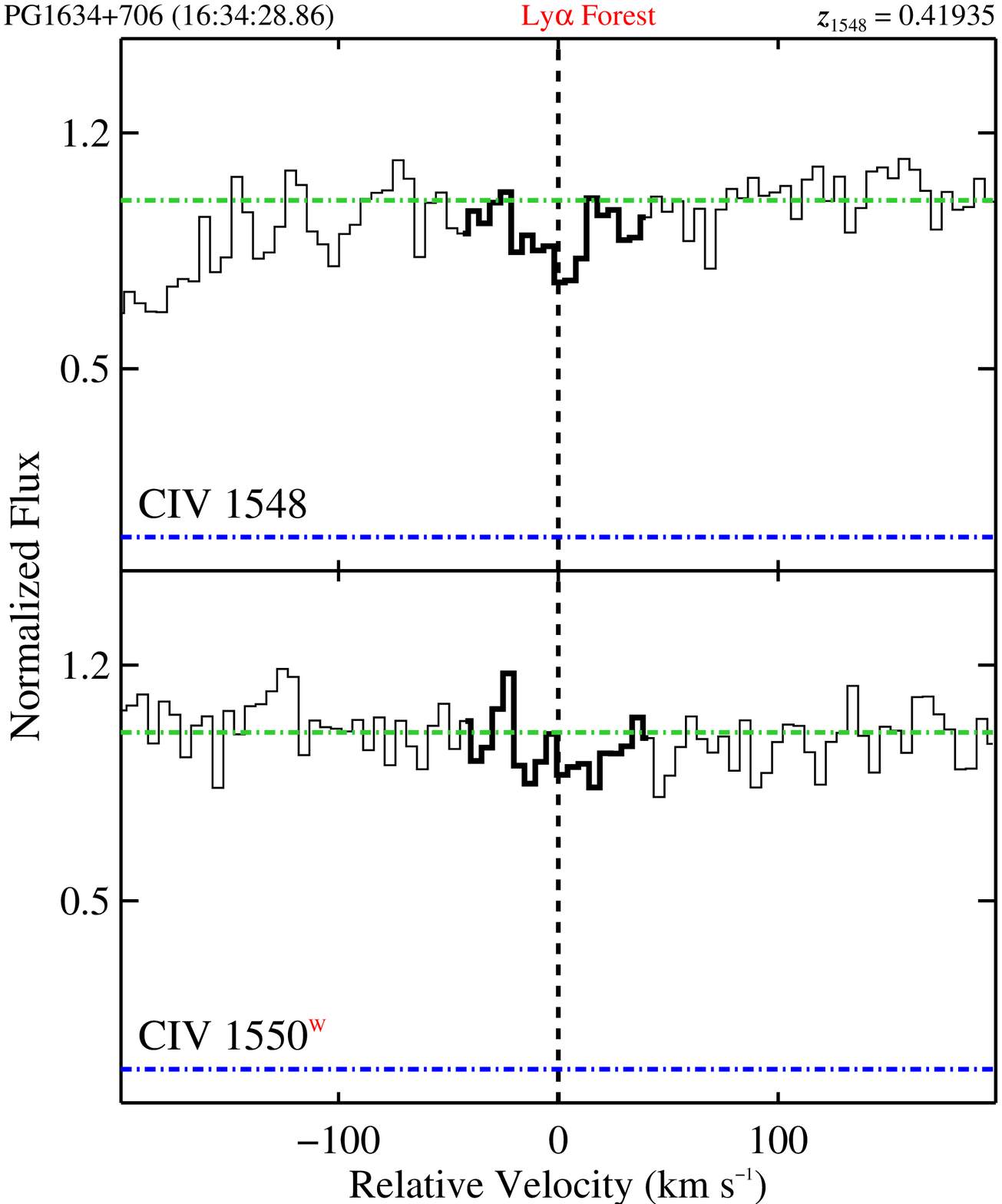} 
    \end{array}$                                      
  \end{center}
  \caption{G = 2 velocity plots (continued) }
\end{figure}
\addtocounter{figure}{-1}

\begin{figure}[!hbt]
  \begin{center}$
    \begin{array}{cc}
      \includegraphics[width=0.45\textwidth]{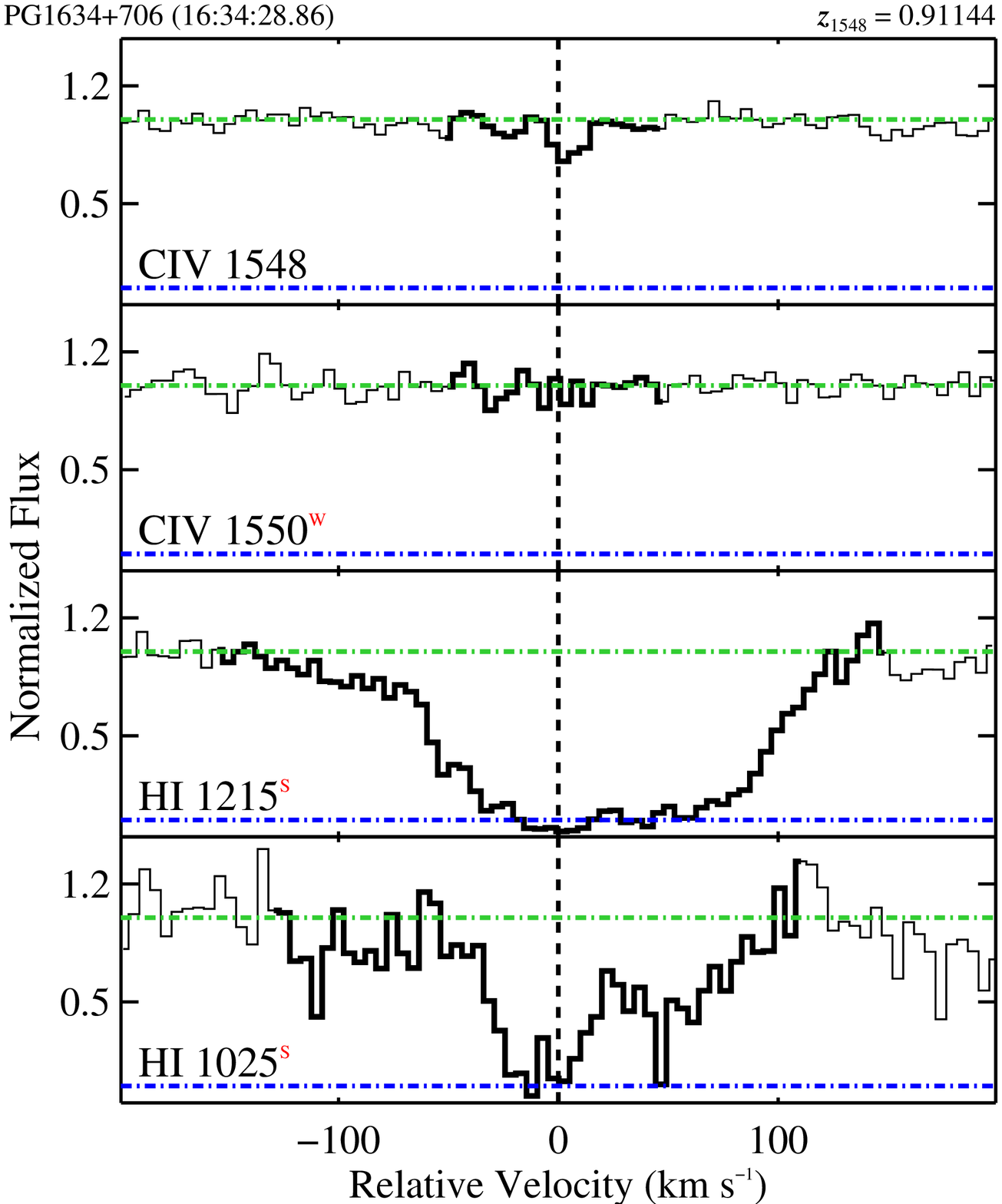} & 
      \includegraphics[width=0.45\textwidth]{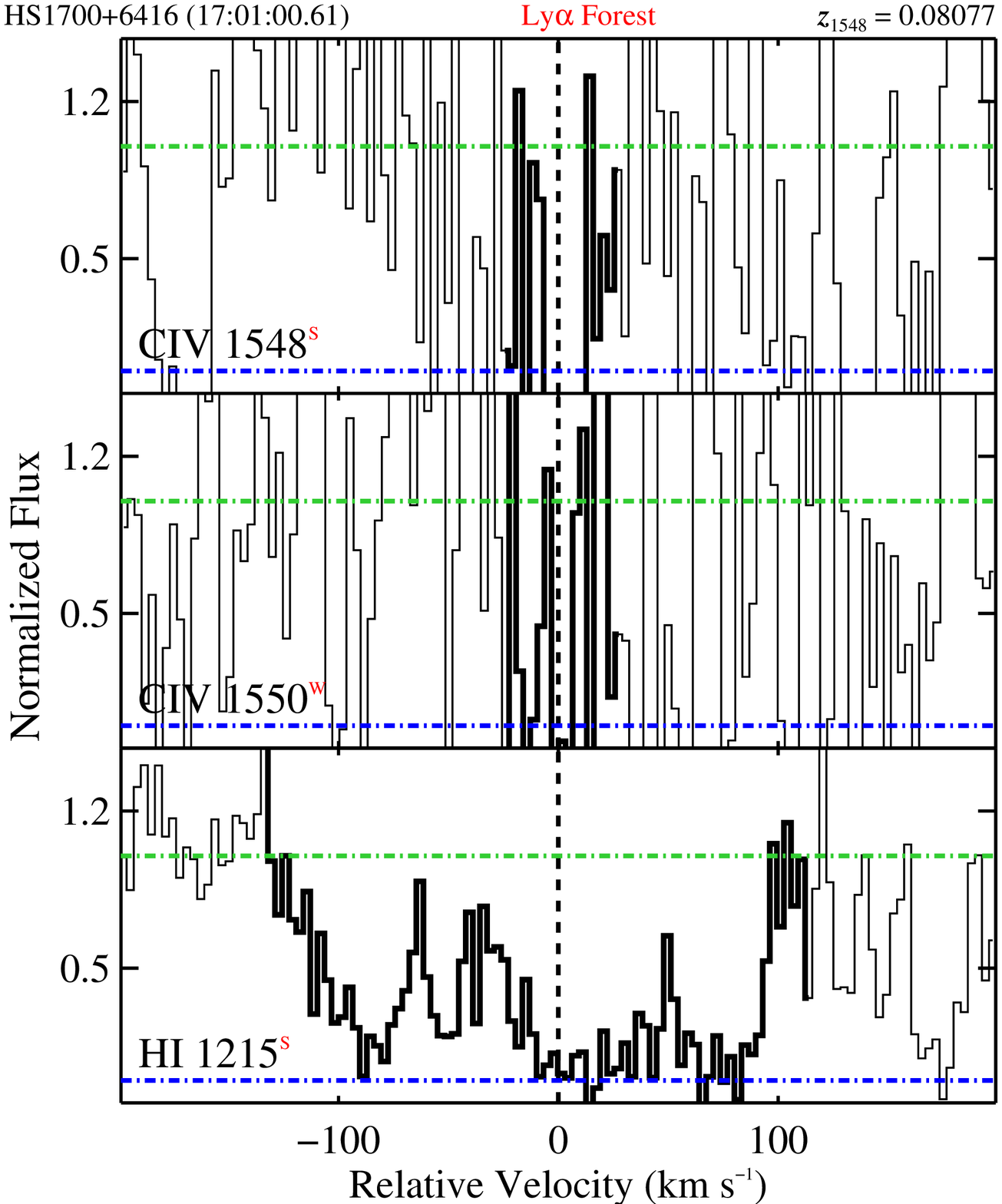} \\
      \includegraphics[width=0.45\textwidth]{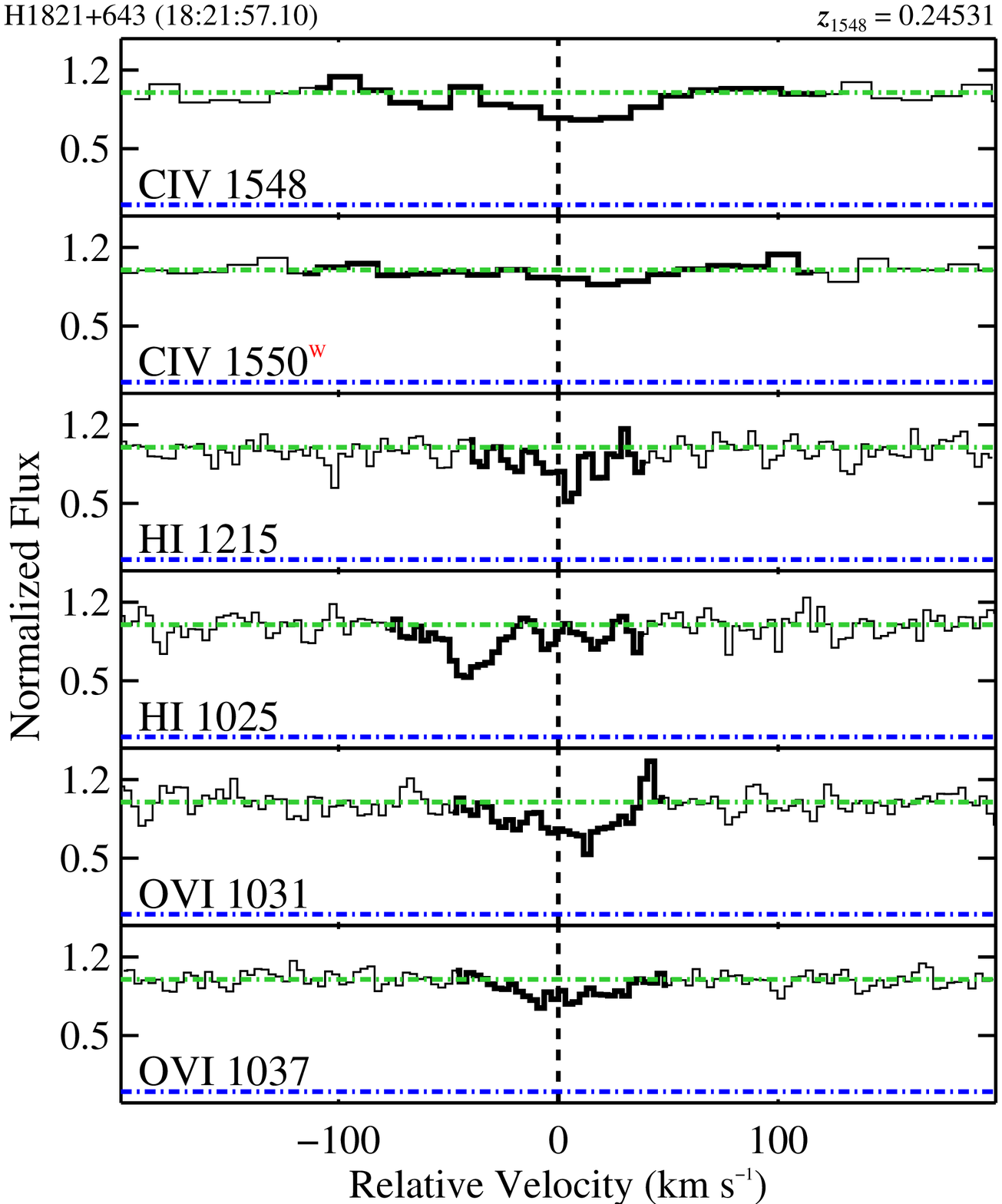} & 
    \end{array}$                                      
  \end{center}
  \caption{G = 2 velocity plots (continued) }
\end{figure}

\clearpage

\section{Maximum-Likelihood Analysis}\label{appdx.maxL}

We used the maximum likelihood method to fit the frequency
distributions with a power-law function. For this discussion, we will
focus primarily on deriving the maximum-likelihood function for the
column densities, where one must account for lower limits due to
saturation effects.  Saturation obviously does not occur for the
equivalent width measurements, and the maximum-likelihood function
$\mathcal{L}$ derived here reduces to a formalism applicable to
\EWr. The following derivation draws heavily from
\citet{storrielombardietal96}.

Assuming Poisson counting statistics, the probability of detecting
$\Num(\N{i})$ absorbers with column densities $\N{i}$ is determined by
the Poisson probability distribution:
\begin{equation}
  P(\Num(\N{i});\mu_{i}) = e^{-\mu_{i}} \frac{\displaystyle
    \mu_{i}^{\Num(\N{i})}}{\displaystyle \Num(\N{i})!}
  {\rm ,} \label{eqn.poiss}
\end{equation}
where $\mu_{i}$ is the expected number, based on the parent
distribution.  The likelihood function is defined as the product of
the probability of observing each absorber with \N{i}:
\begin{equation}
  \mathcal{L} = \prod_{i}^{\Num} P(\Num(\N{i});\mu_{i})  = \prod_{i}^{\Num} e^{-\mu_{i}} \frac{\displaystyle
  \mu_{i}^{\Num(\N{i})}} {\displaystyle \Num(\N{i})!}
\end{equation}
Assume that the expected
number $\mu_{i}$ depends on the frequency distribution:
\begin{equation}
\mu_{i} = \ff{\N{i}} \,\DX{\N{i}} \,\Delta \N{i} 
\end{equation}
(see Equation \ref{eqn.fndef}). Then, consider the limit where
the ``volume'' $\DX{\N{i}}\,\Delta \N{i}$ contains at most one
absorber.  With $\Num = p$ (single-detection) $+ g$ (no-detection)
volumes:
\begin{equation}
\ln \mathcal{L} =  -\int_{\N{min}}^{\N{max}} \ff{\N{i}}
\DX{\N{i}} \ud \N{i} + \sum_{i}^{p} \ln \big(\ff{\N{i}}\DX{\N{i}}\big) {\rm ,}
\end{equation}
in the limit where $\Delta \N{i} \rightarrow 0$ (\DX{\N{i}}\ is
essentially a weight and cannot shrink to zero for all \N{i}).

However, our absorbers include column density measurements where we
have only a lower limit on $\N{i} \ge \N{sat}$. We need a probability
for detecting $q$ saturated absorbers above the saturation limit
$\N{sat}=10^{14.3}\cm{-2}$, $P(q; \mu_{i}^{'})$. Thus, the likelihood
function becomes:
\begin{equation}
  \mathcal{L} = \bigg( \prod_{i}^{\Num} e^{-\mu_{i}} \frac{\displaystyle
  \mu_{i}^{\Num(\N{i})}} {\displaystyle \Num(\N{i})!} \bigg)\
P(q; \mu_{i}^{'}) {\rm .} \label{eqn.maxldef}
\end{equation}
The mean number of saturated absorbers is expected to be:
\begin{equation}
\Num_{\rm sat} = \int_{\N{sat}}^{\infty} \ff{\N{i}}
\DX{\N{i}} \ud \N{i}
\end{equation}
Let $\mu_{i}^{'} = \Num_{\rm sat}$, the expected number of saturated
absorbers from the integral of \ff{\N{i}} between \N{sat} and
infinity. Thus, taking the natural logarithm and expanding Equation
\ref{eqn.maxldef} reduces to:
\begin{eqnarray}
\ln \mathcal{L} & = & -\int_{\N{min}}^{\N{max}} \ff{\N{i}}
\DX{\N{i}} \ud \N{i} -\int_{\N{sat}}^{\infty} \ff{\N{i}}
\DX{\N{i}} \ud \N{i} \\
& & + \sum_{i}^{p} \ln \big(\ff{\N{i}}\DX{\N{i}} \big)
 + q\,\ln \bigg(\int_{\N{sat}}^{\infty} \ff{\N{i}}
\DX{\N{i}} \ud \N{i} \bigg) {\rm .} \nonumber
\end{eqnarray}
Substituting the power-law form of \ff{\NCIV}\ (see Equation
\ref{eqn.fn}) provides the maximum-likelihood function used in the
current study.

\section{Adjusting \Cthr\ Mass Density}\label{appdx.adjOmCIV}

In order to compare \OmCIV\ from the current study to other studies,
we had to account for differences in cosmology and range of \NCIV\
included in the measurement. Changing the adopted Hubble constant
$H_{0}$ is a simple matter of scaling \OmCIV\ by the ratio of the old
$H_{0}$ to the new (see Equation \ref{eqn.omciv}; recall that
$\rho_{c,0} \propto H_{0}^{2}$ and \DXp\ is independent of $H_{0}$).
Less simple is adjusting for changes in $\Omega_{\rm M}$ and
$\Omega_{\Lambda}$, since that enters the computation of \OmCIV\ in
the estimate of the co-moving pathlength $\DXp$ (see Equation
\ref{eqn.x}). Previous authors have frequently scaled the {\it summed}
\OmCIV\ by the ratio of the new pathlength to the old as follows:
\begin{equation}
  \frac{\displaystyle \OmCIV{}_{,{\rm new}}}{\displaystyle \OmCIV{}_{,{\rm old}}} =  
  \frac{\displaystyle H_{0, {\rm old}}
    \frac{\ud X}{\ud z}(\langle z \rangle;\Omega_{\rm M,old},
    \Omega_{\Lambda {\rm ,old}})}
  {\displaystyle H_{0, {\rm new}}
    \frac{\ud X}{\ud z}(\langle z \rangle;\Omega_{\rm M,new}, \Omega_{\Lambda {\rm ,new}})}
    {\rm ,}  \label{eqn.convert}
\end{equation}
where $\ud X/\ud z$ is the derivative of Equation \ref{eqn.x}
evaluated at the median redshift $\langle z \rangle$ and with the
appropriate cosmology. 

This adjustment is an approximation because $\DXp$ is the sum of parts
of spectra that satisfy the redshift and column density constraints
(see Equation \ref{eqn.dx}). Therefore, to correctly adjust for
cosmology, each snippet $\delta X$ must be calculated in the desired
cosmology and then added together to re-estimate \DXp. Most published
studies do not provide the necessary information to do this, which is
why the above equation is the standard adjustment. However, the
approximation works well when the redshift bin is small. For values
quoted in the text and Figures \ref{fig.omciv} and \ref{fig.omcivfit},
we use Equation \ref{eqn.convert} to adjust for differences in
cosmology.

For most of the studies, the adjustment is small since their adopted
cosmology and ours were similar. The results of \citet{songaila01} and
\citet{pettinietal03} were decreased by $\approx55\%$, since
they adopted an Einstein-de Sitter cosmology (EdS) with
$H_{0}=65\kms\,{\rm Mpc}^{-1}$, $\Omega_{\rm M}=1$, and
$\Omega_{\Lambda}=0$. Since we cannot correctly compute their results
precisely for our adopted cosmology, we calculated our \OmCIV\ in
their cosmology, by re-constructing $\DX{\NCIV}$ and re-calculating
the summed \OmCIV. The difference between the values quoted in Table
\ref{tab.fn} and our EdS \OmCIV\ adjusted by Equation
\ref{eqn.convert} is $<8\%$ for the full sample. Thus, Equation
\ref{eqn.convert} adequately adjusts for differences in cosmology when
applied to summed values of \OmCIV.\footnote{Using Equation
  \ref{eqn.convert} to adjust our EdS, {\it integrated} \OmCIV\ values
  introduces errors up to $25\%$.}

The column density limits used in the summed \OmCIV\ estimation was
the other factor we considered when adjusting the values from other
studies. As mentioned in \S\ \ref{subsec.omciv}, the high column
density doublets contain the majority of the mass and dominate the
\Cthr\ mass density. Since there has been no break in the frequency
distribution, we limited the integrated \OmCIV\ measurement to $13 \le
\logCIV \le 15$. However, \citet{scannapiecoetal06} estimated $\OmCIV
= (7.54\pm2.16)\times10^{-8}$ by summing their observed doublets with
$12 \le \logCIV \le 16$. To fairly compare their result with the
current survey, their measured value was decreased by 45\%.
This factor was determined by assuming their summed \OmCIV\ scaled as
the integrated \OmCIV: \begin{equation}
  \frac{\displaystyle \OmCIV{}_{,{\rm new}}}{\displaystyle
  \OmCIV{}_{,{\rm old}}} =  \frac{\N{max,{\rm new}}^{2+\aff{N}} -
  \N{min,{\rm new}}^{2+\aff{N}}} {\N{max,{\rm old}}^{2+\aff{N}} -
  \N{min,{\rm old}}^{2+\aff{N}}} {\rm ,} \label{eqn.scaleOmCIV}
\end{equation}
where the new column density limits are $13 \le \logCIV \le 15$, to
match the current work, and \aff{N}\ is the value measured by the
``old'' study (also see Equation \ref{eqn.omciv_int}). For
\citet{scannapiecoetal06}, the best-fit $\aff{N}=-1.8$. We also made
 adjustments ($<30\%$) to \citet{songaila01},
\citet{pettinietal03}, \citet{boksenbergetal03ph}, and
\citet{danforthandshull08} since $\N{min} < 10^{13}\cm{-2}$ in their
studies and, for most, $\N{max} < 10^{15}\cm{-2}$, which increases
the significance of the lower column density absorbers.

We did {\it not} extrapolate these studies and change \N{max}. These
surveys were surely sensitive to doublets with $\logCIV > 14$, but
they did not survey a sufficiently large pathlength to encounter them,
since they are rare. The affect of pathlength is accounted for in the
summed \OmCIV\ (Equation \ref{eqn.omciv_sum}), where it essentially
weights the sum of the column densities. Therefore, for these studies,
the summed \OmCIV\ values occasionally reflect the fact that the
strongest absorbers are rare, though they dominate the \Cthr\ mass
density.

The result from \citet{ryanweberetal09} could have been increased by a
factor of 1.4, assuming $\aff{N} = -1.8$ (less for a shallower power
law); they were only sensitive to doublets with $13.8 \le \logCIV \le
15$. However, we did not adjust their value since we did not know the
appropriate slope and did not include the highest redshift
measurements in the linear regressions to \OmCIV\ over $t_{age}$.

The total adjustment for cosmology and/or column density limits for
the $z>1$ studies cited in this paper are as follows: $\approx40\%$,
\citet{songaila01}; $38\%$, \citet{pettinietal03}; $81\%$,
\citet{boksenbergetal03ph}; $45\%$, \citet{scannapiecoetal06}; $92\%$,
\citet{danforthandshull08}; and $105\%$, \citet{beckeretal09}.

\section{Detailed Comparison with \citet{milutinovicetal07} and 
\citet{danforthandshull08}}\label{appdx.comp}

\citet{milutinovicetal07} and \citet{danforthandshull08} surveyed
archival \stis\ E230M and E140M spectra, respectively, for
intergalactic absorption lines.  It is useful, therefore, to compare
their results with our own as a consistency check.
\footnote{\citet{fryeetal03} was a conference proceeding, which we do
  not discuss in detail. To summarize, they surveyed nine sightlines
  with \stis\ E140M spectra. We agreed with seven of their nine
  identified \ion{C}{4} doublets. Though we agreed well with their
  $\dNCIVdz$ over equivalent width limit,
  their reported \OmCIV\ value was almost double
  our value for the $z < 0.6$ sample. However, this is likely due to
  small number statistics, since we agreed ($<1\sigma$) when we
  focused just on the E140M data (see Table \ref{tab.fn}).}

In general, our search algorithms and sample selection were consistent
with \citet{milutinovicetal07} and \citet{danforthandshull08}, who
used different procedures.
The current survey, however, has the advantage of analyzing the
largest sample of $\zciv < 1$ sightlines in a consistent fashion.

\subsection{\citet{milutinovicetal07}}

\citet{milutinovicetal07} focused on eight sightlines with \stis\
E230M observations and compiled comprehensive line lists for each
sightline. There were 24 \ion{C}{4} systems in their survey. We agreed
with 19 of these, including the associated transitions (\eg \Lya,
\ion{Si}{2} 1260), though we did not search for all of the transitions
that they did (\eg \ion{C}{2} 1334, \ion{Fe}{2} 1608). We counted the
DLA $\zciv = 0.92677$ \ion{C}{4} doublet towards PG1206+459 as one
system, while \citet{milutinovicetal07} divided it into two. Three of
the remaining five systems are acknowledged to be questionable
identifications by \citet{milutinovicetal07} and were not included in
our G = 1+2 group at all.
Of the other two, the reported doublet at $\zciv = 0.7760$ towards
PG1248+401 does not show up in our survey because the 1548 line was
blended and the 1550 would not be detected with
$\EWlin{1550}\ge3\sigEWr$. The other unconfirmed doublet at $\zciv =
0.9903$ towards PG1634+706 was associated with the \ion{Mg}{2}
absorber, which was the motive for observing the sightline (see \S\
\ref{subsec.losbias}). This doublet would lie in the
highest-wavelength order of the E230M spectrum, which we excluded due
to questionable data quality flags.

Considering the opposite comparison, we identified 29 \ion{C}{4}
doublets in the eight sightlines that \citet{milutinovicetal07} also
analyzed. Of these, 11 were identified as other transitions by
\citet{milutinovicetal07}. In nine cases, the absorption lines we list
as \ion{C}{4} doublets were either: listed as tentative \Lya\ lines; not
identified; or not listed at all. The two remaining disputed cases are
the ones we identify as $\zciv=0.48472$ towards PG1241+176 and
$\zciv=0.55277$ towards PG1248+401, both G = 2 doublets. For the
PG1241+176 lines, \citet{milutinovicetal07} identifies our 1548 line
as \ion{O}{6} 1037 at $z=1.2142$ and 1550 as a tentative \Lya\ line.
The 1548 profile is consistent with being blended, so we conclude that
both identifications are valid. For the PG1248+401 lines,
\citet{milutinovicetal07} lists our 1548 line as \Lya\ absorption,
with no associated transitions, and does not list our 1550 line, which
we detect at $>3\sigEWr$. We stand by our identification and accept
the possibility that the $\zciv=0.55277$ doublet is \Lya\
contamination, as discussed in \S\ \ref{subsec.lya}.

\subsection{\citet{danforthandshull08}}

\citet{danforthandshull08} focused on 28 sightlines with \stis\ E140M
observations and supplementary \fuse\ spectra. They conducted a
\Lya-targeted survey, where they identified \Lya\ absorbers first,
then searched for other, associated intergalactic transitions (\eg
\Lyb, \ion{C}{3}). They detected 24 $\zciv<0.12$ \ion{C}{4} doublets
with at least one line detected at $\ge4\sigEWr$ so long as $R_{W}$
was consistent (see \S\ \ref{subsec.flag}). For a total pathlength
$\Dz=2.42$, they measured $\dNCIVdz=10^{+4}_{-2}$ for $\EWr \ge
30\mA$. They also fit a power law to \dNCIVdz\ in bins of $\Delta
\logCIV = 0.2$; the best-fit exponent was $\aff{N} = -1.79\pm0.17$.
They measured a summed $\OmCIV = (7.78\pm1.47)\times10^{-8}$ for
$12.83 \le \logCIV \le 14.13$.

The current survey included the same 28 sightlines though not all of
the \fuse\ data, since some had S/N too low for continuum fitting. Of the
24 doublets in \citet{danforthandshull08}, we detected nine also and
could neither agree or disagree with three, which were detected in the
zeroth order of the \stis\ E140M spectra, which we did not include due
to questionable data quality flags. The zeroth order covered $1711\Ang
< \lambda < 1729\Ang$ (or $0.105 < z < 0.115$), and the sensitivity of
the E140M spectra drops rapidly at $\lambda > 1700\Ang$. Therefore, we
would not likely have included any doublets in the zeroth order in our
analysis, since both lines were not detected with $\EWr\ge3\sigma$ in
\citet{danforthandshull08}.

Of the remaining 11 doublets, two towards NGC7469 were excluded from
our sample because they were within $3000\kms$ of the background
source. The other nine either did not look like absorption lines or
the \ion{C}{4} 1548 line did not have $\EWlin{1548}\ge3\sigEWr$ in our
co-added and normalized spectra.
As discussed in \citet{cookseyetal08}, \citet{danforthandshull08}
cited questionably small errors ($<5\%$) for their rest equivalent
widths, which led them to consider more lines to be detected at a
given significance level.

We detected 12 \ion{C}{4} doublets in the E140M spectra, that includes
three doublets not cited in \citet{danforthandshull08}. For two cases,
we identified the absorption lines as G = 2 \ion{C}{4} doublets with
only the 1548 line detected with $\EWlin{1548}\ge3\sigEWr$:
$\zciv=0.02616$ towards 3C249.1\footnote{\citet{danforthandshull08}
  identified the
  $\ge3\sigEWr$ absorption line (that we list as \ion{C}{4} 1548) as
  \Lya\ at $z=0.30788$, which was their default identification for
  single lines.} and $\zciv=0.08077$ towards HS1700+6416. These
doublets were not included in our analyses. The last case was the G =
1, $\zciv=0.00574$ doublet towards QSO--123050+011522, which was also
associated with a \ion{Si}{4} doublet (another target absorption lines
of their study). They did not detect this doublet because the \Lyb\
line was contaminated by a spurious artifact in the \fuse\ spectra,
but they now agree with our identification (C. Danforth, private communication)

\end{document}